\documentclass[journal]{IEEEtran}
\IEEEoverridecommandlockouts

\usepackage{cite}
\usepackage{amsmath,amssymb,amsfonts}
\usepackage{algorithmic}
\usepackage{graphicx}
\usepackage{textcomp}
\usepackage{xcolor}
\usepackage{float}
\usepackage{overpic}
\usepackage{multirow}
\usepackage{subfigure}
\usepackage[colorlinks,
linkcolor=red,
anchorcolor=red,
citecolor=green]{hyperref}
\def\BibTeX{{\rm B\kern-.05em{\sc i\kern-.025em b}\kern-.08em
    T\kern-.1667em\lower.7ex\hbox{E}\kern-.125emX}}

\begin{document}

	\title{DA-DRN: Degradation-Aware Deep Retinex \\Network for Low-Light Image Enhancement}

	\author{Xinxu~Wei,
	Xianshi~Zhang\IEEEauthorrefmark{1}, Shisen~Wang, Cheng~Cheng, Yanlin~Huang, 
	Kaifu~Yang,
	\\and Yongjie~Li,~\IEEEmembership{Senior Member,~IEEE}
	\thanks{This work was supported by Guangdong Key R\&D Project (\#2018B030338001) and Natural Science Foundations of China (\#61806041, \#62076055). (Corresponding author: Xian-Shi~Zhang, email: zhangxianshi@uestc.edu.cn)}%
	\thanks{Xinxu~Wei, Xianshi~Zhang, Cheng~Cheng, Yanlin~Huang, Kaifu~Yang, Yongjie~Li are with the MOE Key Lab for Neuroinformation, University of Electronic Science and Technology of China (UESTC), Chengdu 610054, China.}}

	\maketitle


\maketitle

\begin{abstract}
Images obtained in real-world low-light conditions are not only low in brightness, but they also suffer from many other types of degradation, such as color distortion, unknown noise, detail loss and halo artifacts. In this paper, we propose a Degradation-Aware Deep Retinex Network (denoted as DA-DRN) for low-light image enhancement and tackle the above degradation. Based on Retinex Theory, the decomposition net in our model can decompose low-light images into reflectance and illumination maps and deal with the degradation in the reflectance during the decomposition phase directly. We propose a Degradation-Aware Module (DA Module) which can guide the training process of the decomposer and enable the decomposer to be a restorer during the training phase without additional computational cost in the test phase. DA Module can achieve the purpose of noise removal while preserving detail information into the illumination map as well as tackle color distortion and halo artifacts. We introduce Perceptual Loss to train the enhancement network to generate the brightness-improved illumination maps which are more consistent with human visual perception. We train and evaluate the performance of our proposed model over the LOL real-world and LOL synthetic datasets, and we also test our model over several other frequently used datasets without Ground-Truth (LIME, DICM, MEF and NPE datasets). We conduct extensive experiments to demonstrate that our approach achieves a promising effect with good rubustness and generalization and outperforms many other state-of-the-art methods qualitatively and quantitatively.
Our method only takes 7 ms to process an image with 600x400 resolution on a TITAN Xp GPU.   
\end{abstract}



\begin{figure}
	\subfigure[Input (1.67/26.19)]{
		\begin{minipage}[b]{0.3\linewidth}
			\includegraphics[width=2.8cm]{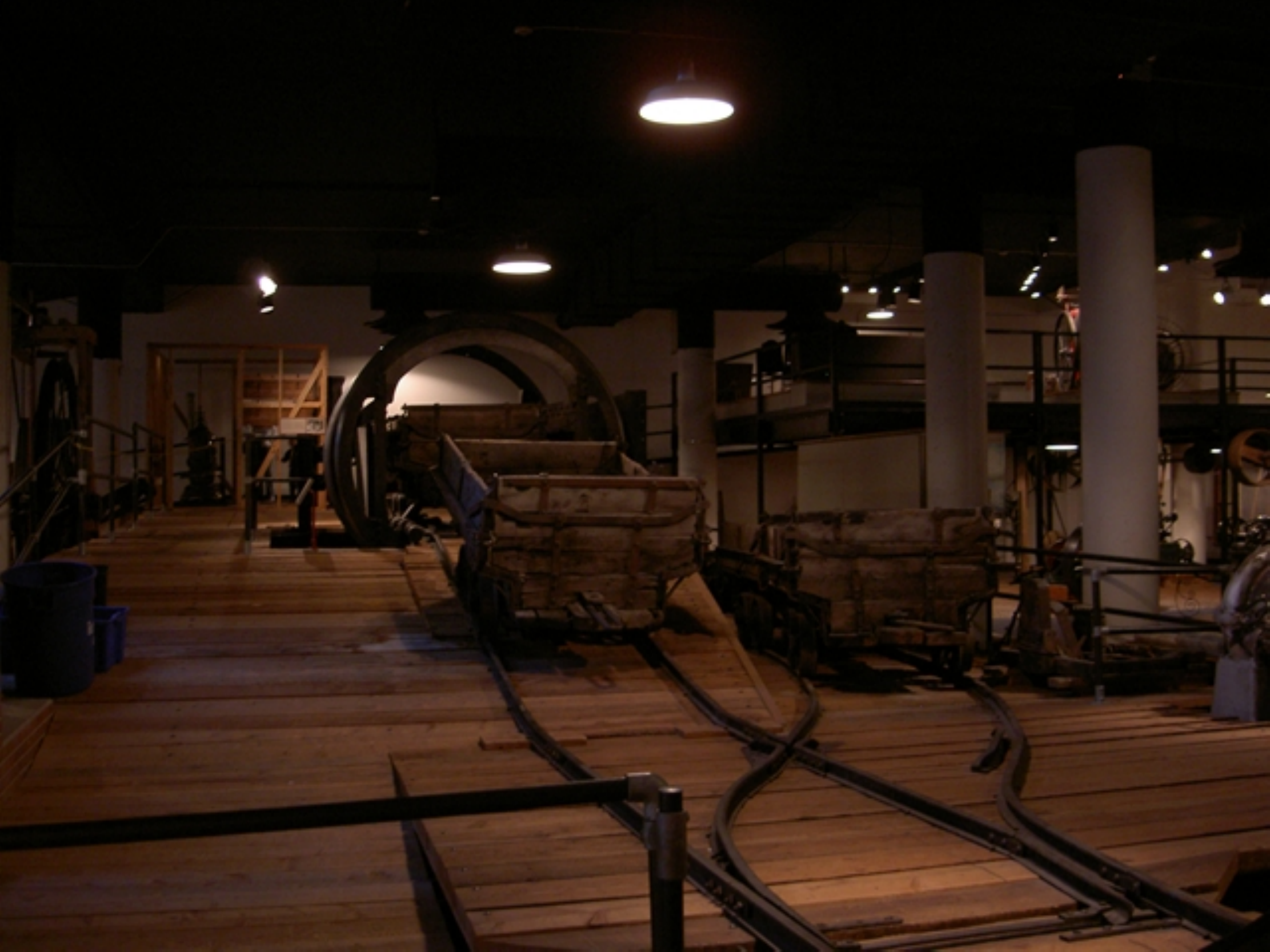}\vspace{-3pt}
	\end{minipage}}
\vspace{-4pt}
	\subfigure[GLAD (1.45/34.52)]{
		\begin{minipage}[b]{0.3\linewidth}
			\includegraphics[width=2.8cm]{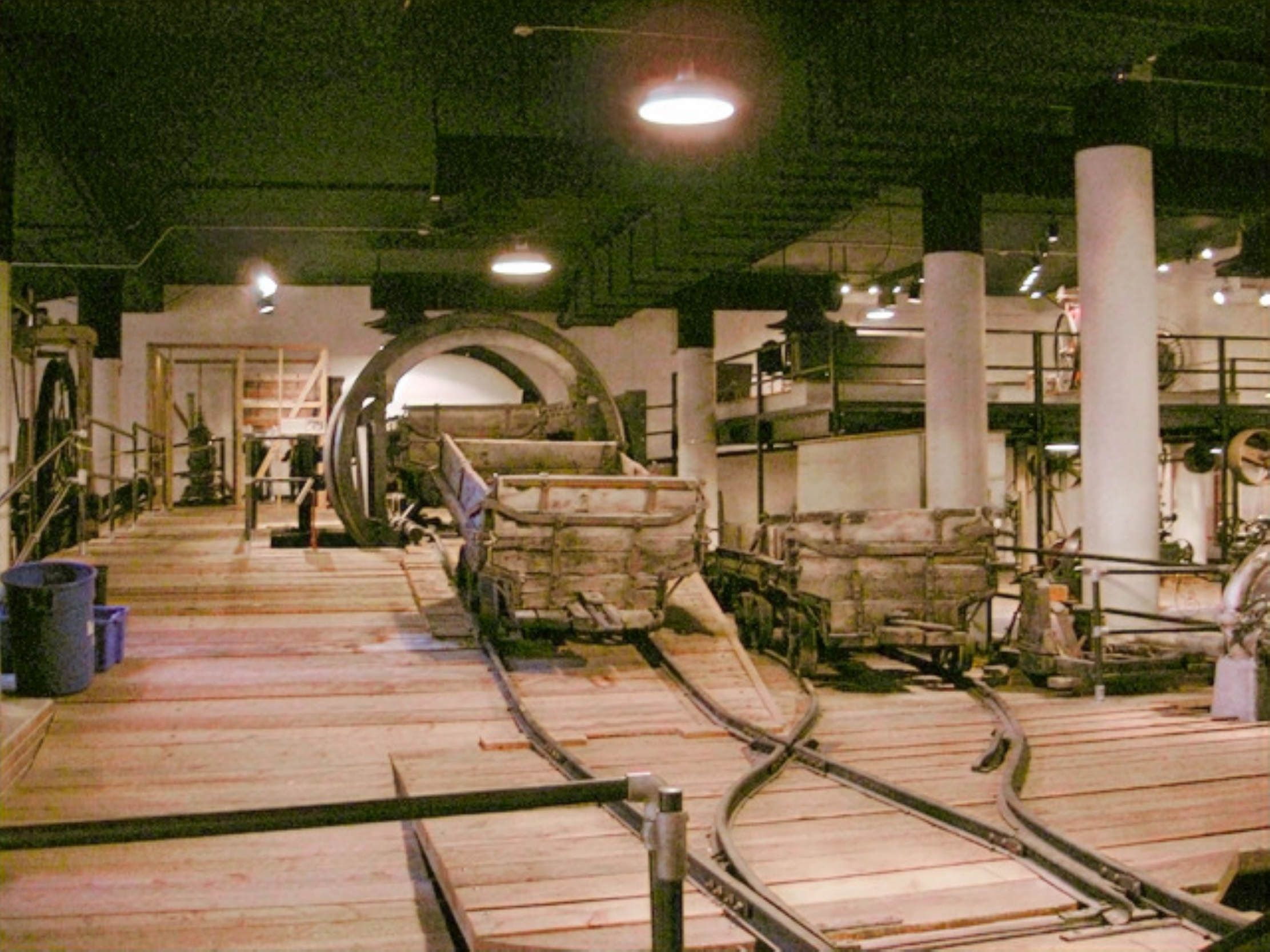}\vspace{-3pt}
	\end{minipage}}
\vspace{-4pt}
	\subfigure[Ours \textbf{(1.15/57.17)}]{
		\begin{minipage}[b]{0.3\linewidth}
			\includegraphics[width=2.8cm]{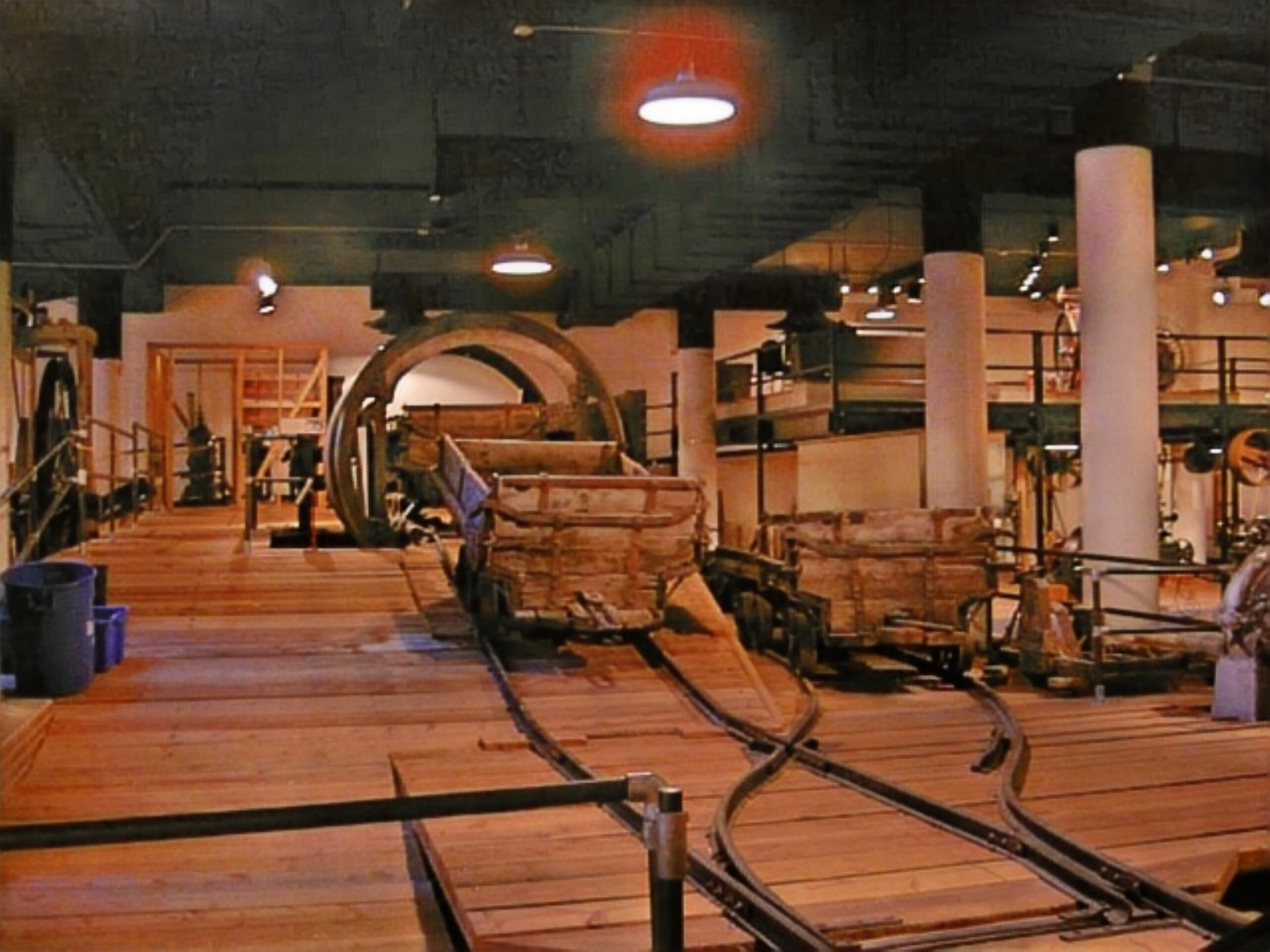}\vspace{-3pt}
	\end{minipage}}
\vspace{1pt}
\\
	\subfigure[Input (1.42/62.25)]{
		\begin{minipage}[b]{0.3\linewidth}
			\includegraphics[width=2.8cm]{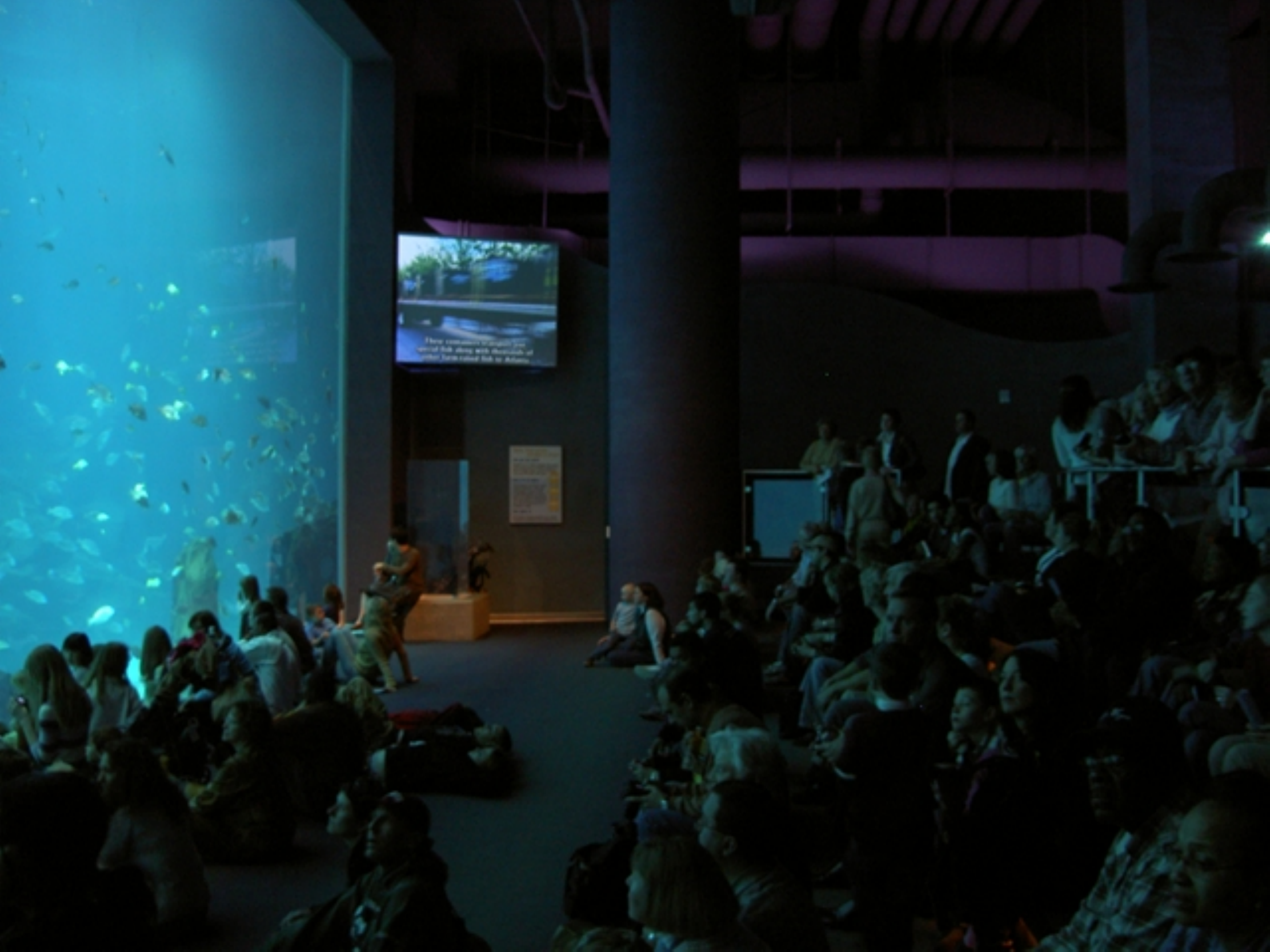}\vspace{-3pt}
	\end{minipage}}
\vspace{-4pt}
	\subfigure[GLAD (1.35/25.65)]{
	\begin{minipage}[b]{0.3\linewidth}
		\includegraphics[width=2.8cm]{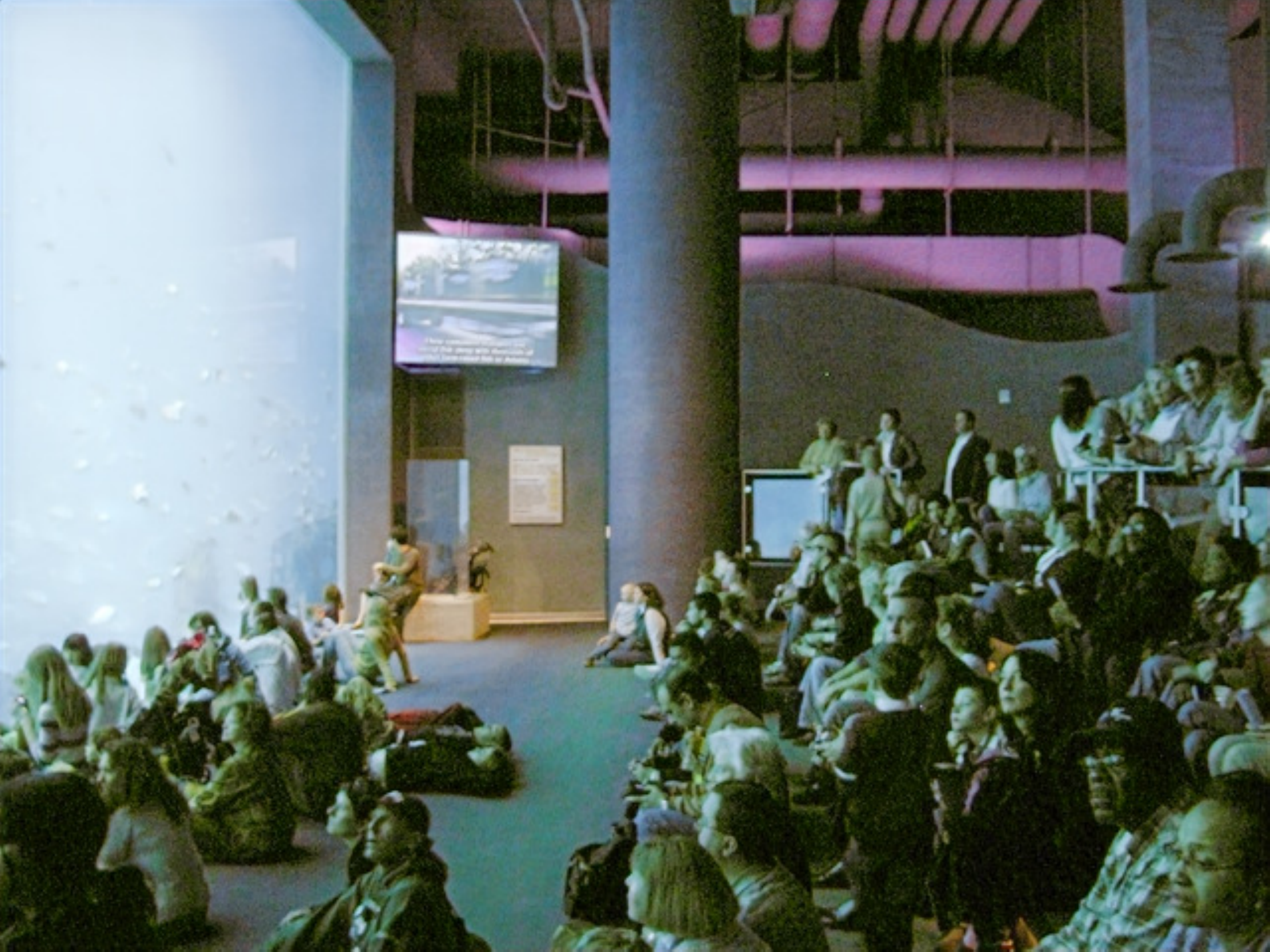}\vspace{-3pt}
	\end{minipage}}
\vspace{-4pt}
	\subfigure[Ours \textbf{(1.06/75.15)}]{
		\begin{minipage}[b]{0.3\linewidth}
			\includegraphics[width=2.8cm]{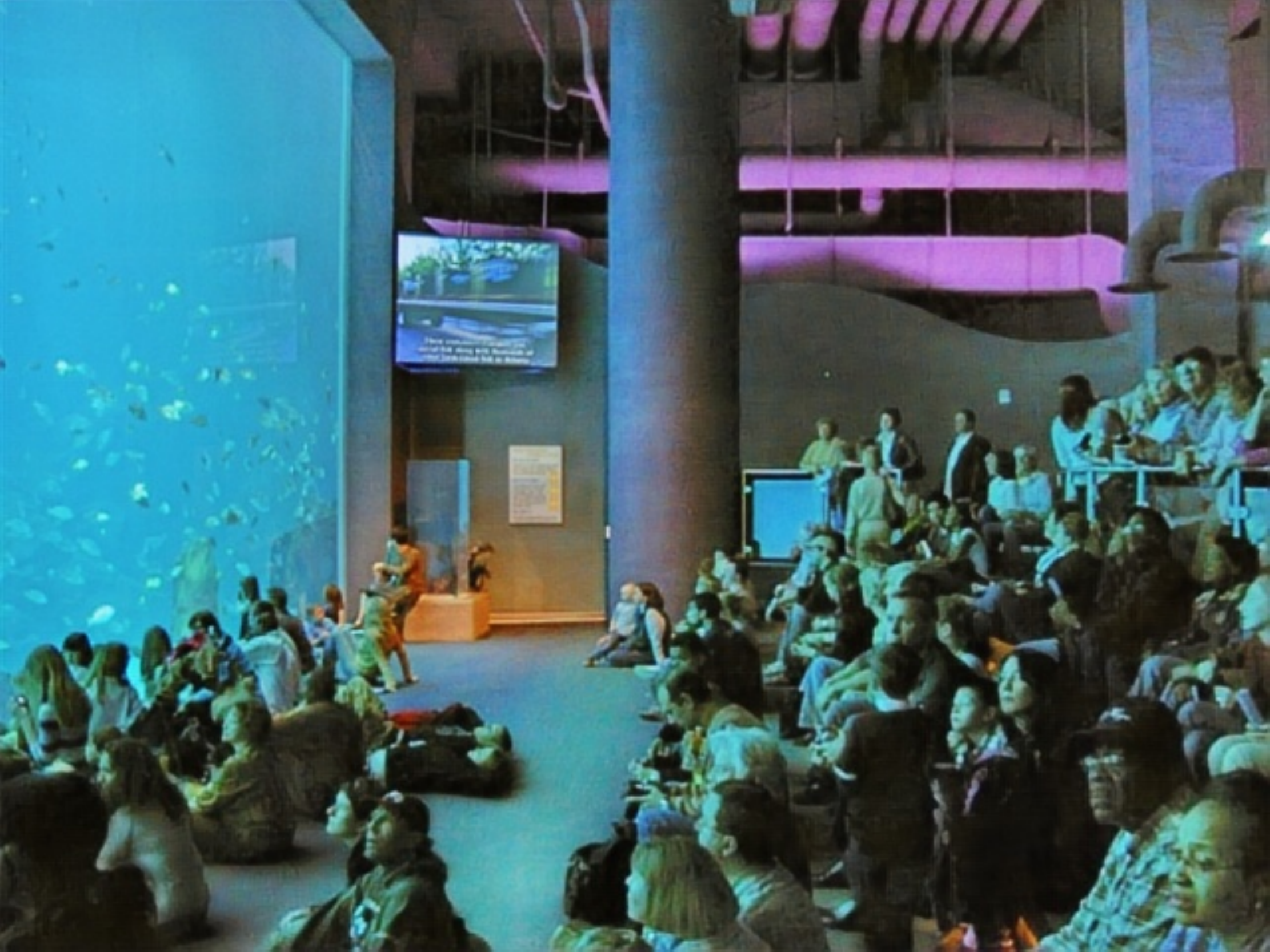}\vspace{-3pt}
	\end{minipage}}
\vspace{3pt}
	\caption{Some low-light image examples and their corresponding enhanced results using our DA-DRN. There are noticeable noise and color distortion in the results of GLAD\cite{wang2018gladnet}. Our results are noise-free and color-corrected. *(Noise Level\cite{liu2013single} / Colorfulness\cite{hasler2003measuring}). *The low value of noise level and colorfulness, the less the noise and color information of the image.}
	\label{pre}
\end{figure}

\section{Introduction}
\IEEEPARstart{L}{ow-light} image enhancement is a very challenging low-level computer vision task because, while enhancing the brightness, we also need to tackle color distortion, suppress amplified noise, preserve detail information and eliminate halo artifacts. 
Images captured in insufficient lighting conditions often suffer from several types of degradation, such as poor visibility, low contrast, color distortion and severe ISO 
noise, which have negative effects on other computer vision tasks, such as image recognition \cite{lecun1998gradient} \cite{simonyan2014very} \cite{he2016deep}, object detection \cite{redmon2016you} \cite{girshick2014rich} \cite{girshick2015fast} 
and image segmentation \cite{chen2017deeplab} \cite{he2017mask} \cite{zhao2017pyramid}. Therefore, there is a huge demand for low-light image enhancement. 
In recent years, a great number of approaches have been proposed and achieved remarkable results in low-light image enhancement, but, to the best of our knowledge, there are few successful methods for simultaneously dealing with all the degradation contained in low-light images. \\ 
It is still a very challenging task to cope with so many problems simultaneously. 
Although many existing methods can improve the brightness of low-light images, some of them cause serious color distortion after increasing the brightness\cite{wang2018gladnet}\cite{zhang2019kindling}\cite{jobson1997multiscale}\cite{guo2016lime}\cite{wei2018deep}\cite{jiang2021enlightengan}. For example, as shown in Fig.\ref{pre}, the results of GLAD\cite{wang2018gladnet} is under-saturated in terms of colorfulness, while as shown in Fig.\ref{DICM}, the colorfulness of RetinexNet\cite{wei2018deep} is over-saturated, both of them suffered from severe color distortion. In addition, some methods do not suppress noise when enhancing brightness, resulting in serious noise in the image after enhancement \cite{jobson1997multiscale}\cite{wang2019rdgan}\cite{guo2020zero}. Other methods \cite{lv2018mbllen}\cite{li2018structure}\cite{ren2018joint}, in order to eliminate noise, cause the enhanced and denoised image to be too smooth, resulting in blurred edges, missing too much detail information and causing color dostortion. 
For example, as shown in Fig.\ref{748_r}, KinD \cite{zhang2019kindling} uses a simple U-Net to decompose low-light images. Noise hidden in the dark regions is amplified in reflectance map, so a deeper U-Net is used in KinD to remove noise in reflectance, but this introduces color distortion because color information is lost when the noise-free reflectance map is reconstructed in the up-and-down sampling structure of the deep U-Net.
Degradation can affect not only our visual effects but other visual tasks as well.
Therefore, how to deal with all the degradation is a problem worth studying.\\
Based on the assumption of Retinex Theory \cite{land1977retinex}, a natural image can be decomposed into two parts: reflectance and illumination. The reflectance part contains detail and color information of the original image. The illumination part carries the information about the intensity and distribution of light in the original image, and it should be smooth enough and not contain any high-frequency information such as details and noise.
The decomposition task concentrates the high-frequency components, including the noise and detail information, on the reflectance part, whereas the illumination part is mainly composed of low-frequency components, so there is almost no noise in the illumination part and we only need to denoise the reflectance part.\\
Many previous methods\cite{jobson1997multiscale}\cite{wang2018gladnet}\cite{jiang2021enlightengan} do not consider denoising the enhanced image. Some methods use other state-of-the-art denoising methods\cite{dabov2006image}\cite{zhang2017beyond}\cite{guo2019toward} as post-processing operations for enhanced images, which can slow down the speed of the model considerably. In other methods\cite{zhang2019kindling}, a denoising network or algorithm is specially designed to achieve noise-free in the enhanced image.\\
In this paper, we improve RetinexNet \cite{wei2018deep} and propose a novel framework based on Retinex Theory \cite{land1977retinex} to achieve brightness enhancement, denoising, color distortion correction, details preservation and halo artifacts removal simultaneously and efficiently. The proposed framework consists of two U-Nets \cite{ronneberger2015u} and a convolutional neural network (CNN). In the decomposition phase, we use a U-Net to decompose low-light images into reflectance and illumination map. According to \cite{xiong2020unsupervised}, the up-and-down sampling structure of U-Net can be used as a decomposition network and inspired by\cite{lore2017llnet}\cite{guo2019toward}\cite{zhang2019kindling}, as shown in Table.\ref{unet_abla}, a deep U-Net with up-and-down sampling structure can achieve the function of denoising. So unlike previous methods\cite{wei2018deep}\cite{zhang2019kindling}, we adopt a deeper U-Net as our decomposition net for avoiding the amplification of noise hidden in the dark regions. More details about the choice of the decomposer will be explained in the methodology section.\\
We propose a Degradation-Aware Module (denoted as DA Module), which uses a CNN for the awareness of degradation and a loss function (DA Loss) for tackling all the degradation in the reflectance map. Different from previous methods, instead of designing a network specifically to deal with amplified noise in the reflectance map after the decomposition, we use DA Module to guide the training of the decomposer and enable the decomposer to be the restorer as well, which can generate degradation-free reflectance map directly during the decomposition stage. In the enhancement stage, we use a U-Net to reconstruct the illumination and introduce Perceptual Loss\cite{johnson2016perceptual} to constrain the supervised training process in order to generate the results which are in line with human visual perception better. More details about the DA Module will be explained in the following sections.
We highlight the contributions of this paper as follows:
\begin{itemize}
	\item Based on the Retinex theory, we improve RetinexNet \cite{wei2018deep} and propose an end-to-end Degradation-Aware Deep Retinex Network (DA-DRN) for low-light image decomposition and enhancement.\\ 
	\item We propose a Degradation-Aware (DA) Module for the awareness of all the degradation in reflectance and guide the decomposer to tackle them simultaneously during the decomposition phase directly.\\
	\item By adjusting the coefficient of Degradation-Aware (DA) Loss, we can flexibly adjust the colorfulness of the image.\\
	\item We introduce Perceptual Loss to train the enhancement network to enhance the brightness of illumination by directly learning the mapping of low/normal-light illumination map.\\
	\item Extensive comparisons and ablation experiments are conducted on commonly used datasets to demonstrate the superiority of our method qualitatively and quantitatively, including its fast computation speed.
\end{itemize}

\section{Related Works}
In recent years, many effective methods have been developed for the tasks of low-light image enhancement.
According to the algorithm principle, low-light image enhancement methods can be roughly grouped into three categories: Histogram Equalization(HE)-based enhancement methods, Retinex-based enhancement methods and Learning-based enhancement methods.

\subsubsection{HE-based Enhancement Methods}
There are many improved methods based on HE. Dynamic histogram equalization (DHE) divides the histogram of the image into sub-blocks and uses HE to stretch the contrast for each sub-block. Adaptive histogram equalization (AHE) \cite{pizer1987adaptive} changes image contrast by calculating the histogram of multiple local areas of the image and redistributing the brightness. AHE is more suitable for improving the local contrast and details of the image, but the noise is amplified. Contrast limited adaptive histogram equalization (CLAHE) \cite{pizer1990contrast} restricts the histogram of each sub-block and well controls the noise brought by AHE. \\
\subsubsection{Retinex-based Enhancement Methods}
Methods based on Retinex Theory \cite{land1977retinex} decompose the image into reflectance and illumination. Such methods maintain the consistency of the reflectance, increase the brightness of the illumination and then take the pixel-wise product to enhance the low-light image. Single Scale Retinex (SSR) \cite{jobson1997properties} aims to restore the brightness of illumination after Retinex decomposition. Multi-Scale Retinex (MSR) \cite{jobson1997multiscale} combines the filtering results of multiple scales based on SSR. MSRCR adds a color recovery factor to tackle the color distortion caused by contrast enhancement in local areas of the image. NPE \cite{wang2013naturalness} strikes a balance between contrast enhancement and naturalness of estimated illumination. MF \cite{fu2016fusion} is a fashion-based method for contrast enhancement of the illumination. SRIE \cite{fu2016weighted} is a weighted variational model to estimate reflectance and illumination. BIMEF \cite{ying2017bio} provides a bio-inspired dual-exposure fusion algorithm to provide accurate contrast and lightness enhancement, which obtains results with less contrast and lightness distortion. LIME \cite{guo2016lime} estimates a structure-aware illumination map with structure prior and uses BM3D \cite{dabov2006image} for the post-processing denoising operation. Dong at el. \cite{dong2011fast} proposed an efficient algorithm by inverting an input low-lighting video and applying an optimized image de-haze algorithm on the inverted video. CRM \cite{ying2017new} uses the camera response model for the enhancement process. RRM \cite{li2018structure} is a Robust Retinex Model that considers noise mapping to improve the enhancement performance of low-light intensity noise images. LECARM \cite{ren2018lecarm} is a novel enhancement framework using the response characteristics of cameras to tackle color and lightness distortions. JED \cite{ren2018joint} is a joint low-light enhancement and denoising strategy. 

\subsubsection{Learning-based Enhancement Methods}
With the advent of deep learning, a great number of state-of-the-art methods have been developed for low-light image enhancement. LLNet \cite{lore2017llnet} is a deep auto-encoder model for enhancing lightness and denoising simultaneously. LLCNN \cite{tao2017llcnn} is a CNN-based method utilizing multi-scale feature maps and SSIM loss for low-light image enhancement. MSR-Net \cite{shen2017msr} is a feedforward CNN with different Gaussian convolution kernels to simulate the pipeline of MSR for directly learning end-to-end mapping between dark and bright images. GLADNet \cite{wang2018gladnet} is a global illumination-aware and detail-preserving network that calculates global illumination estimation. LightenNet \cite{li2018lightennet} serves as a trainable CNN by taking a weakly illuminated image as the input and outputting its illumination map. MBLLEN \cite{lv2018mbllen} uses multiple subnets for enhancement and generates the output image through multi-branch fusion. RetinexNet \cite{wei2018deep} decomposes low-light input into reflectance and illumination and enhances the lightness over illumination. EnlightenGan \cite{jiang2021enlightengan} trains an unsupervised generative adversarial network (GAN) without low/normal-light pairs. KinD \cite{zhang2019kindling} first decomposes low-light images into a noisy reflectance and a smooth illumination and then uses a U-Net to recover reflectance from noise and color distortion. RDGAN \cite{wang2019rdgan} proposes a Retinex decomposition based GAN for low-light image enhancement. 
RetinexDIP \cite{zhao2021retinexdip} provides a unified deep framework using a novel ”generative” strategy for Retinex decomposition. Zhang et al. \cite{zhang2020self} presented a self-supervised low-light image enhancement network, which is only trained with low-light images. Zero-DCE \cite{guo2020zero} estimates the brightness curve of the input image without any paired or unpaired data during training.

\section{Motivation and Analysis}
\begin{figure}
	\subfigure[Framework of RetinexNet\cite{wei2018deep}]{
		\begin{minipage}[b]{0.45\linewidth}
			\includegraphics[width=4.3cm]{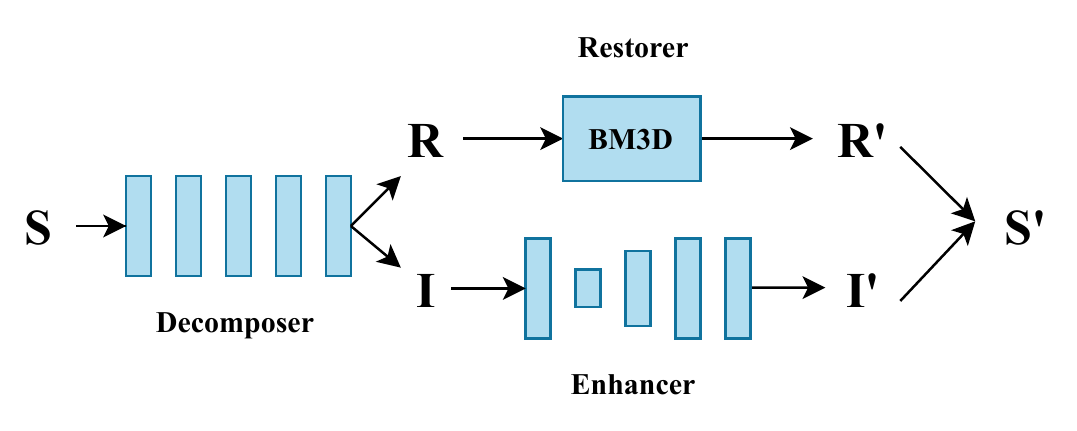}\vspace{-3pt}
	\end{minipage}}
	\subfigure[Framework of KinD\cite{zhang2019kindling}]{
		\begin{minipage}[b]{0.45\linewidth}
			\includegraphics[width=4.3cm]{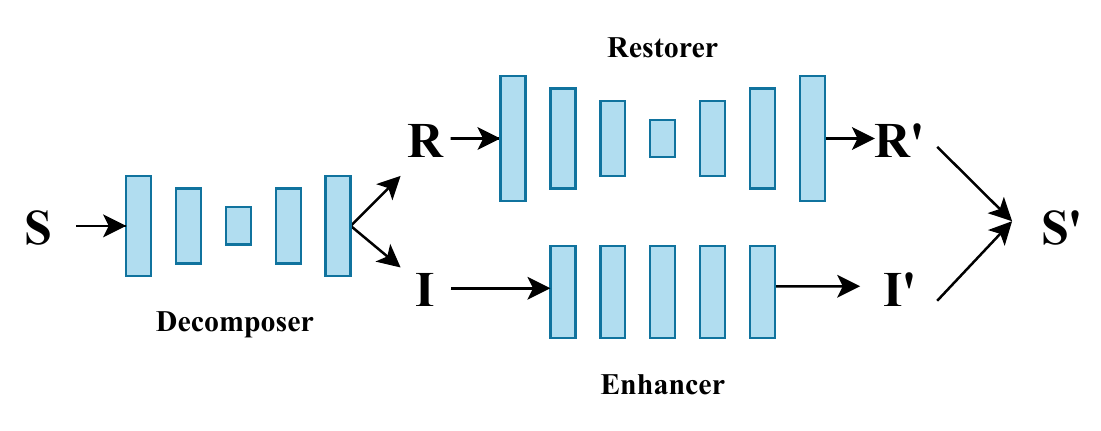}\vspace{-3pt}
	\end{minipage}}
\\
	\subfigure[Training Phase of DA-DRN]{
		\begin{minipage}[b]{0.45\linewidth}
			\includegraphics[width=4.3cm]{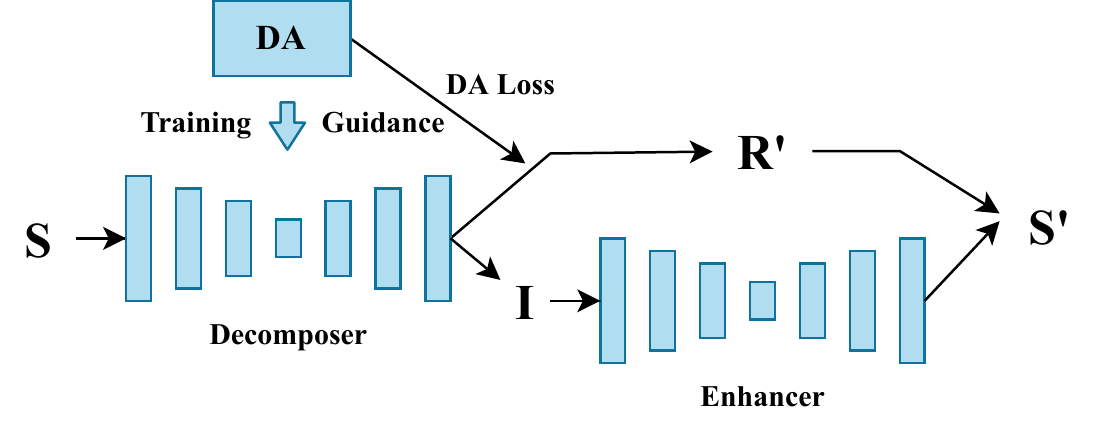}\vspace{-3pt}
	\end{minipage}}
		\subfigure[Test Phase of DA-DRN]{
		\begin{minipage}[b]{0.45\linewidth}
			\includegraphics[width=4.3cm]{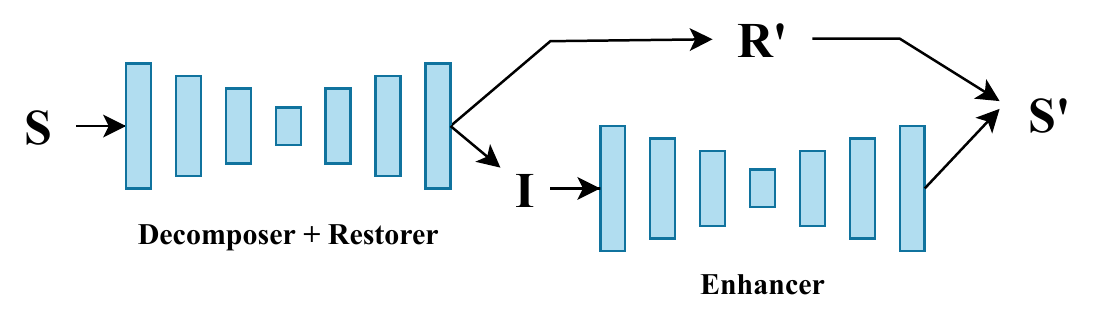}\vspace{-3pt}
	\end{minipage}}
	\vspace{4pt}
	\caption{Comprison with other Retinex-based deep learning frameworks. DA is only used in training phase to guide the training of decomposer and enable the decomposer the ability of restoration. ($S$, $R$, $I$ : Low-light input, Reflectance with degradation, Illumination with low lightness)  ($S^{'}$, $R^{'}$, $I^{'}$ : Normal-light input, Reflectance without degradation, Illumination with normal lightness)}
	\label{model_abla}
\end{figure}

Based on the assumption of Retinex Theory \cite{land1977retinex}, a natural image (S) can be decomposed into two components: Reflectance (R) and Illumination (I). * represents a pixel-wise product operator.
\begin{align}
	S=S^{'}+D=R^{'}*I^{'} + D
\end{align}
where $S$ and $S^{'}$ mean Low-light Input Image and Normal-light Ground-Truth. $R^{'}$ and $I^{'}$ represent Reflectance map without degradation and Illumination map without degradation. $D$ means a set of degradation. 
\begin{align}
	D=N+LL
\end{align}
where $N$ and $LL$ represent Noise and Low Lightness, respectively.\\
Reflectance is usually a three-channel image that contains color and high-frequency components, such as noise and detail information. Illumination is a very smooth single-channel image that only contains low-frequency components, such as the intensity and distribution of lumination. In the process of decomposition, illumination is smooth enough to be regarded as noise-free since it only contains low-frequency information. However, noise hidden in the dark is amplified in the reflectance, which results in a very low signal-to-noise ratio (SNR) for the reflectance.\\ 
When Retinex decomposition is conducted, degradation will change. Here we give a detailed analysis and formula derivation of the degradation model in Retinex decomposition. \\
As shown in Formula.\ref{f3}, when a low-light image is decomposed into Reflectance and Illumination, noise hidden in the darkness will be amplified in the reflectance, in addition, color distortion and unpleasant halo artifacts occur; illumination map now is an image with low lightness and poor visibility:
\begin{align}
	S=R*I=(R^{'} + AN + CD + HA)*(I^{'} + LL)
	\label{f3}
\end{align}
where $R$, $I$, $AN$, $CD$ and $HA$ represent Reflectance map with degradation, Illumination map with degradation, Amplified Noise, Color Distortion and Halo Artifacts, respectively.\\
When we conduct denoising on $R$, although noise can be removed, details will be lost in the reflectance; when we conduct enhancement on $I$, over/under-exposure may occur:
\begin{align}
	R=R^{'} +& AN + CD + HA \xrightarrow{Denoising} R^{'} + CD + HA + DL \\
	\notag 
	&I=I^{'} + LL \xrightarrow{Enhancement} I^{'} + OE/UE
	\label{f4}
\end{align}
where $DL$, $OE$ and $UE$ represent Details Loss, Over-Exposure and Under-Exposure, respectively.\\
Previous approaches\cite{guo2016lime}\cite{wei2018deep}\cite{zhang2019kindling} used additional well-designed denoisers\cite{dabov2006image}\cite{guo2019toward} or an embedded denoiser to denoise the reflectance. However, there may be some problems such as color distortion and loss of high-frequency details in the reflectance map after applying extra denoisers. Furthermore, additional denoisers can significantly reduce the forward inferencing speed of the whole pipeline.\\
In the enhancement stage, the brightness of the illumination is enhanced, but if the intensity of brightness and the distribution of light are not restored correctly, the result will be overexposed or underexposed. The color information of the image depends not only on the reflectance but also on the brightness information of the illumination. Incorrectly predicted illumination maps can also result in color distortion in the final enhanced results.\\
A variety of degradation may arise after enhancing the brightness of the low-light image. Many of the previous methods used multiple sub-methods or sub-networks to tackle some of these problems in numerous steps \cite{guo2016lime}\cite{wei2018deep}\cite{zhang2019kindling}, which can slow down the speed of the pipeline. Unlike previous methods, as shown in Fig.\ref{model_abla}, we aim to enhance the lightness of low-light images without introducing extra restorers to deal with dagradations in the reflectance map. We use Degradation-Aware (DA) Module for the awareness of the degradation in reflectance and guide the decomposer to eliminate them directly during the decomposition stage rather than designing multiple sub-networks to deal with these problems individually. We are inspired by RetinexNet \cite{wei2018deep} and overcome the shortcomings of RetinexNet.

\begin{figure*}[htbp]
	\centering
	\includegraphics[width=17cm]{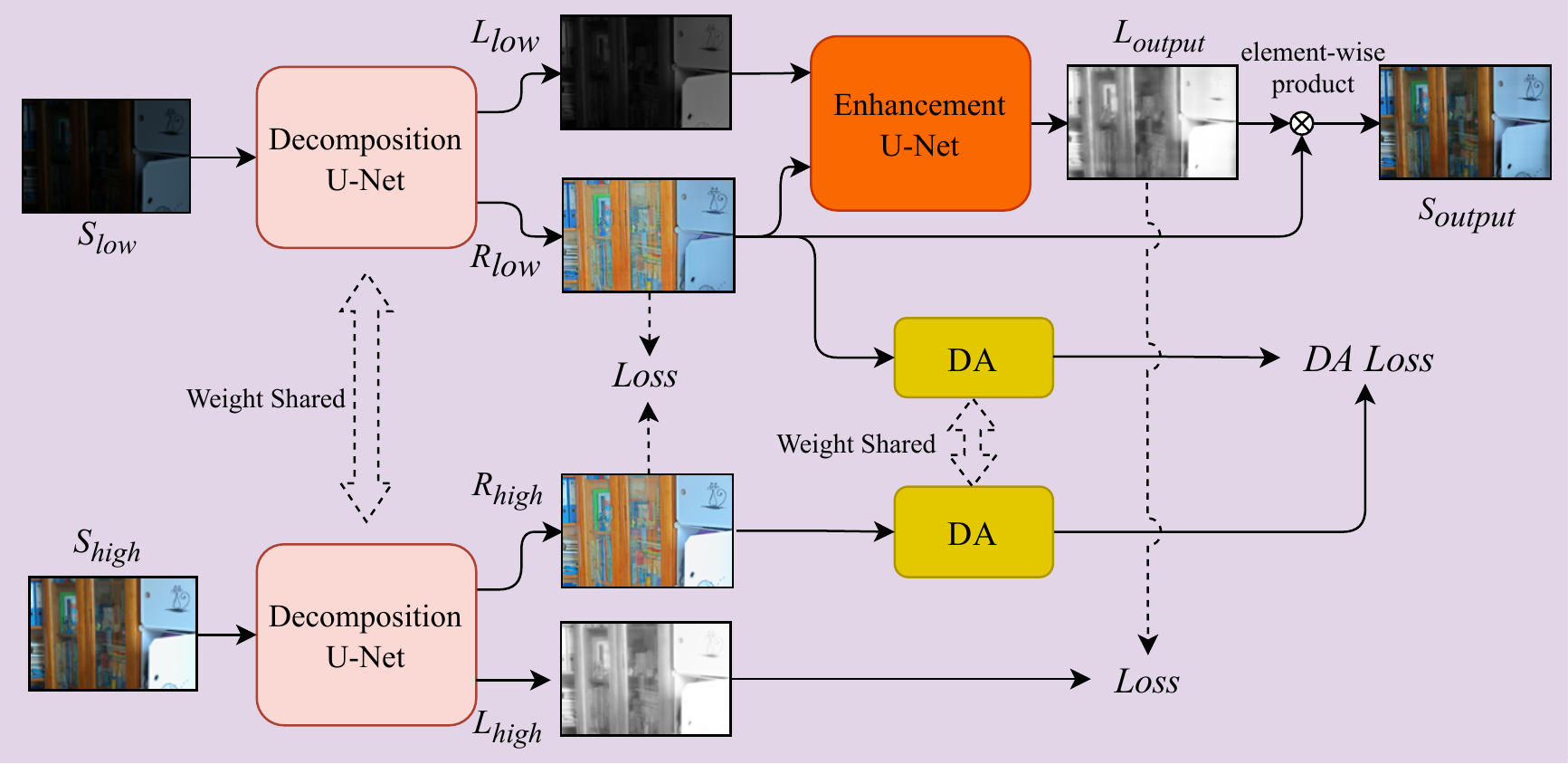}
	\includegraphics[width=17cm]{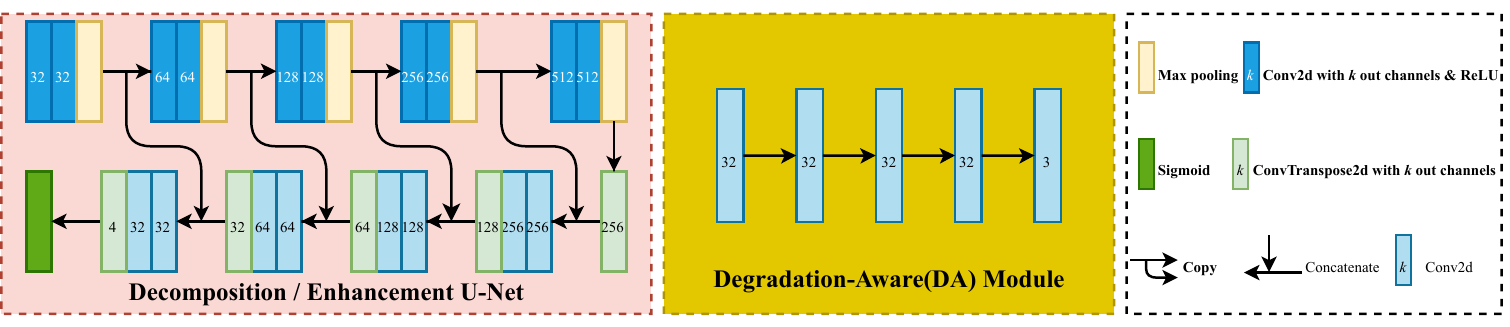}
	\caption{The network architecture of DA-DRN. Both the Decomposition Net and Enhancement Net adopt the U-Net structure(The reason can be found in the following sections). The DA Module is a CNN that consists of several stacked convolutional blocks.}
	\label{model}
\end{figure*}

\section{Methodology}
As shown in Fig.\ref{model}, inspired by Retinex Theory, our proposed method adopts two U-Nets as decomposition and enhancement networks, respectively. In the decomposition stage, we use a deep U-Net to decompose the low-light image into two components (i.e., reflectance and illumination) and propose a Degradation-Aware (DA) Module for guiding the decomposer to tackle degradation in the reflectance directly during the decomposition stage. In the enhancement stage, we introduce Perceptual Loss\cite{johnson2016perceptual} to train a deep U-Net as our enhancement net. In the following sections, we will explain our network architecture and loss functions in detail.

\begin{table}[htbp]
	\centering 
	\caption{Quantitative comparison of \textbf{Reflectance (R)} decomposed by normal/low-light images of several methods in Fig. \ref{748_r}. “↑” indicates the higher the better, “↓” indicates the lower the	better. The best results
		are highlighted in bold.}
	\begin{tabular}{ c c|c|c|c }
		\hline
		Methods & Reflectance & Noise Level↓ & PSNR↑ & DeltaE↓ \\ \hline\hline
		\multirow{2}{*}{RetinexNet\cite{wei2018deep}} & Normal & 3.4678 &- &- \\ 
		& Low & 36.5095 & 14.7454 & 20.9036 \\ \hline
		\multirow{3}{*}{KinD\cite{zhang2019kindling}} & Normal & 5.3774 &- &- \\
		& Restorer & 1.7625 & 21.9833 & 8.9493 \\ 
		& Low & 13.9367 & 13.3297 & 25.2839 \\ \hline
		\multirow{2}{*}{Ours w/o DA} & Normal & 0.9443 &- &- \\
		& Low & 3.1812 & 22.7766 & 8.6116 \\ \hline
		\multirow{2}{*}{Ours with DA} & Normal & \textbf{0.9265} & - & - \\
		& Low & \textbf{2.1764} & \textbf{25.0536} & \textbf{5.5829} \\ \hline
		\label{748_r_abla}
	\end{tabular}
\end{table}

\subsection{Decomposition Network}
\begin{figure*}
	\centering
	
	\subfigure[R of RetinexNet\cite{wei2018deep}]{
		\begin{minipage}[b]{0.23\textwidth}
			\includegraphics[width=4.2cm]{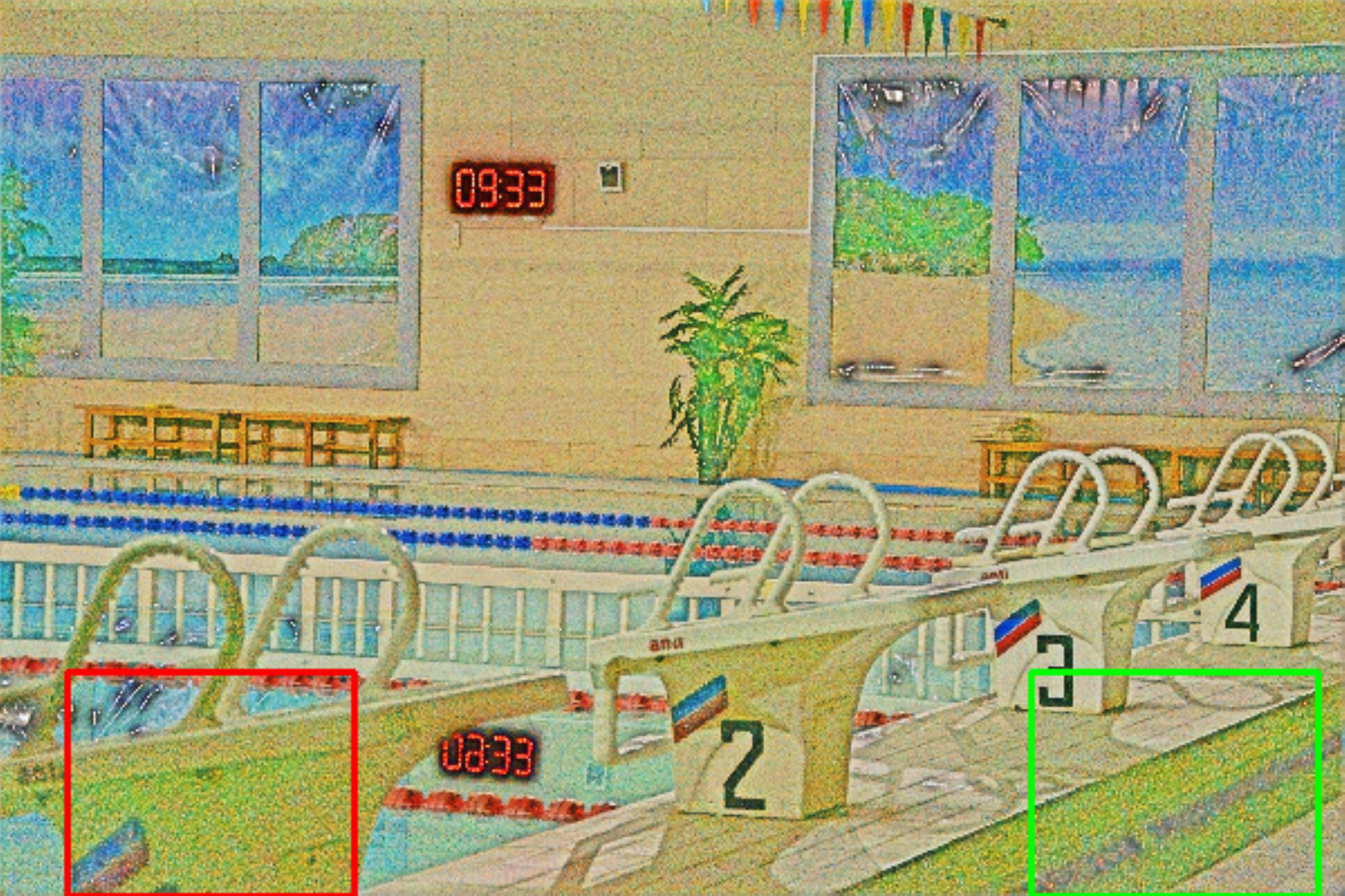}\vspace{1.5pt} \\
			\includegraphics[width=2.02cm]{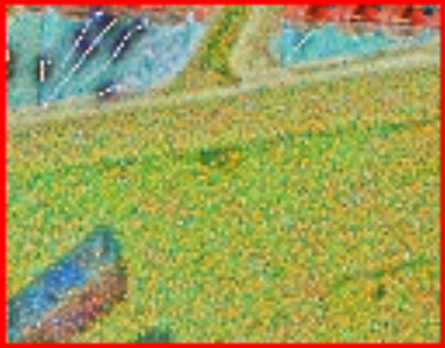}
			\includegraphics[width=2.02cm]{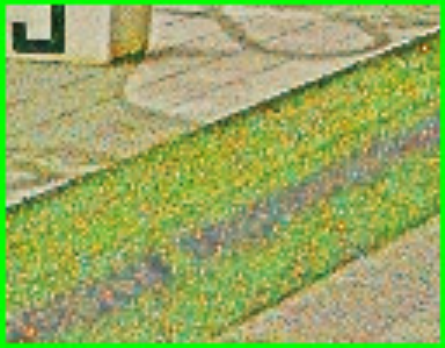}
		\end{minipage}
	}\hspace{-5pt}
	\subfigure[R of KinD\cite{zhang2019kindling}]{
		\begin{minipage}[b]{0.23\textwidth}
			\includegraphics[width=4.2cm]{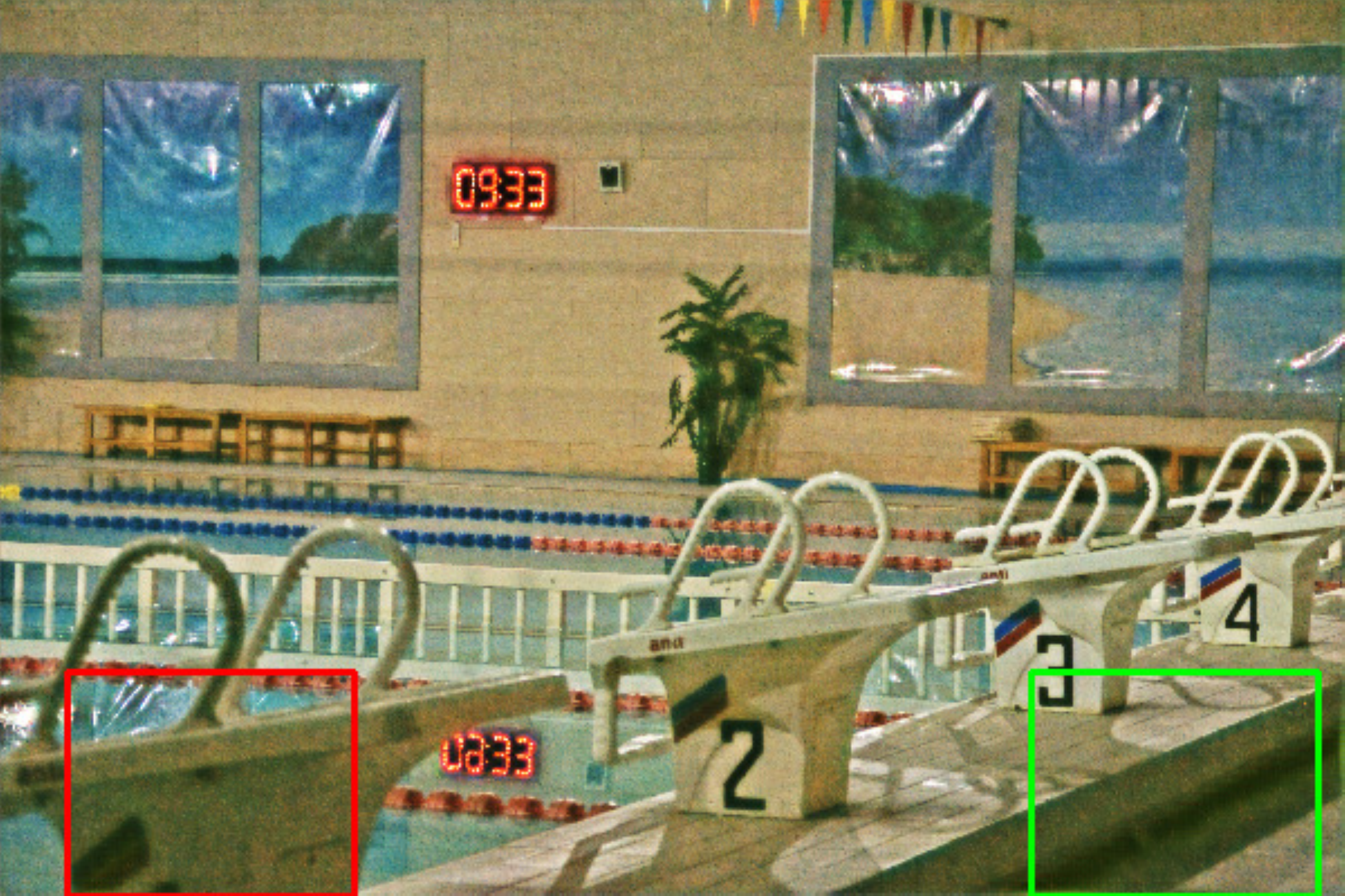}\vspace{1.5pt} \\
			\includegraphics[width=2.02cm]{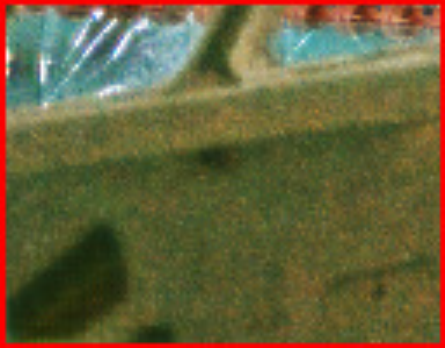}
			\includegraphics[width=2.02cm]{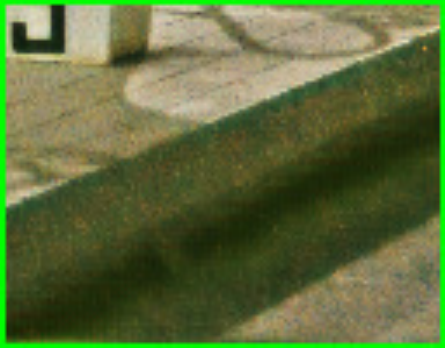}
		\end{minipage}
	}\hspace{-5pt}
	\subfigure[R of DA-DRN w/o DA]{
		\begin{minipage}[b]{0.23\textwidth}
			\includegraphics[width=4.2cm]{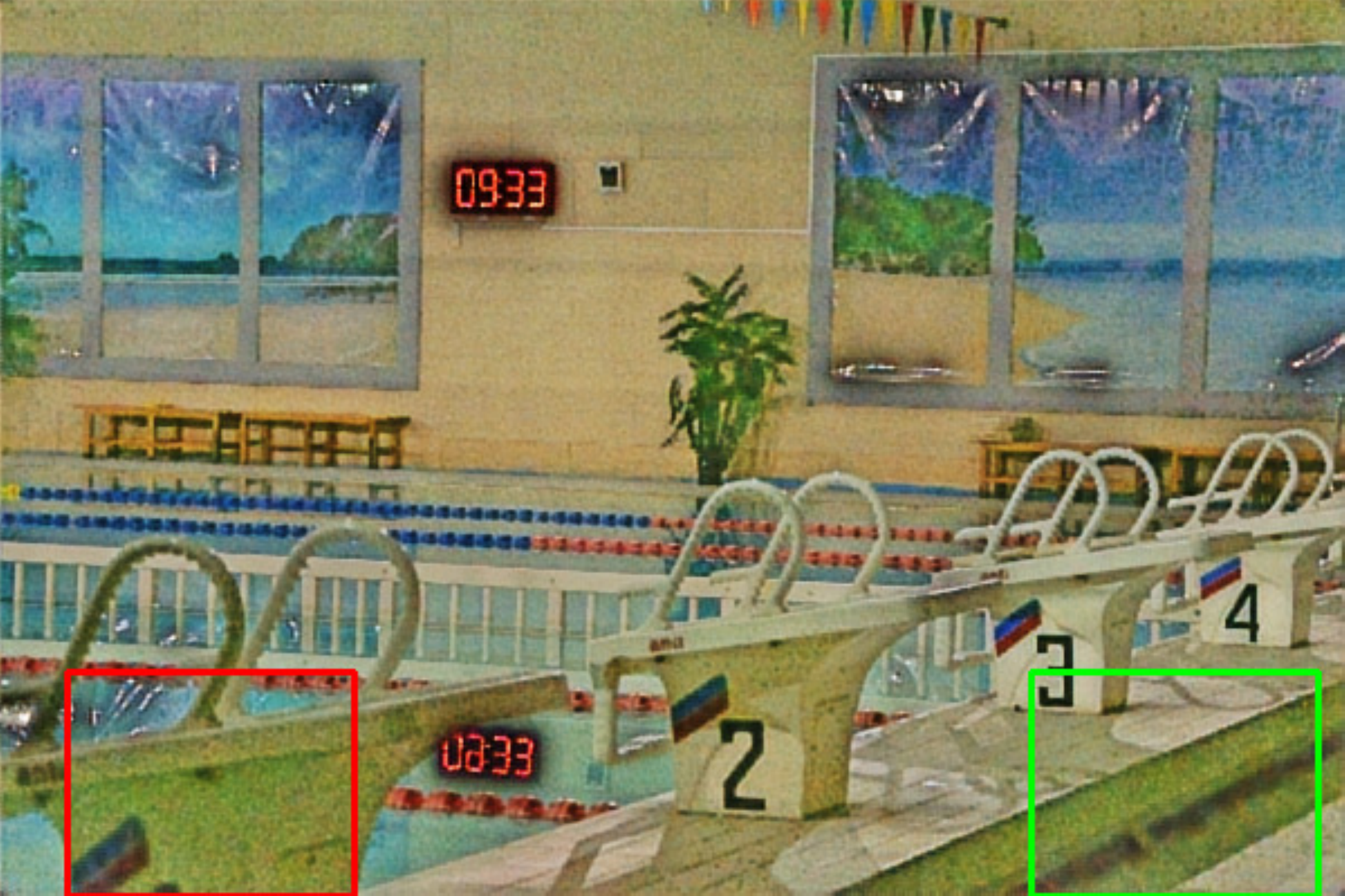}\vspace{1.5pt} \\
			\includegraphics[width=2.02cm]{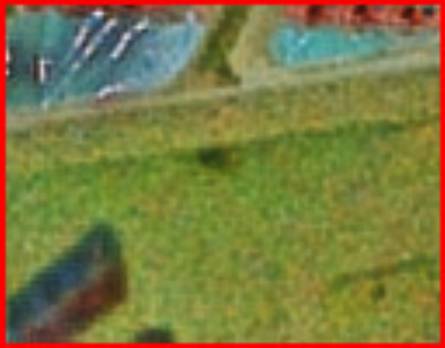}
			\includegraphics[width=2.02cm]{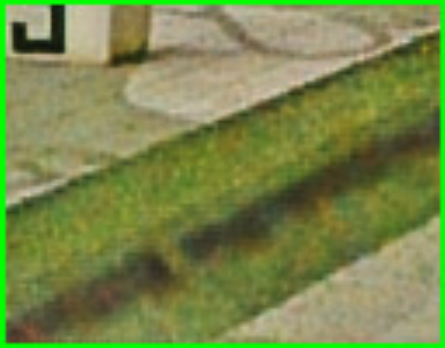}
		\end{minipage}
	}\hspace{-5pt}
	\subfigure[R of DA-DRN with DA]{
		\begin{minipage}[b]{0.23\textwidth}
			\includegraphics[width=4.2cm]{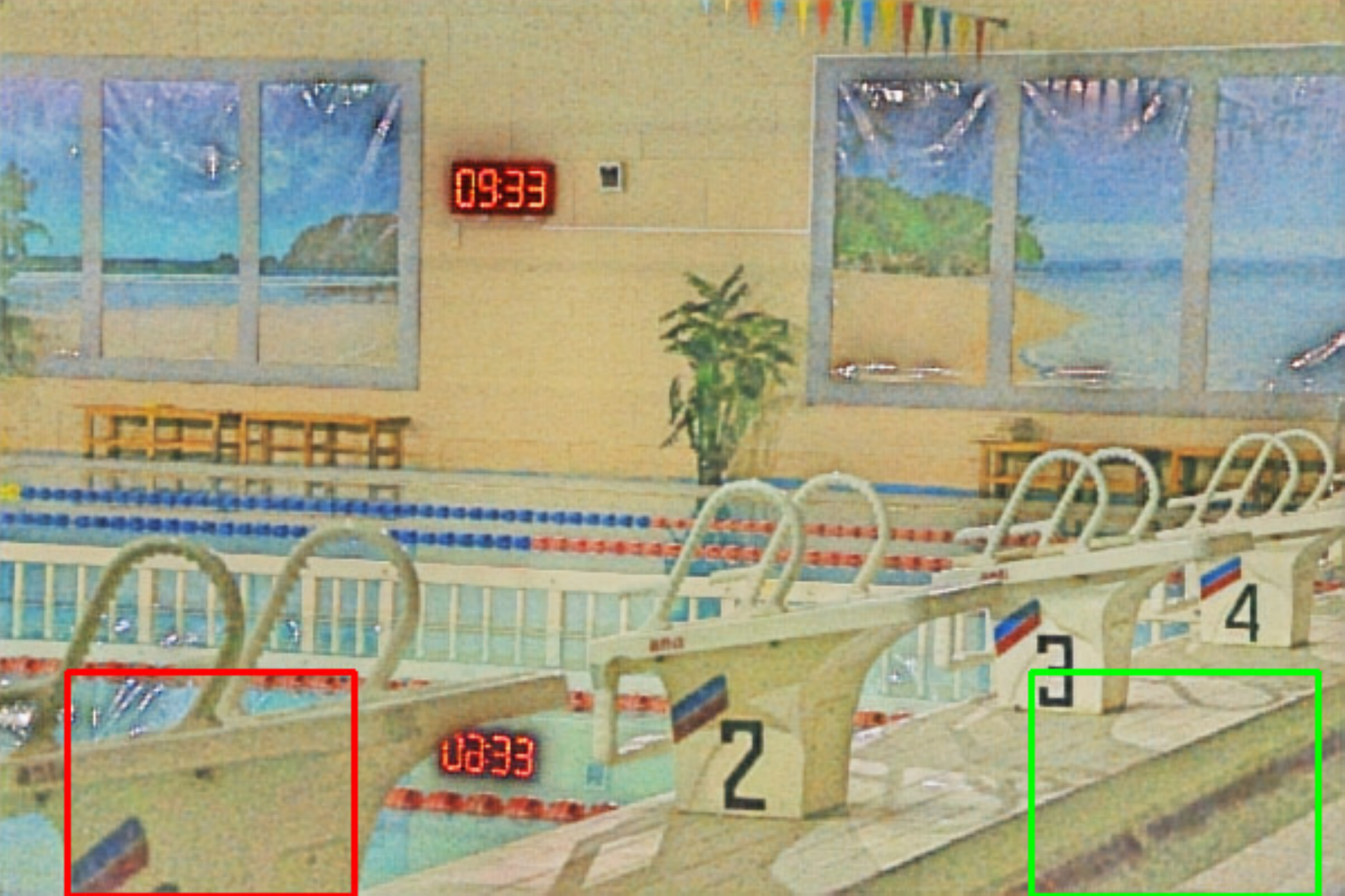}\vspace{1.5pt} \\
			\includegraphics[width=2.02cm]{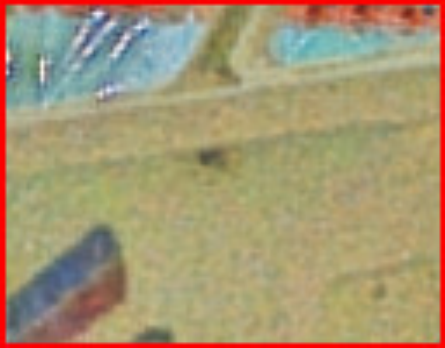}
			\includegraphics[width=2.02cm]{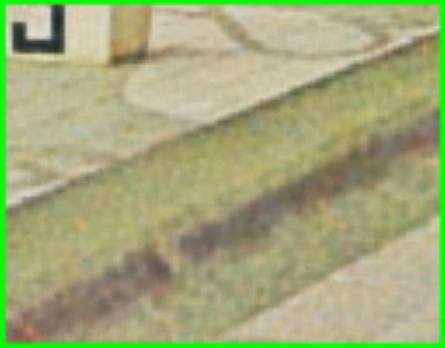}
		\end{minipage}
	}
	\caption{Comparison with other Retinex-based methods in terms of Reflectance(R). The first row: Reflectance of Normal-light Ground Truth Input. The second row: Reflectance of Low-light Input. Quantitative comparison of Reflectances in terms of Noise Level, PSNR and DeltaE can be seen in Table.\ref{748_r}}
	\label{748_r}
\end{figure*}

As shown in Fig.\ref{model_abla}, RetinexNet \cite{wei2018deep} uses a Plain CNN without up-and-down sampling structure as the decomposition net to decouple the low-light image into the reflectance and illumination, but as shown in Fig.\ref{748_r} and Table.\ref{748_r_abla}, this Plain CNN may amplify the noise and introduce color distortion in the reflectance. KinD \cite{zhang2019kindling} uses a shallow U-Net as the decomposition net, but as we can seen in Fig.\ref{748_r} and Table.\ref{748_r_abla}, the noise is still amplified in the reflectance, even the noise in the reflectance map, which is decomposed from normal-light Ground-truth images, is also amplified. In the reflectance maps obtained by decomposition of the normal-light Ground-Truth images through the decomposers of RetinexNet\cite{wei2018deep} and KinD\cite{zhang2019kindling}, the noise levels are 3.4678 and 5.3774, By contrast, the noise level of their counterparts decomposed through our decomposer is only 0.9443. In addition, there are also serious color distortion and unpleasant halo artifacts. Therefore, inspired by the restoration net in \cite{zhang2019kindling}, we use a deeper U-Net as our decomposition net because according to\cite{lore2017llnet}\cite{guo2019toward}\cite{zhang2019kindling}\cite{wang2018gladnet}, as shown in Table.\ref{unet_abla}, the up-and-down sampling structure of a deeper U-Net has the function of denoising and keeping the noise from being amplified during the decomposition stage.

\begin{table}[htbp] 
	\centering \caption{Comparison of \textbf{Reflectance (R)} maps in Fig.\ref{pic_04} decomposed by different decomposer network architectures.}
	\begin{tabular}{  c | c | c  }
		\hline Decomposer Network &Kernel Settings  & Noise Level↓  \\
		\hline
		\hline  
		Plain CNN &$<$64 64 64 64 64$>$ & 1.1962  \\
		CNN + U-Net &$<$64 64 64 128 64 64 64$>$  & 0.7751 \\
		Shallow U-Net &$<$32 64 32$>$ & 0.4287 \\
		U-Net &$<$32 64 128 64 32$>$  & 0.3962 \\
		Deep U-Net &$<$32 64 128 256 128 64 32$>$  & 0.3762\\ 
		\hline
		\hline 
		Deep U-Net with DA &$<$32 64 128 256 128 64 32$>$  & \textbf{0.2156}\\
		\hline\end{tabular}\vspace{0cm}
	\label{unet_abla}
\end{table}

\begin{figure*}
	
	
	\subfigure[R of Plain CNN \cite{wei2018deep}]{
		\begin{minipage}[b]{0.185\textwidth}
			\includegraphics[width=3.5cm]{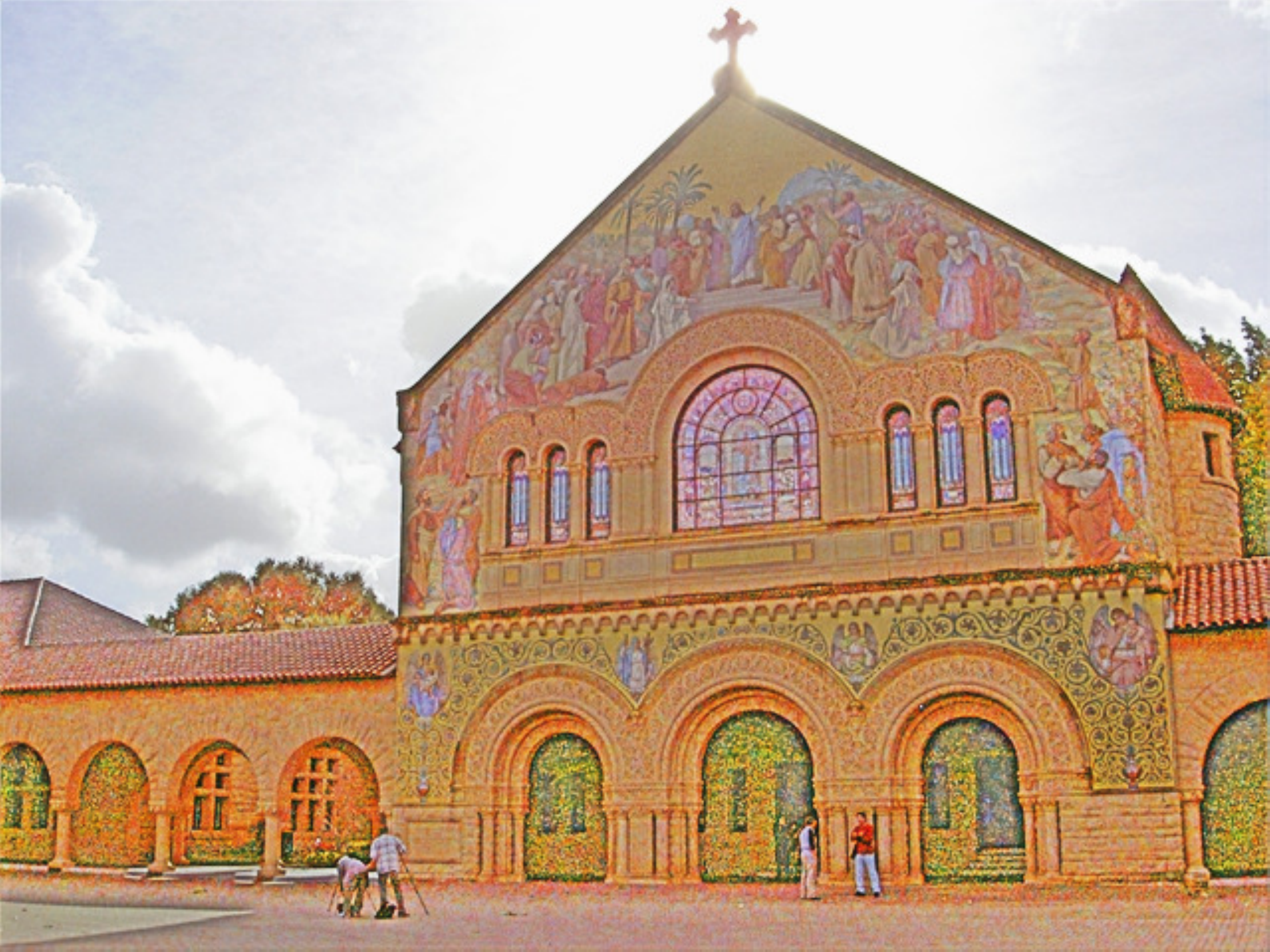}\vspace{-10pt} \\
		\end{minipage}
	}\hspace{-5pt}
	\subfigure[R of CNN+U-Net]{
		\begin{minipage}[b]{0.185\textwidth}
			\includegraphics[width=3.5cm]{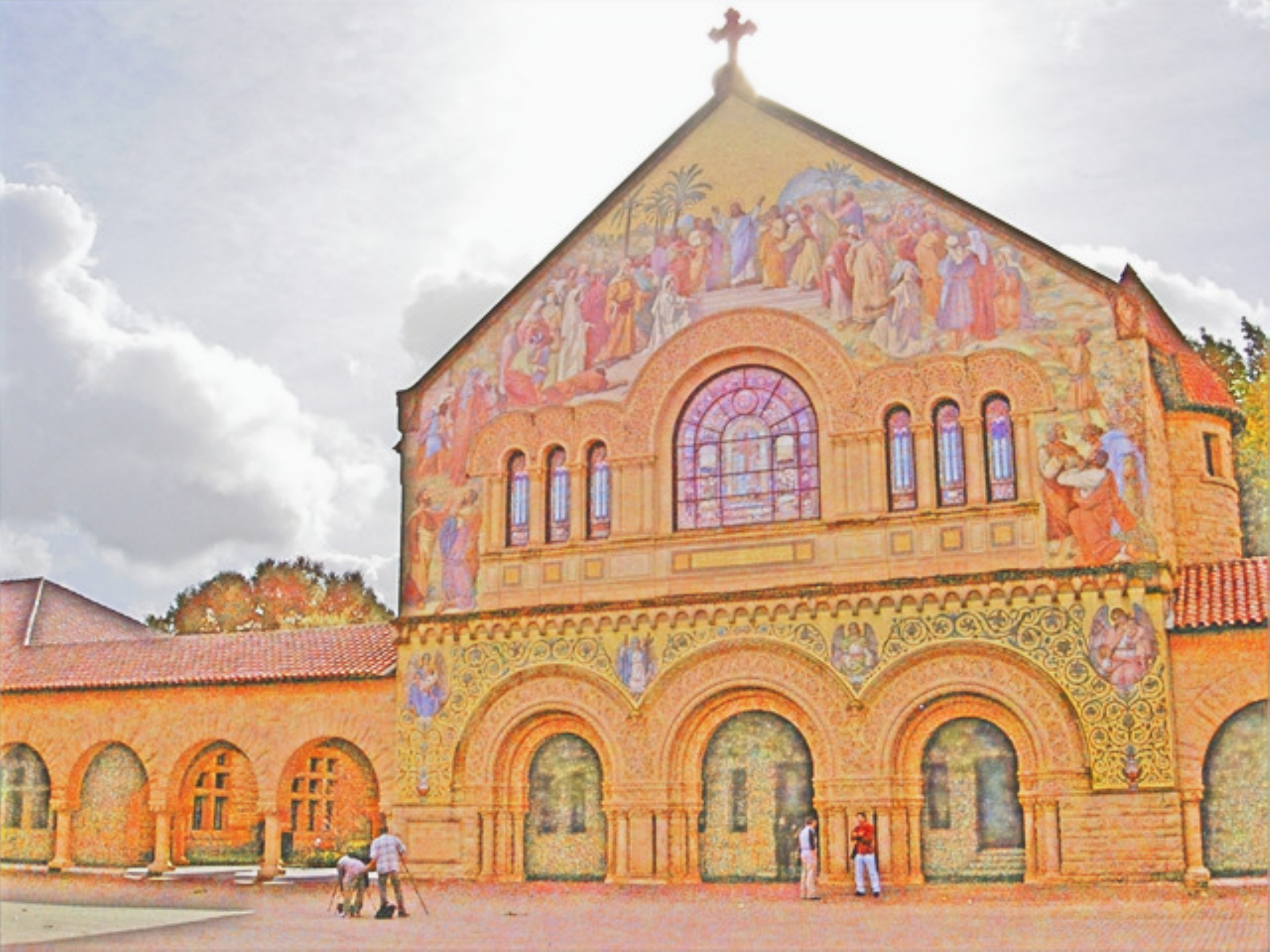}\vspace{-10pt} \\
		\end{minipage}
	}\hspace{-5pt}
	\subfigure[R of U-Net]{
		\begin{minipage}[b]{0.185\textwidth}
			\includegraphics[width=3.5cm]{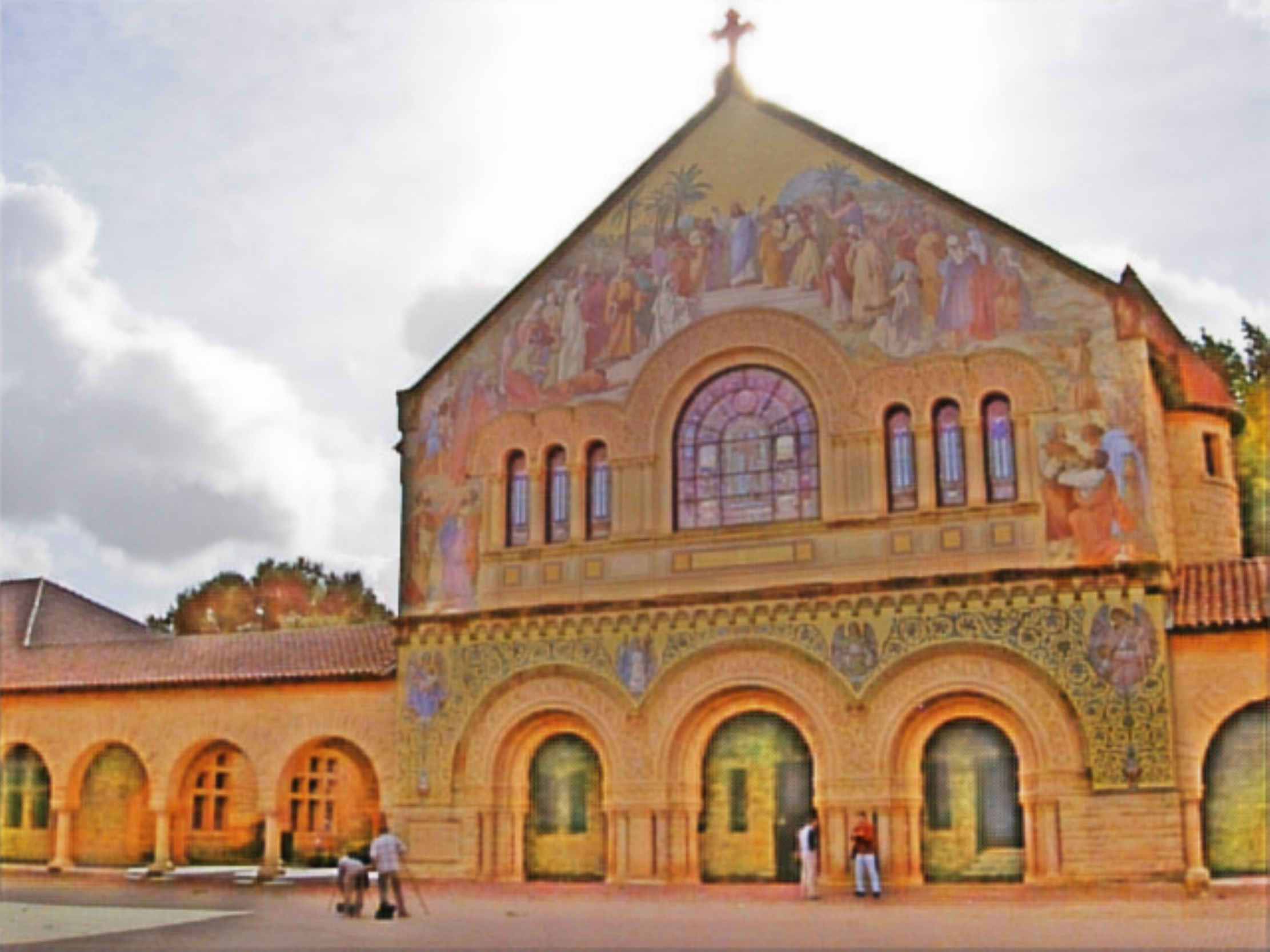}\vspace{-10pt} \\
		\end{minipage}
	}\hspace{-5pt}
	\subfigure[R of Deep U-Net]{
		\begin{minipage}[b]{0.185\textwidth}
			\includegraphics[width=3.5cm]{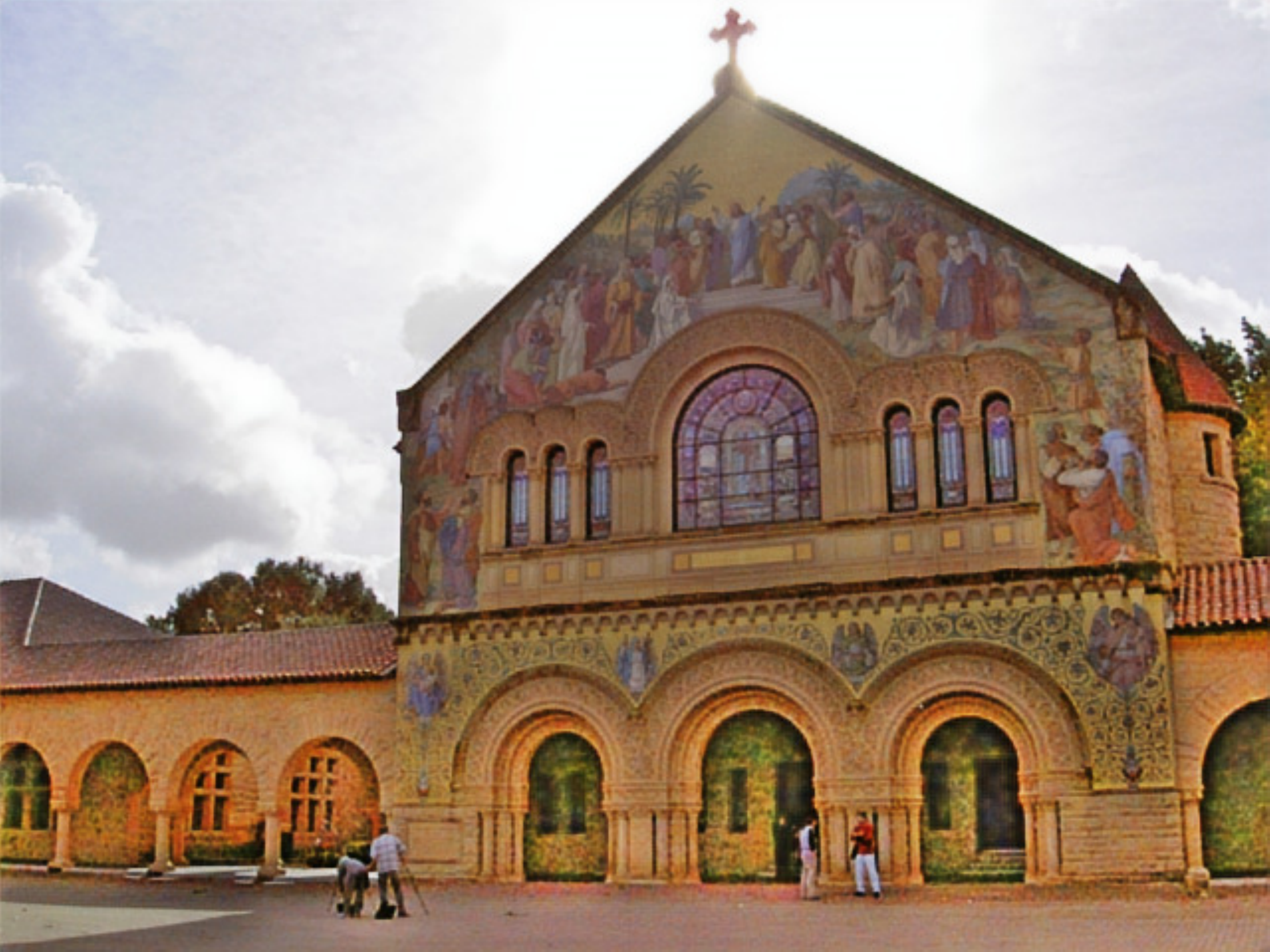}\vspace{-10pt} \\
		\end{minipage}
	}\hspace{-5pt}
	\subfigure[R of Deep U-Net with DA]{
		\begin{minipage}[b]{0.185\textwidth}
			\includegraphics[width=3.5cm]{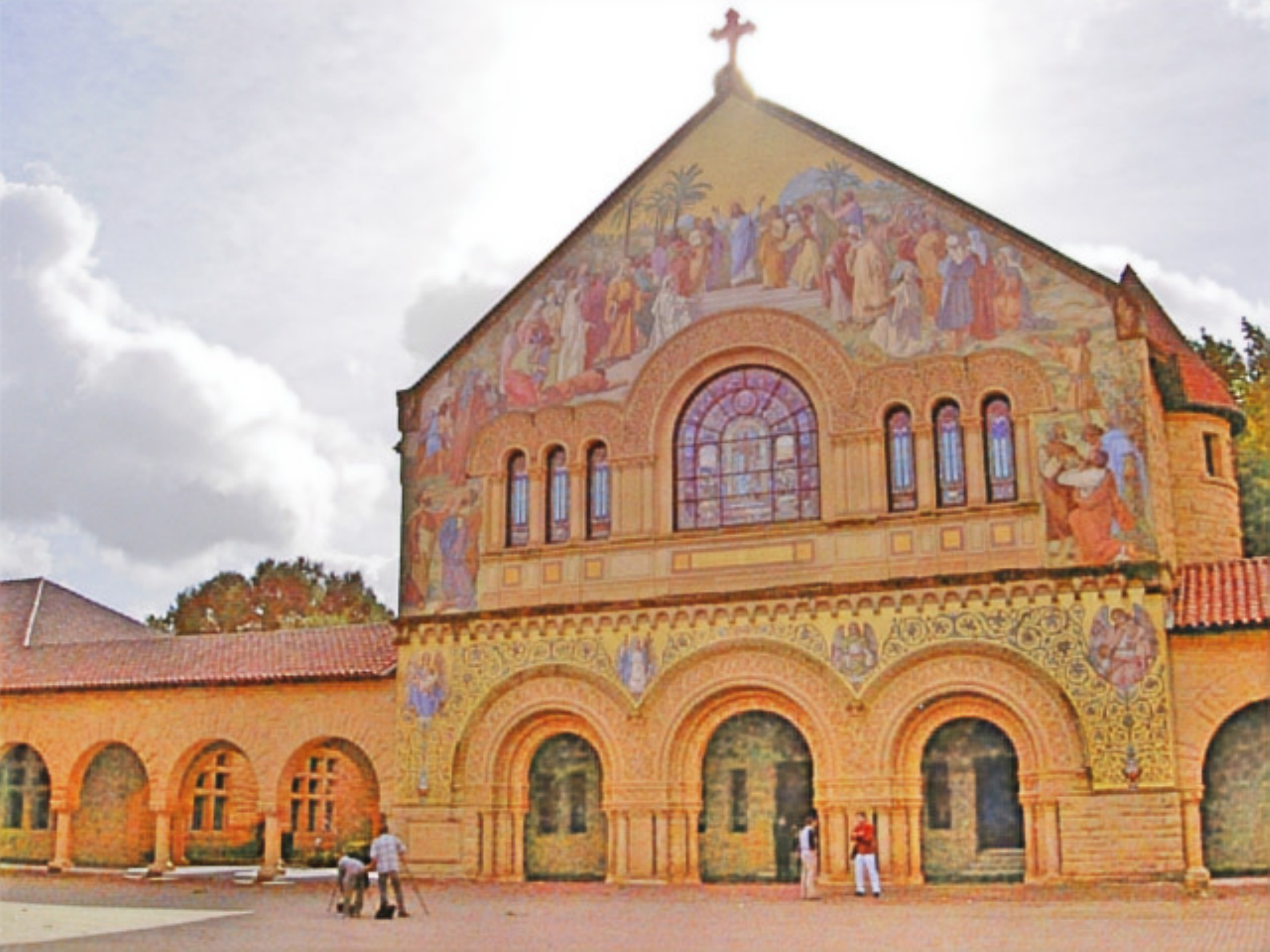}\vspace{-10pt} \\
		\end{minipage}
	}
	\caption{Visual comparison of different reflectance (R) maps decomposed by different decomposer network architecture. Quantitative comparison of them can be seen in Table.\ref{unet_abla}}
	\label{pic_04}
\end{figure*}

However, unpleasant noise, color distortion and halo artifacts still exist in the reflectance, so we introduce the Degradation-Aware (DA) Module for the awareness of the degradation in the reflectance and guide the decomposer to tackle them during the decomposition phase instead of using an extra denoiser for the restoration of the reflectance map. DA can guide the training process of the decomposition net and enable the trained decomposer the ability of restoration, which can tackle the noise, color distortion and halo artifacts during the decomposition stage directly.\\
Compared with KinD\cite{zhang2019kindling}, as shown in Table.\ref{unet_abla} and Fig.\ref{pic_04}, we use a deeper U-Net with more parameters as our decomposer in order to suppress and prevent noise hidden in the darkness from being amplified as well as train the decomposition net to be the decomposer and restorer with the guidance of DA.\\
In the decomposition, inspired by\cite{wei2018deep}\cite{zhang2019kindling}, we also use the reconstruction loss $L_{rc}$ proposed in RetinexNet \cite{wei2018deep} to constrain the process of decomposition.
\begin{equation}
L_{rc}^{D}=\sum_{i}\sum_{j}\lambda_{ij}\left \|R_{i}\ast I_{j} - S_{j}  \right \|_{1}
\label{lrc}
\end{equation}
where $*$ denotes the pixel-wise product and $i,j$ denote the low- and high-frequency components, respectively. $R_{high}$, $R_{low}$ and $I_{high}$, $I_{low}$ represent the reflectance and the illumination decomposed by Ground-Truth and low-light input, respectively. When $i=j$, $\lambda_{ij}=1$; otherwise, $\lambda_{ij}=0.001$. More details can be found in \cite{wei2018deep}.\\
We use $L1$ Loss as the equal loss $L_{equal}$ to constrain the reflectance to be as similar as possible to that decomposed by Ground-Truth. 
\begin{equation}
L_{equal}=\left \| R_{low} - R_{high} \right \|_{1}
\end{equation}
Previous Retinex-based methods \cite{wei2018deep}\cite{zhang2019kindling} use the smoothness loss to maintain the spatial smoothness of illumination so that the illumination contains only low-frequency information, such as intensity and distribution of brightness, and all of the high-frequency information, such as details and noise, are retained in the reflectance map. We also use the smoothness loss $L_{smooth}$ proposed in RetinexNet \cite{wei2018deep} to keep the illumination map spatially smooth. 
\begin{equation}
L_{smooth}=\sum_{i}\left \| \triangledown _{h}I_{i}\ast e^{-\lambda \ast \triangledown_{h}R_{i} } \right \|_{1}  +  \sum_{i}\left \| \triangledown _{v}I_{i}\ast e^{-\lambda \ast \triangledown_{v}R_{i} } \right \|_{1}
\label{sml}
\end{equation}
where $i=low,high$ and $\triangledown_{h}$, $\triangledown_{v}$ denote the gradients in the horizontal and vertical directions. According to RetinexNet \cite{wei2018deep}, we also set the weight coefficient $\lambda$ to 10. $\lambda$ is a parameter which can regulates the smoothness of the generated illumination map. \\
We introduce Degradation-Aware (DA) Module and DA Loss to giude the training process of the decomposer aiming to enable the decomposer to be the restorer as well and the reflectance map without degradation can be directly genreated by the decomposer trained with DA.
So the total decomposition loss is as follows: 
\begin{equation}
L_{decom}=1.0L_{rc}^{D} + 0.1L_{smooth} +  0.01L_{equal} + 1.0L_{DA}
\end{equation}
The details of Degradation-Aware (DA) Module and DA Loss will be explained in next section.\\
According to the previous methods \cite{guo2016lime}\cite{wei2018deep}\cite{wang2019rdgan}\cite{zhang2019kindling}, the illumination map is smooth enough and contains only low-frequency information, and all of the noise after decomposition appears in the reflectance map. However, as shown in Fig.\ref{pic_tv_abla}, the illumination map of DA-DRN contains some high-frequency details information and even some noise, and with the coefficient of TV Loss $\lambda_{TV}$ in Formula.\ref{lda} gets larger, more and more details are transferred from reflectance into illumination map.

\begin{figure*}
	\flushleft
	
	\subfigure[Ground-Truth]{
		\begin{minipage}[b]{0.155\textwidth}
			\includegraphics[width=2.8cm]{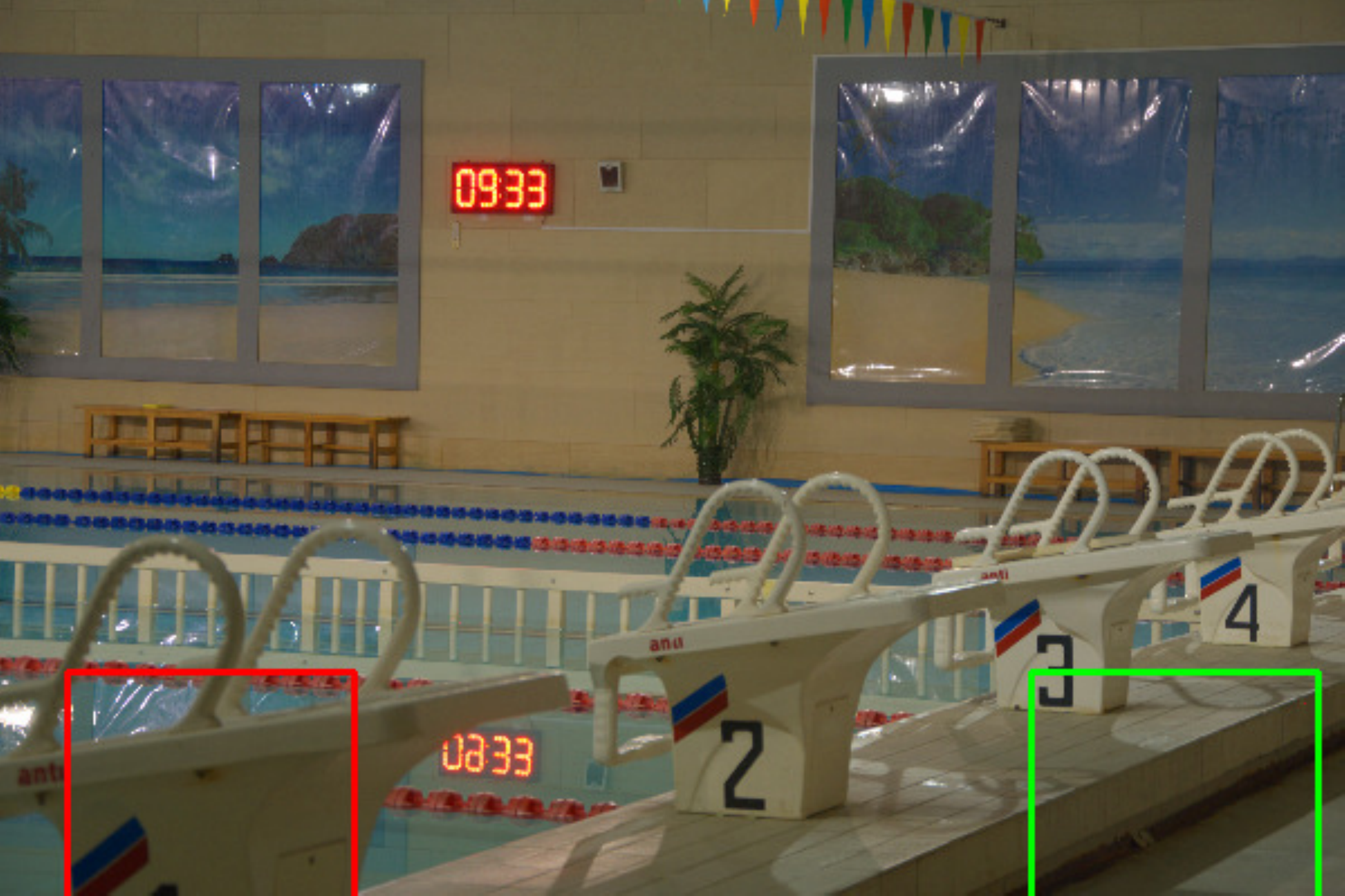}\vspace{1pt} \\
			\includegraphics[width=1.35cm]{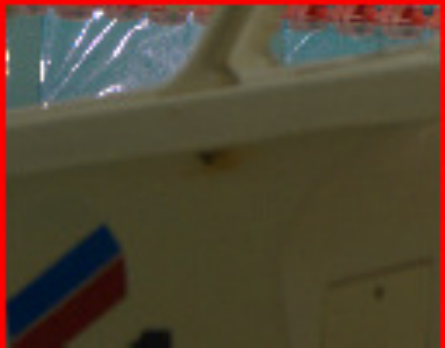}
			\includegraphics[width=1.35cm]{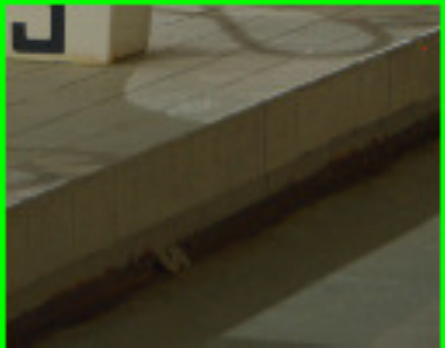}\vspace{5pt}
		\end{minipage}
	}\hspace{-5pt}
	\subfigure[EnlightenGan\cite{jiang2021enlightengan}]{
		\begin{minipage}[b]{0.155\textwidth}
			\includegraphics[width=2.8cm]{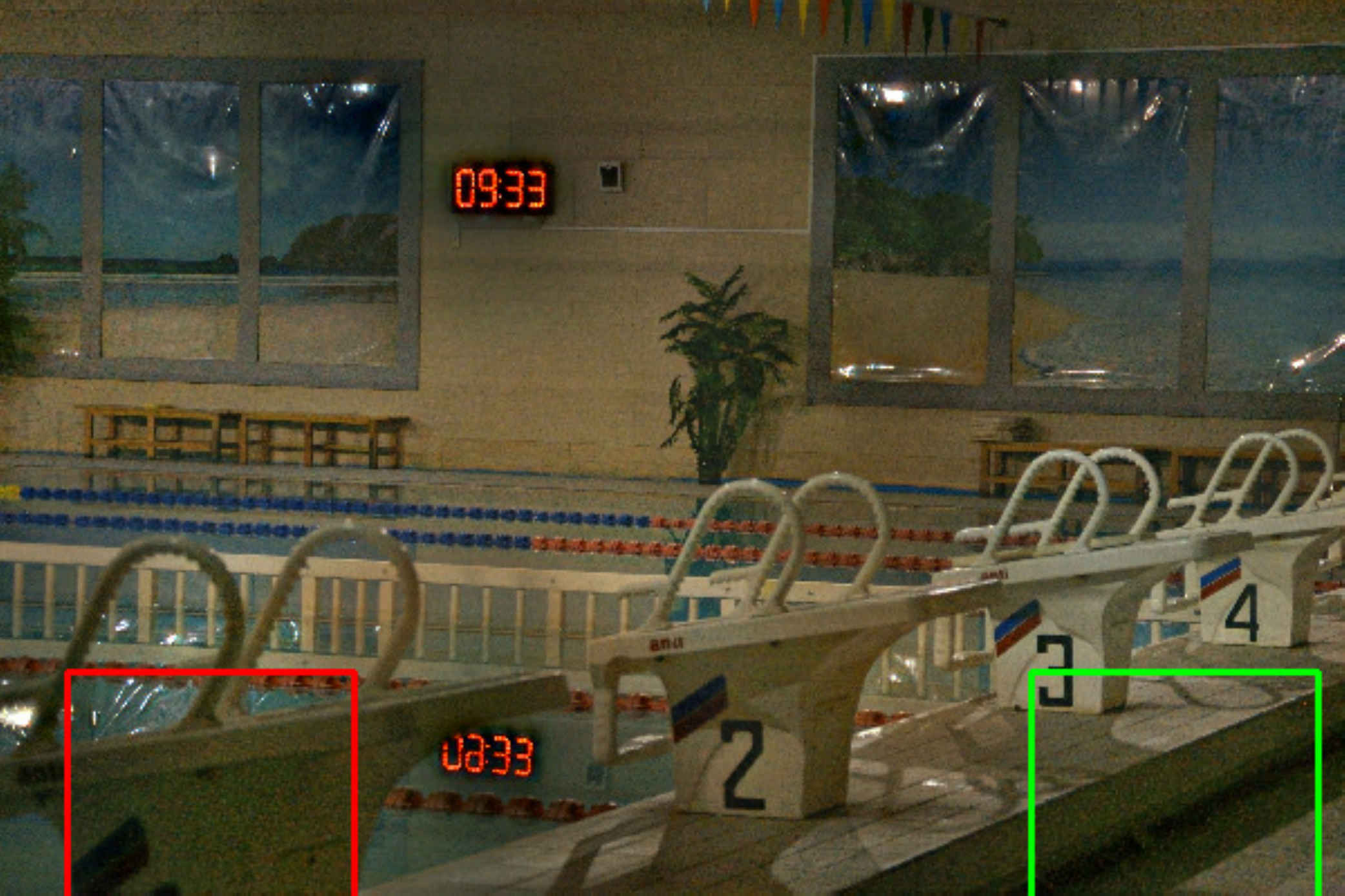}\vspace{1pt} \\
			\includegraphics[width=1.35cm]{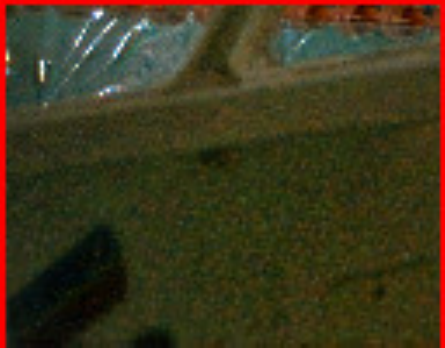}
			\includegraphics[width=1.35cm]{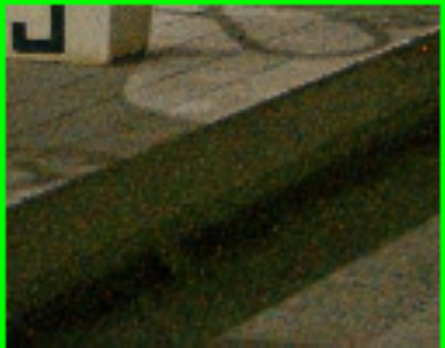}\vspace{5pt}
		\end{minipage}
	}\hspace{-5pt}
	\subfigure[RetinexNet\cite{wei2018deep}]{
		\begin{minipage}[b]{0.155\textwidth}
			\includegraphics[width=2.8cm]{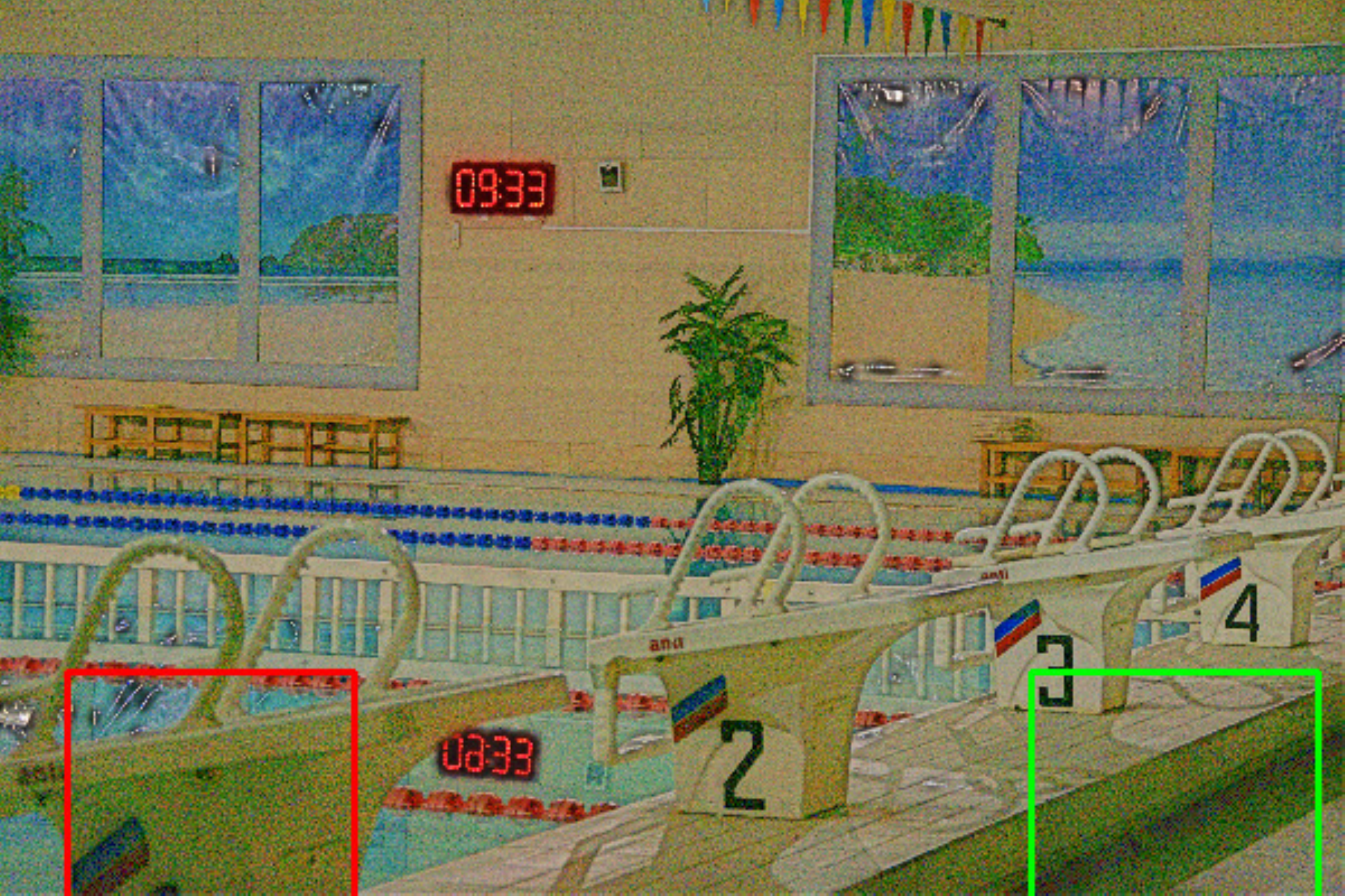}\vspace{1pt} \\
			\includegraphics[width=1.35cm]{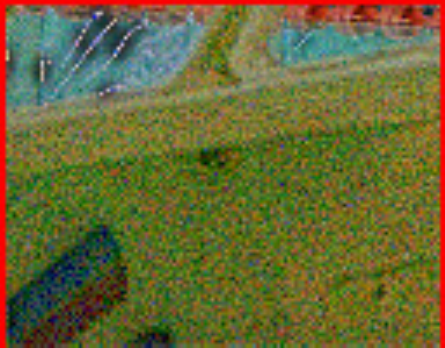}
			\includegraphics[width=1.35cm]{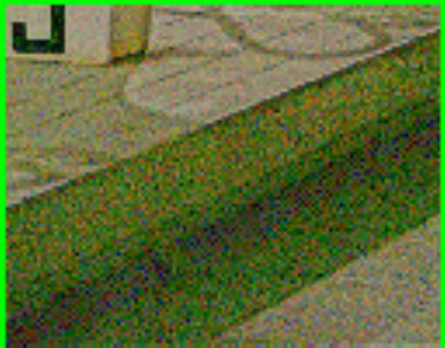}\vspace{5pt}
		\end{minipage}
	}\hspace{-5pt}
	\subfigure[KinD\cite{zhang2019kindling}]{
		\begin{minipage}[b]{0.155\textwidth}
			\includegraphics[width=2.8cm]{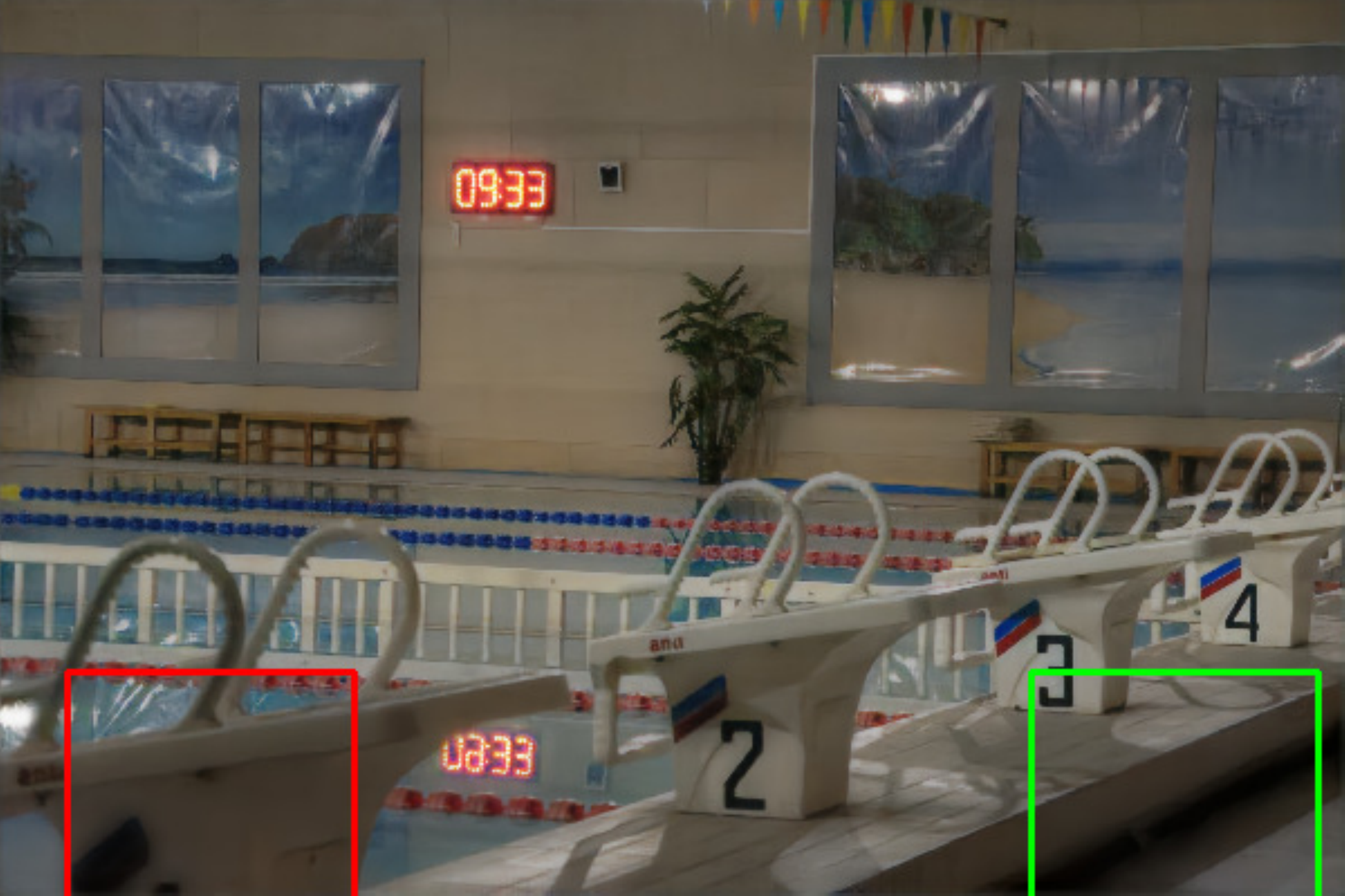}\vspace{1pt} \\
			\includegraphics[width=1.35cm]{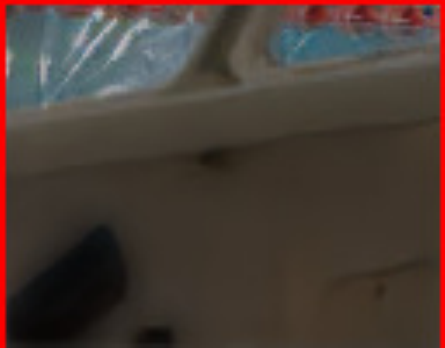}
			\includegraphics[width=1.35cm]{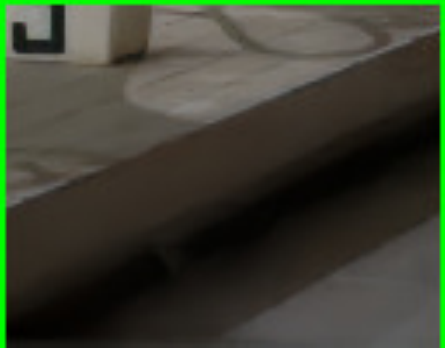}\vspace{5pt}
		\end{minipage}
	}\hspace{-5pt}
	\subfigure[DA-DRN w/o DA]{
		\begin{minipage}[b]{0.155\textwidth}
			\includegraphics[width=2.8cm]{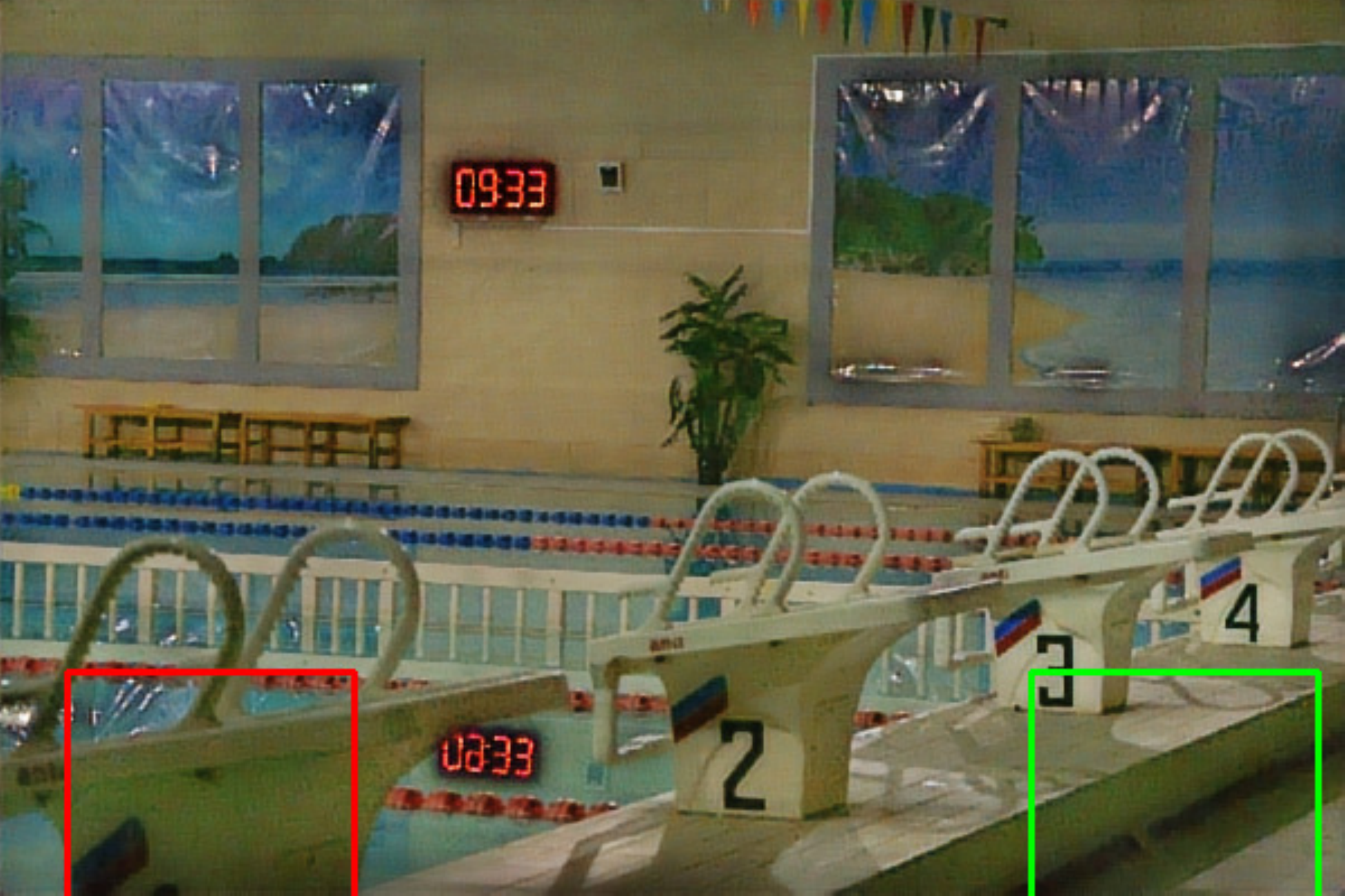}\vspace{1pt} \\
			\includegraphics[width=1.35cm]{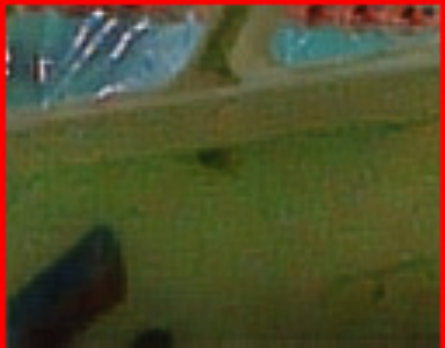}
			\includegraphics[width=1.35cm]{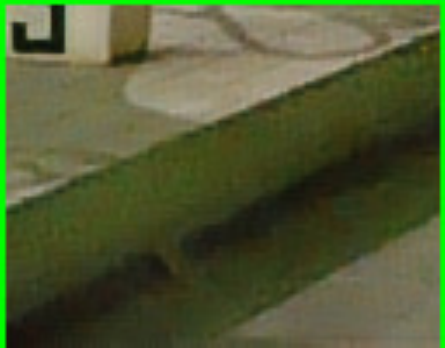}\vspace{5pt}
		\end{minipage}
	}\hspace{-5pt}
	\subfigure[DA-DRN with DA]{
		\begin{minipage}[b]{0.155\textwidth}
			\includegraphics[width=2.8cm]{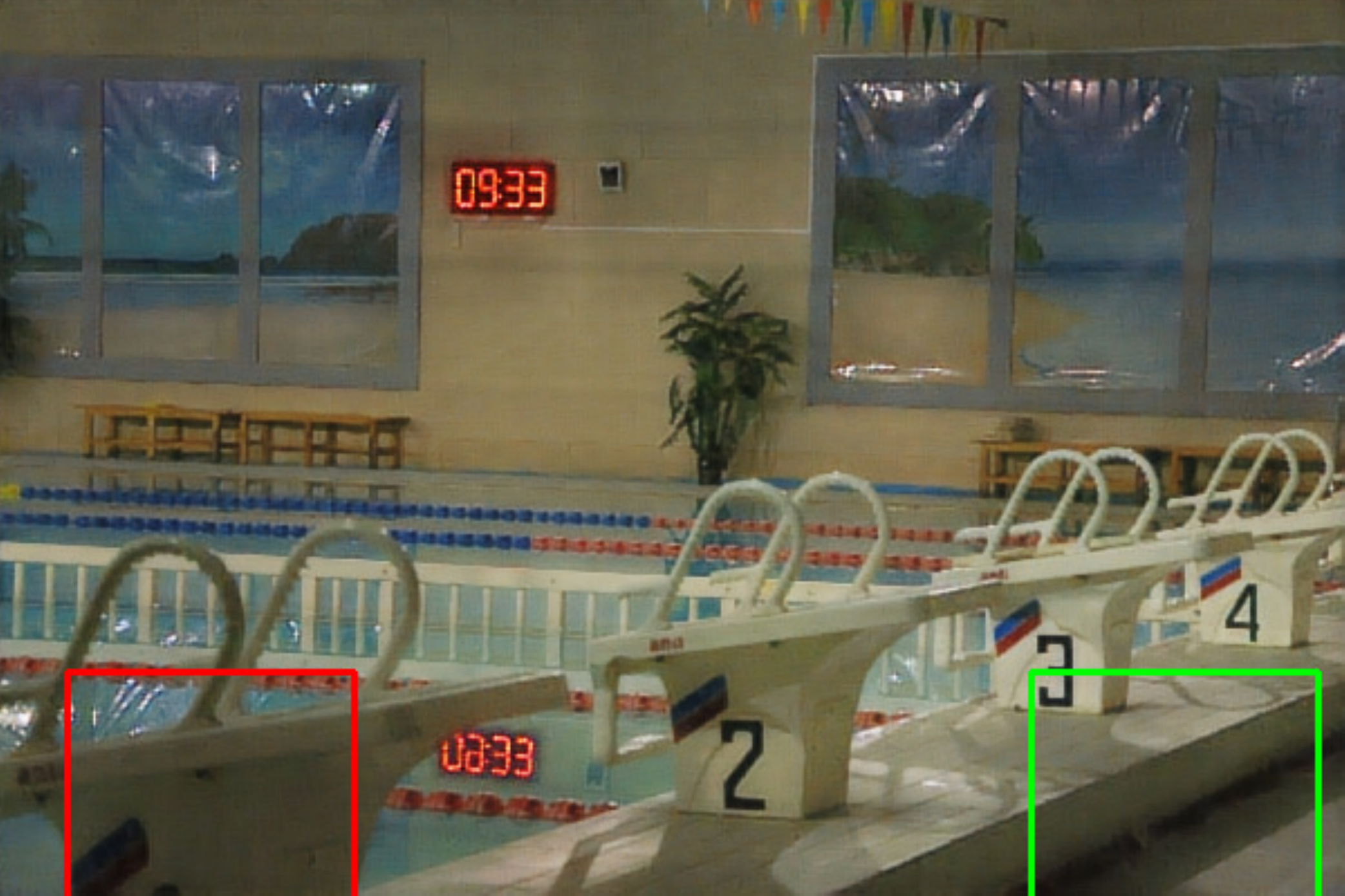}\vspace{1pt} \\
			\includegraphics[width=1.35cm]{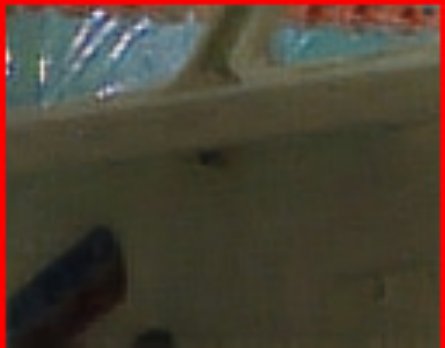}
			\includegraphics[width=1.35cm]{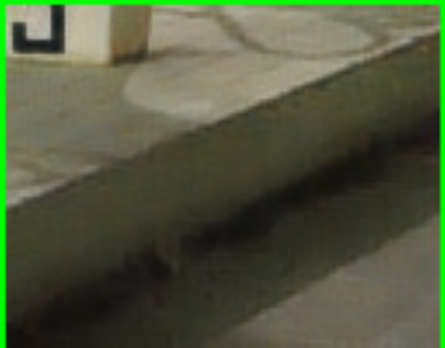}\vspace{5pt}
		\end{minipage}
	}
	\caption{Visual comparison of the final results generated by different methods on LOL real-world dataset.}
	\label{748}
\end{figure*}

\begin{figure*}
	\subfigure{
		\begin{minipage}[b]{0.9\linewidth}
			\includegraphics[width=17.5cm]{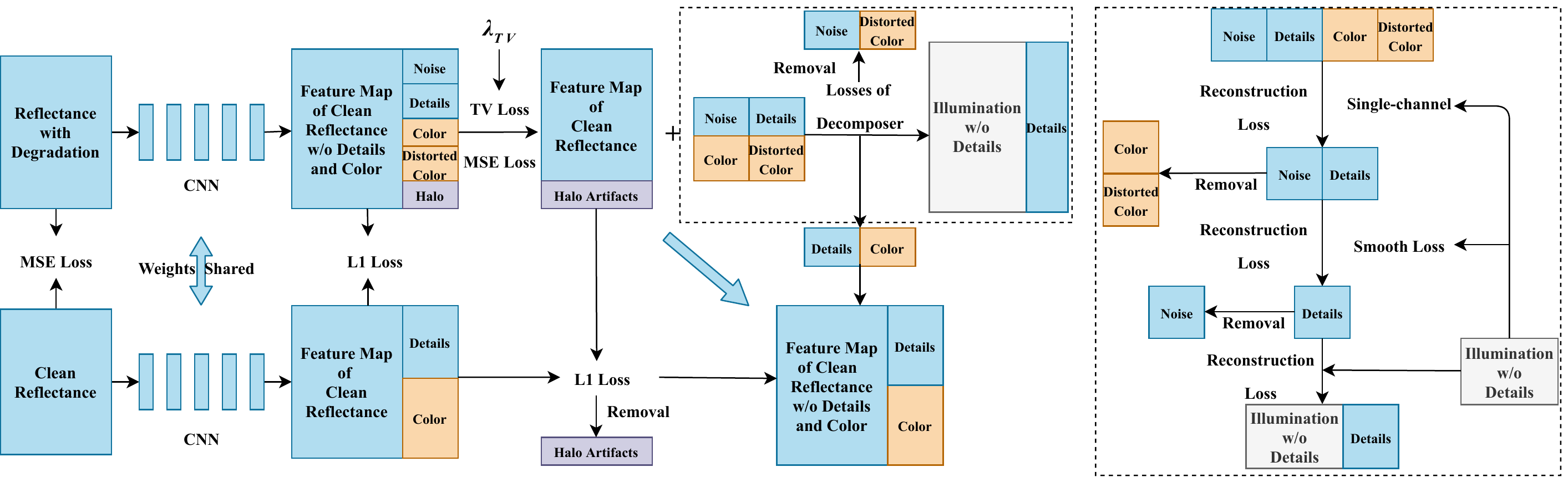}
	\end{minipage}}
	\vspace{-3pt}
	\caption{The schematic of DA Module. The dotted box shows the process of reconstructing the details information into the illumination map. More details of the reconstruction process can be seen in the right dotted box in the figure. Clean reflectance means the reflectance without noise, color distortion and artifacts.}
	\label{da}
\end{figure*}


\begin{table}[htbp] 
	\centering \caption{Ablation study of \textbf{DA Loss} and quantitative comparison of \textbf{Reflectance (R)} decomposed by our method
		and other state-of-the-art methods on \textbf{LOL REAL-WORLD evaluation dataset}.}
	\begin{tabular}{ c | c  c  c}
		\hline Methods &PSNR↑  & SSIM↑ & DeltaE↓ \\
		\hline
		\hline  
		Baseline & 21.0210  & 0.7763   &11.6729 \\
		Baseline w/o DA & 16.2574  & 0.7140   &18.6383 \\
		\hline
		\hline
		DA w/o TV and MSE &19.5902  & 0.7358  &12.1806\\
		DA w/o MSE Loss &18.5617  & 0.7350  &14.1855\\
		DA w/o TV Loss &17.7151  & 0.7045  &14.1923\\
		DA w/o L1 Loss & 20.1673 & 0.7562  &12.5627\\
		\hline
		\hline
		RetinexNet\cite{wei2018deep} & 18.4610 & 0.5223   &13.5327 \\
		KinD\cite{zhang2019kindling} & 13.3871 & 0.5144   &25.0592 \\
		KinD\cite{zhang2019kindling} + Restorer & 19.3947 & 0.8049   &10.6695 \\
		\hline\end{tabular}\vspace{0cm}
	\label{loss_abla}
\end{table}

\begin{figure*}
	\centering
	
	
	\subfigure[DA-DRN w/o DA]{
		\begin{minipage}[b]{0.155\textwidth}
			\includegraphics[width=2.8cm]{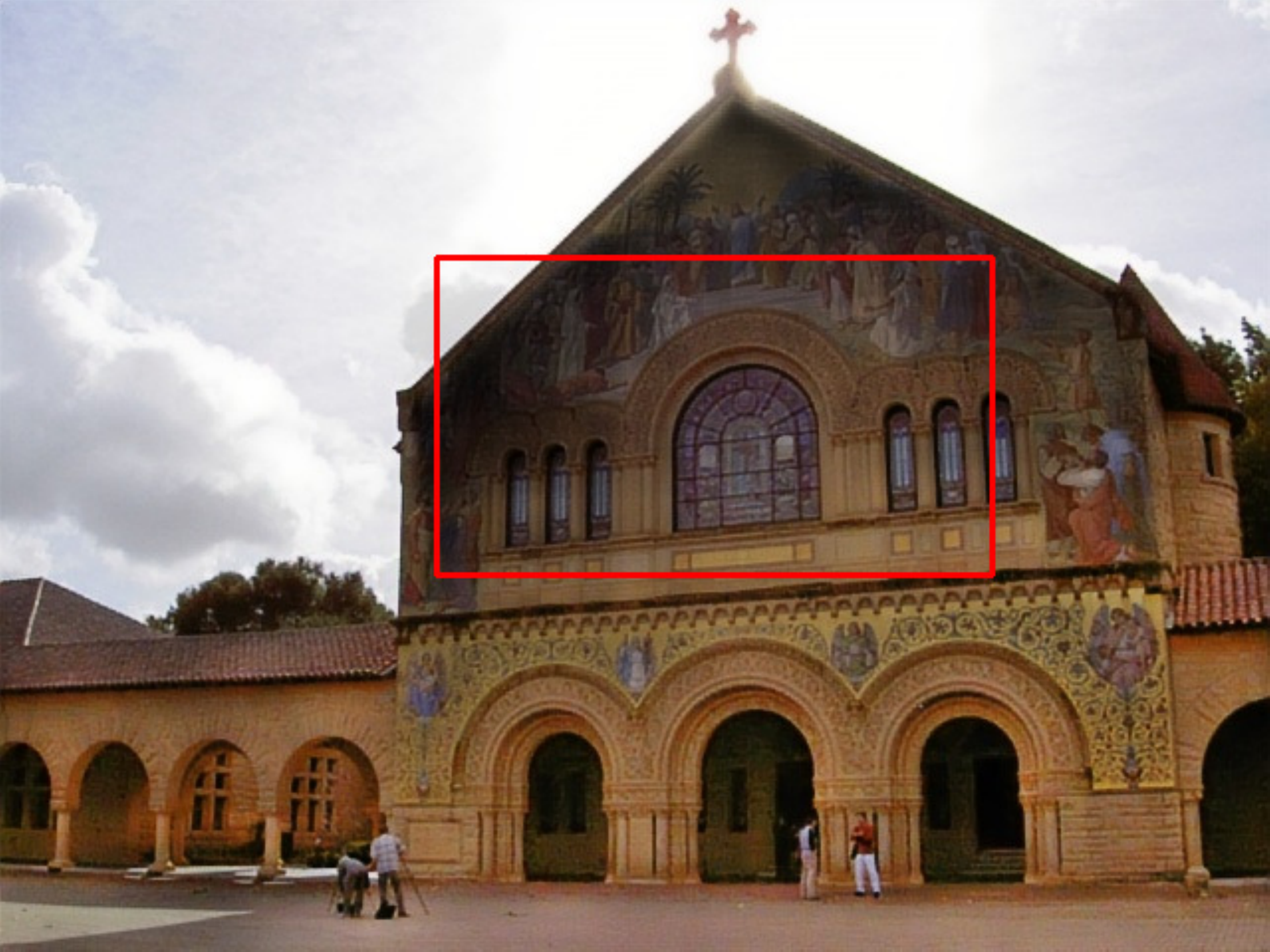}\vspace{2pt}
			\includegraphics[width=2.8cm]{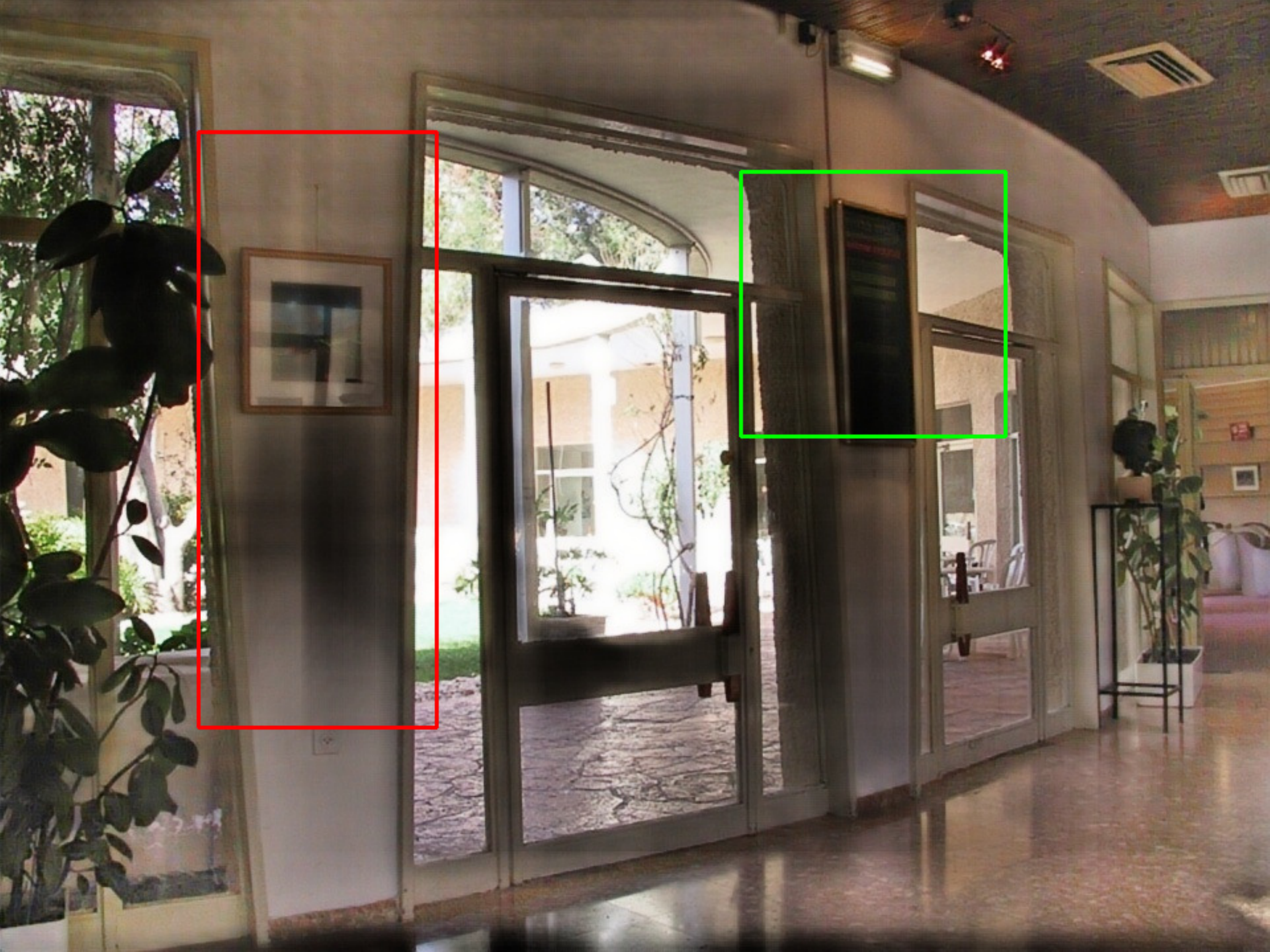}\vspace{2pt} \\
			\includegraphics[width=2.8cm]{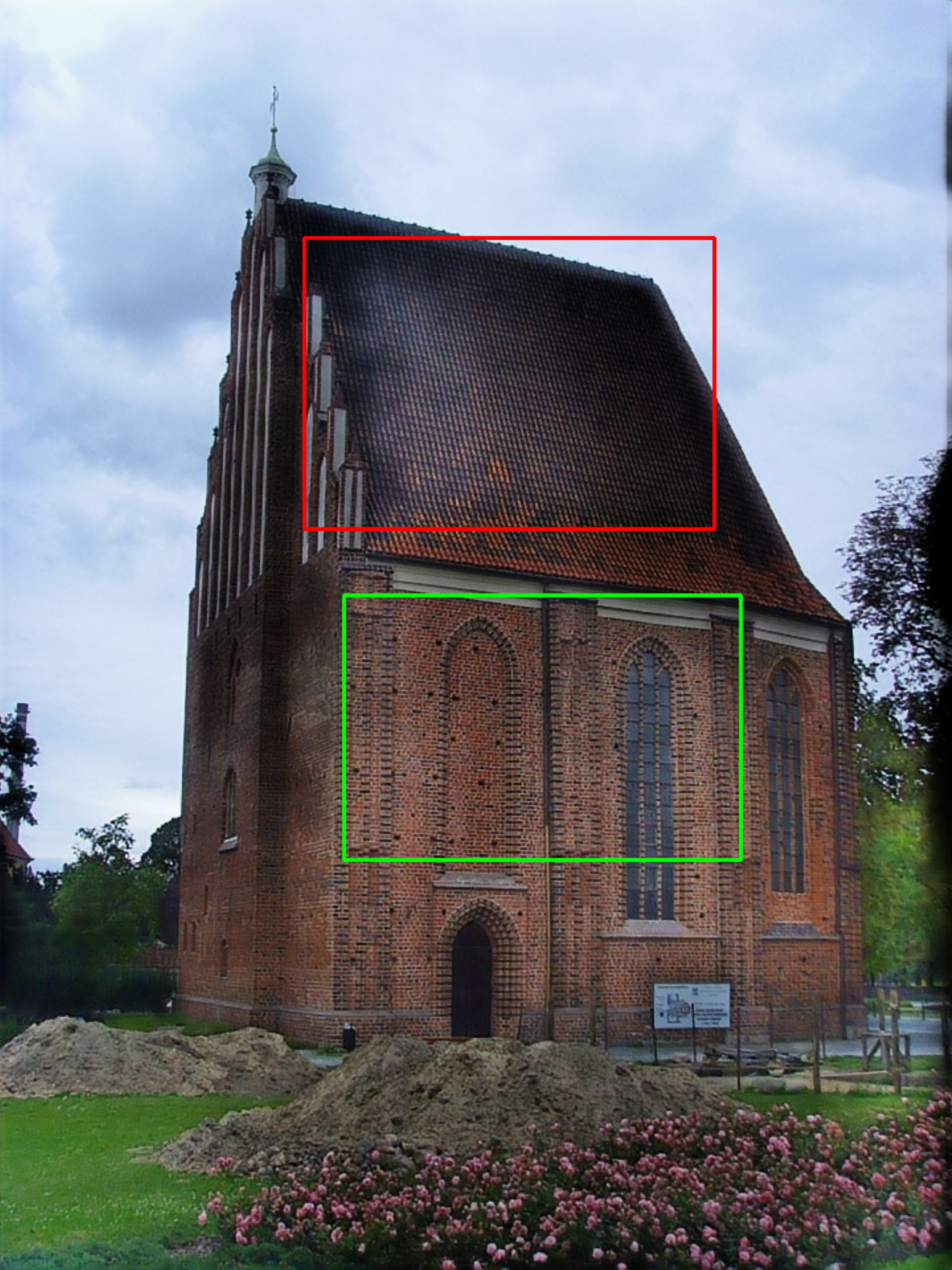}
		\end{minipage}
	}\hspace{-5pt}
	\subfigure[DA w/o TV and MSE]{
		\begin{minipage}[b]{0.155\textwidth}
			\includegraphics[width=2.8cm]{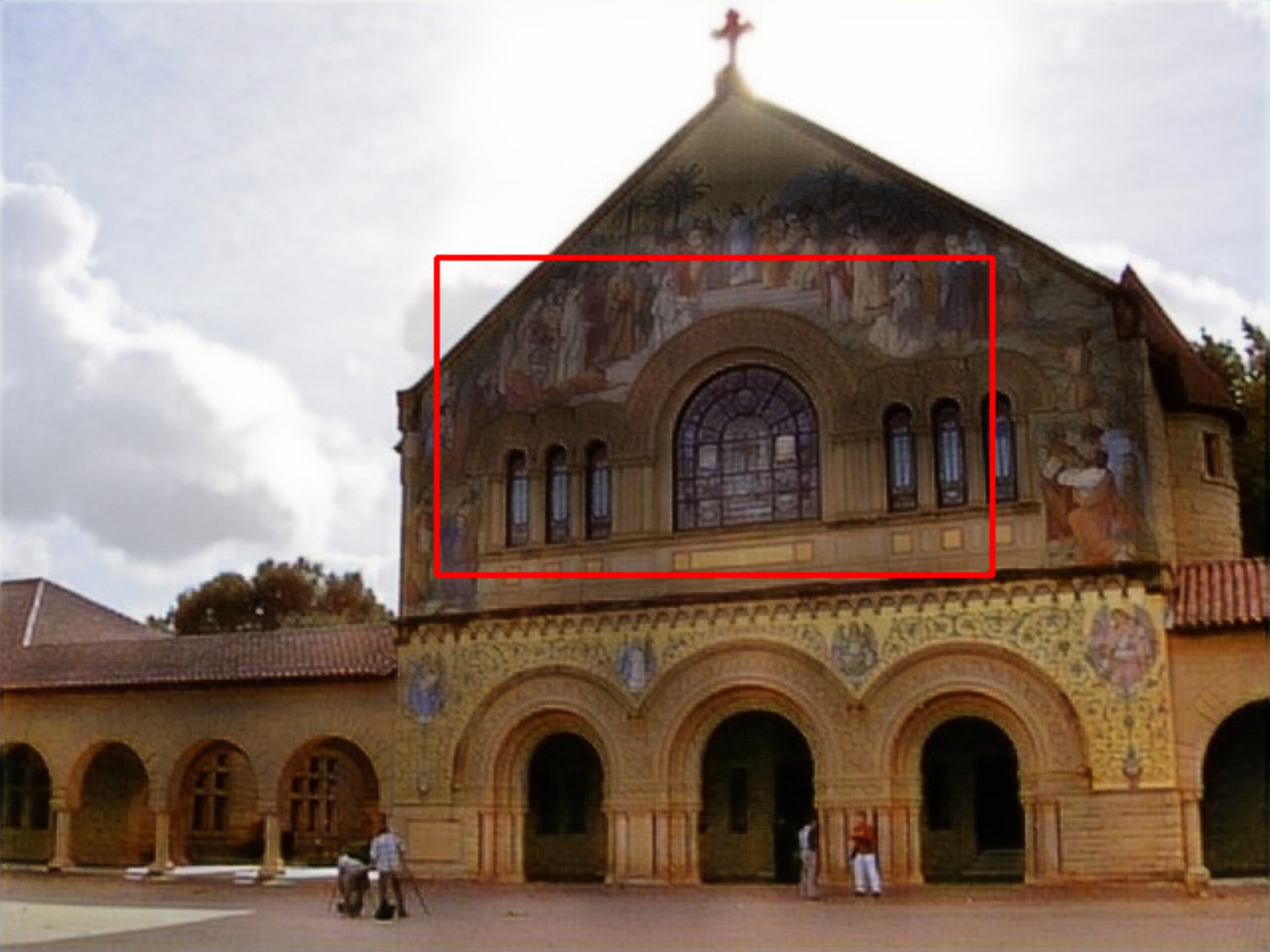}\vspace{2pt}
			\includegraphics[width=2.8cm]{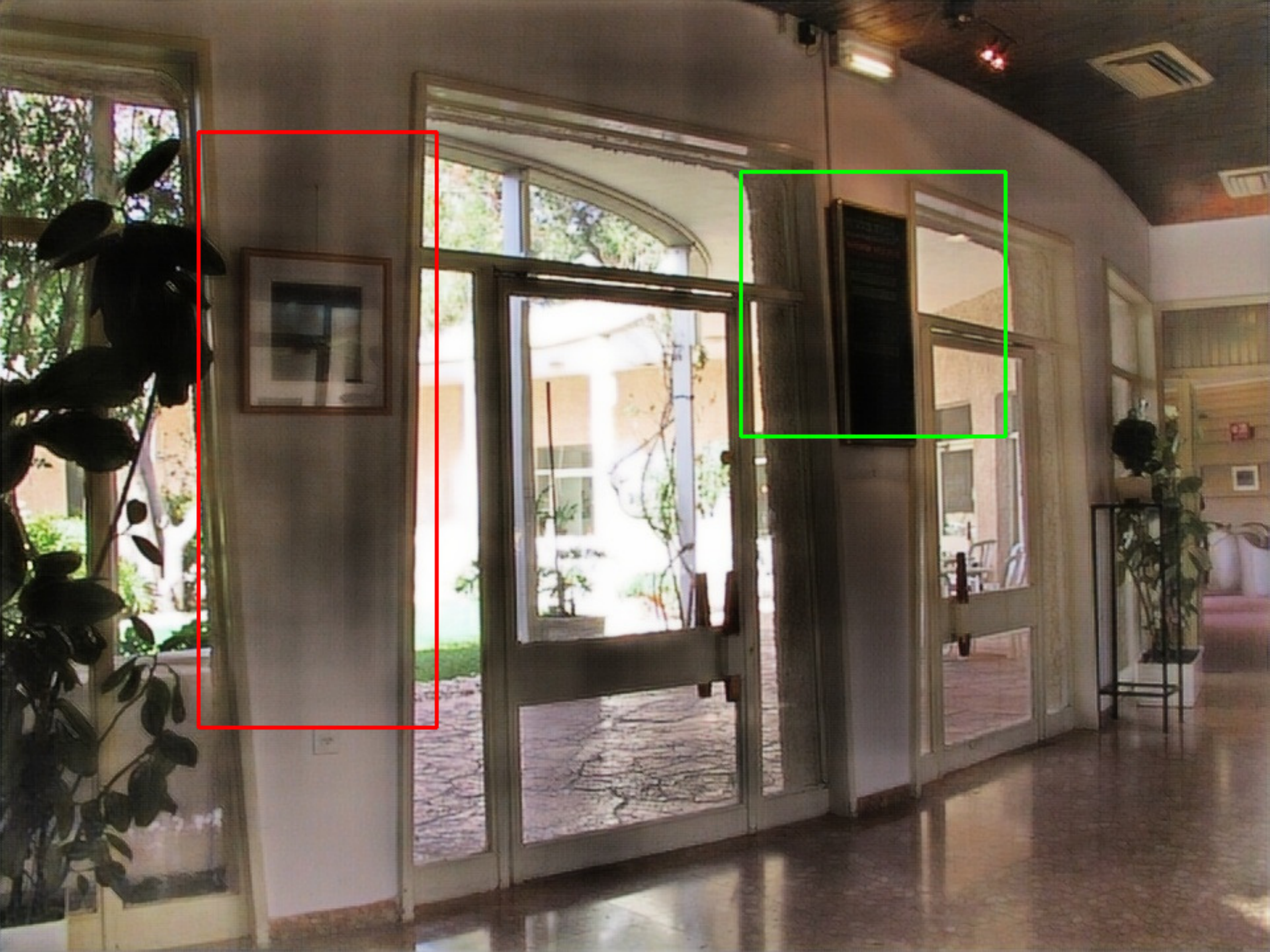}\vspace{2pt} \\
			\includegraphics[width=2.8cm]{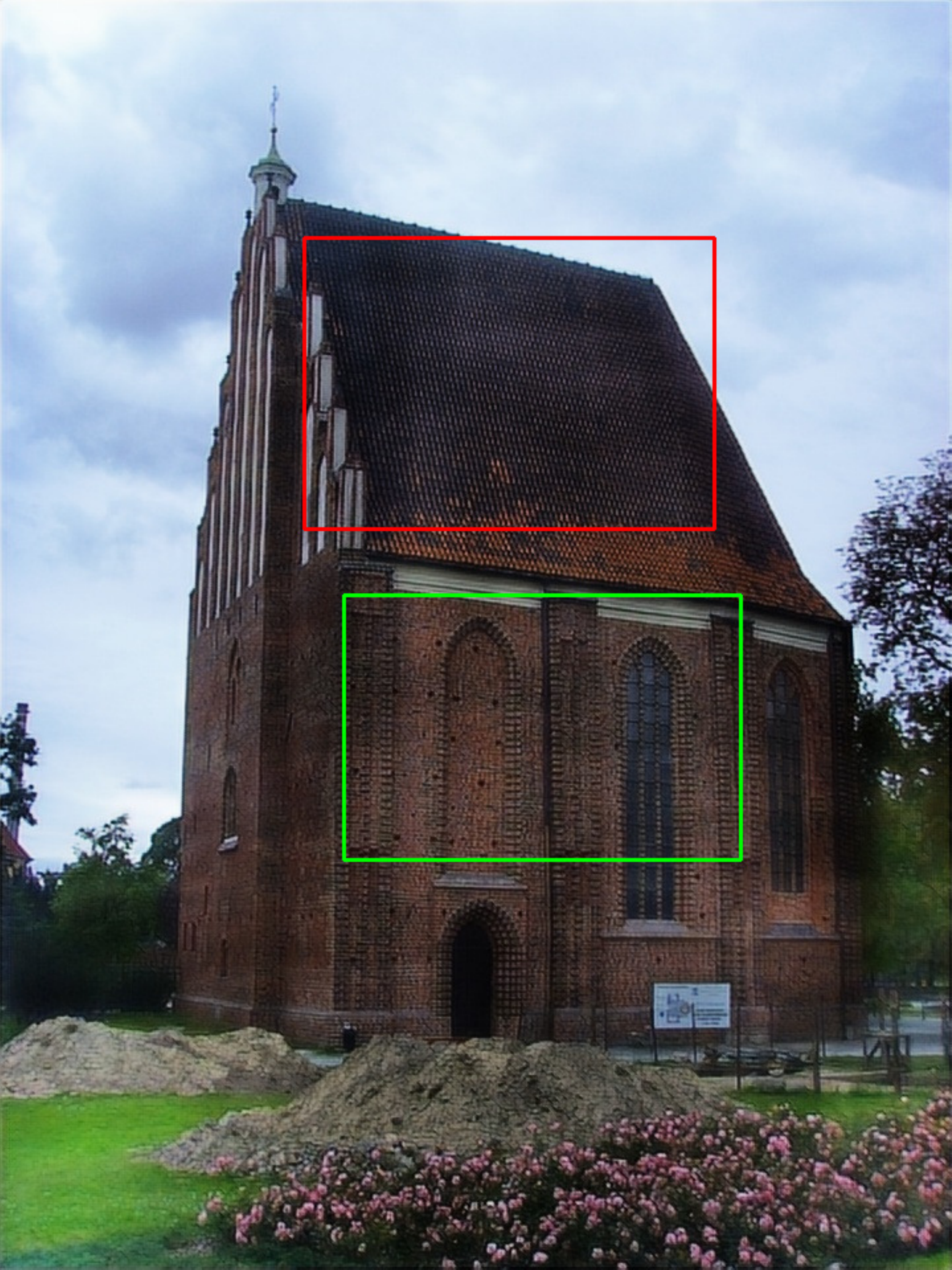}
		\end{minipage}
	}\hspace{-5pt}
	\subfigure[DA w/o TV Loss]{
		\begin{minipage}[b]{0.155\textwidth}
			\includegraphics[width=2.8cm]{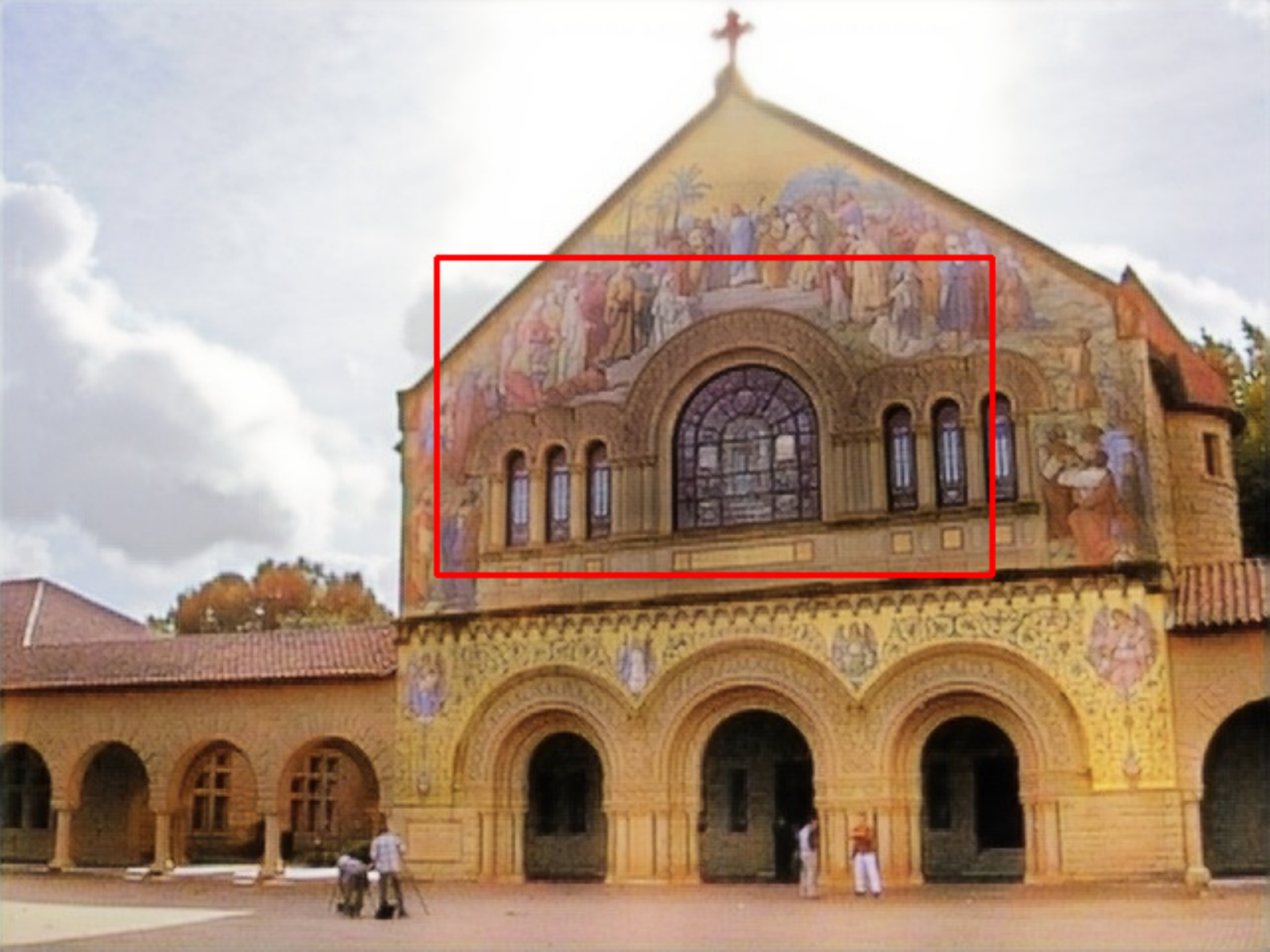}\vspace{2pt}
			\includegraphics[width=2.8cm]{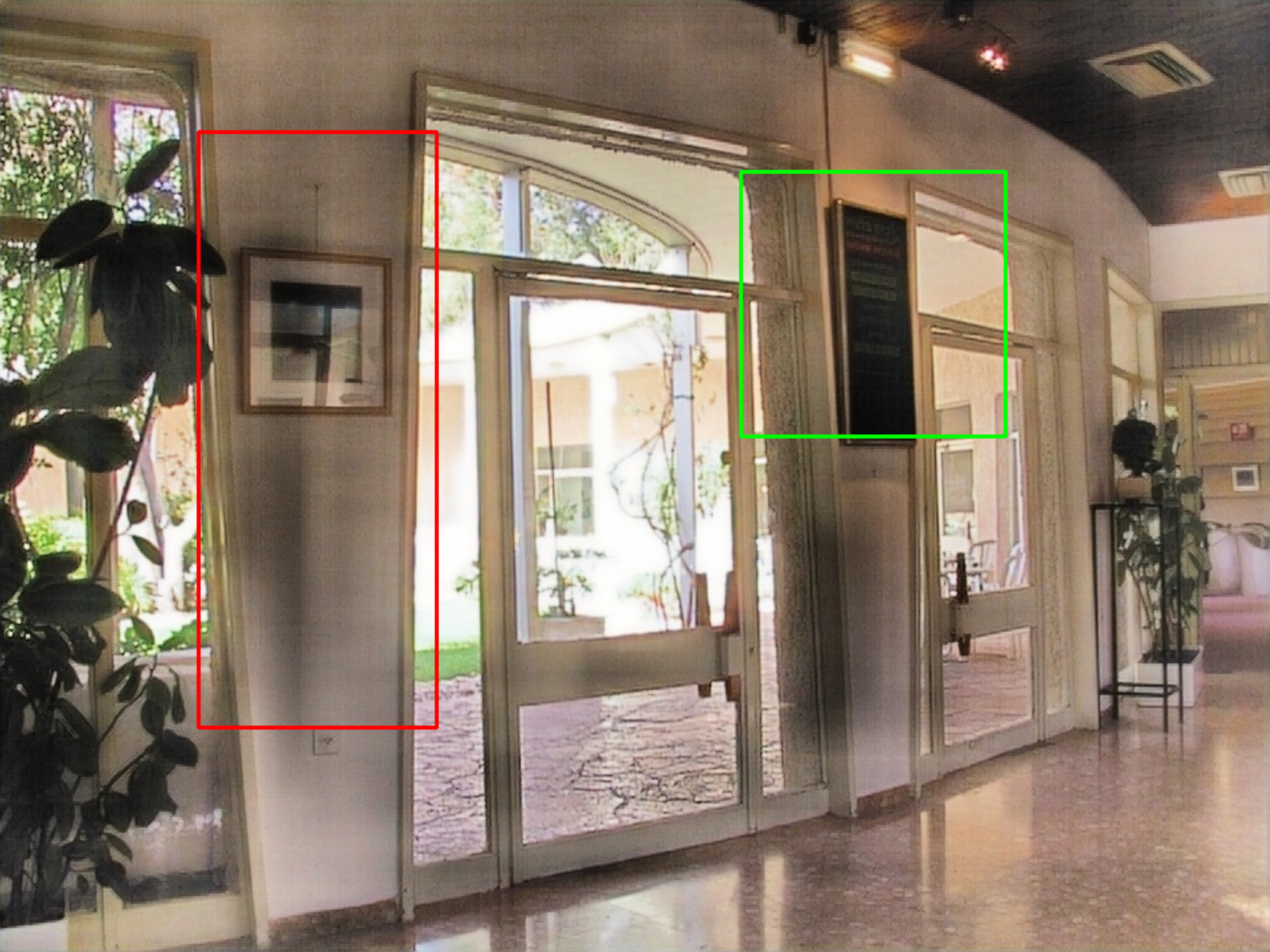}\vspace{2pt} \\
			\includegraphics[width=2.8cm]{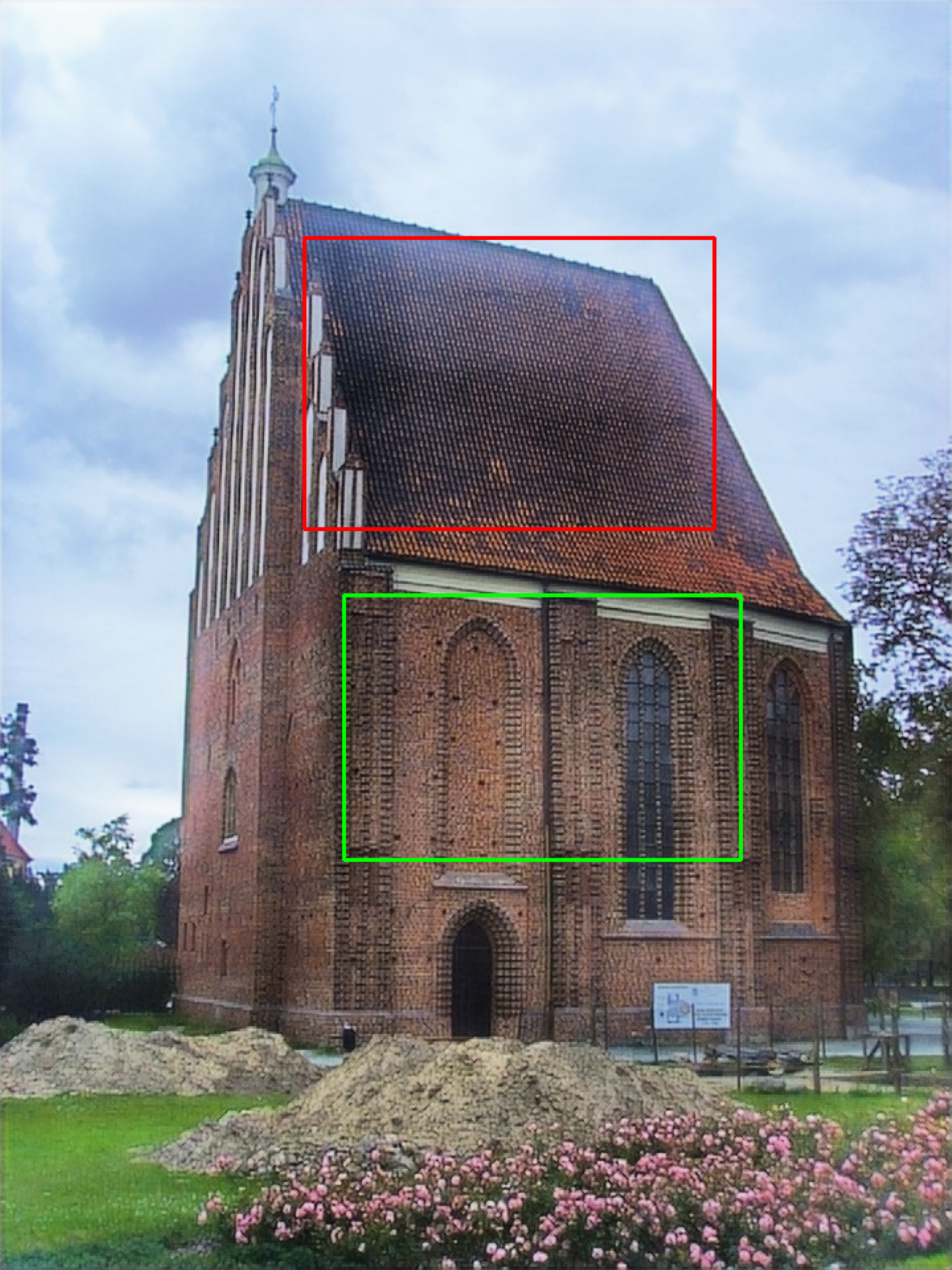}
		\end{minipage}
	}\hspace{-5pt}
	\subfigure[DA w/o MSE Loss]{
		\begin{minipage}[b]{0.155\textwidth}
			\includegraphics[width=2.8cm]{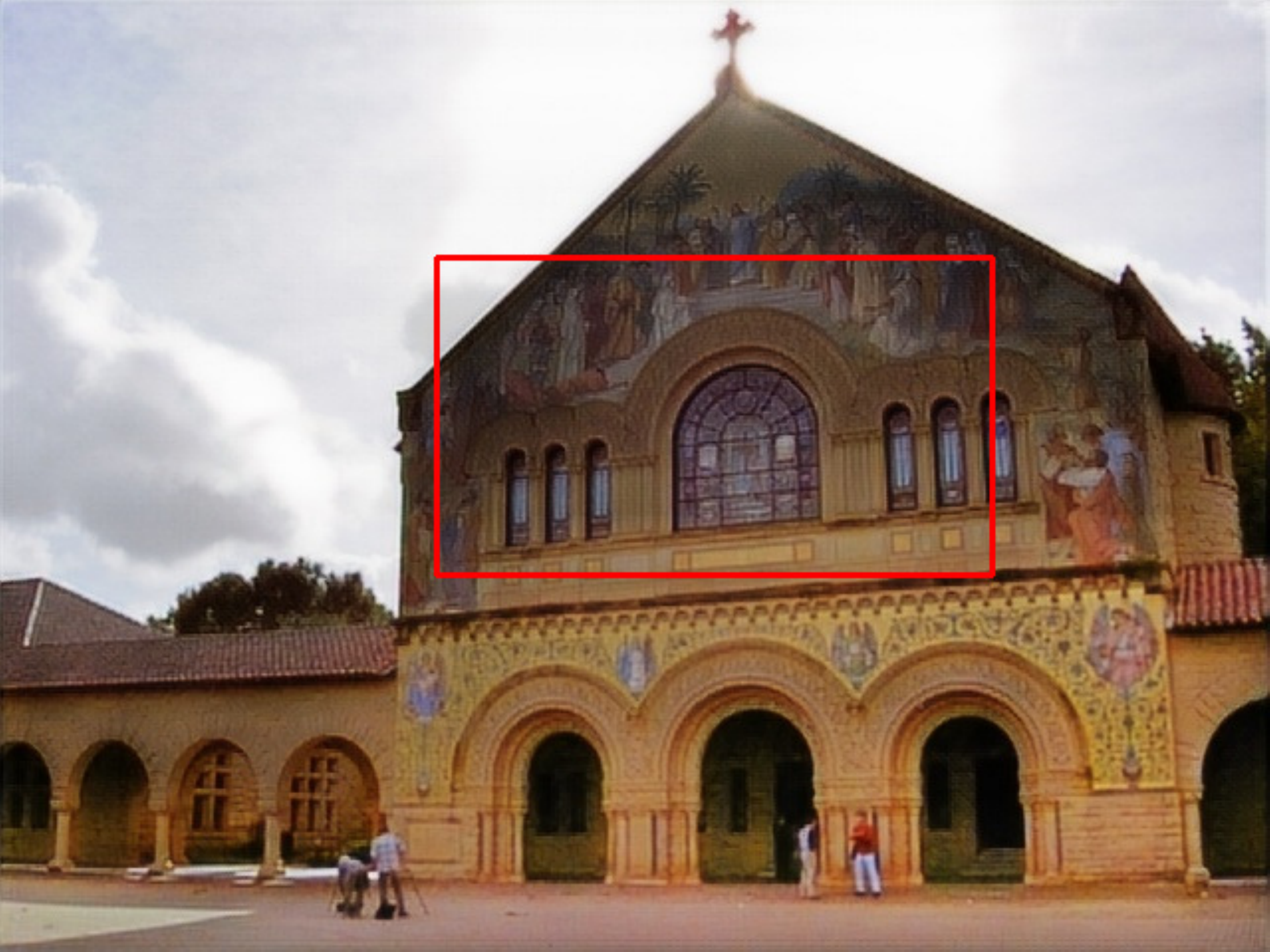}\vspace{2pt}
			\includegraphics[width=2.8cm]{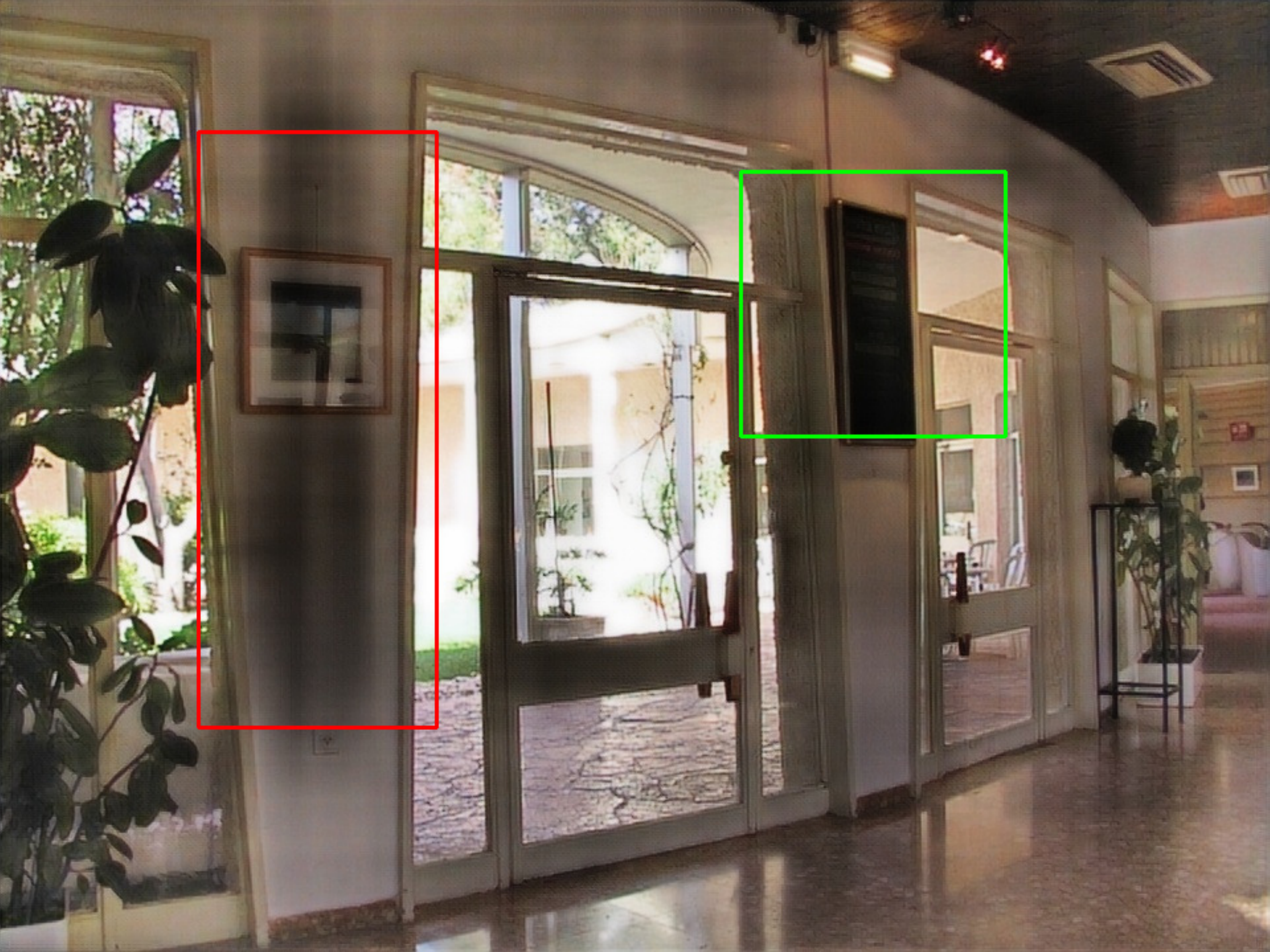}\vspace{2pt} \\
			\includegraphics[width=2.8cm]{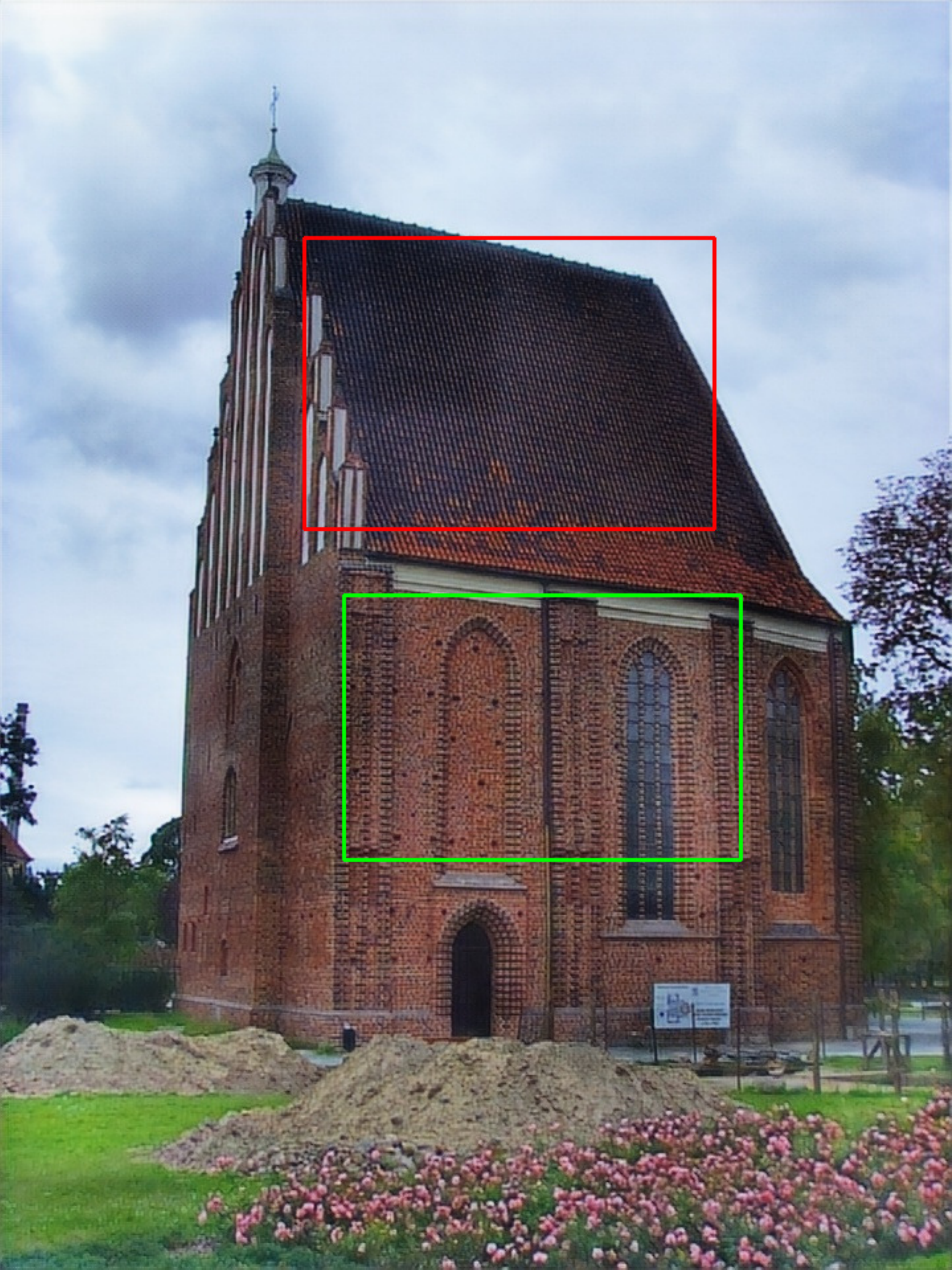}
		\end{minipage}
	}\hspace{-5pt}
	\subfigure[DA w/o L1 Loss]{
		\begin{minipage}[b]{0.155\textwidth}
			\includegraphics[width=2.8cm]{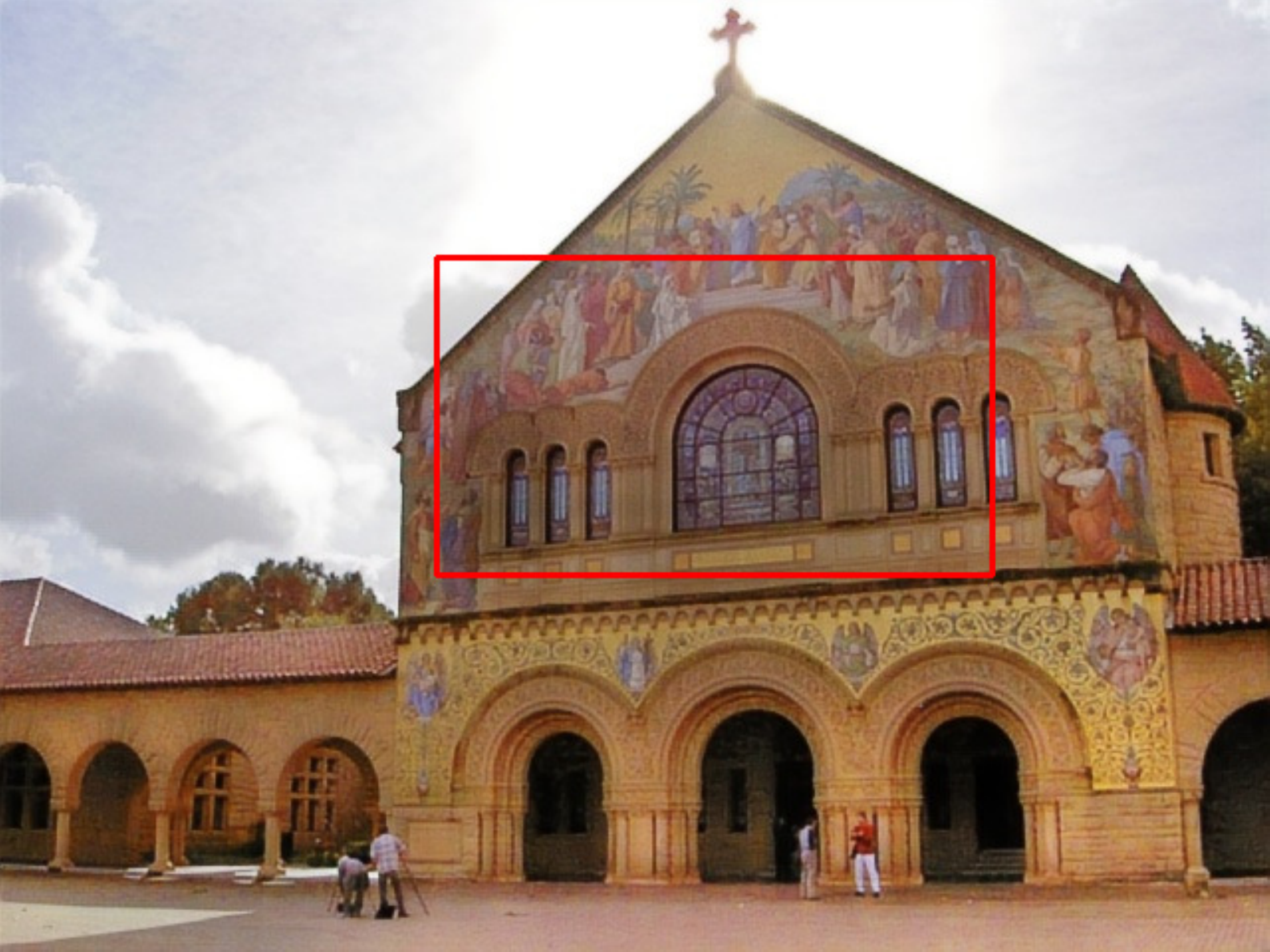}\vspace{2pt}
			\includegraphics[width=2.8cm]{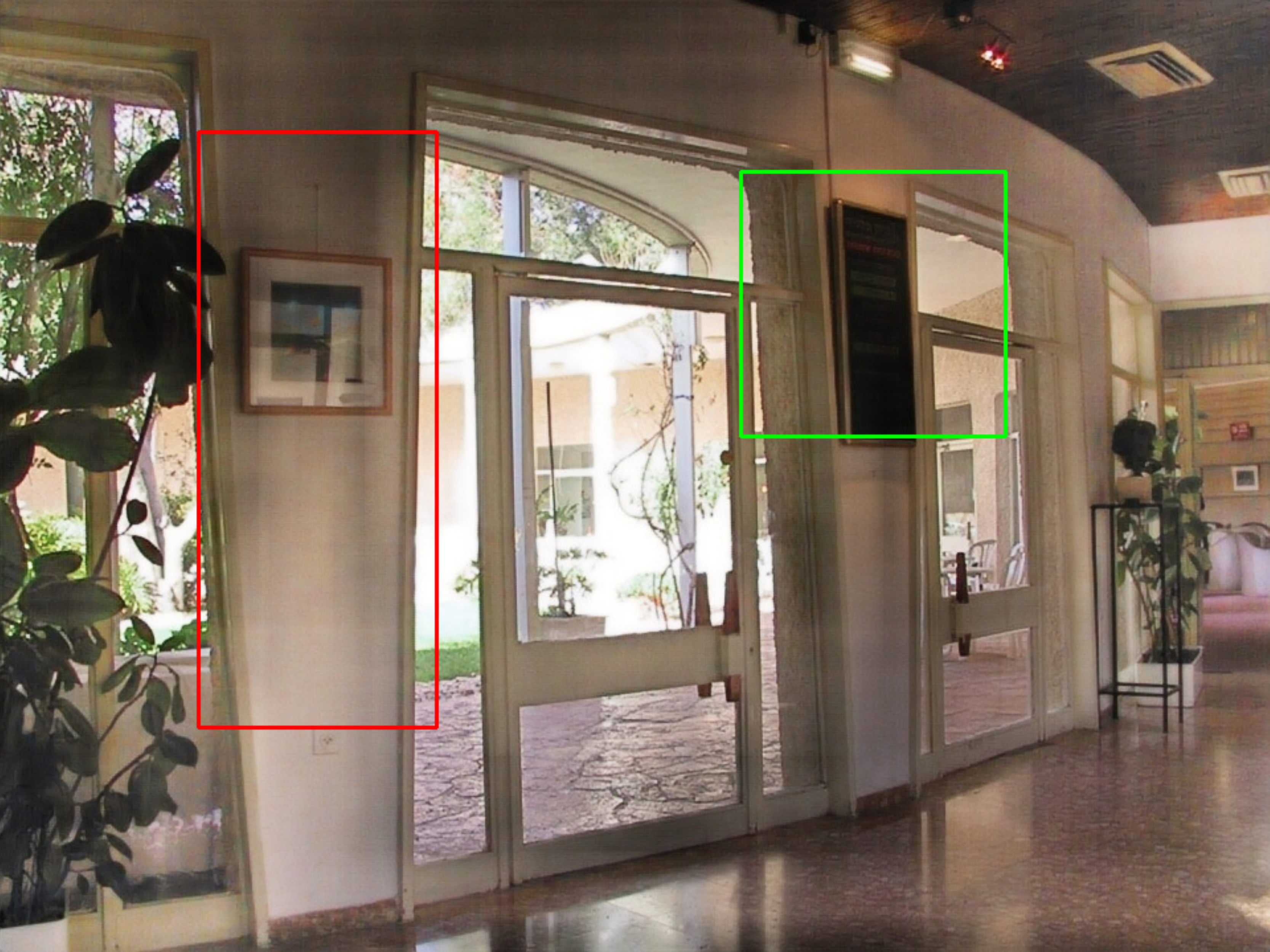}\vspace{2pt} \\
			\includegraphics[width=2.8cm]{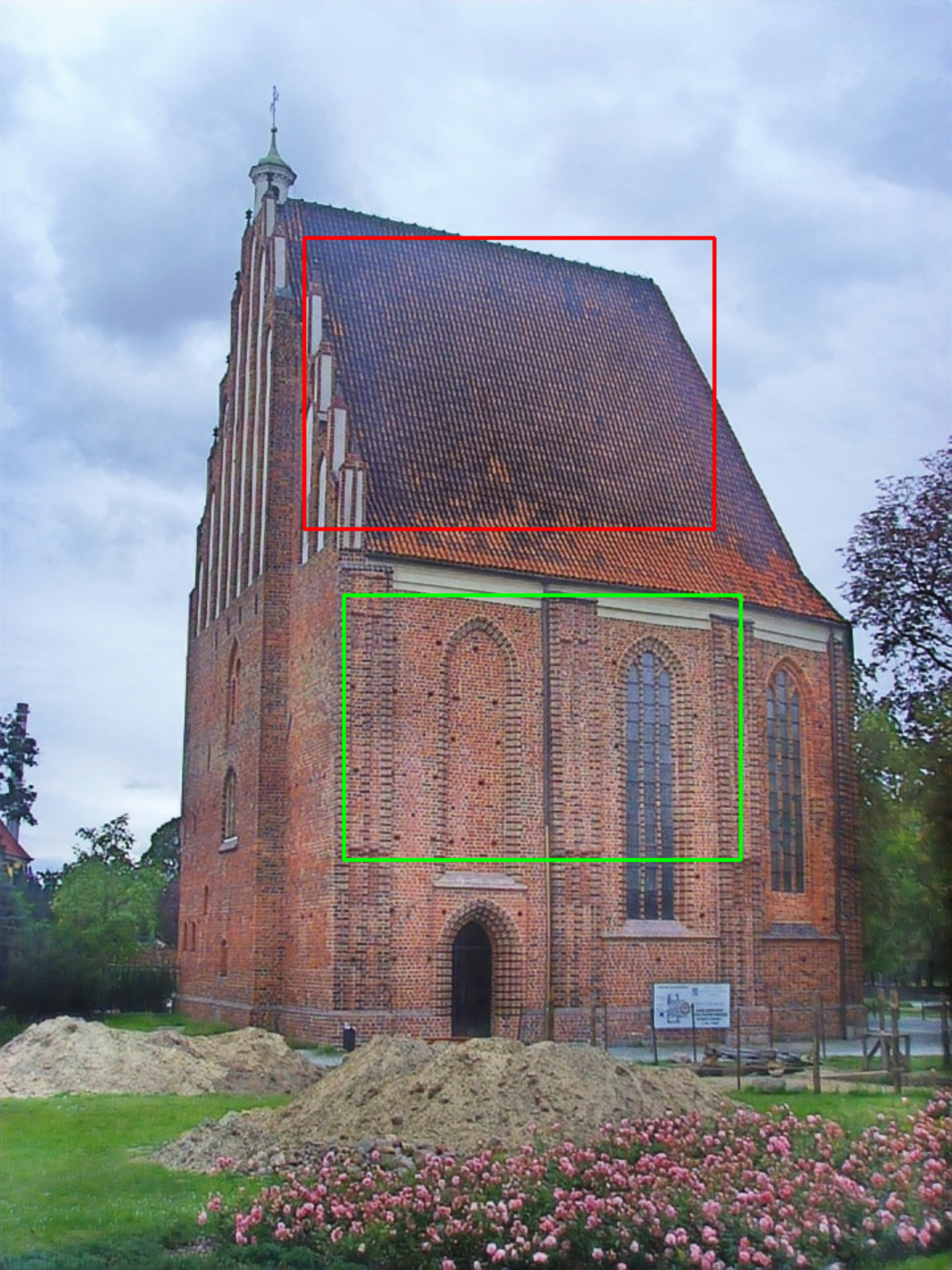}
		\end{minipage}
	}\hspace{-5pt}
	\subfigure[DA-DRN Baseline]{
		\begin{minipage}[b]{0.155\textwidth}
			\includegraphics[width=2.8cm]{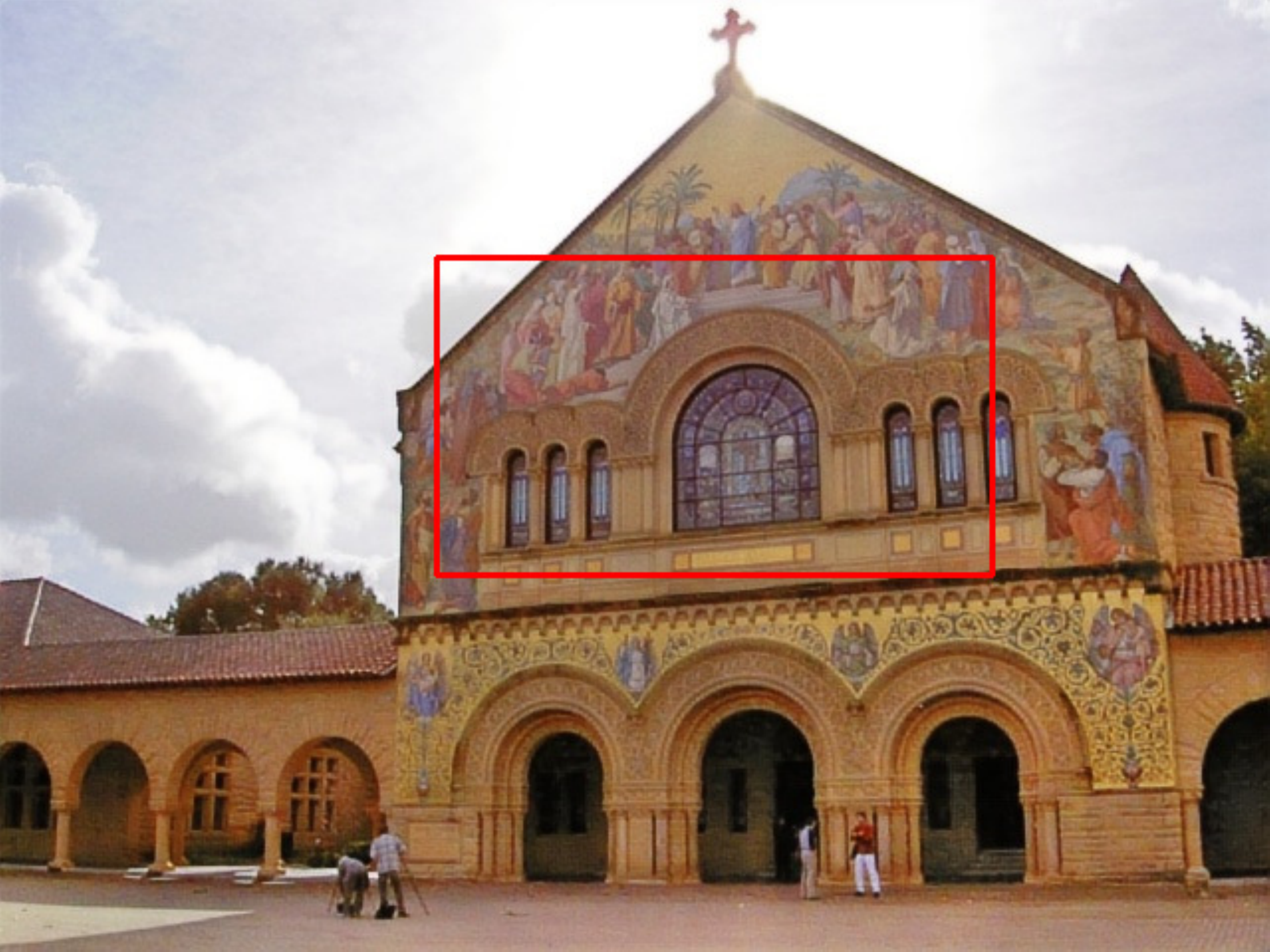}\vspace{2pt}
			\includegraphics[width=2.8cm]{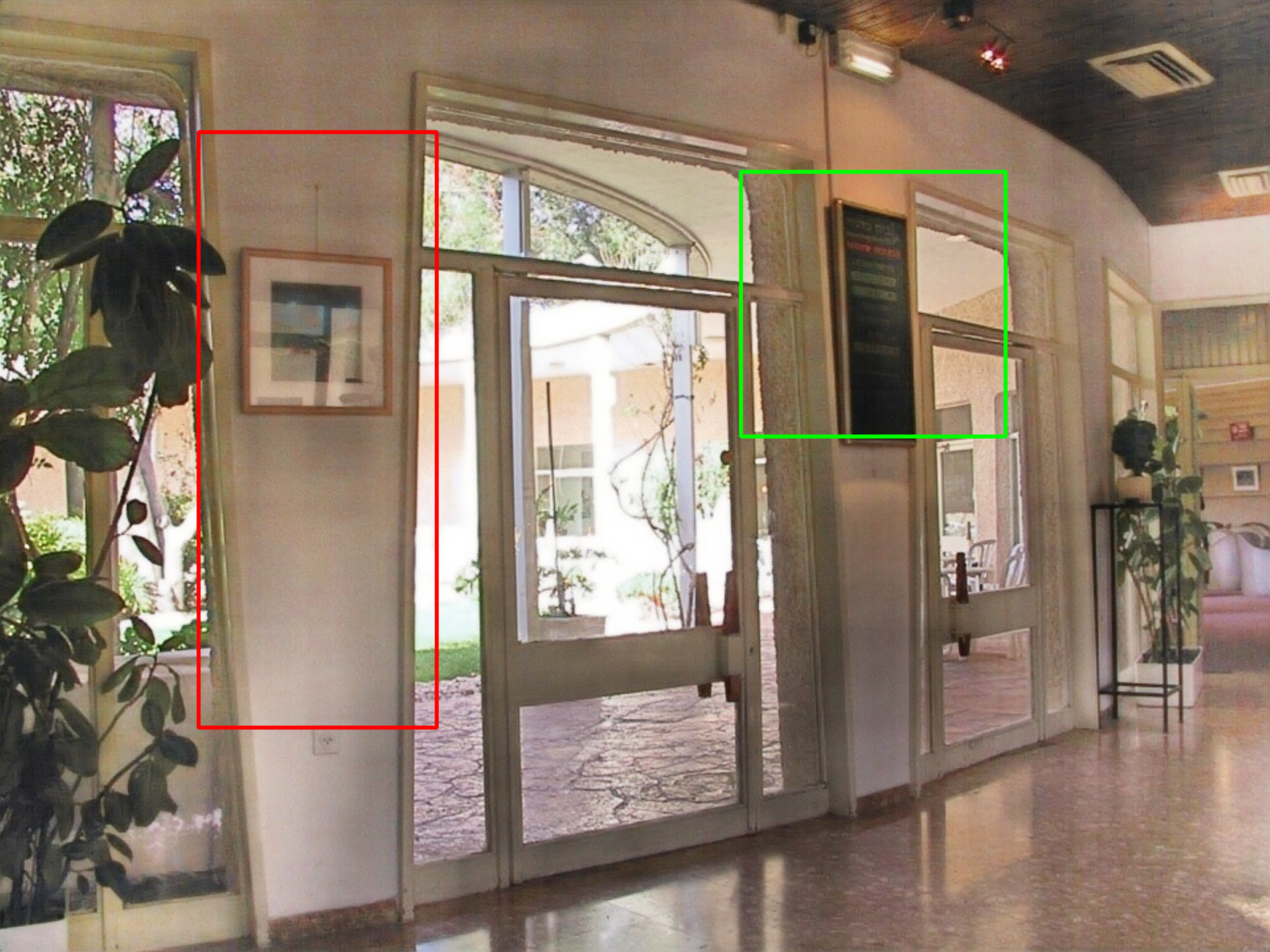}\vspace{2pt} \\
			\includegraphics[width=2.8cm]{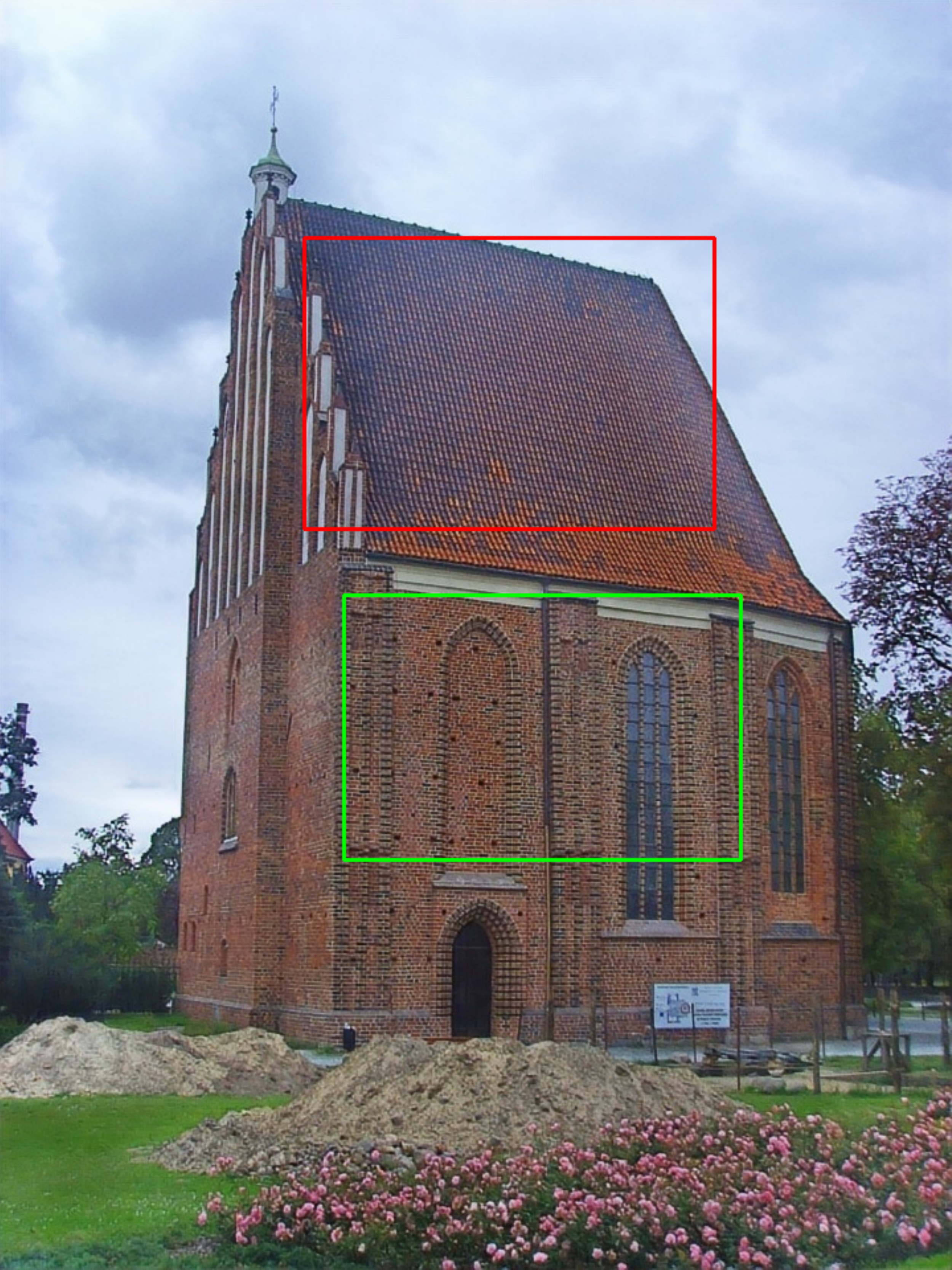}
		\end{minipage}
	}
	\caption{Visual comparison from the ablation study of Degradation-Aware (DA) Module and the loss functions in DA.}
	\label{da_abla_dicm}
\end{figure*}

\subsection{Degradation-Aware Module}
\begin{figure*}
	\centering
	
	
	\subfigure[KinD\cite{zhang2019kindling}]{
		\begin{minipage}[b]{0.155\textwidth}
			\includegraphics[width=2.8cm]{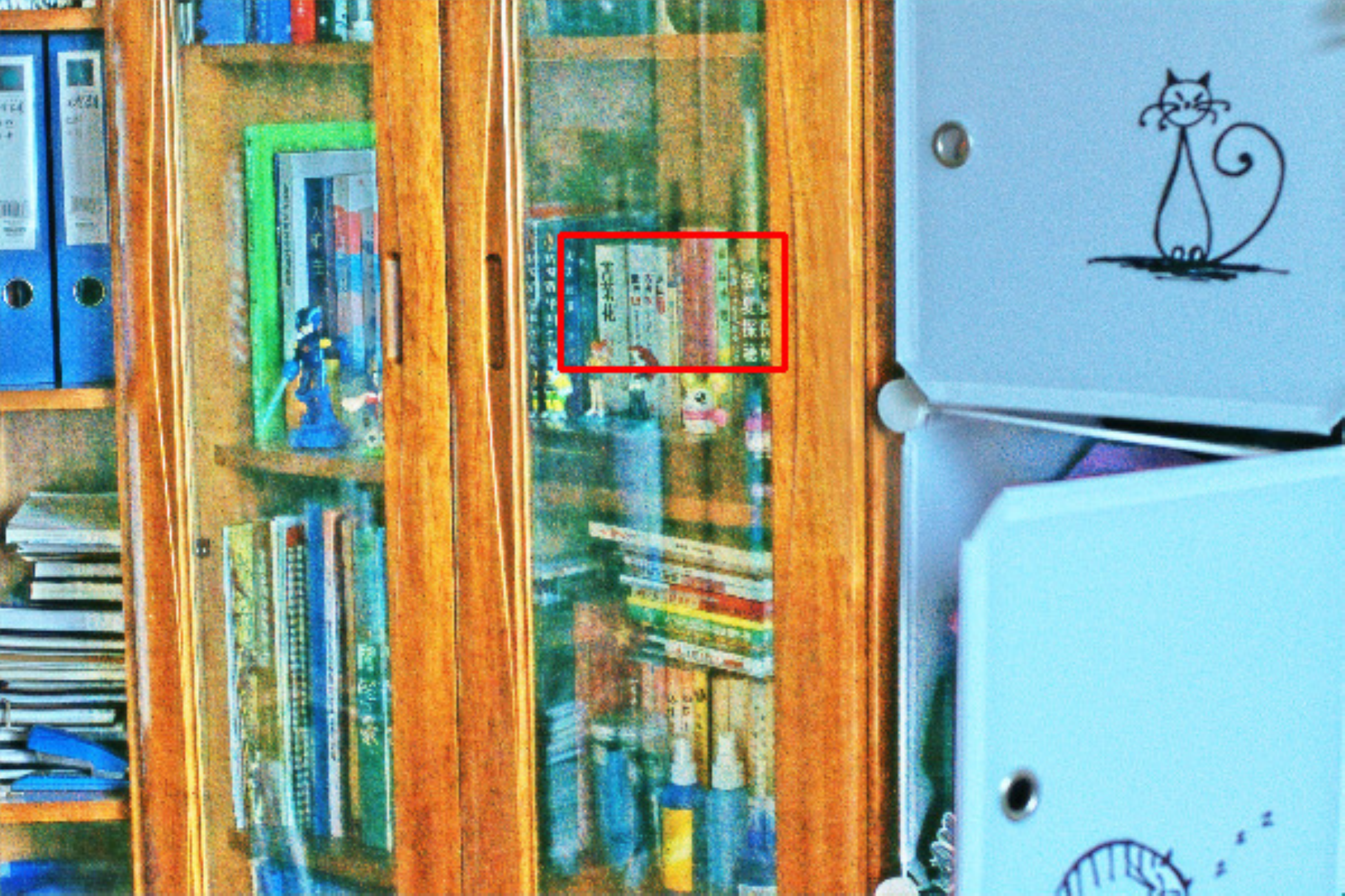}\vspace{2pt} \\
			\includegraphics[width=2.8cm]{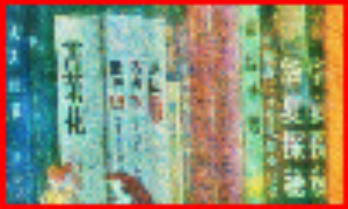}\vspace{5pt}
			\includegraphics[width=2.8cm]{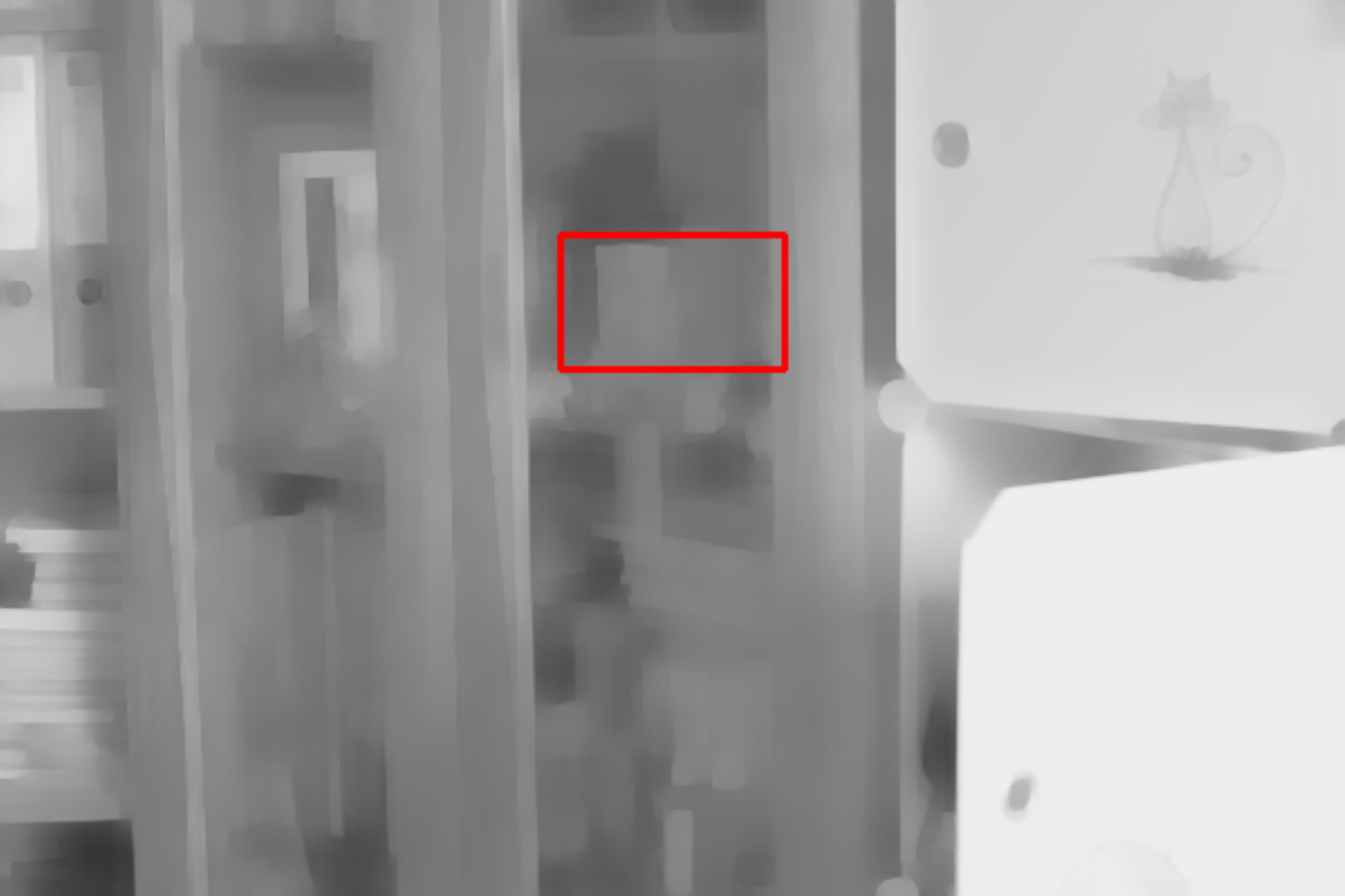}\vspace{2pt} \\
			\includegraphics[width=2.8cm]{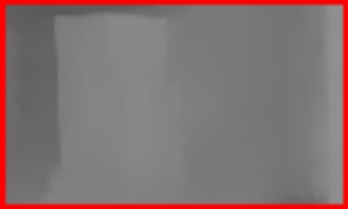}
		\end{minipage}
	}\hspace{-5pt}
	\subfigure[RetinexNet\cite{wei2018deep}]{
		\begin{minipage}[b]{0.155\textwidth}
			\includegraphics[width=2.8cm]{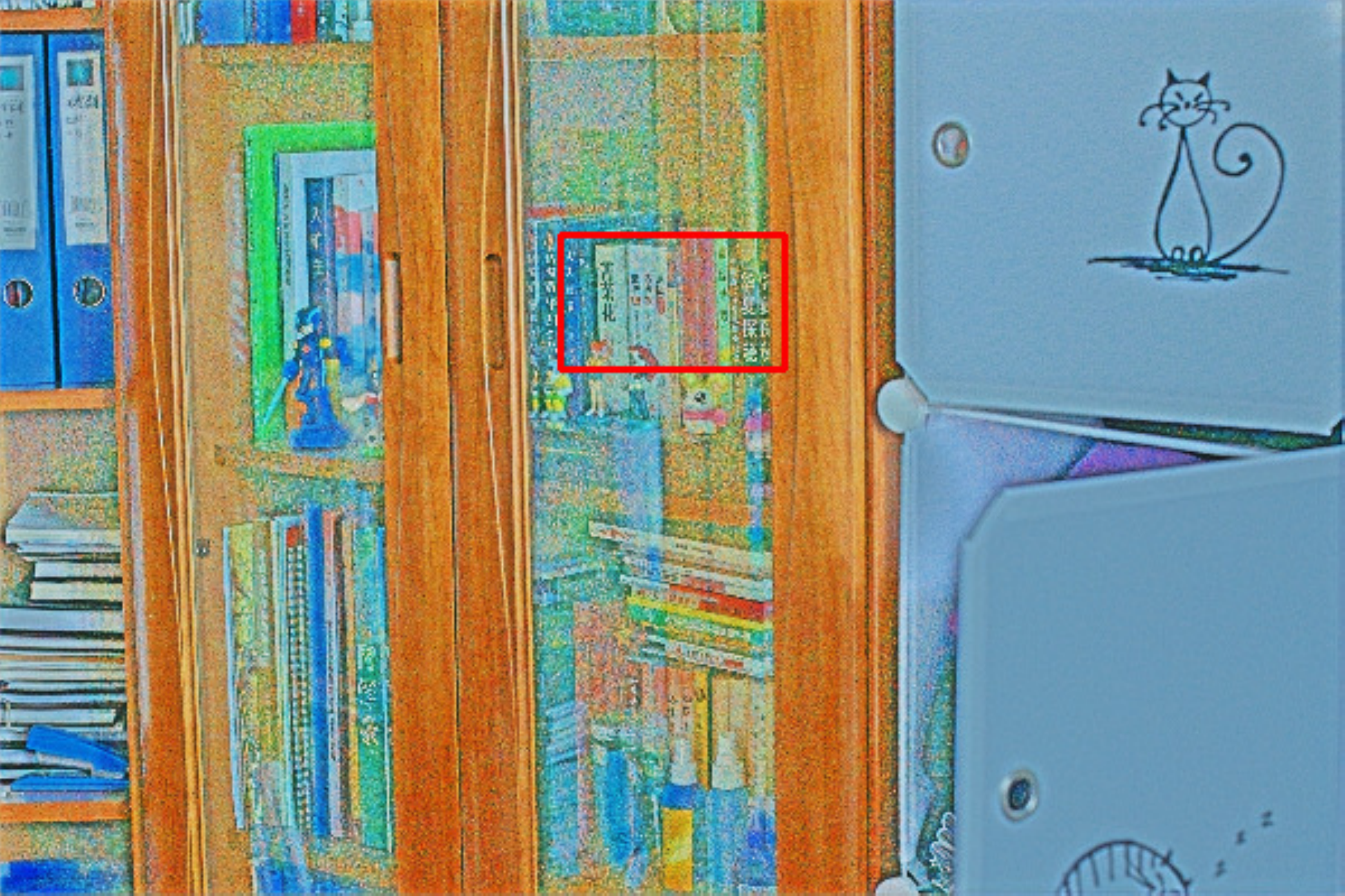}\vspace{2pt} \\
			\includegraphics[width=2.8cm]{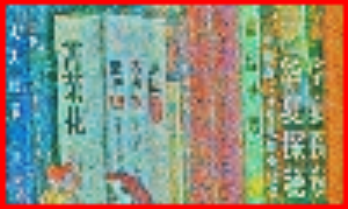}\vspace{5pt}
			\includegraphics[width=2.8cm]{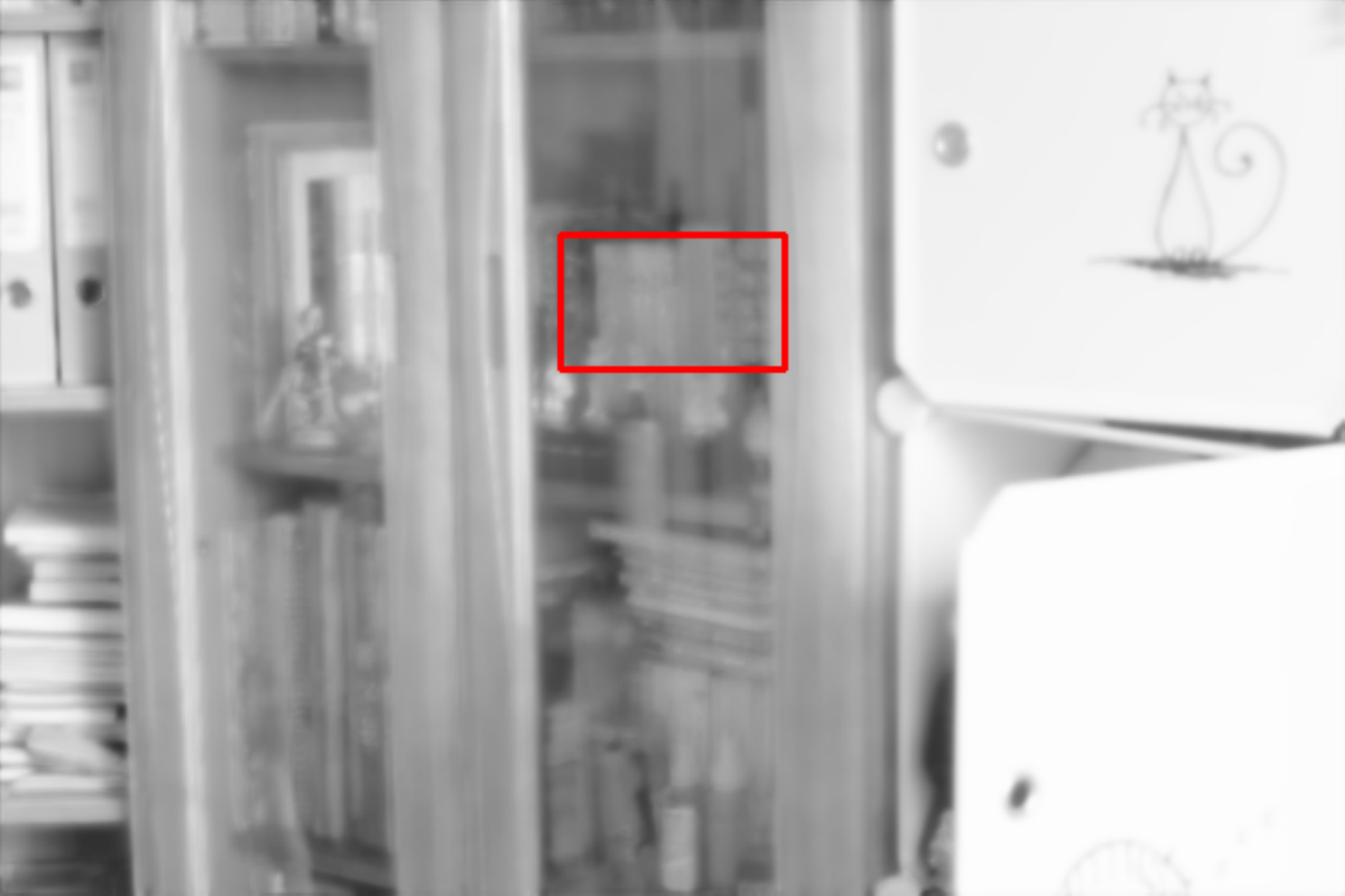}\vspace{2pt} \\
			\includegraphics[width=2.8cm]{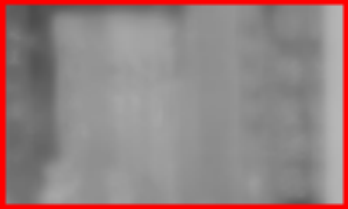}
		\end{minipage}
	}\hspace{-5pt}
	\subfigure[$\lambda_{TV}$=0.1 in DA]{
		\begin{minipage}[b]{0.155\textwidth}
			\includegraphics[width=2.8cm]{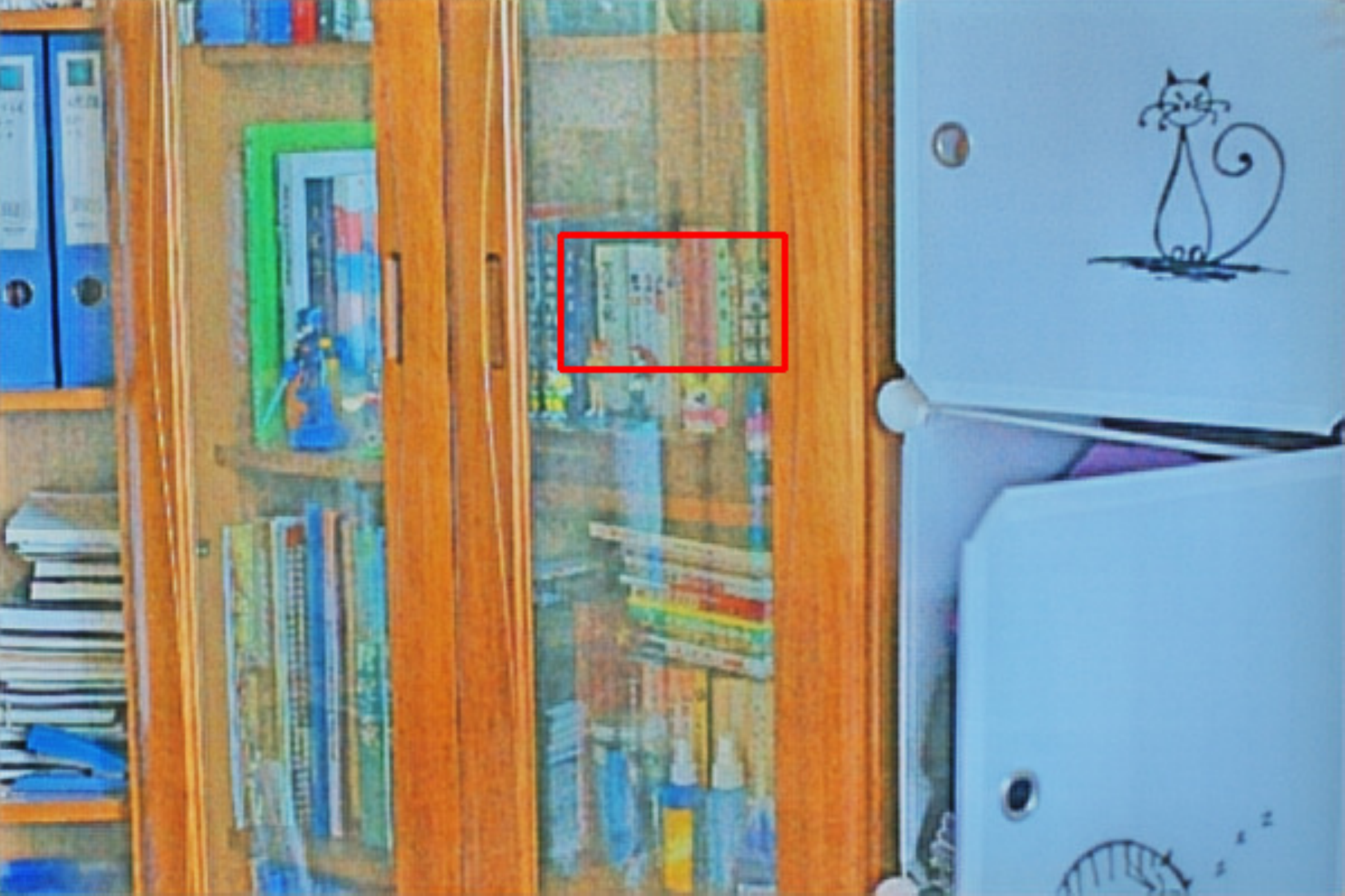}\vspace{2pt} \\
			\includegraphics[width=2.8cm]{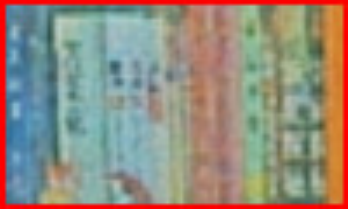}\vspace{5pt}
			\includegraphics[width=2.8cm]{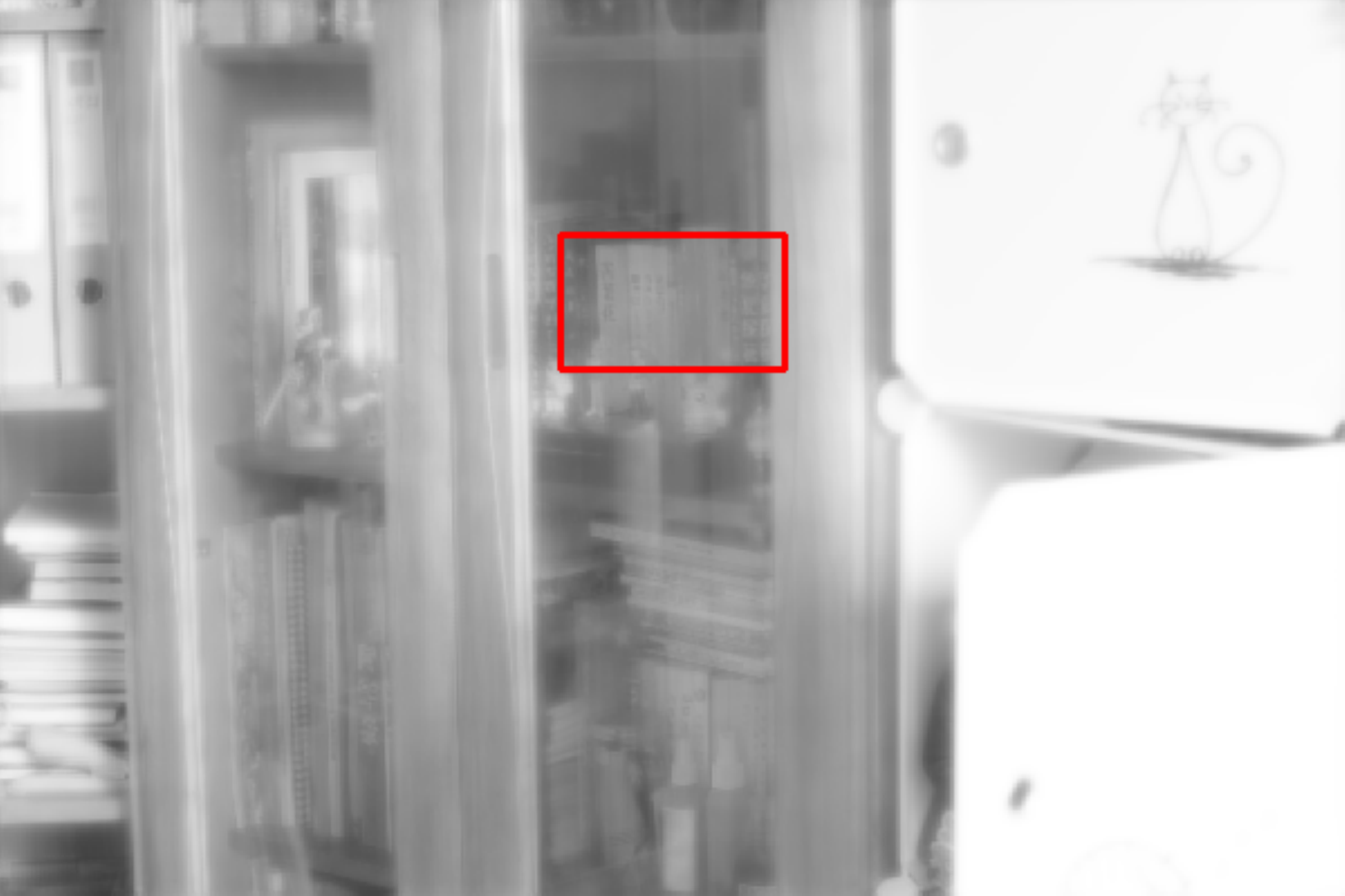}\vspace{2pt} \\
			\includegraphics[width=2.8cm]{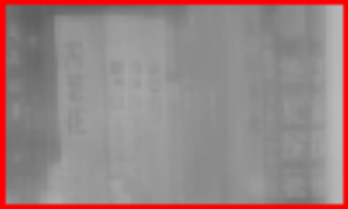}
		\end{minipage}
	}\hspace{-5pt}
	\subfigure[$\lambda_{TV}$=0.2 in DA]{
		\begin{minipage}[b]{0.155\textwidth}
			\includegraphics[width=2.8cm]{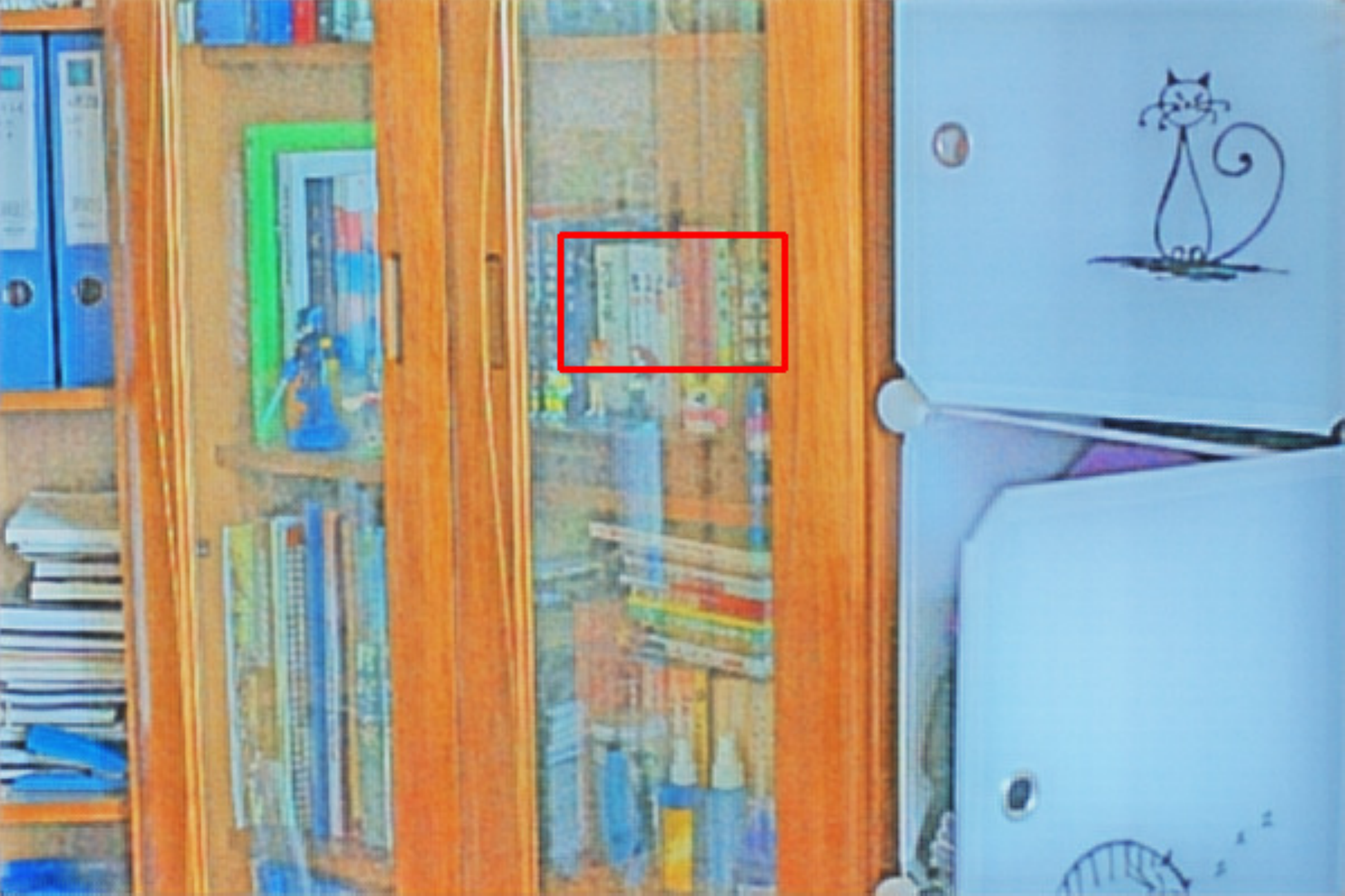}\vspace{2pt} \\
			\includegraphics[width=2.8cm]{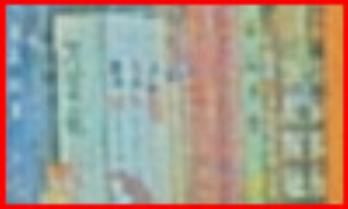}\vspace{5pt}
			\includegraphics[width=2.8cm]{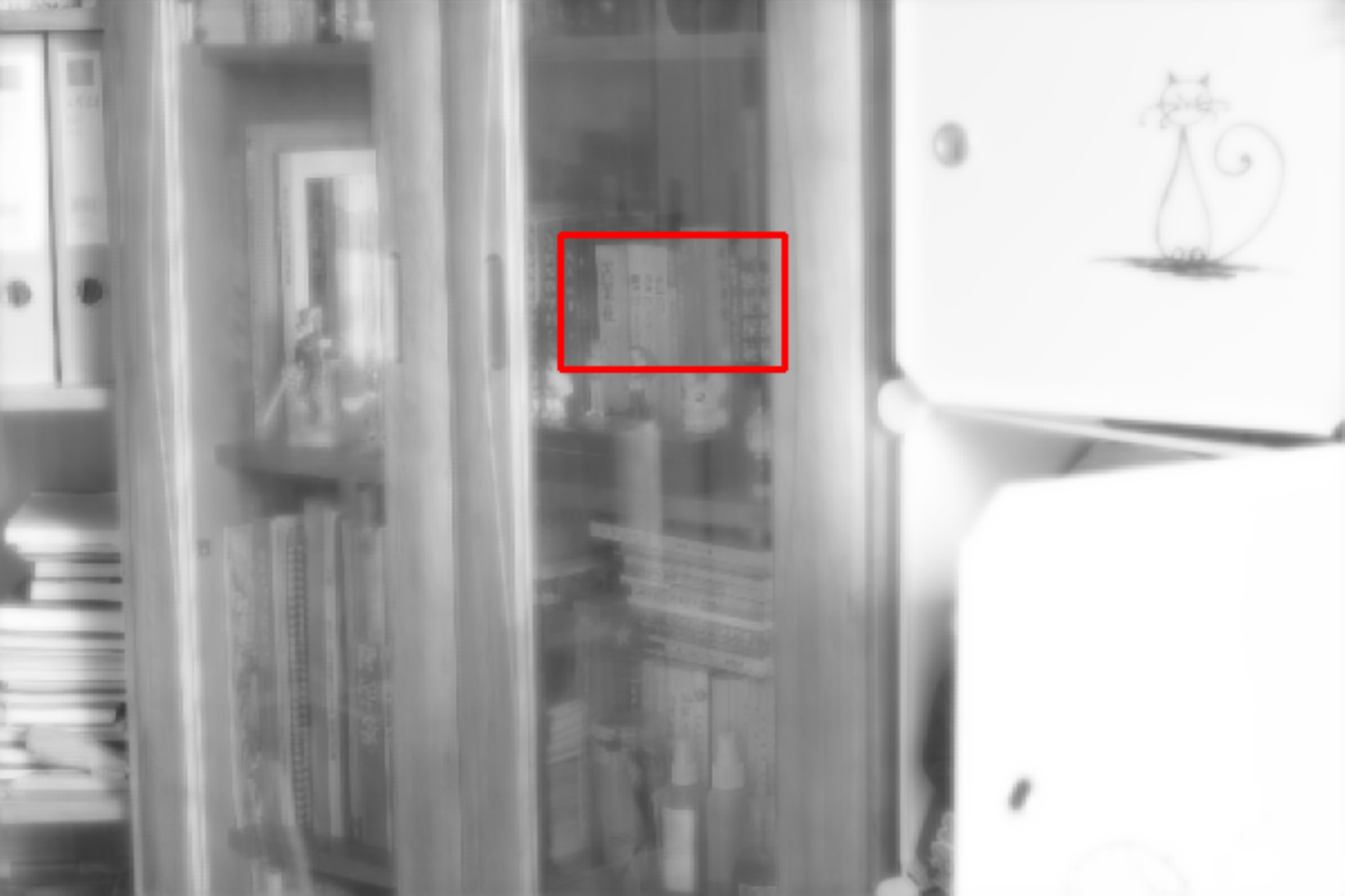}\vspace{2pt} \\
			\includegraphics[width=2.8cm]{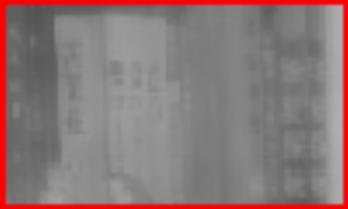}
		\end{minipage}
	}\hspace{-5pt}
	\subfigure[$\lambda_{TV}$=0.4 in DA]{
		\begin{minipage}[b]{0.155\textwidth}
			\includegraphics[width=2.8cm]{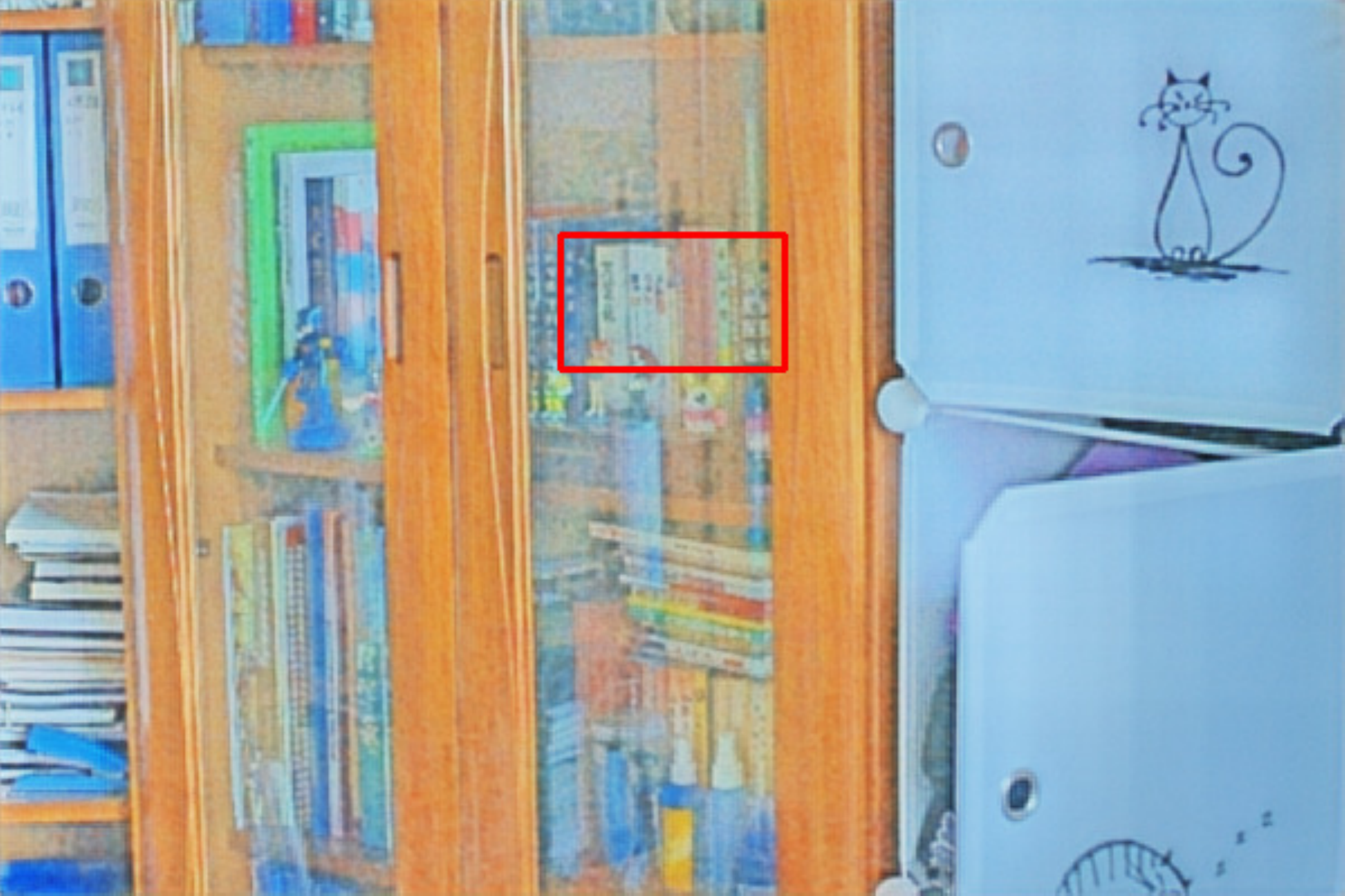}\vspace{2pt} \\
			\includegraphics[width=2.8cm]{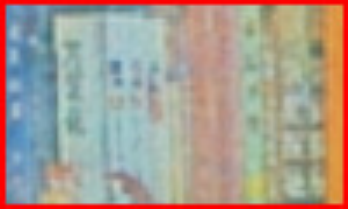}\vspace{5pt}
			\includegraphics[width=2.8cm]{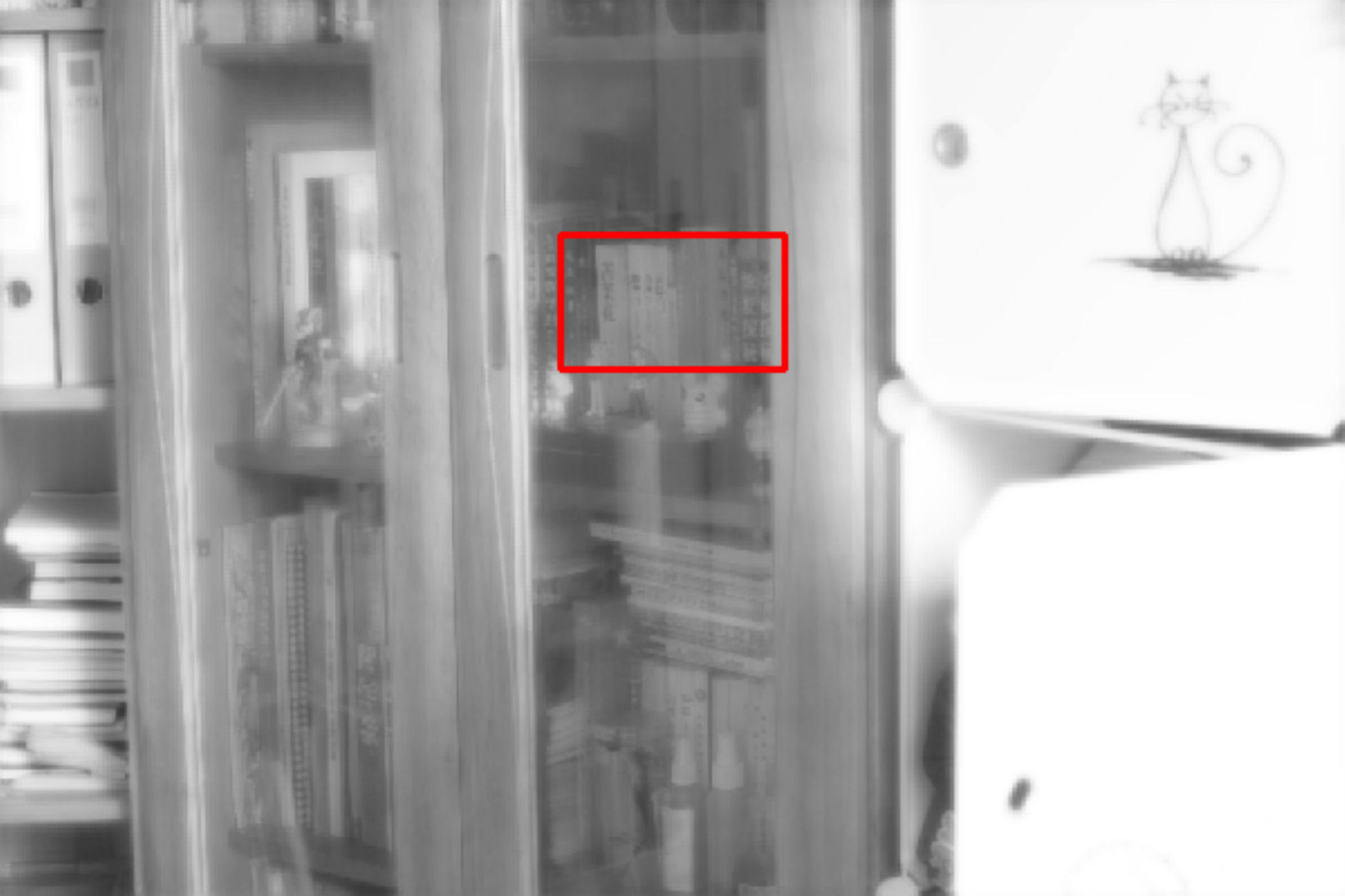}\vspace{2pt} \\
			\includegraphics[width=2.8cm]{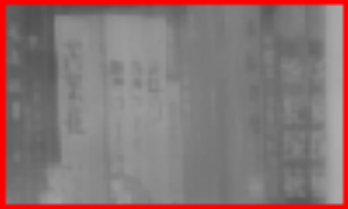}
		\end{minipage}
	}\hspace{-5pt}
	\subfigure[$\lambda_{TV}$=0.7 in DA]{
		\begin{minipage}[b]{0.155\textwidth}
			\includegraphics[width=2.8cm]{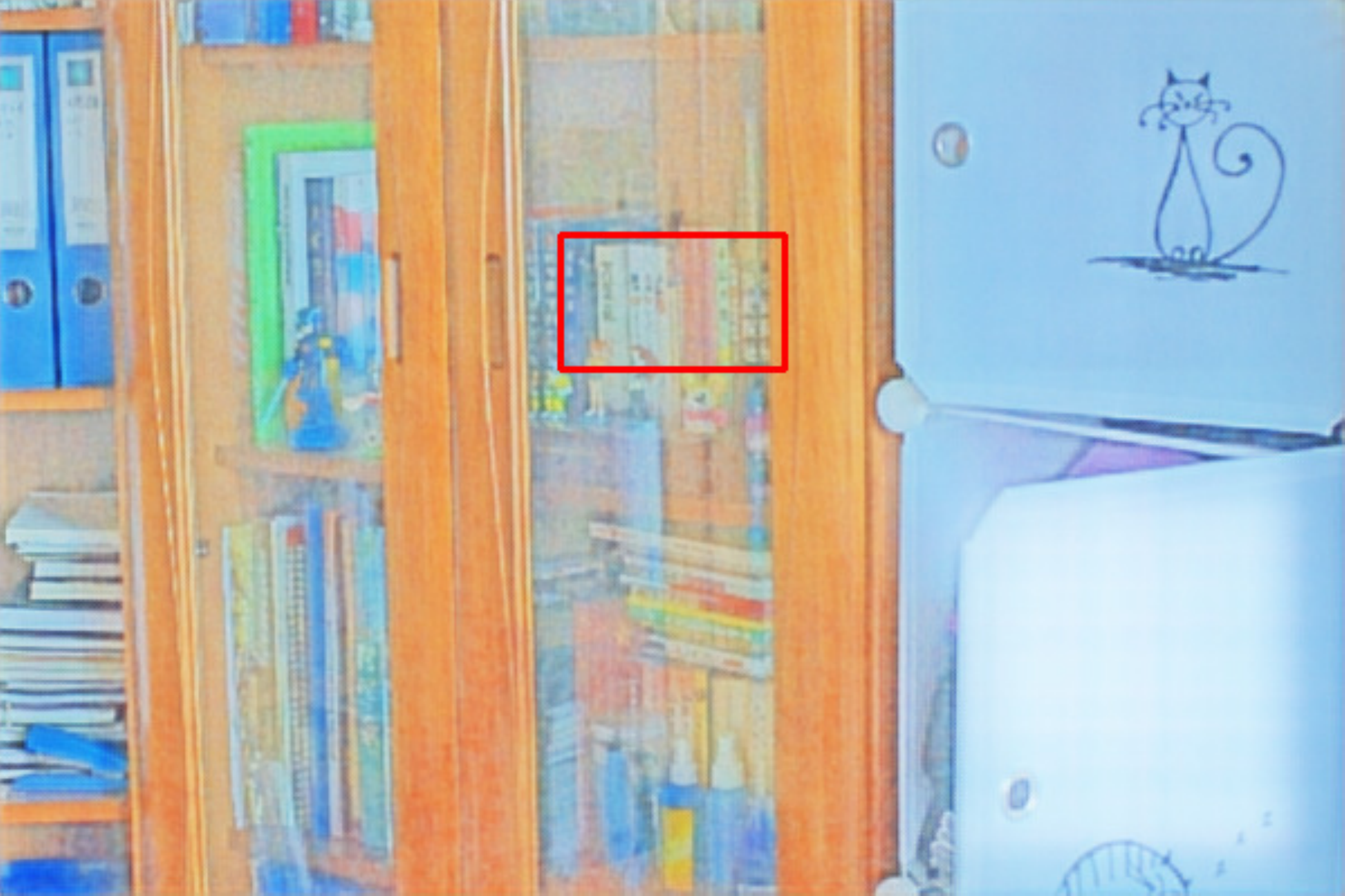}\vspace{2pt} \\
			\includegraphics[width=2.8cm]{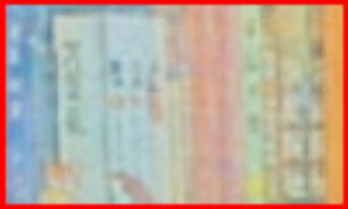}\vspace{5pt}
			\includegraphics[width=2.8cm]{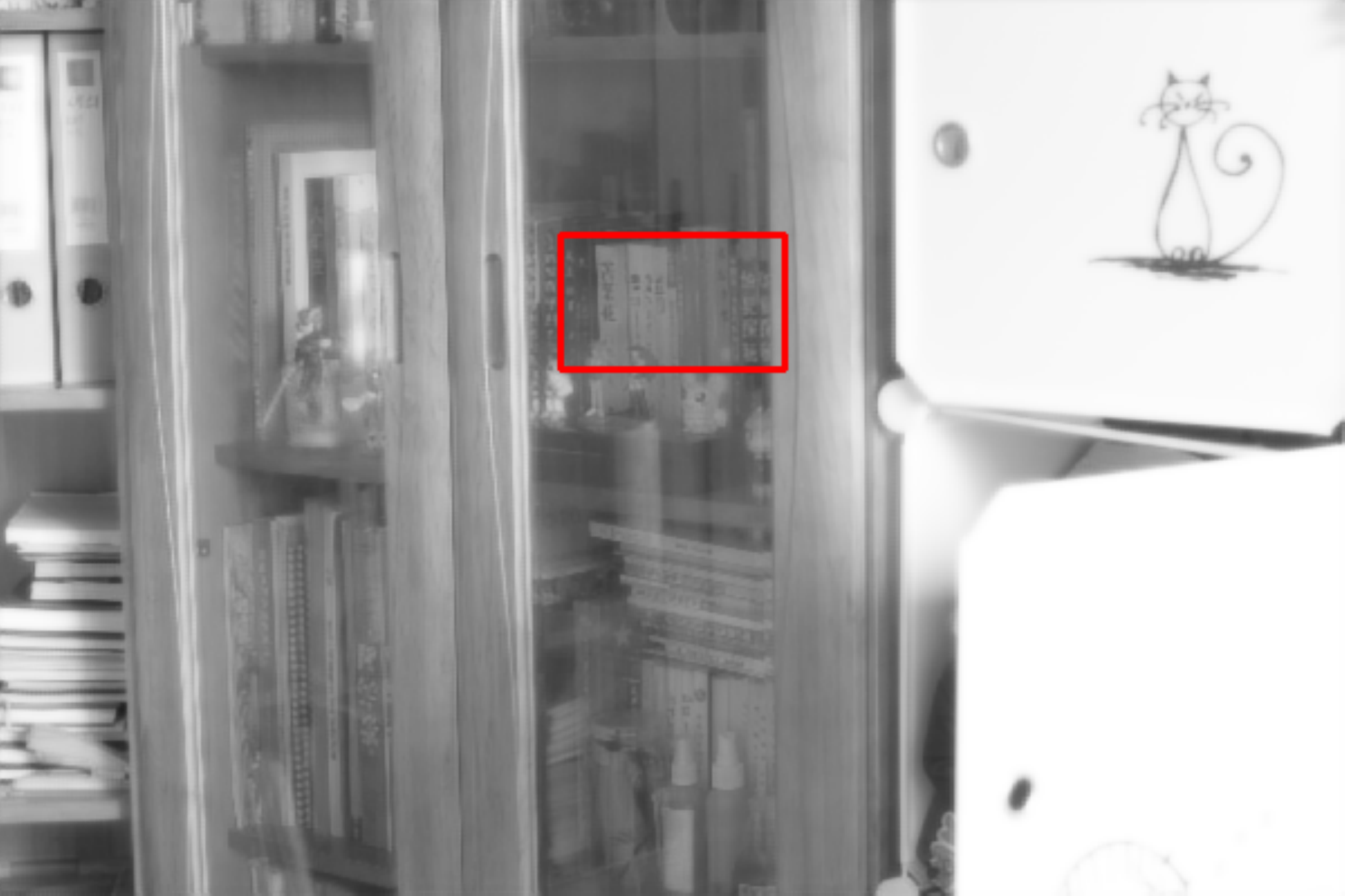}\vspace{2pt} \\
			\includegraphics[width=2.8cm]{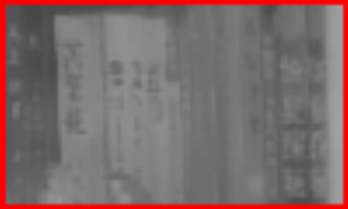}
		\end{minipage}
	}
	\caption{Visual comparison and ablation study of the \textbf{reflectance(R)} maps genreated by different Retinex-based methods and different coefficients of TV Loss($\lambda_{TV}$).}
	\label{pic_tv_abla}
\end{figure*}

Through experiments, we found that directly using TV loss to smooth the reflectance is not good for denoising and has no effect on the elimination of color distortion and halo artifacts. In order to restore the reflectance map without introducing additional networks or restorers, which may slow down the processing of the pipeline and cause color distortion again in the reflectance map, we propose the Degradation-Aware (DA) Module and DA Loss for the awareness and removal of degradation in the reflectance as well as details preservation.\\
As shown in Fig.\ref{748_r}, there are barely noise and color distortion in the reflectance decomposed by normal-light Ground-Truth image, and it is colorful and full of high-frequency details information. On the contrary, the reflectance decomposed by low-light image is full of amplified noise, color distortion and halo artifacts. We propose Degradation-Aware (DA) Module to guide the training of the decomposer, which can enable the decomposer to directly generate the reflectance without amplified noise, color distortion and halo artifacts.\\
The schematic of DA Module is shown in Fig.\ref{da}. We use a CNN to extract the feature maps of the reflectance of the normal/low-light images for the awareness and extraction of the feature of degradation, even though the convolution operation is the local operator, we can construct CNN by stacking multiple convolutional layers to enlarge the receptive field so as to obtain the global feature information of the image, such as color information and texture. Hence, CNN in DA Module has the ability to capture the feature map of global noise and color information of the image. Before feeding the reflectance maps into the CNN, we apply MSE Loss between the normal-light and low-light reflectance maps for the pre-denoising.
MSE Loss is as follows:
\begin{equation}
	L_{MSE}=\left \| R_{low} - R_{high}\right \|_{2}^{2}
\end{equation}
And then, we use CNN to extract the feature map of the reflectance maps decomposed by low/normal-light images.
Together with MSE Loss, we then use Total Variation (TV) Loss to remove the high-frequency information (noise and details) and distorted color information from the feature map of the reflectance decomposed by low-light image ($R_{low}$). 
Total variation loss (TV loss) \cite{guo2019toward} is as follows:
\begin{equation}
	L_{TV}^{low}=\left \| \triangledown _{h}F (R_{low})] \right \|_{2}^{2}+\left \| \triangledown _{v}F (R_{low})] \right \|_{2}^{2}
\end{equation}
where $F$ means the output feature map of the CNN in DA Module, and $\triangledown_{h}$ and $\triangledown_{v}$ denote the gradients in the horizontal and vertical directions. \\
According to \cite{zhao2016loss} and \cite{wang2018gladnet}, we use MSE loss to denoise as well as constrain the similarity of $R_{low}$ and $R_{high}$ while using TV loss to smooth the reflectance map. 
Now the details information in the $R_{low}$ is erased along with the noise, and some color information will also be erased, therefore, the corresponding $R_{low}$ is noise-free and less colorful than the $R_{low}$ before feeding into CNN.\\
Next we adopt $L1$ Loss on the feature map of normal/low-light reflectance for removing halo artifacts and restoring details and corrected color information as much as possible. 
$L1$ Loss is as follows:
\begin{equation}
	L_{1}=\left \| F (R_{low})- F (R_{high})\right \|_{1}
\end{equation}
where $F$ represents the output feature map of CNN in DA.
The total DA loss $L_{DA}$ is as follows: 
\begin{equation}
	L_{DA}=\lambda_{TV}(0.05L_{TV}^{low} + 1.0L_{MSE}) + 0.1L_{1}
	\label{lda}
\end{equation}
During the training process, the coefficient of TV and MSE Loss ($\lambda_{TV}$) is a parameter that can be adjusted to suit images with different noise levels and degradation. We set $\lambda_{TV}$ to 0.2 for the training on the LOL real-world dataset and 0.05 for the LOL synthetic dataset because the latter is not as so noisy, dark and degenerate as the former one. This makes our method more flexble to different datasets with different degrees of degradation and noise. \\
Although the details are removed from the generated noise-free reflectance, as shown in Fig.\ref{pic_tv_abla}, details information can be reconstructed into the single-channel illumination map with the help of loss functions in the decomposition phase. And as shown in Fig.\ref{pic_tv_abla}, when we increase the coefficient of TV Loss ($\lambda_{TV}$), more high-frequency information is erased from the reflectance, which means a higher PSNR, at the same time, more details information can be reconstructed into the illumination map. In this way, we can remove the noise while preserving the details information. This makes our model more flexible for different datasets with different degrees of noise and other degradation.\\
Now we discuss how the details information can be reconstructed into the single-channel illumination map and how the noise can be removed.
As the schematic shown in the right dotted box in the figure, high-frequency components (noise and details information) and color information are removed from reflectance map with the function of TV and MSE Loss, because of the constraint of reconstruction loss ($L_{rc}$) \ref{lrc} in the decomposition phase, the information removed from reflectance map has tendency to be reconstructed into the illumination map, but the illumination map is a single-channel image without color information, only the high-frequency information (noise and details information) can be reconstructed into the illumination map.\\
In the decomposition phase, Smooth Loss\ref{sml} is applied on the illumination map, so when the high-frequency components (noise and details) are reconstructed in the illuminance map, noise with higher frequency will be mostly filtered out by Smooth Loss, and most of the details with relatively lower frequency will be reconstructed in the illuminance map. In this way, we can denoise the reflectance and preserve details information into the illumiantion map. However, there is still a small amount of noise leakage into the illumination map with details information, which requires our enhanced network to have a certain noise removal ability.  

\begin{table*}[htbp] 
	\centering 
	\caption{Quantitative ablation study of the \textbf{reflectance(R)} maps genreated by different Retinex-based methods and different coefficients of TV Loss($\lambda_{TV}$).}
	\begin{tabular}{  c | c  c | c  c  c  c  c  c  c}
		\hline  Reflectance  & KinD\cite{zhang2019kindling}   & RetinexNet\cite{wei2018deep}  & $\lambda_{TV}$=0.1  &$\lambda_{TV}$=0.2   & $\lambda_{TV}$=0.3  & $\lambda_{TV}$=0.4 & $\lambda_{TV}$=0.5  &$\lambda_{TV}$=0.6 & $\lambda_{TV}$=0.7\\
		\hline
		\hline  
		Noise Level↓ & 12.1900  & 12.5234 & 0.8679 & 0.8574 & 0.8506 &0.8477   &0.7921  & 0.7739 & 0.7642\\
		PSNR↑  & 15.9720 & 14.8130 & 17.9139 &19.3568 & 19.6542 & 19.9428  &20.1469 & 21.1157 &21.4358\\
		SSIM↑ & 0.6626 & 0.6631 & 0.8443 & 0.8519 & 0.8413 &0.8446   &0.8421  & 0.8339 & 0.8236\\
		DeltaE↓  & 19.5286 & 17.7442 & 11.6357 &9.0756 & 10.1457 & 11.1745  &12.3876 & 12.4631 & 12.7662 \\
		Colorfulness  & 76.7577 & 69.8642 & 65.3096 &65.0251 & 62.6727 & 60.1175  &60.0732 & 59.9943 &59.9407\\
		\hline\end{tabular}\vspace{0cm}
	\label{tv_abla_tb}
\end{table*}

\begin{figure*}
	
	
	\subfigure[Noise Level\cite{liu2013single}]{
		\begin{minipage}[b]{0.185\textwidth}
			\includegraphics[width=3.9cm]{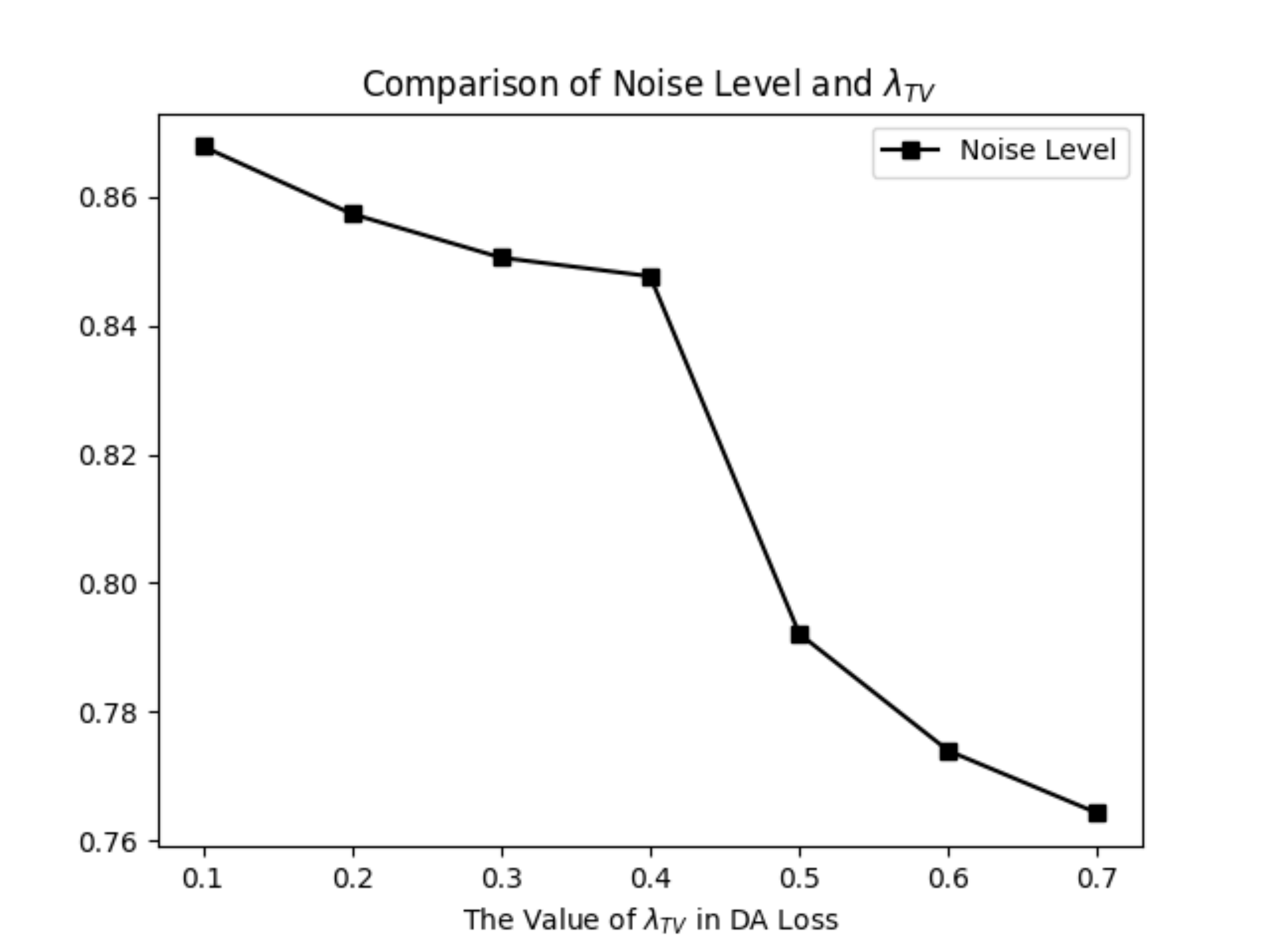}\vspace{-10pt} \\
		\end{minipage}
	}\hspace{-5pt}
	\subfigure[PSNR]{
		\begin{minipage}[b]{0.185\textwidth}
			\includegraphics[width=3.9cm]{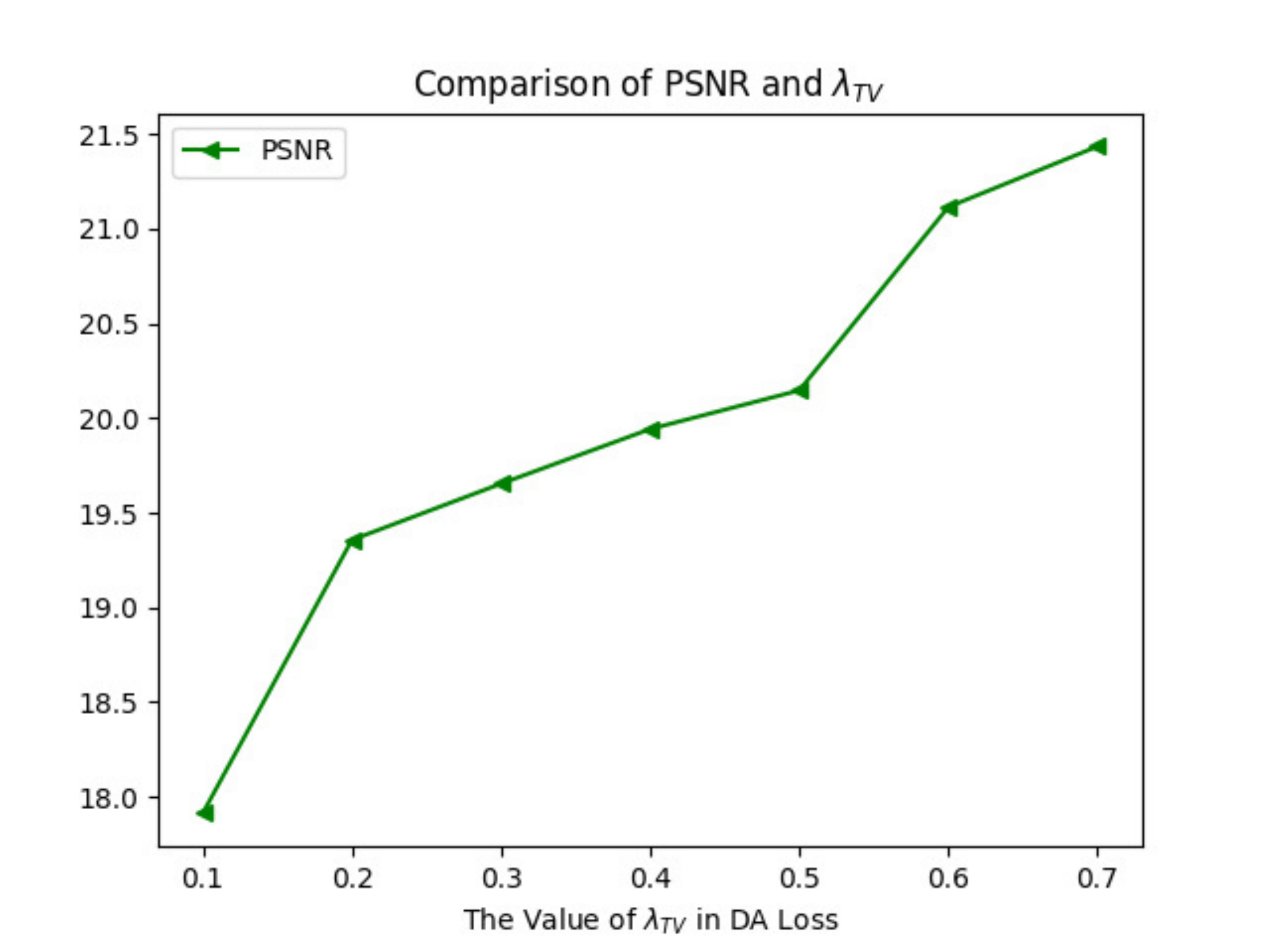}\vspace{-10pt} \\
		\end{minipage}
	}\hspace{-5pt}
	\subfigure[SSIM\cite{wang2004image}]{
		\begin{minipage}[b]{0.185\textwidth}
			\includegraphics[width=3.9cm]{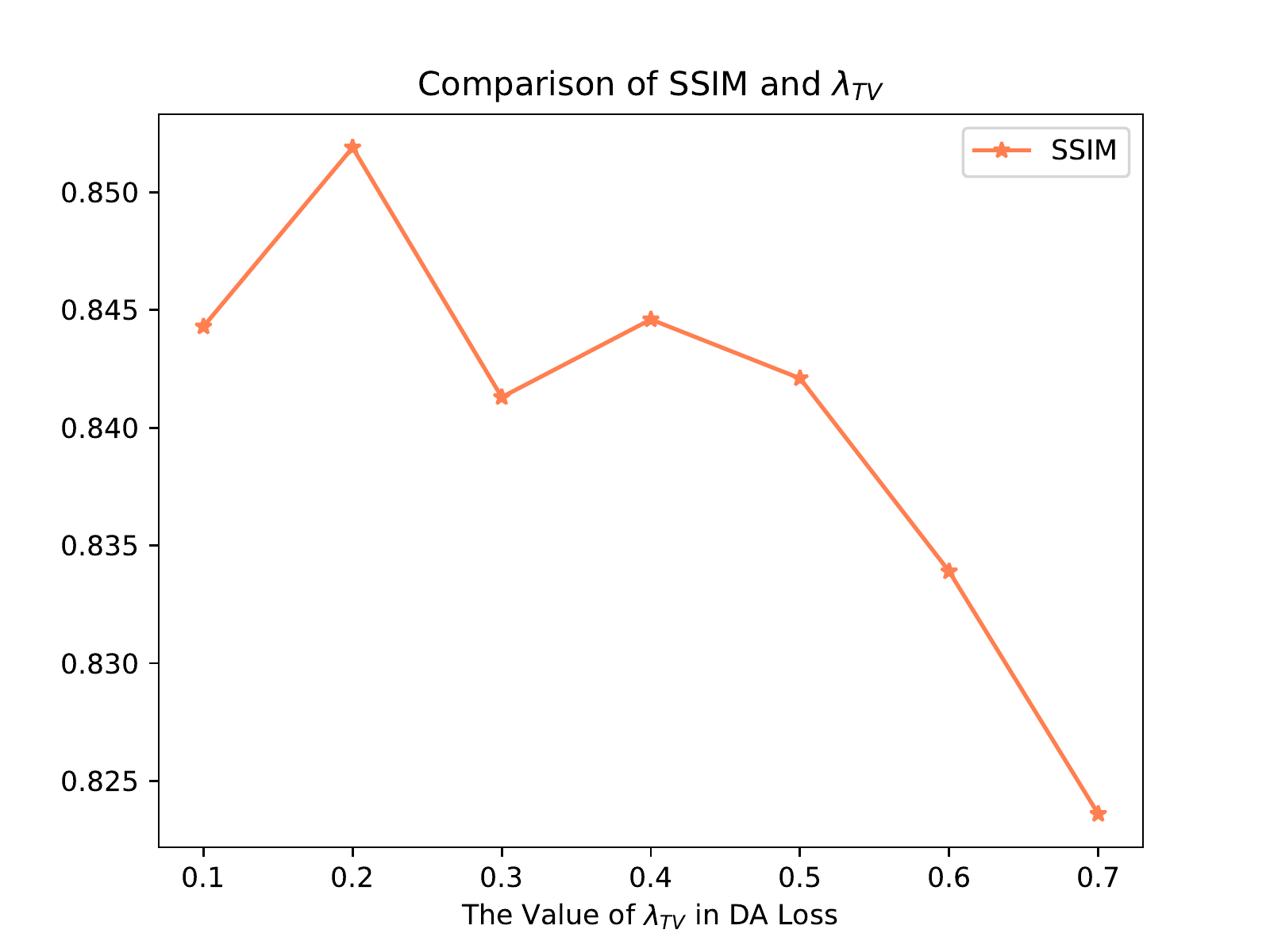}\vspace{-10pt} \\
		\end{minipage}
	}\hspace{-5pt}
	\subfigure[DeltaE\cite{sharma2005ciede2000}]{
		\begin{minipage}[b]{0.185\textwidth}
			\includegraphics[width=3.9cm]{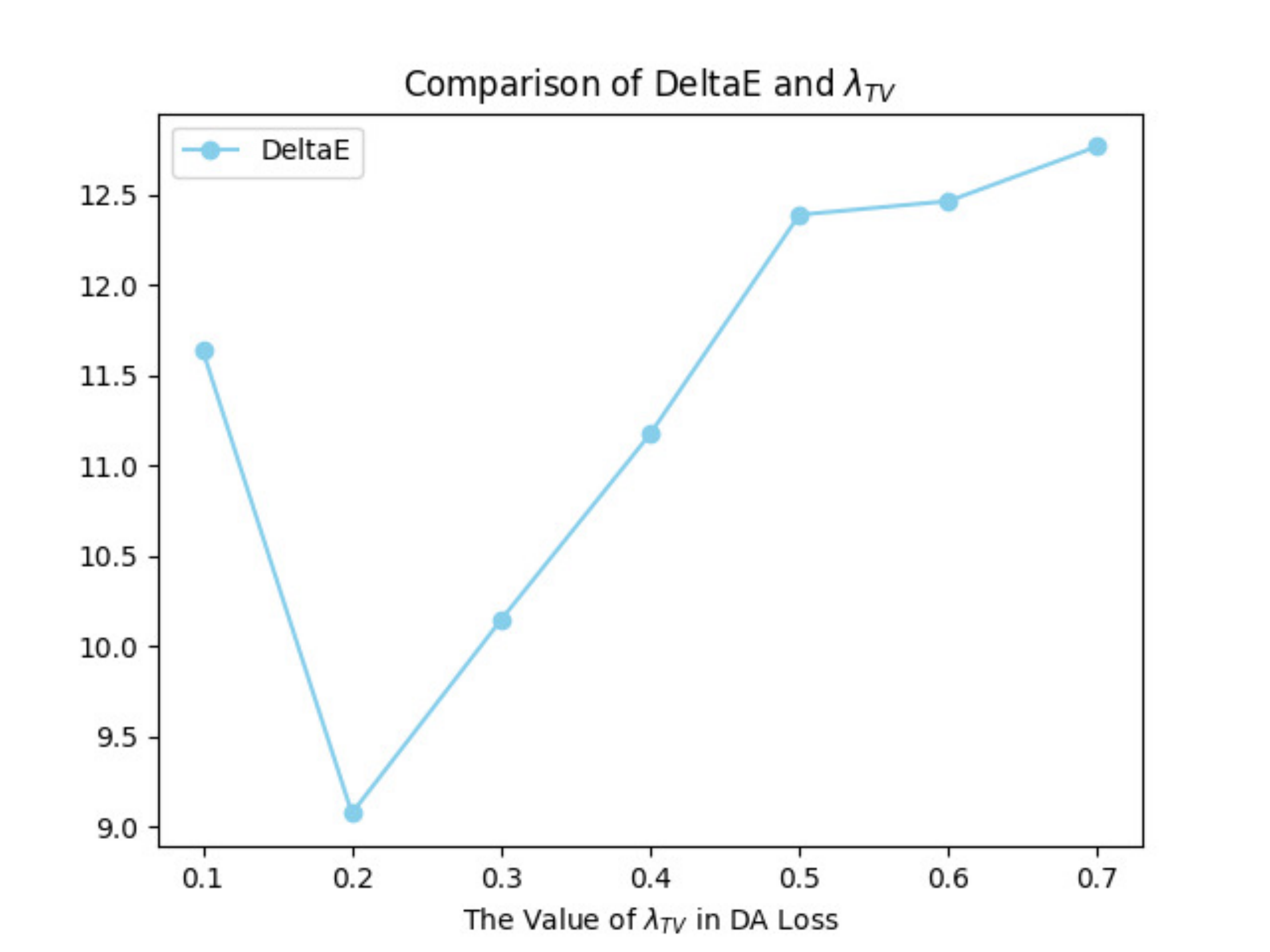}\vspace{-10pt} \\
		\end{minipage}
	}\hspace{-5pt}
	\subfigure[Colorfulness\cite{hasler2003measuring}]{
		\begin{minipage}[b]{0.185\textwidth}
			\includegraphics[width=3.9cm]{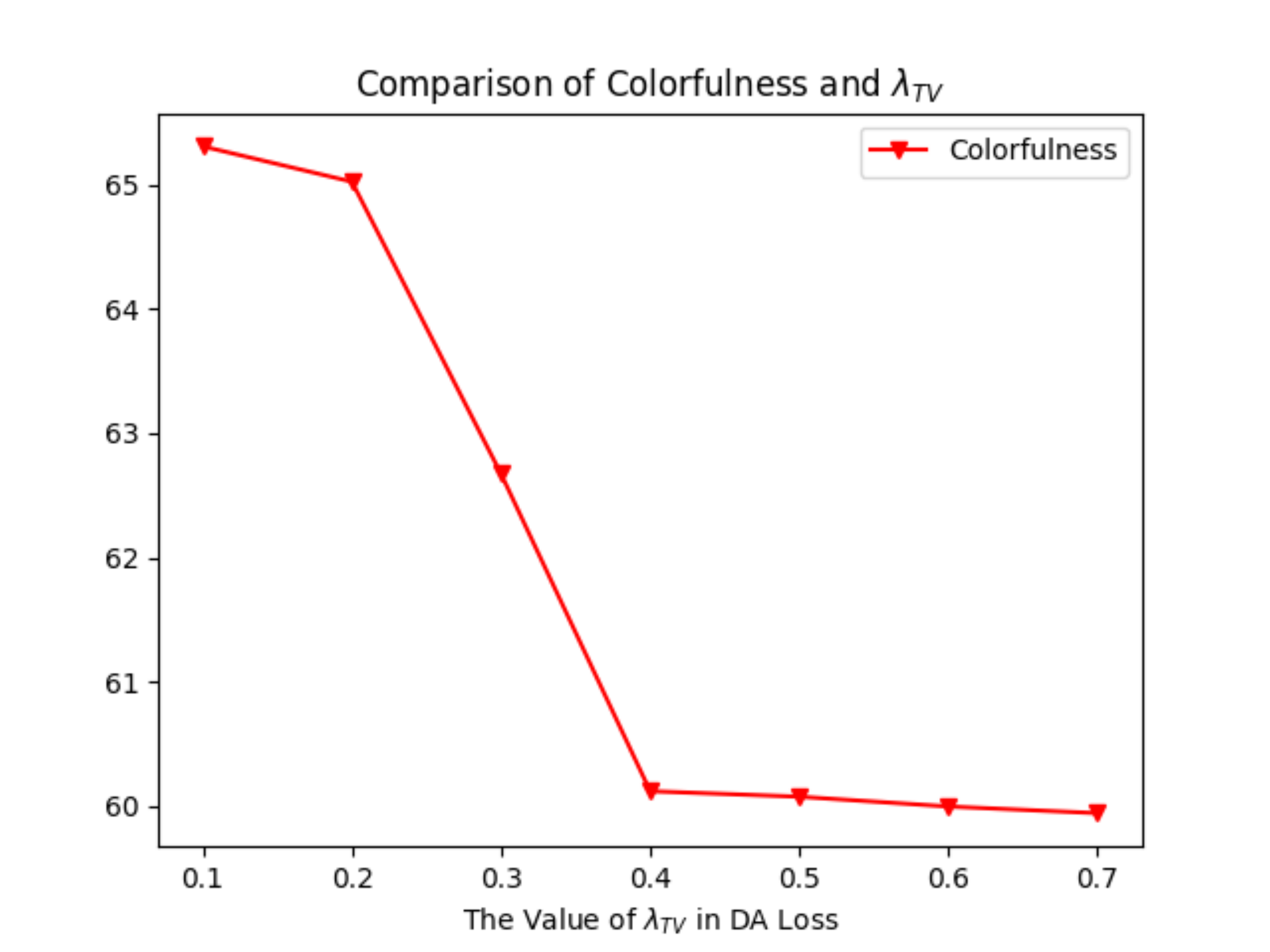}\vspace{-10pt} \\
		\end{minipage}
	}
	\caption{Comparison in graphs of the \textbf{reflectance (R)} maps genreated by different Retinex-based methods and different coefficients of TV Loss($\lambda_{TV}$).}
	\label{line_tv_abla}
\end{figure*}

\subsection{Enhancement Network}
We use a deep U-Net as our enhancer. Because when the details information is reconstructed into the illumination map, most of the noise is removed by Smooth Loss, however, through the experiment, we found that some noise was still transferred into the illumination map. If we use plain CNN to enhance the brightness like the enhancer in KinD\cite{zhang2019kindling}, as shown in Table.\ref{relinet_abla}, the small amount of noise in the illumination map will be amplified. So following the choice of decomposer network architecture, we use deep U-Net as our enhancer for avoiding the residual noise being amplified. 
\begin{table}[htbp] 
	\centering \caption{Comparison of the final enhanced results generated by different choices of enhancer network architectures. The input of these different enhancer are generated by the same decomposer network(Deep U-Net with DA)}
	\begin{tabular}{ c | c  c  c}
		\hline Enhancer Network  &PSNR↑  & SSIM↑ & DeltaE↓ \\
		\hline
		\hline  
		Plain CNN in KinD\cite{zhang2019kindling}   & 19.2341  & 0.7669   &14.7876 \\
		Enhancer in RetinexNet\cite{wei2018deep}  & 19.6362  & 0.7917  &14.0004\\
		U-Net &19.4824  & 0.7734  &14.1133\\
		Deep U-Net & \textbf{20.7282} & \textbf{0.7939}  & \textbf{12.9350}\\
		\hline\end{tabular}\vspace{0cm}
	\label{relinet_abla}
\end{table}
In addition, as shown in Table.\ref{relinet_abla}, although all the enhancers have the same decomposer (Deep U-Net with DA), the results of different enhancers have different values in terms of DeltaE (an indicator of color distortion and bias), which mean the lightness condition in the illumination map can affect the color of the results genreated by the pixel-wise product of the reflectance and the illumination map. As shown in Table.\ref{relinet_abla}, deep U-Net can estimate the illumination map more accurately, which leads to higher SSIM (better performance on structural similarity) and lower DeltaE (better performance on color distortion restoration).\\
The effectiveness of our method depends primarily on the contribution of our enhancement loss function, especially the Perceptual Loss. The total enhancement loss function in the enhancement stage is as follows:
\begin{equation}
L_{eh}=1.0 L_{rc}^{E} + 1.0 L_{bri} + 1.0 L_{per}  + 1.0 L_{grad}
\end{equation}
We use reconstruction loss $L_{rc}^{E}$ to constrain the training process of enhancer and regularize the final output after the pixel-wise product to be as close as possible to the Ground-Truth. The reconstruction loss is as follows:
\begin{equation}
L_{rc}^{E}=\left \|R_{low}\ast I_{output} - S_{high}  \right \|_{1}
\end{equation}
where $R_{low}$, $I_{output}$ and $S_{high}$ respectively represent the reflectance maps decomposed from the low-light image, the brightness-improved illumination maps generated by our enhancement net and the normal-light input Ground-Truth.\\ 
We use $L1$ loss as our brighten loss function $L_{bri}$ to enhance the brightness of the illumination decomposed from low-light images and constrain the output brightness-improved illumination to be as similar as possible to its counterpart from Ground-Truth. The brighten loss function is as follows:
\begin{equation}
L_{bri}=\left \| I_{output} - I_{high} \right \|_{1}
\end{equation}
where $I_{output}$ and $I_{high}$ denote the brightness-improved output of the enhancement net and the decomposed illumination map of the normal-light Ground-Truth, respectively.\\
Perceptual loss is widely used in super-resolution and style tranfer \cite{johnson2016perceptual}. Because we transfer some of the image details information from reflectance to illumination map, we use perceptual loss to preserve the texture information during illumination reconstruction in the enhancement network. In addition, perceptual loss can make our outputs more consistent with human visual perception. 
Perceptual loss $L_{per}$ is as follows:
\begin{equation}
L_{per}=\frac{1}{CHW}\left \| F (R_{low}\ast I_{output})-F (S_{high}) \right \|_{2}^{2}
\end{equation}
where $F$ is the 31st feature map obtained by the VGG16 \cite{simonyan2014very} network pre-trained on ImageNet database. C,H,W represents the number of channels, the height and the width of the input image, respectively.
\begin{table}[htbp] 
	\centering \caption{Quantitative ablation study of the \textbf{Perceptual Loss} in the enhancement stage on the \textbf{LOL real-world evaluation} dataset.}
	\begin{tabular}{ c | c  c  c  c}
		\hline Methods  &PSNR↑  & SSIM↑ & LPIPS↓  & FSIM↑  \\
		\hline
		\hline  
		Baseline   & 20.7282  & 0.7939   &0.3124  &0.9458 \\
		w/o $L_{per}$  & 18.2658  & 0.7535  &0.3520  &0.9154 \\
		\hline
		\hline Methods  &UQI↑  & SERE↑ & SAM↑  & DeltaE↓ \\
		\hline
		\hline
		Baseline   & 54.1478  & 88.2747  &0.9378  &12.9350 \\
		w/o $L_{per}$  & 52.8766  & 87.4807  &0.8946  &16.3991\\
		\hline
\end{tabular}\vspace{0cm}
	\label{perloss}
\end{table}

\begin{figure}
	\flushleft
	
	\subfigure[Input]{
		\begin{minipage}[b]{0.15\textwidth}
			\includegraphics[width=2.8cm]{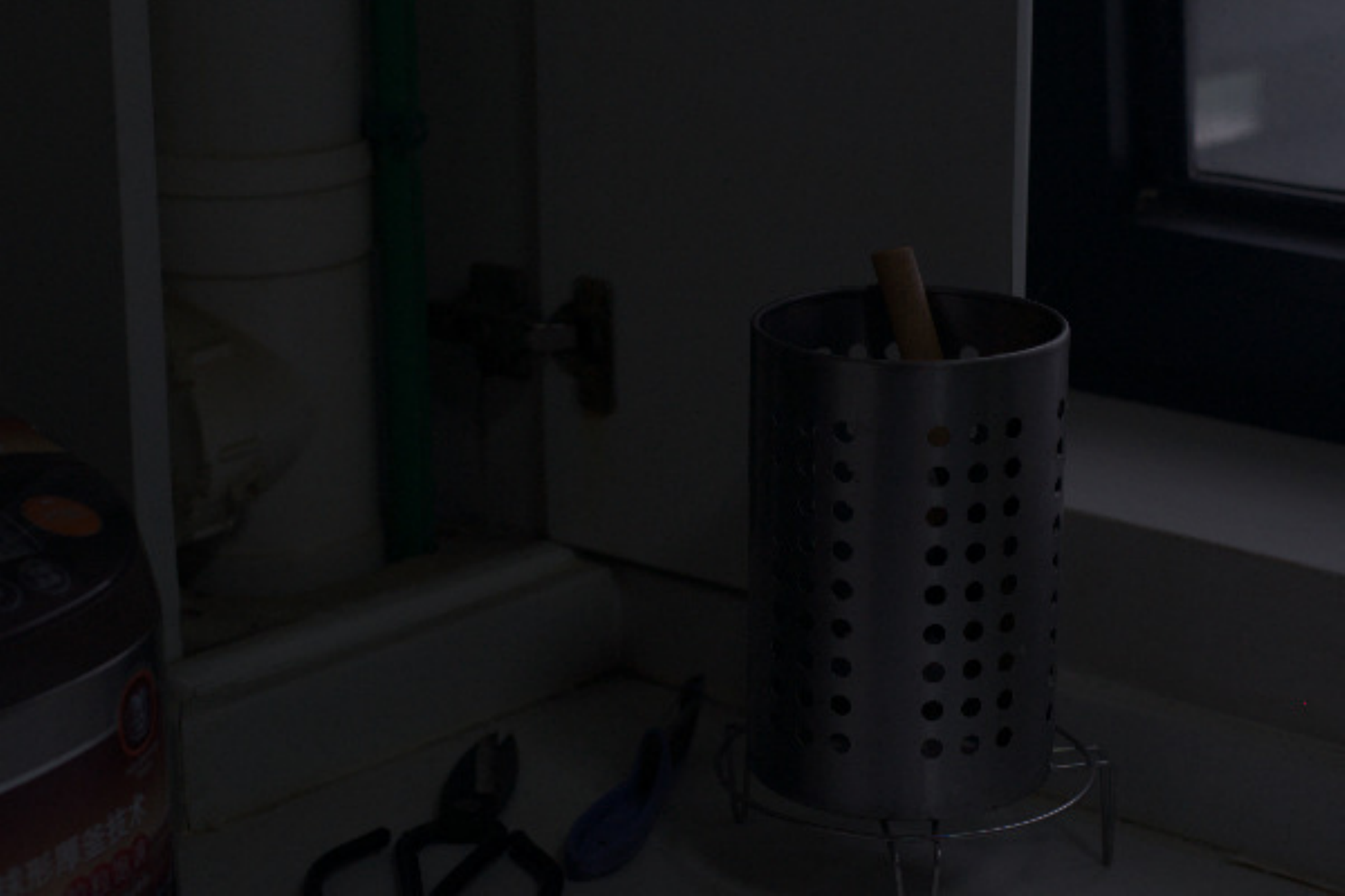}\vspace{2pt} \\
			\includegraphics[width=2.8cm]{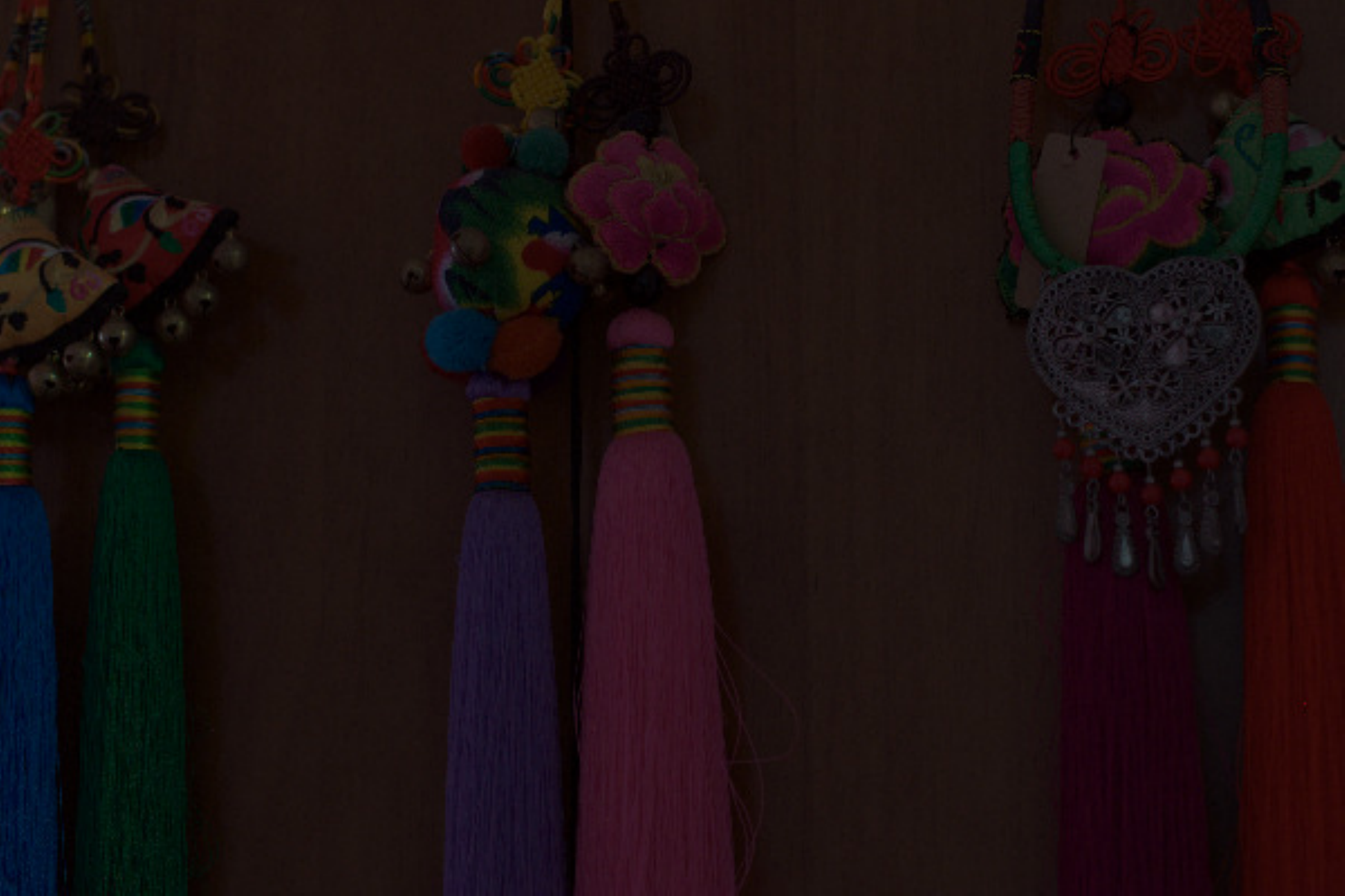}\vspace{-15pt} \\
		\end{minipage}
	}\hspace{-5pt}
	\subfigure[w/o Perceptual Loss]{
		\begin{minipage}[b]{0.15\textwidth}
			\includegraphics[width=2.8cm]{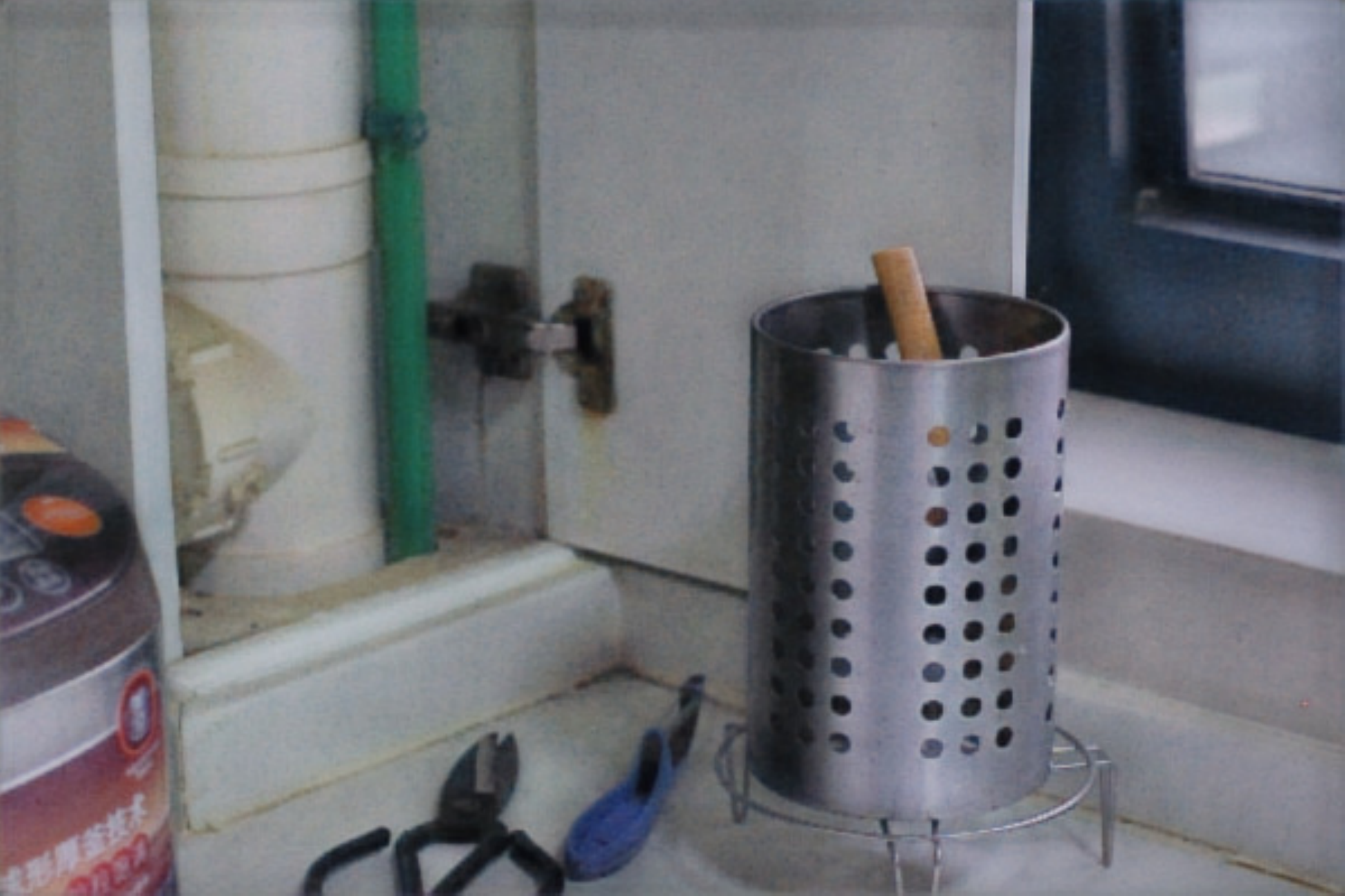}\vspace{2pt} \\
			\includegraphics[width=2.8cm]{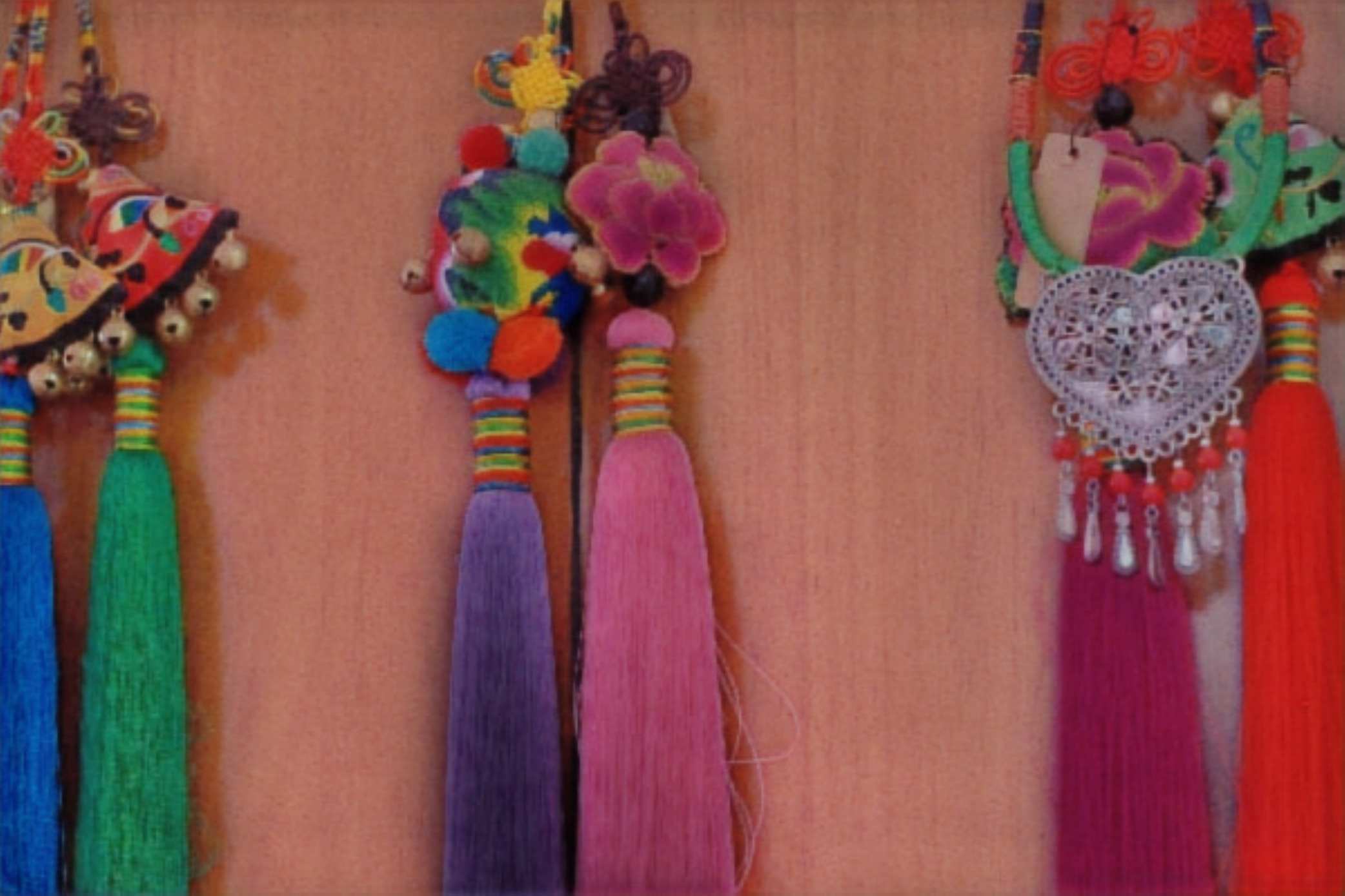}\vspace{-15pt} \\
		\end{minipage}
	}\hspace{-5pt}
	\subfigure[DA-DRN Baseline]{
		\begin{minipage}[b]{0.15\textwidth}
			\includegraphics[width=2.8cm]{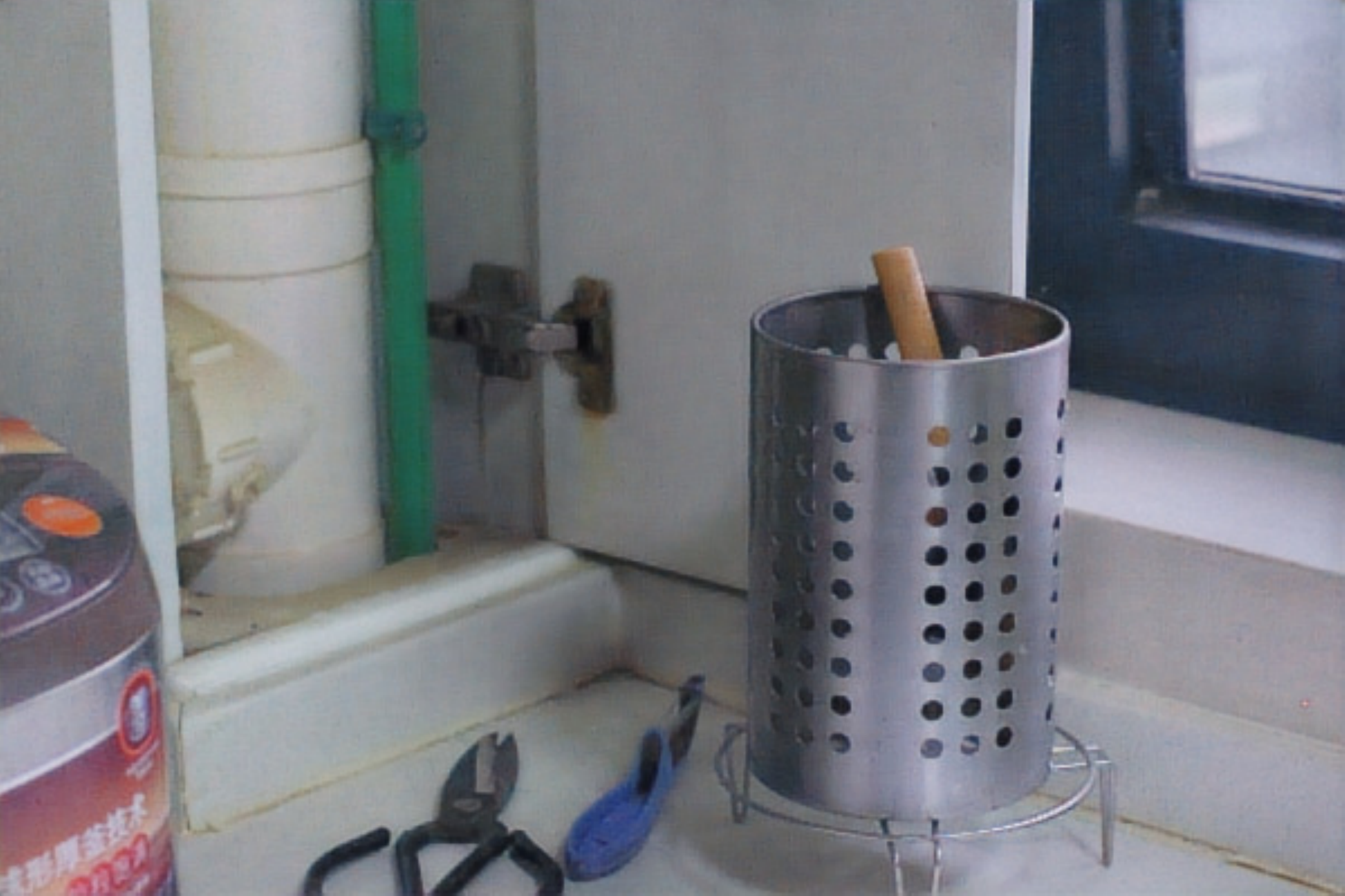}\vspace{2pt} \\
			\includegraphics[width=2.8cm]{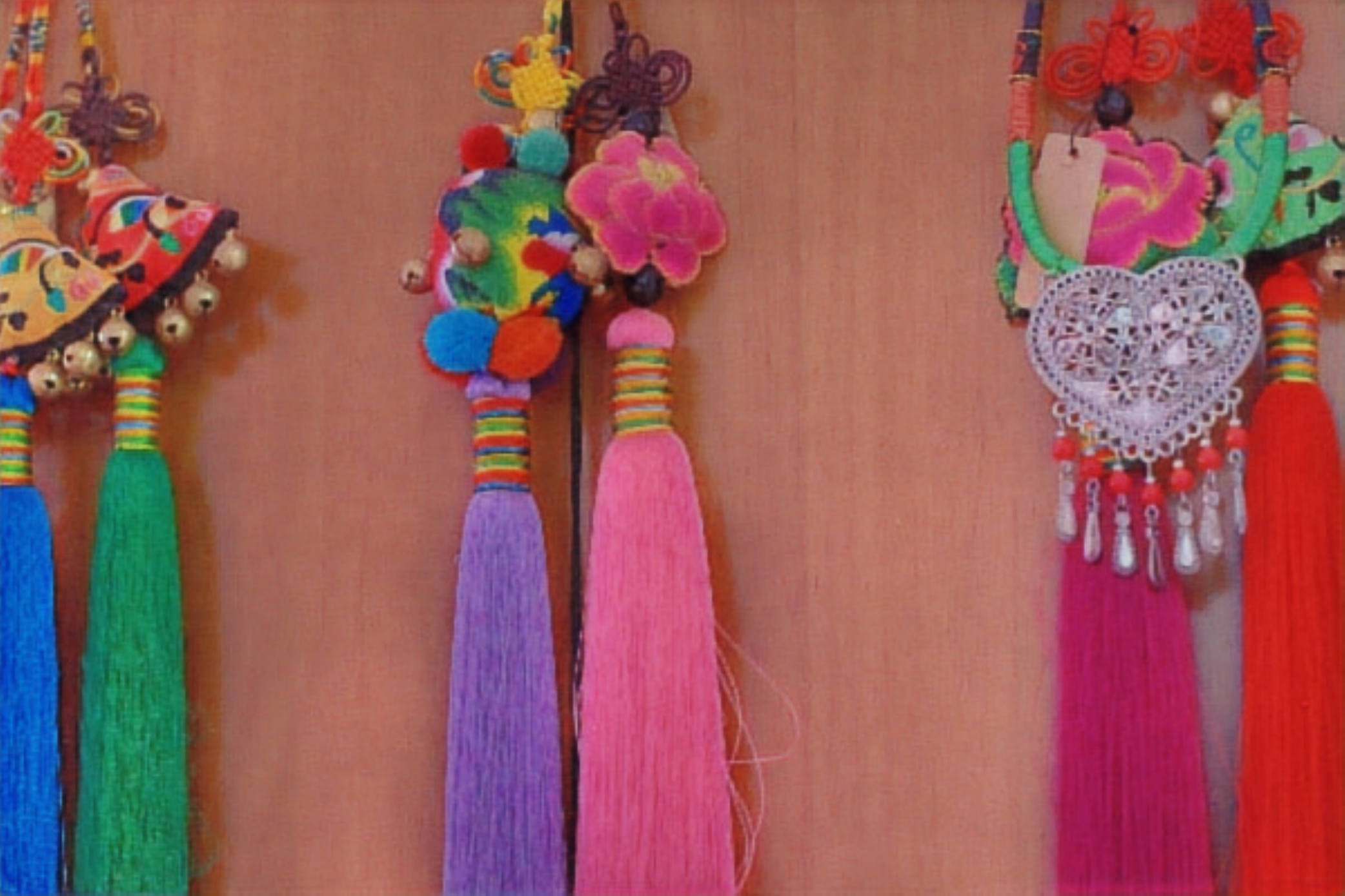}\vspace{-15pt} \\
		\end{minipage}
	}\hspace{-5pt}
	\caption{Ablation study and visual comparison of Perceptual Loss in the enhancement stage on the \textbf{LOL real-world evaluation} dataset.}
	\label{pic_perloss}
\end{figure}

We find that the obtained output is rather rough, and the details and edges information are still lost partly. Therefore, we use gradient loss $L_{grad}$ to maintain the balance of the sharpness and smoothness of the output.
\begin{equation}
\begin{aligned}
L_{grad}=\left \| \triangledown_{h}(R_{low}\ast I_{output})-\triangledown_{h}(S_{high})  \right \|_{1}
\\+ \left \| \triangledown_{v}(R_{low}\ast I_{output})-\triangledown_{v}(S_{high})  \right \|_{1}
\end{aligned}
\end{equation}
where $\triangledown_{h}$, $\triangledown_{v}$ denote the gradients in the horizontal and vertical directions, respectively. $S_{high}$ is the normal-light Ground-Truth.\\
The channel number of the enhancement net is four. Like RetinexNet \cite{wei2018deep}, we also concatenate the illumination and reflectance of the low-light image decomposed by the decomposition net as the four-channel input of the enhancement net. 
Because both the illumination and the reflectance contain high-frequency details and texture information after decomposition, concatenating the illumination and reflectance as the input of our enhancement net can help the enhancement net reconstruct the details and texture information and better estimate the illumination map. 
In addition, the output of the enhancement net is the corresponding brightness-improved single-channel illumination map. Finally, we multiply the output illuminantion map with the noise-removed reflectance map at the pixel-wise level and obtain the final output, which is noise-free, detail-preserved, color-restored and brightness-improved. According to Retinex Theory\cite{land1977retinex}, the three channels of the reflectance map share the same illumination map, so we copy the single-channel illumination map three times in order to make it a  three-channel illumination map and then multiply by the reflectance map.

\section{Experimental Evaluation}
\subsection{Implementation details}
We train our model on the LOL \cite{wei2018deep} real-world and synthetic training datasets individually and evaluate it on the LOL real-world and synthetic validation datasets. In addition, we test our model on four popular test datasets: LIME \cite{guo2016lime}, DICM \cite{lee2013contrast}, MEF\cite{ma2015perceptual} and NPE\cite{wang2013naturalness} datasets. The LOL dataset consists of two parts: the real-world dataset and the synthetic dataset. The LOL real-world dataset contains 500 low-light/normal-light image pairs captured by adjusting the exposure time and ISO of the camera, the lightness of them is very low and the degradation are severe. We use 485 image pairs as the training dataset and the remaining 15 image pairs as the validation dataset. The LOL synthetic dataset contains 1000 low-light/normal-light image pairs obtained by artificially synthesizing, and over/under-exposure conditions exist in the synthetic dataset. We use 900 image pairs as the training set and the remaining 100 image pairs as the validation set. We use the Adam \cite{kingma2014adam} optimizer to optimize the training of the model and set the training batch-size to four and the patch-size of random crop to 384x384. We use the PyTorch framework to build our model on a PC with an Nvidia TITAN XP GPU and an Intel Core i7-9700 3.00GHz CPU.\\
We evaluate our method on the LOL \cite{wei2018deep} validation dataset and test it on several widely used datasets, including LIME \cite{guo2016lime}, DICM \cite{lee2013contrast}, MEF \cite{ma2015perceptual} and NPE\cite{wang2013naturalness} datasets. We adopt PSNR, SSIM \cite{wang2004image}, LPIPS \cite{zhang2018unreasonable}, FSIM \cite{zhang2011fsim},  UQI \cite{wang2002universal}, Signal to Reconstruction Error Ratio (SRER), Root-MSE (RMSE) and Spectral Angle Mapper (SAM)\cite{de2000spectral} as the quantative metrics to measure the performance of our method. In addition, we use the mean and median value of Angular Error \cite{hordley2004re} and their average as well as  DeltaE\cite{sharma2005ciede2000} as the metrics of color distortion to calculate the color bias between our results and Ground-Truth. 

\begin{figure}[htbp]
	\includegraphics[width=9cm]{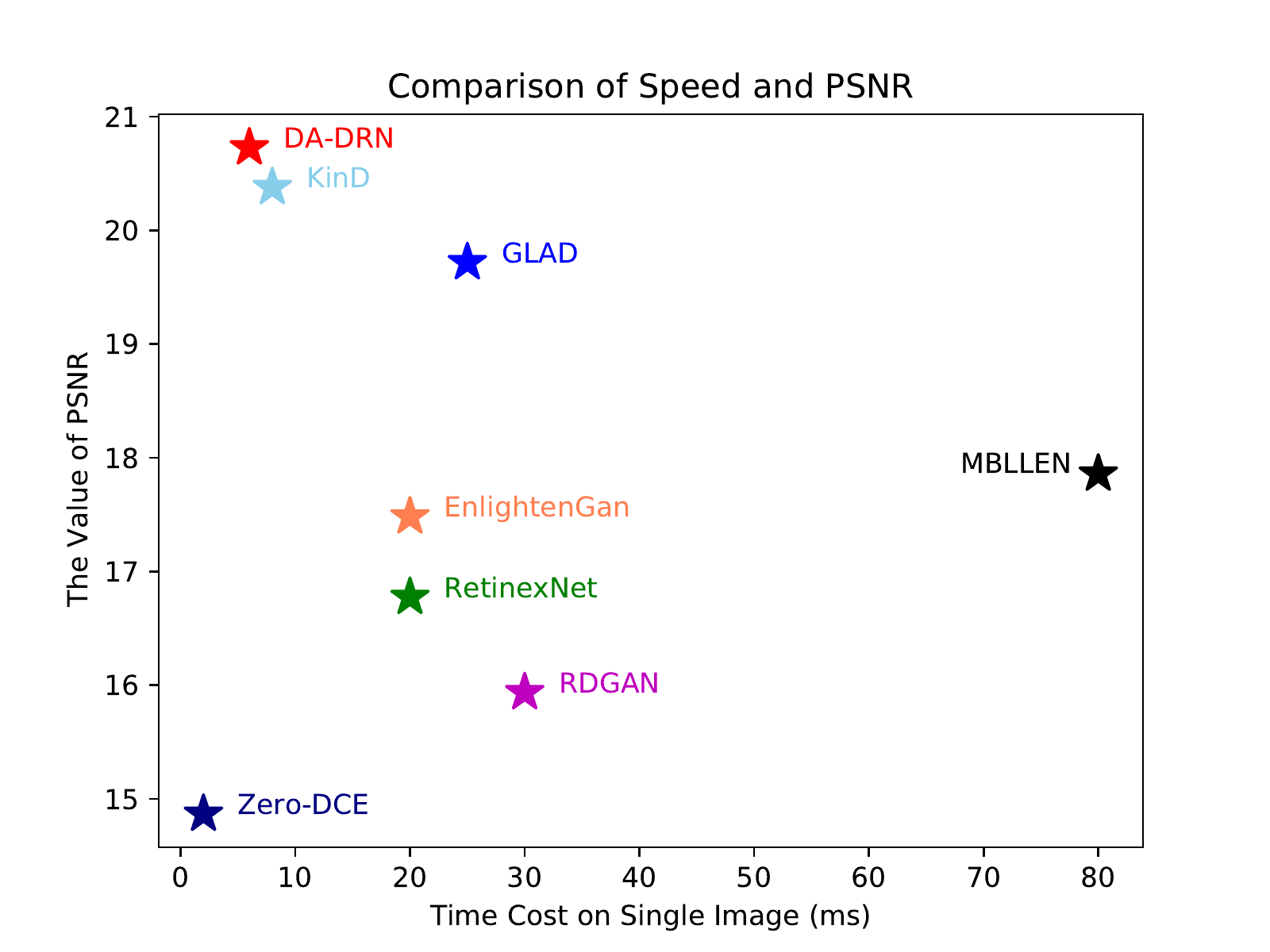}
	\caption{Runtime cost and performance comparison of our method and other state-of-the-art deep learning methods on the LOL Real-world dataset.}
	\label{speed}
\end{figure}

\begin{table*}[htbp] 
	\centering \caption{Quantitative comparison of several metrics between our method
		and other state-of-the-art methods on \textbf{LOL REAL-WORLD} dataset. Mean, Median and Avg represent the mean and median values of the Angular Error and the average value of them, respectively. “↑” indicates the higher the better, “↓” indicates the lower the	better. \color{red}Red: the best, \color{blue}Blue: the second best.}
	\begin{tabular}{  c | c  c  c  c  c  c  c  c | c  c  c  c }
		\hline  TD Methods  & PSNR↑   & SSIM↑  & LPIPS↓  &FSIM↑   & UQI↑  & SRER↑  & RMSE↓  & SAM↑   & Mean↓   & Median↓   &Avg↓  &DeltaE↓ \\
		\hline
		\hline  Input & 7.7733 & 0.1914 & 0.4173 &0.7190   &0.0622  & 47.5772 & 0.0264 & 76.5801  &3.8061 & 3.9728 & 3.8895 & 76.5837\\
		MSRCR \cite{jobson1997multiscale}  & 13.1728 & 0.4615 & 0.4404 &0.8450   &0.7884 & 50.4277 & 0.0147 & 87.5276 & 3.7421 & 4.5877 & 4.1649 & 27.4496\\
		BIMEF \cite{ying2017bio}  & 13.8752 & 0.5949 & 0.3673 &0.9263   &0.7088 & 50.6598 & 0.0141 & 86.6089 & 3.4004 & 3.5187 & 3.4595 & 33.8820\\
		LIME \cite{guo2016lime}  & 16.7586 & 0.4449 & 0.4183 &0.8549   &0.8805 & 52.1989 & 0.0094 & 86.9102 
		& 3.2096 & 4.0825 & 3.6460  & 21.1816\\
		Dong \cite{dong2011fast} & 16.7165 & 0.4783 & 0.4226 &0.8886   & 0.8078 & 52.1309 & 0.0101 & 86.2800 & 3.3499 & 4.1481 & 3.7490 & 25.3349\\
		SRIE \cite{fu2016weighted}  & 11.8552 & 0.4954 & 0.3657 &0.9085   &0.5033 & 49.6370 & 0.0173 & 85.0271 & 3.4488 & 4.0751 & 3.7620 & 44.3194\\
		MF \cite{fu2016fusion}  & 16.9662 & 0.5075 & 0.4092 &0.9236   &0.8572 & 52.2864 & 0.0099 & 86.5465 & 3.3277 & 3.9810 & 3.6544 & 24.5488\\
		NPE \cite{wang2013naturalness}  & 16.9697 & 0.4839 & 0.4156 &0.8964   &0.8943 & 52.2944 & 0.0093 & 87.0226 & 3.5588 & 4.2505 & 3.9046 & 22.6374\\
		LECARM \cite{ren2018lecarm}  & 14.4099 & 0.5448 & 0.3687 &0.9288   &0.6406 & 50.9764 & 0.0133 & 85.1824 & 3.4091 & 3.9668 & 3.6979 & 34.5143\\
		CRM \cite{ying2017new}  & 17.2032 & 0.6229 & 0.3748 & \color{blue}0.9456   &0.8441 & 52.4903 & 0.0099 & 87.0542 & 3.4396 & 3.6790 & 3.5993 & 23.7405\\
		\hline
		\hline  DL Methods  & PSNR↑   & SSIM↑   & LPIPS↓  &FSIM↑  & UQI↑  & SRER↑  & RMSE↓  & SAM↑   & Mean↓   & Median↓   &Avg↓  &DeltaE↓ \\
		\hline
		\hline
		MBLLEN \cite{lv2018mbllen}  & 17.8583 & 0.7247 & 0.3672 &0.9262   &0.8261 & 52.7664 & 0.0086 & 86.1212 & 3.2716 & 4.4620 & 3.8669 & 21.5774\\
		RetinexNet \cite{wei2018deep}  & 16.7740 & 0.4249 & 0.4670 &0.8642   &0.9110 & 52.2075 & 0.0094 & \color{blue}88.2461 & 3.7501 & 4.4975 & 4.3589 & 21.3550\\
		GLAD \cite{wang2018gladnet}  & 19.7182 & 0.6820 & 0.3994 & 0.9329   &0.9204 & 53.7990 & 0.0070 & 88.2170  & 3.3110 & 3.8021 & 3.5565 & 16.0393\\
		RDGAN \cite{wang2019rdgan}  & 15.9363 & 0.6357 & 0.3985 &0.9276   &0.8296 & 51.7681 & 0.0114 & 87.4576 & 4.3899 & 5.3027 & 4.8463 & 26.3796\\
		Zero-DCE \cite{guo2020zero}  & 14.8671 & 0.5623 & 0.3852 &0.9276   &0.7205 & 51.2269 & 0.0126 & 85.9968 & 4.1051 & 4.6860 & 4.3955 & 31.4451\\
		EnlightenGan \cite{jiang2021enlightengan}  & 17.4828 & 0.6515 & 0.3903 &0.9226  &0.8499 & 52.5934 & 0.0095 & 87.7195 & 4.5296 & 5.2536 & 4.8916 & 21.9113\\
		KinD \cite{zhang2019kindling}  & \color{blue}20.3792 & \color{red}0.8056 & \color{red}0.2711 &0.9397  &\color{blue}0.9250 & \color{blue}54.1233 & \color{blue}0.0066 & 87.5607 & \color{blue}2.2947 & \color{blue}2.6376 & \color{blue}2.4662 & \color{blue}13.9618\\
		\hline 
		\hline 
		DA-DRN w/o DA & 18.4446 & 0.7605 & 0.3514 &0.9261   &0.9259 & 53.0684 & 0.0078 & 88.0573 & 2.8645 & 3.9083 & 3.3864 & 16.0604\\
		DA-DRN  & \color{red}20.7282 & \color{blue}0.7939 & \color{blue}0.3126 &\color{red}0.9458   &\color{red}0.9378  &  \color{red}54.1478 & \color{red}0.0061 & \color{red}88.2747 &\color{red}2.1638 & \color{red}2.3149 & \color{red}2.2394 & \color{red}12.9350\\
		\hline\end{tabular}\vspace{0cm}
	\label{lol_real}
\end{table*}

\begin{table*}[htbp] 
	\centering \caption{Quantitative comparison of several metrics between our method
		and other state-of-the-art methods on \textbf{LOL SYNTHETIC} dataset. Mean, Median and Avg represent the mean and median values of the Angular Error and the average value of them, respectively. “↑” indicates the higher the better, “↓” indicates the lower the	better. \color{red}Red: the best, \color{blue}Blue: the second best.}
	\begin{tabular}{  c | c  c  c  c  c  c  c  c | c  c  c  c }
		\hline  TD Methods  & PSNR↑   & SSIM↑  & LPIPS↓  &FSIM↑   & UQI↑  & SRER↑  & RMSE↓  & SAM↑   & Mean↓   & Median↓   &Avg↓  &DeltaE↓ \\
		\hline
		\hline  Input & 10.2533 & 0.4193 & 0.2871 &0.7802   &0.3502  & 48.9243 & 0.0112 & 77.5350  &3.2315 & 3.1539 & 3.1927 & 51.9337\\
		MSRCR \cite{jobson1997multiscale}  & 15.3540 & 0.8070 & 0.2328 &0.9213   &0.6497 & 52.1708 & 0.0113 & 87.6335 & 6.9882 & 8.6362 & 7.8122 & 22.7014\\
		BIMEF \cite{ying2017bio}  & 14.9942 & 0.768 & 0.1831 &0.9115   &0.7850 & 51.3881 & 0.0122 & 85.5286 & 3.9513 & 4.5929 & 4.2721 & 23.9757\\
		LIME \cite{guo2016lime}  & 17.0682 & 0.7606 & 0.2040 &0.8617   &0.8804 & 52.3908 & 0.0092 & 85.8743 
		& 3.2096 & 4.0825 & 3.6460  & 21.1816\\
		Dong \cite{dong2011fast} & 15.5510 & 0.7281 & 0.2511 &0.8549   & 0.8207 & 51.6197 & 0.0109 & 84.9840 & 2.4316 & 2.5599 & 2.4957 & 21.3182\\
		SRIE \cite{fu2016weighted}  & 12.1494 & 0.6252 & 0.2051 &0.8622   &0.6045 & 49.8886 & 0.0163 & 82.1820 & 2.9142 & 2.7892 & 2.8517 & 34.3793\\
		MF \cite{fu2016fusion}  & 15.3212 & 0.7683 & 0.1809 &0.9140   &0.7651 & 51.5230 & 0.0116 & 83.9633 & \color{blue}2.4751 & 2.4093 & 2.4422 & 23.5243\\
		NPE \cite{wang2013naturalness}  & 14.6603 & 0.7724 & 0.1866 &0.9036   &0.7921 & 51.1505 & 0.0123 & 85.1261 & 2.4371 & 2.6856 & 2.5614 & 23.4608\\
		LECARM \cite{ren2018lecarm}  & 16.9126 & 0.7868 & 0.1894 &0.8936   &0.7787 & 52.3143 & 0.0098 & 84.2295 & 3.0717 & 3.2291 & 3.1854 & 22.9359\\
		CRM \cite{ying2017new}  & 14.9942 & 0.7689 & 0.1831 &0.9115   &0.7850 & 51.3881 & 0.0122 & 85.5286 & 3.9513 & 4.5929 & 4.2721 & 23.9757\\
		\hline
		\hline  DL Methods  & PSNR↑   & SSIM↑   & LPIPS↓  &FSIM↑  & UQI↑  & SRER↑  & RMSE↓  & SAM↑   & Mean↓   & Median↓   &Avg↓  &DeltaE↓ \\
		\hline
		\hline
		MBLLEN \cite{lv2018mbllen}  & 14.2620 & 0.6552 & 0.2903 &0.9039   &0.7013 & 50.9951 & 0.0132 & 84.1075 & 2.5991 & 3.1658 & 2.8825 & 27.5349\\
		RetinexNet \cite{wei2018deep}  & \color{blue}17.2025 & 0.7639 & 0.2467 &0.8639   &0.8888 & 52.4594 & 0.0095 & \color{red}88.1026 & \color{blue}1.7625 & 2.7897 & \color{blue}2.2761 & 18.2853\\
		GLAD \cite{wang2018gladnet}  & 16.2292 & 0.8007 & 0.1888 & 0.9378   &0.8406 & 52.1234 & 0.0105 & 86.2192  & 3.6618 & 3.8868 & 3.7743 & 19.4709\\
		RDGAN \cite{wang2019rdgan}  & 18.2270 & \color{blue}0.8368 & 0.1706  &\color{blue}0.9415   &\color{blue}0.8971 & \color{blue}53.1087 & \color{blue}0.0084 & 87.1588 & 3.1857 & 3.4799 & 3.3329 & \color{blue}16.6347\\
		Zero-DCE \cite{guo2020zero}  & 16.5206 & 0.8173 & 0.1772 &0.9256   &0.8150 & 52.2576 & 0.0102 & 85.2074 & 4.0482 & 3.6468 & 3.8475 & 21.8502\\
		EnlightenGan \cite{jiang2021enlightengan}  & 15.2653 & 0.7516 & 0.1754 &0.8947  &0.7953 & 51.4678 & 0.0117 & 85.9107 & 3.0516 & 3.8443 & 3.4479 & 22.0353\\
		KinD \cite{zhang2019kindling}  & 16.2156 & 0.8173 & \color{red}0.1457 &0.9306  &0.8257 & 51.9733 & 0.0102 & 85.5904 & 1.7839 &3.1954  & 2.4897 & 18.7326\\
		\hline 
		\hline 
		DA-DRN w/o DA & 16.5855 & 0.8028 & 0.2425 &0.9213   &0.9019 & 52.1280 & 0.0096 & 86.8376 & 2.2544 & 2.4824 & 2.3684 & 17.1552\\
		DA-DRN  & \color{red}20.5360 & \color{red}0.8388 & \color{blue}0.1691 &\color{red}0.9549   &\color{red}0.9359  &  \color{red}54.1278 & \color{red}0.0063 & \color{blue}87.2474 &\color{red}1.5082 & \color{red}1.7007 & \color{red}1.6045 & \color{red}12.3788\\
		\hline\end{tabular}\vspace{0cm}
	\label{lol_syn}
\end{table*}

\begin{table*}[htbp] 
	\centering \caption{Quantitative comparison of several metrics between our method
		and other state-of-the-art methods on \textbf{LIME, DICM, MEF and NPE} datasets. *(IL-NIQE↓\cite{zhang2015feature} / PIQE↓\cite{venkatanath2015blind} / Noise Level↓\cite{liu2013single}) .“↓” indicates the lower the	better. \color{red}Red: the best, \color{blue}Blue: the second best.}
	\begin{tabular}{  c | c | c | c | c }
		\hline  TD Methods  & LIME   & DICM   & MEF  &NPE \\
		\hline
		\hline  Input & 31.2672 / 42.9709 / 0.7603 & 32.1408 / 29.1601 / 1.2393 & 39.9522 / 44.4448 / 0.8500  & 25.6812 / 33.9167 / 1.2874  \\
		MSRCR \cite{jobson1997multiscale}  & 28.9141 / 40.3416 / 2.8542 & 22.2660 / 38.6280 / 5.9468 & 25.7093 / 44.7977 / 7.6916 &25.7093 / 44.7977 / 7.6916 \\
		BIMEF \cite{ying2017bio}  & 29.3937 / 35.7954 / 1.2679 & 22.9029 / 28.5598 / \textcolor{blue}{0.4283} & 30.2226 / 39.1439 / 3.5885 &31.5463 / 40.0646 / 3.1890 \\
		LIME \cite{guo2016lime}  & 32.6487 / 41.0324 / 2.2140 & 26.9526 / 37.7184 / 4.0929 & 32.8506 / 46.5014 / 8.8760 &31.5463 / 40.0646 / 3.1890 \\
		Dong \cite{dong2011fast} & 30.4354 / 38.4488 / 1.6179 & 27.1374 / 29.8391 / 0.6409 & 32.0582 / 40.4559 / 5.2508 & 27.7929 / 38.1724 / 0.8549\\
		SRIE \cite{fu2016weighted}  & 29.2130 / 34.7993 / 1.0342 & 25.4863 / 29.6566 / 0.3801 & 31.5463 / 40.0646 / 3.1890 & 24.0372 / 37.3384 / 0.8622\\
		MF \cite{fu2016fusion}  & 30.0656 / 36.4530 / 1.6531 & 25.1276 / 29.4292 / 0.8224 & 30.4437 / 40.9272 / 7.1779 & 24.5837 / 38.4197 / 0.9140 \\
		NPE \cite{wang2013naturalness}  & 29.1324 / 36.4812 / 1.5852 & 24.8242 / 32.8927 / 1.4183 & 30.7461 / 42.2345 / 7.8979 & 24.1103 / 37.3556 / 0.9036\\
		LECARM \cite{ren2018lecarm}  & 28.9609 / 35.7329 / 1.5941 & 30.7963 / 30.7964 / 0.4826 & 29.9232 / 37.1885 / 3.1479 & 27.6458 / 39.8107 / 0.8936 \\
		CRM \cite{ying2017new}  &30.0496 / 37.0007 / 1.5248 &\textcolor{blue}{21.4971} / 34.3147 / 0.5682 & 29.5136 / 39.9534 / 4.2715 & \textcolor{blue}{23.7106} / 37.8869 / 1.6866\\
		\hline
		\hline  DL Methods  & LIME   & DICM   & MEF  &NPE \\
		\hline
		\hline
		MBLLEN \cite{lv2018mbllen}  & 33.5665 / 52.2802 / \textcolor{red}{0.4267} & 27.9186 / 39.3463 / 0.4659 & 32.9730 / 47.0705 / 0.6065 & 28.3009 / 45.0218 / 0.9039 \\
		RetinexNet \cite{wei2018deep}  & 33.5524 / 42.7741 / 3.1563 & 29.9806 / 41.6317 / 7.6714 & 26.0972 / 40.9106 / 6.7381 & 29.0044 / 39.3471 / 0.8639 \\
		GLAD \cite{wang2018gladnet}  & 26.9276 / 36.2863 / 1.3145 & 23.5602 / 33.3614 / 1.9812 & 24.5708 / 36.6039 / 2.5051 & 24.7363 / 34.4245 / 0.9378  \\
		RDGAN \cite{wang2019rdgan}  & 27.6757 / \textcolor{blue}{32.8818} / 1.4781  & 21.8475 / 34.9916 / 0.7022  & \textcolor{blue}{24.5320} / \textcolor{red}{28.3491} / 2.0782  &23.7579 / \textcolor{blue}{31.3491} / 0.9940  \\
		Zero-DCE \cite{guo2020zero}  & 28.6369 / 35.8671 / 1.6435 & 25.2122 / \textcolor{blue}{25.9288} / \textcolor{red}{0.4241} & 29.9260 / 36.5886 / 3.7810 & 24.3265 / 37.5726 / 0.9256\\
		EnlightenGan \cite{jiang2021enlightengan}  & \textcolor{blue}{25.3370} / 33.7214 / 0.9571 & 23.9541 / 28.0480 / 1.1907 & 26.6239 / \textcolor{blue}{31.3382} / 0.9139 & 24.3083 / 32.8102 / 0.8947 \\
		KinD \cite{zhang2019kindling}  & 25.3681 / 45.3092 / 0.6575 & 24.4079 / 45.6483 / 0.7183 & 31.1011 / 56.4110 / \textcolor{blue}{0.5167} & 25.0960 / 46.7505 / \textcolor{red}{0.5436} \\
		\hline 
		\hline 
		DA-DRN  & \textcolor{red}{25.1070} / \textcolor{red}{30.8348} / \textcolor{blue}{0.6196} & \textcolor{red}{21.2704} / \textcolor{red}{25.6733} / 0.5855 & \textcolor{red}{24.0172} / 36.2347 / \textcolor{red}{0.5068}  & \textcolor{red}{23.4195} / \textcolor{red}{30.4935} / \textcolor{blue}{0.6231} \\
		\hline\end{tabular}\vspace{0cm}
	\label{wogt}
\end{table*}

\begin{figure*}[htbp]
	\flushleft
	\subfigure[Input]{
		\begin{overpic}[scale=.205]{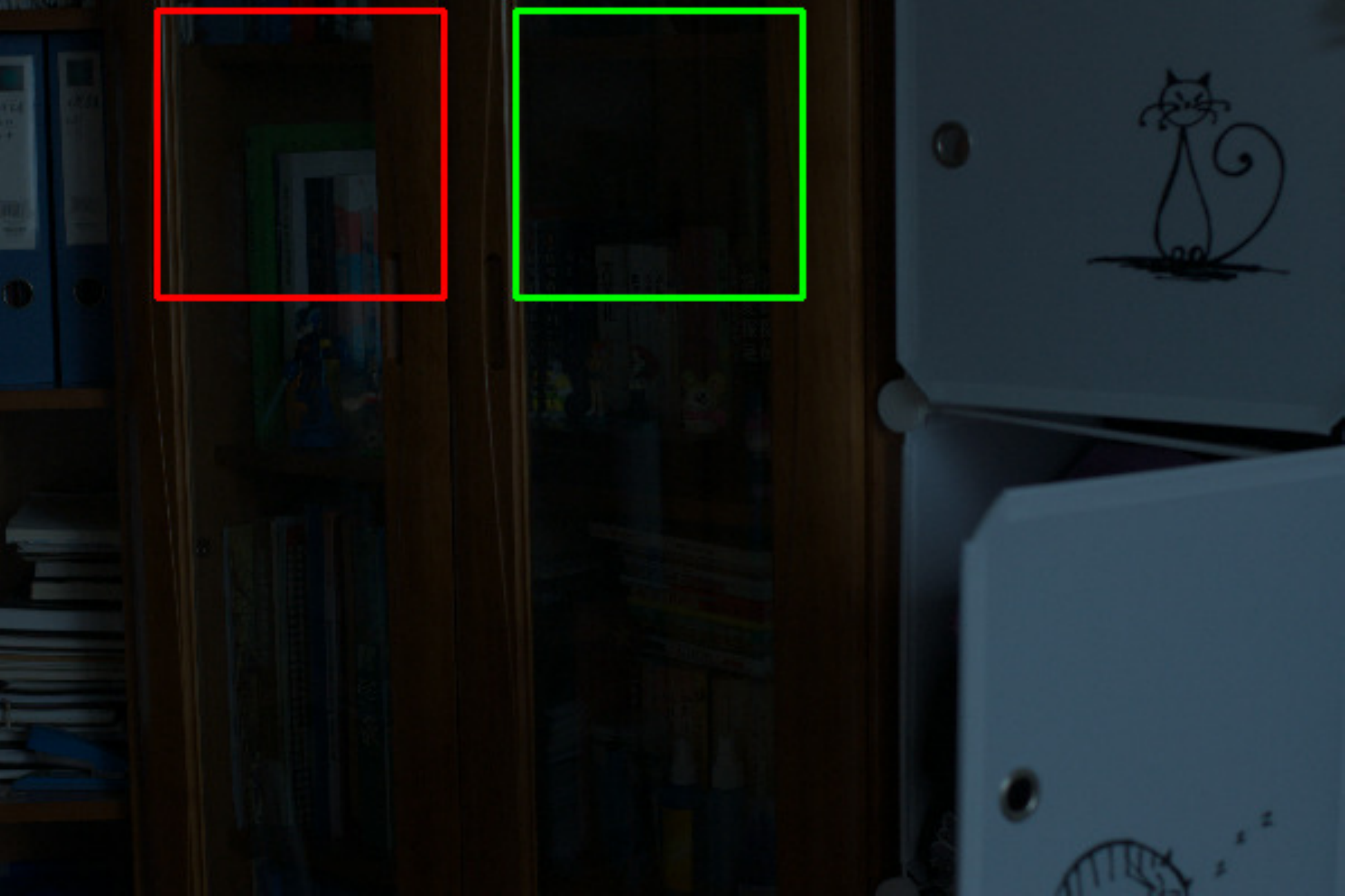}   		\put(70,0){\includegraphics[scale=.28]%
				{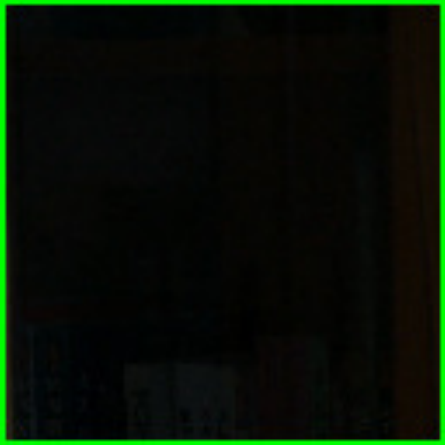}} 
		\end{overpic}
	}%
	\subfigure[MSRCR \cite{jobson1997multiscale}]{
		\begin{overpic}[scale=.205]{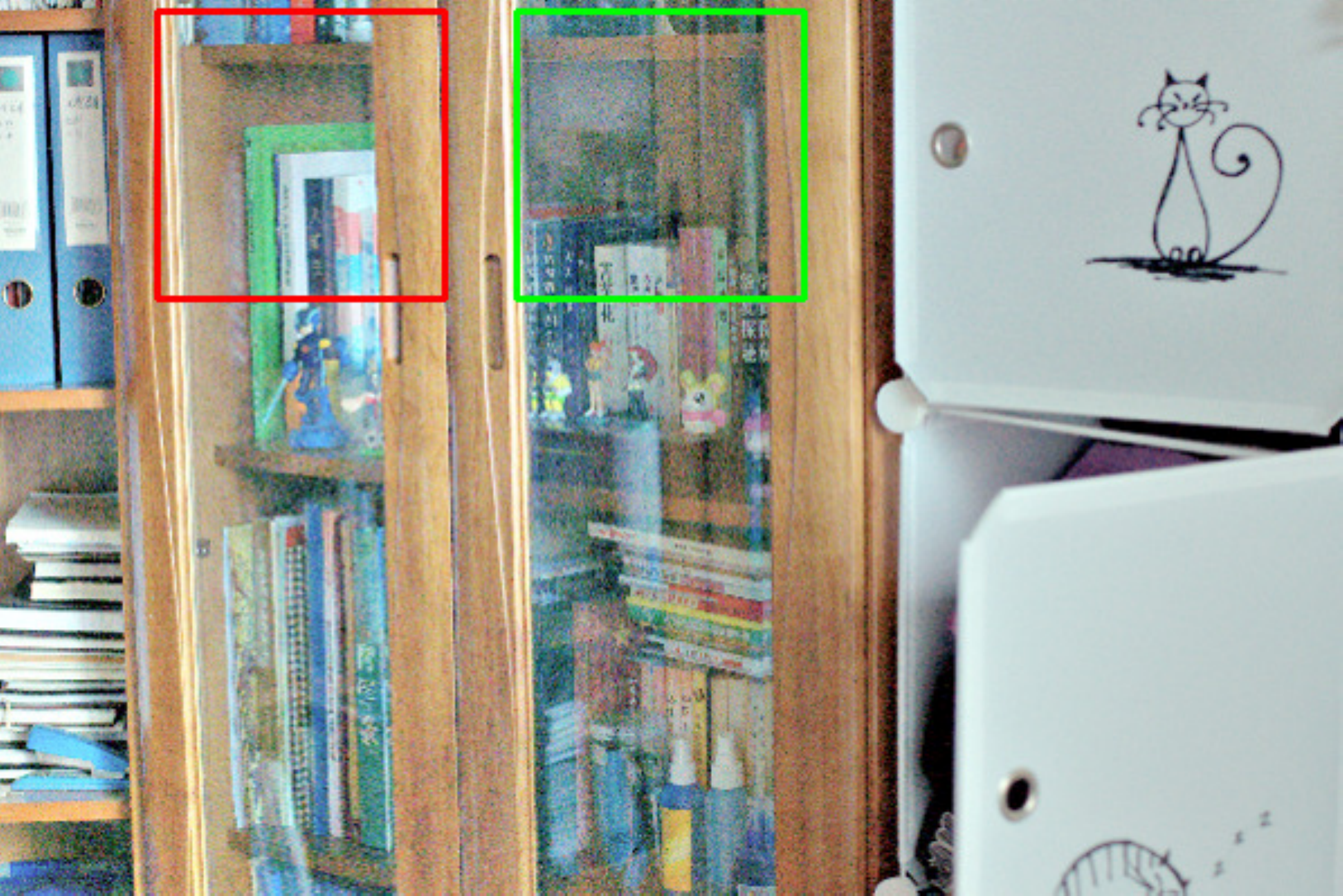}   		\put(70,0){\includegraphics[scale=.28]%
				{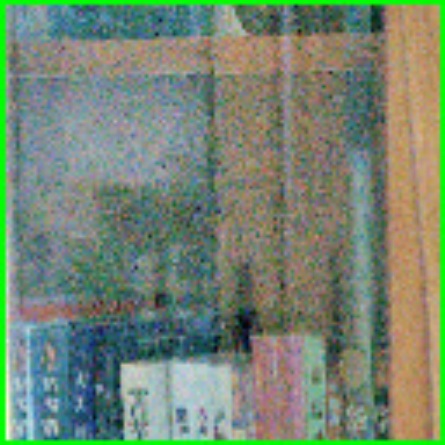}} 
		\end{overpic}
	}%
	\subfigure[BIMEF \cite{ying2017bio}]{
		\begin{overpic}[scale=.205]{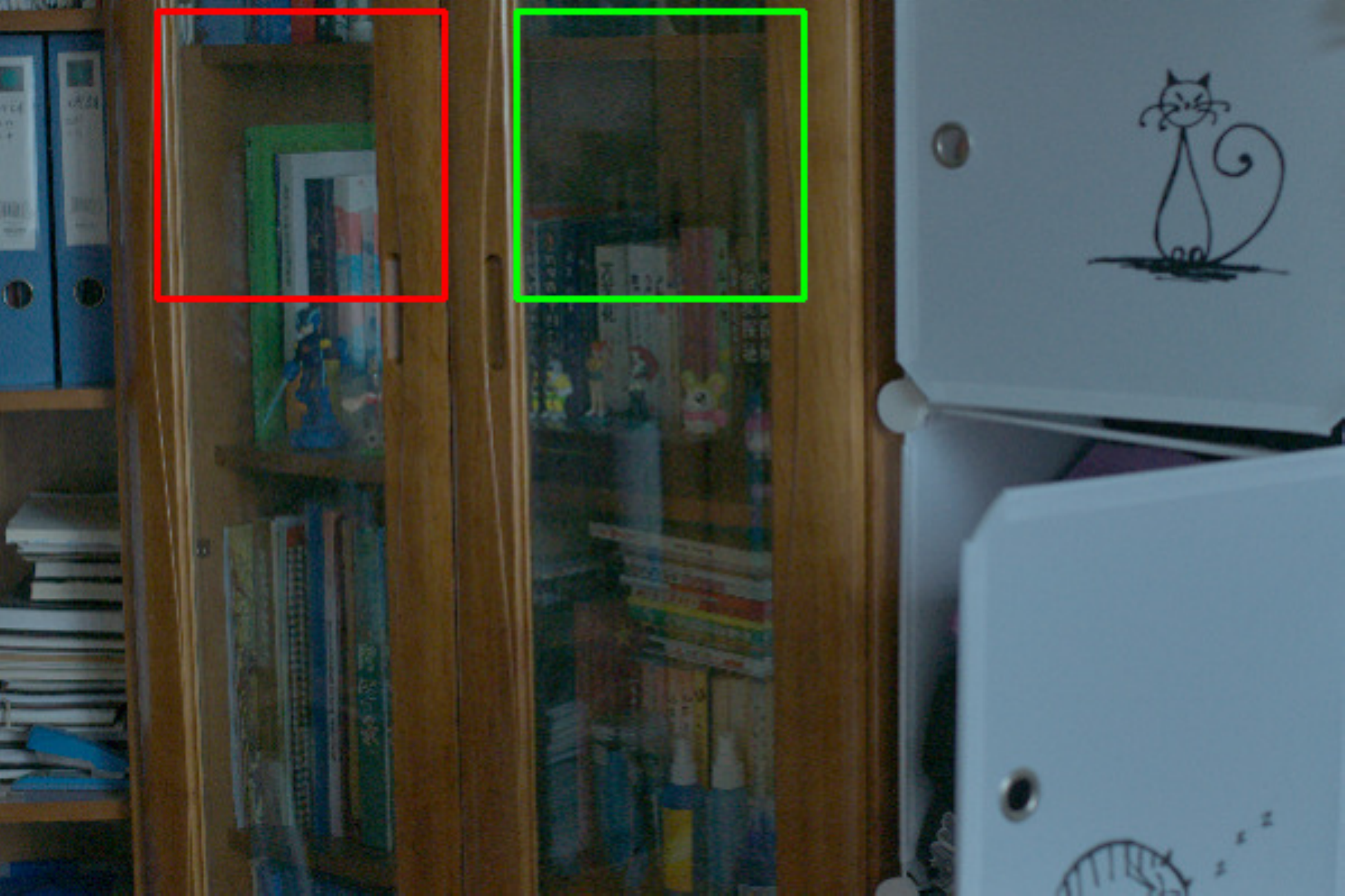}   		\put(70,0){\includegraphics[scale=.28]%
				{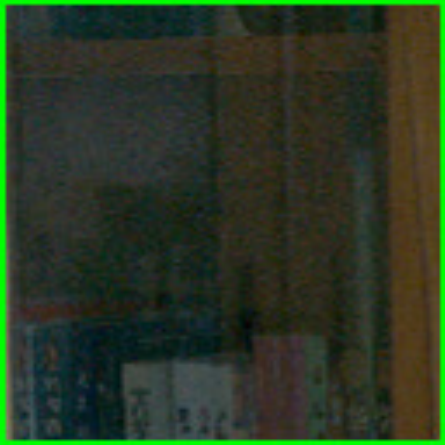}} 
		\end{overpic}
	}%
	\subfigure[LIME \cite{guo2016lime}]{
		\begin{overpic}[scale=.205]{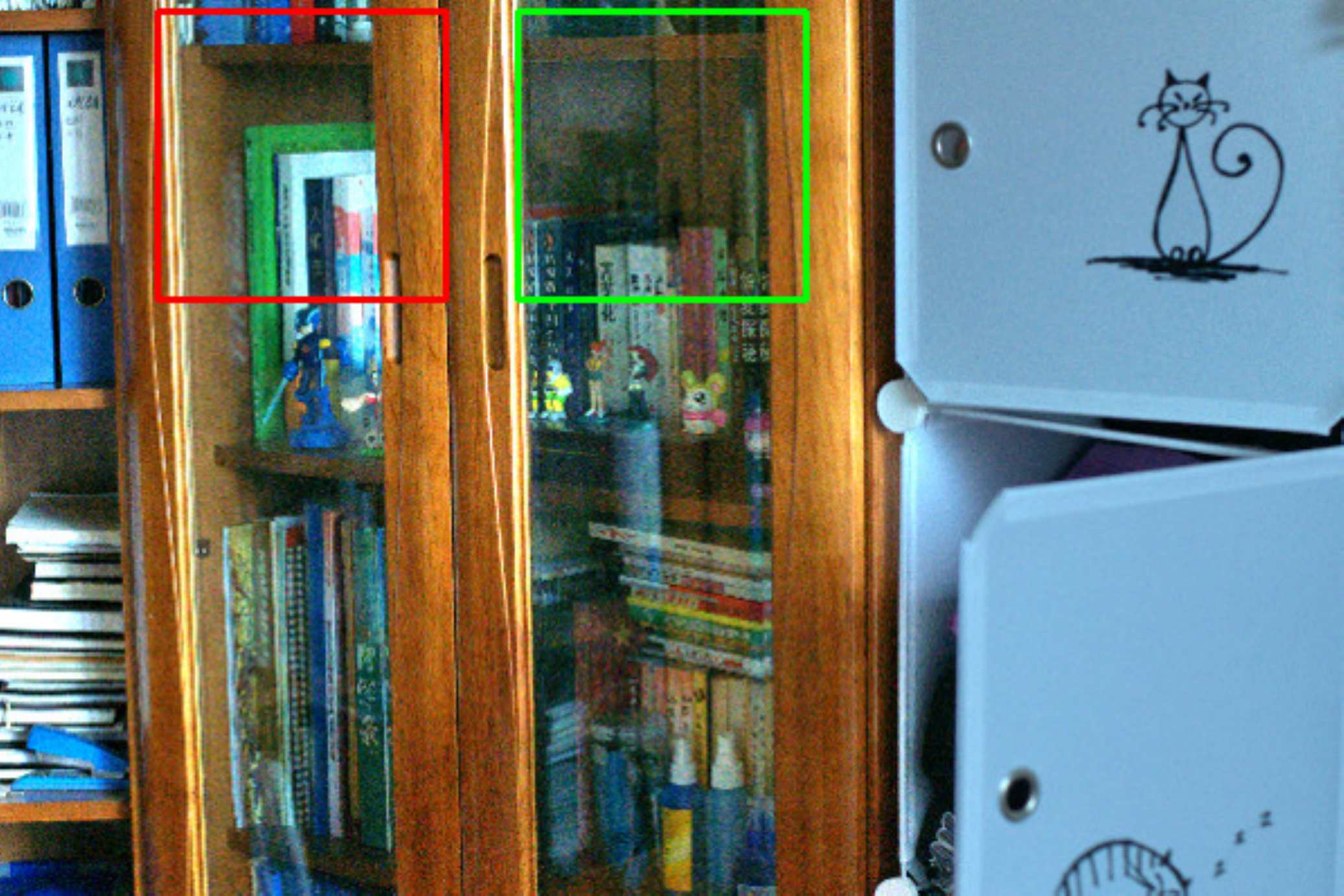}   		\put(70,0){\includegraphics[scale=.28]%
				{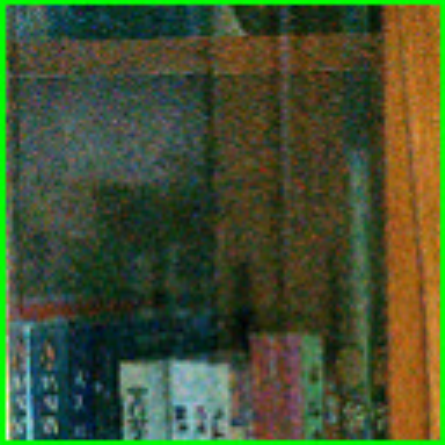}} 
		\end{overpic}
	}%
	
	\subfigure[Dong \cite{dong2011fast}]{
		\begin{overpic}[scale=.205]{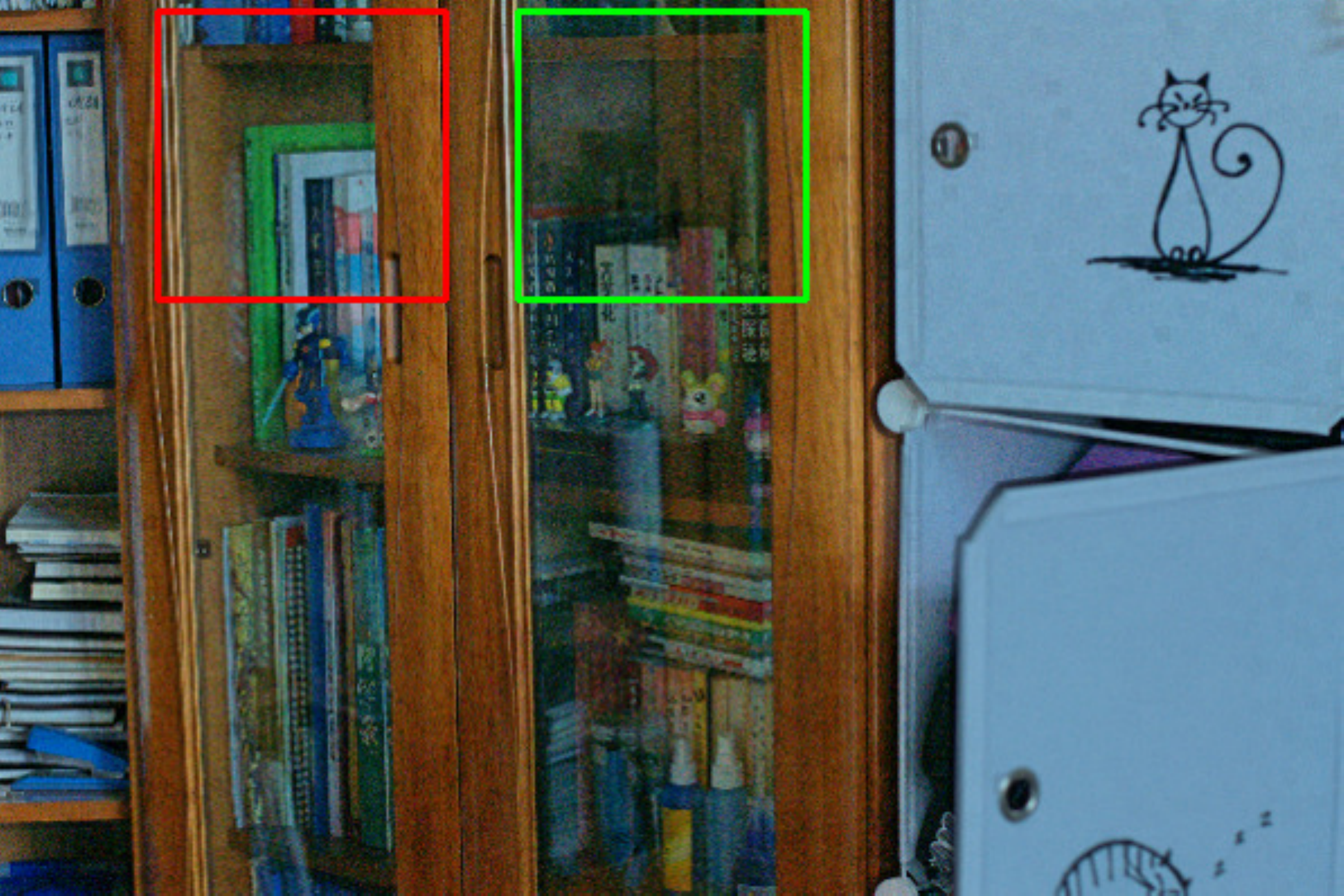}   		\put(70,0){\includegraphics[scale=.28]%
				{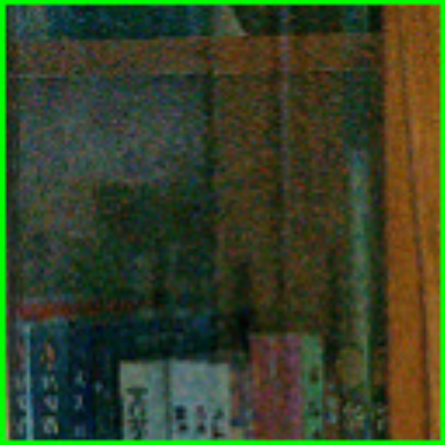}} 
		\end{overpic}
	}%
	\subfigure[SRIE \cite{fu2016weighted}]{
		\begin{overpic}[scale=.205]{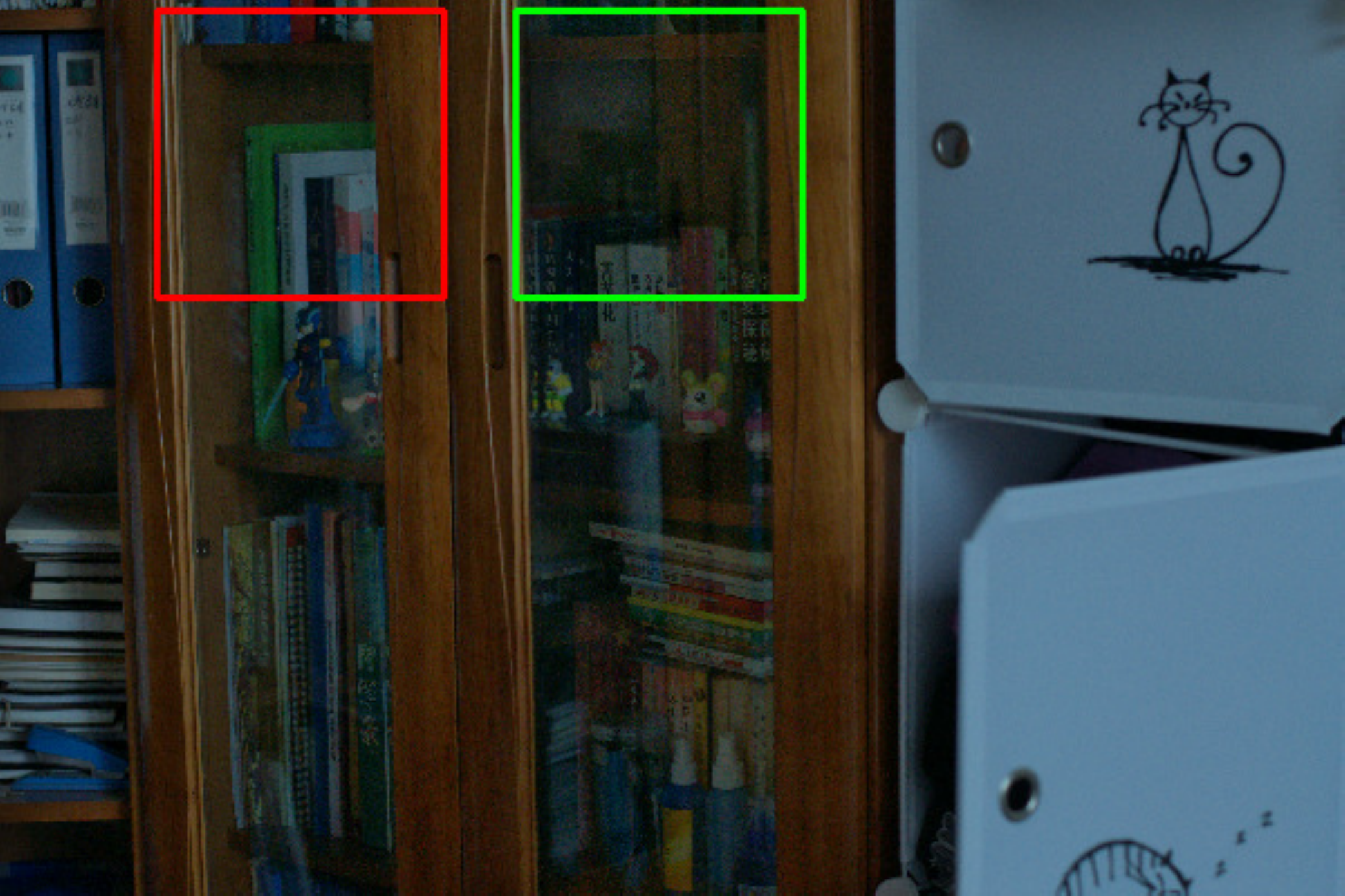}   		\put(70,0){\includegraphics[scale=.28]%
				{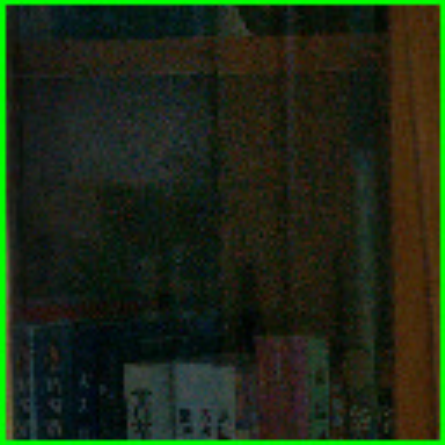}} 
		\end{overpic}
	}%
	\subfigure[MF \cite{fu2016fusion}]{
		\begin{overpic}[scale=.205]{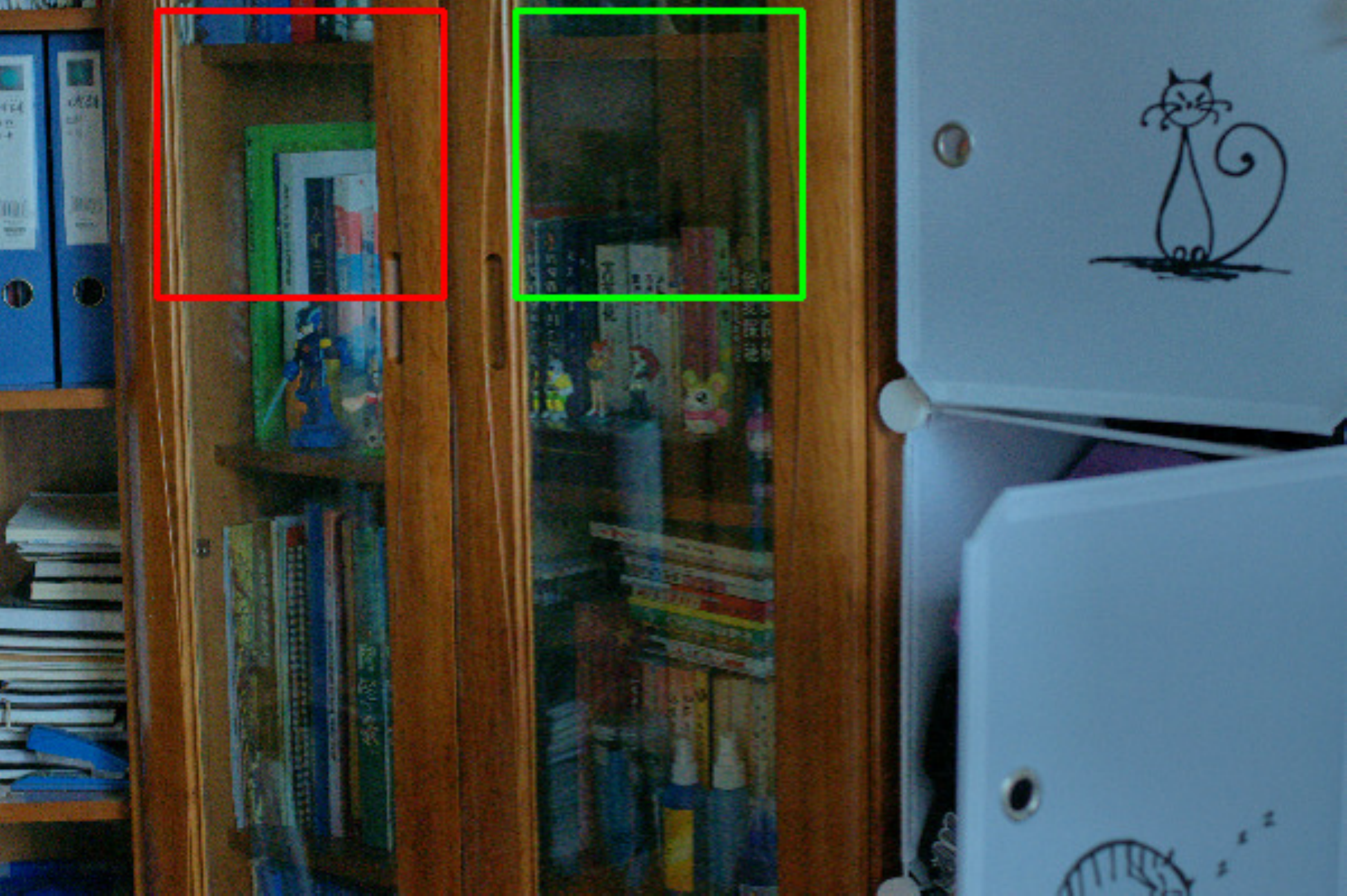}   		\put(70,0){\includegraphics[scale=.28]%
				{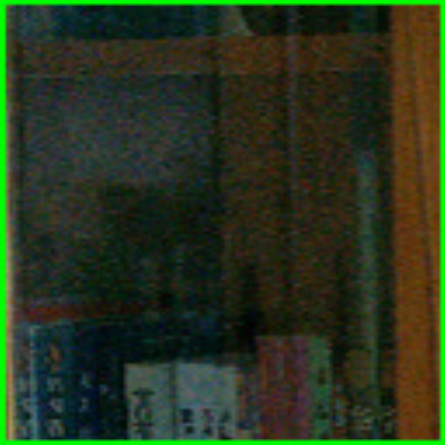}} 
		\end{overpic}
	}%
	\subfigure[NPE \cite{wang2013naturalness}]{
		\begin{overpic}[scale=.205]{pic/1/Dong-eps-converted-to.pdf}   		\put(70,0){\includegraphics[scale=.28]%
				{pic/1/Dong_1-eps-converted-to.pdf}} 
		\end{overpic}
	}%
	
	\subfigure[RRM \cite{li2018structure}]{
		\begin{overpic}[scale=.205]{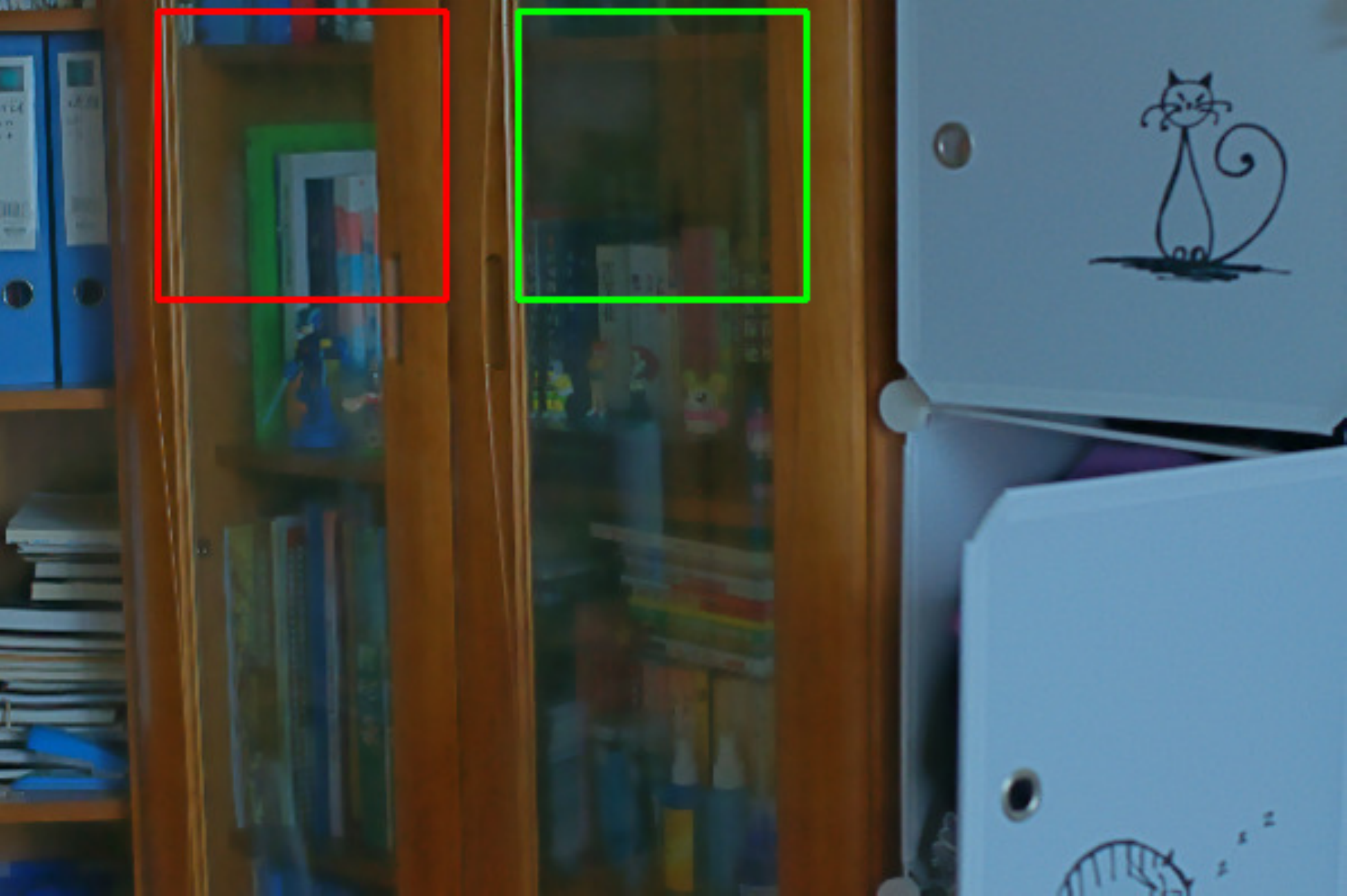}   		\put(70,0){\includegraphics[scale=.28]%
				{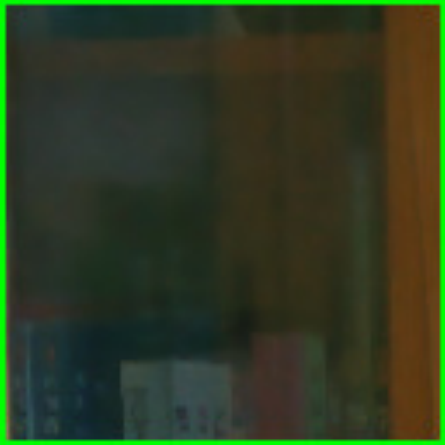}} 
		\end{overpic}
	}%
	\subfigure[MBLLEN \cite{lv2018mbllen}]{
		\begin{overpic}[scale=.205]{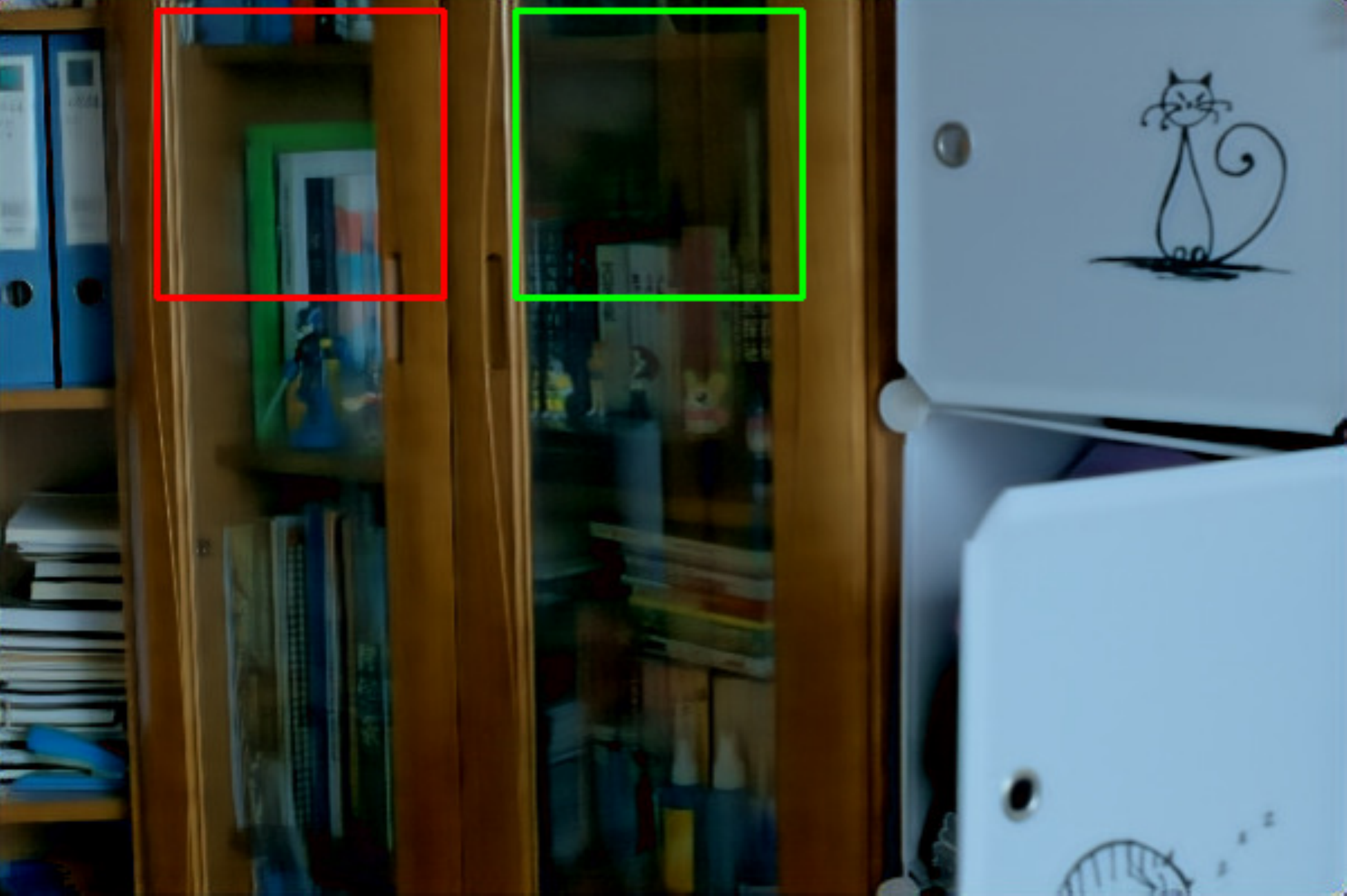}   		\put(70,0){\includegraphics[scale=.28]%
				{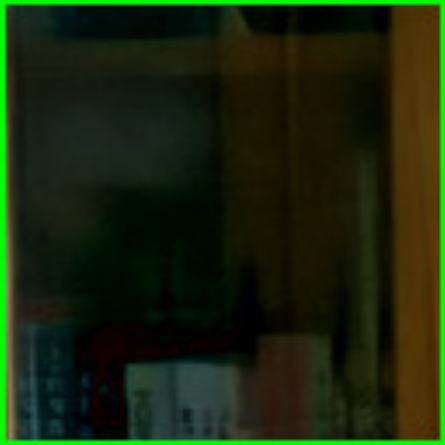}} 
		\end{overpic}
	}%
	\subfigure[RetinexNet \cite{wei2018deep}]{
		\begin{overpic}[scale=.205]{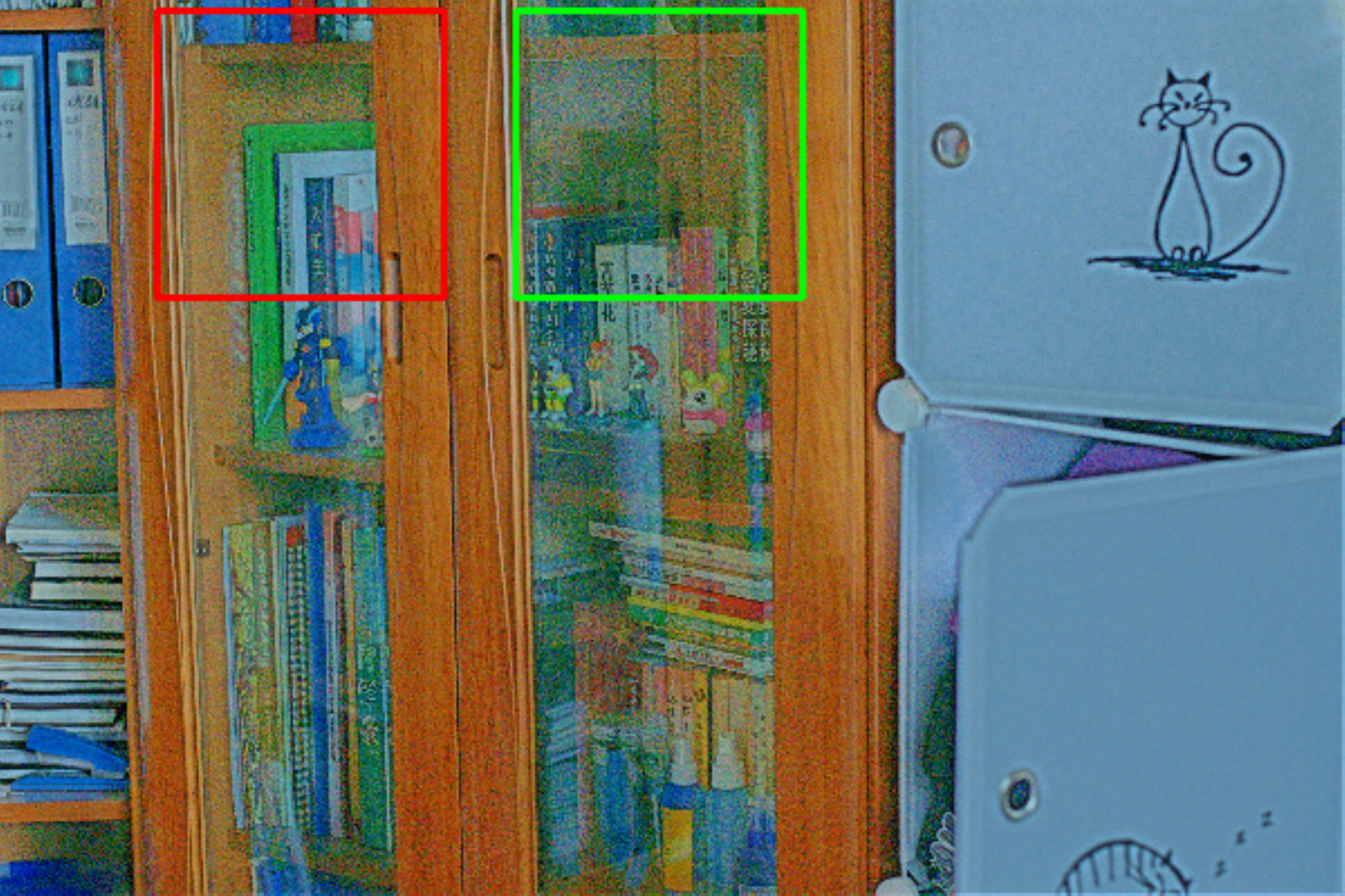}   		\put(70,0){\includegraphics[scale=.28]%
				{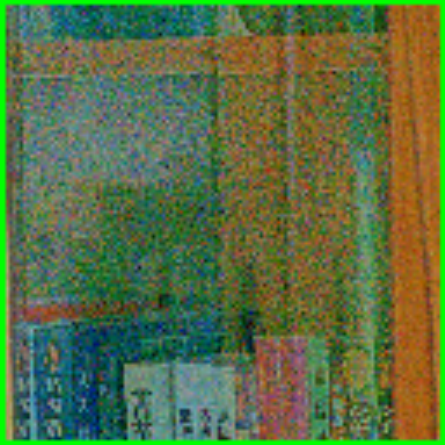}} 
		\end{overpic}
	}%
	\subfigure[GLAD \cite{wang2018gladnet}]{
		\begin{overpic}[scale=.205]{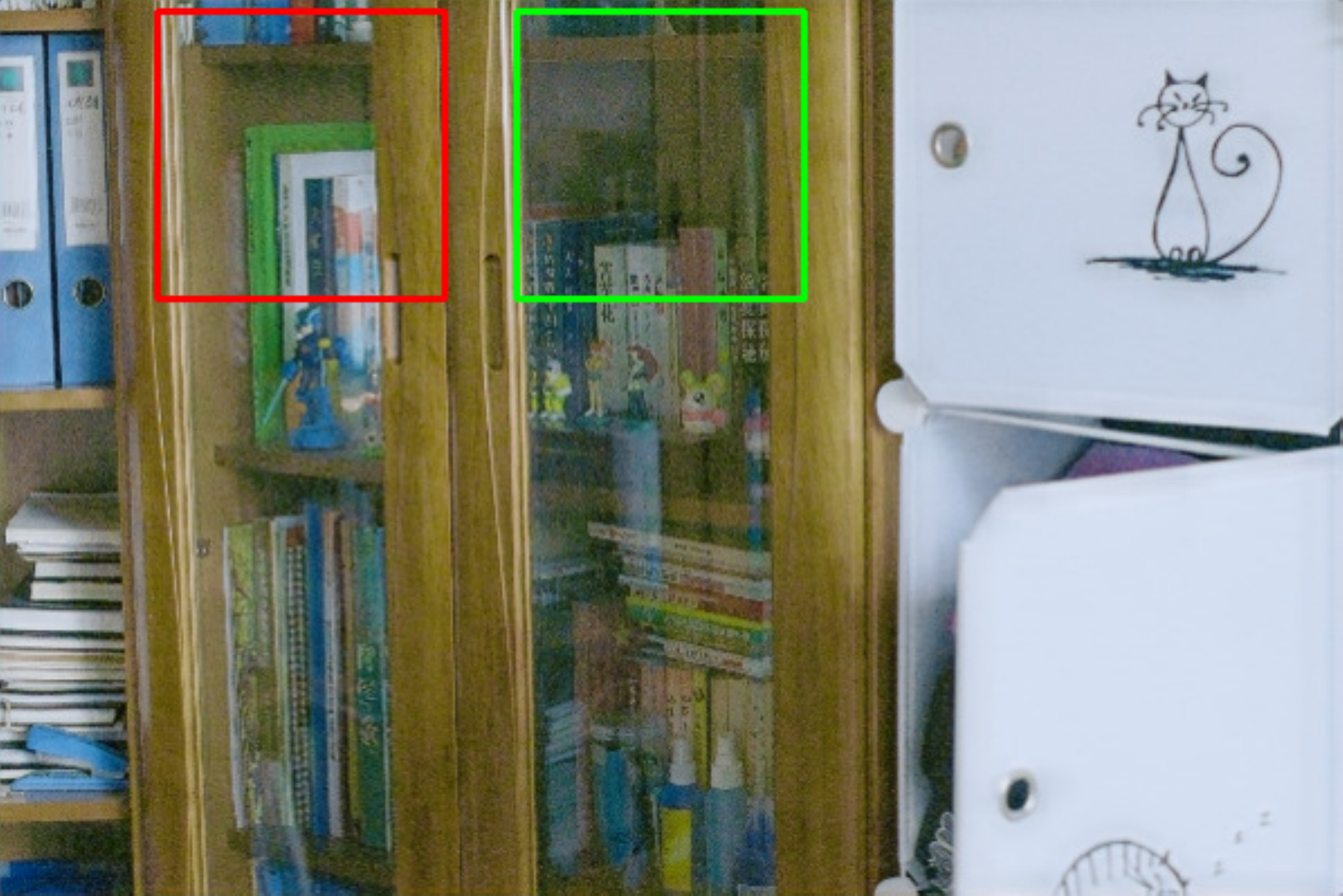}   		\put(70,0){\includegraphics[scale=.28]%
				{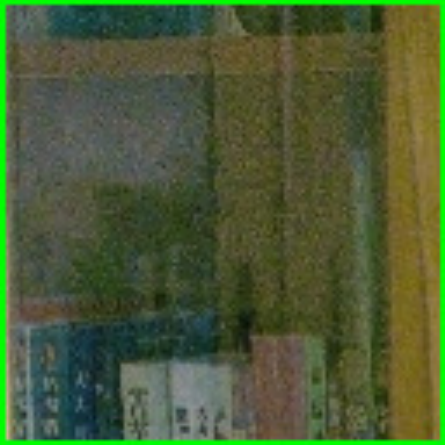}} 
		\end{overpic}
	}%
	
	\subfigure[EnlightenGan \cite{jiang2021enlightengan}]{
		\begin{overpic}[scale=.205]{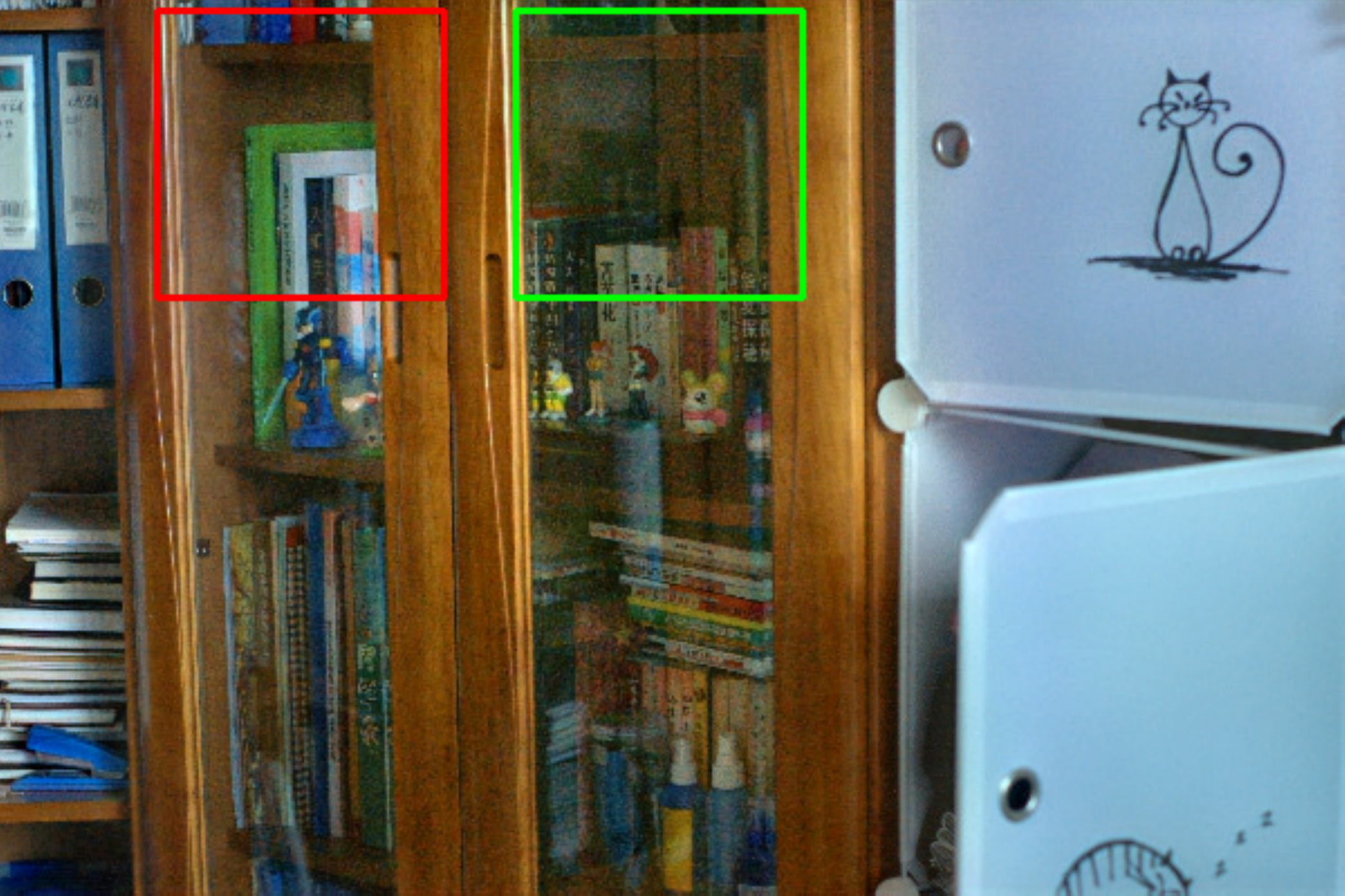}   		\put(70,0){\includegraphics[scale=.28]%
				{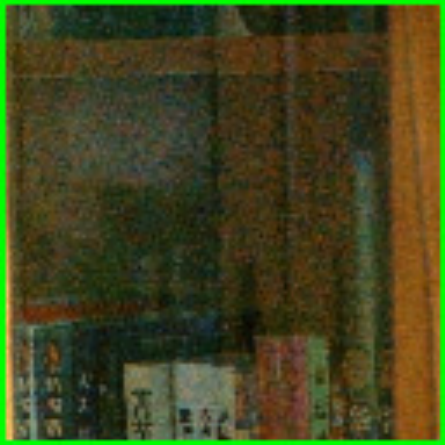}} 
		\end{overpic}
	}%
	\subfigure[Zero-DCE \cite{guo2020zero}]{
		\begin{overpic}[scale=.205]{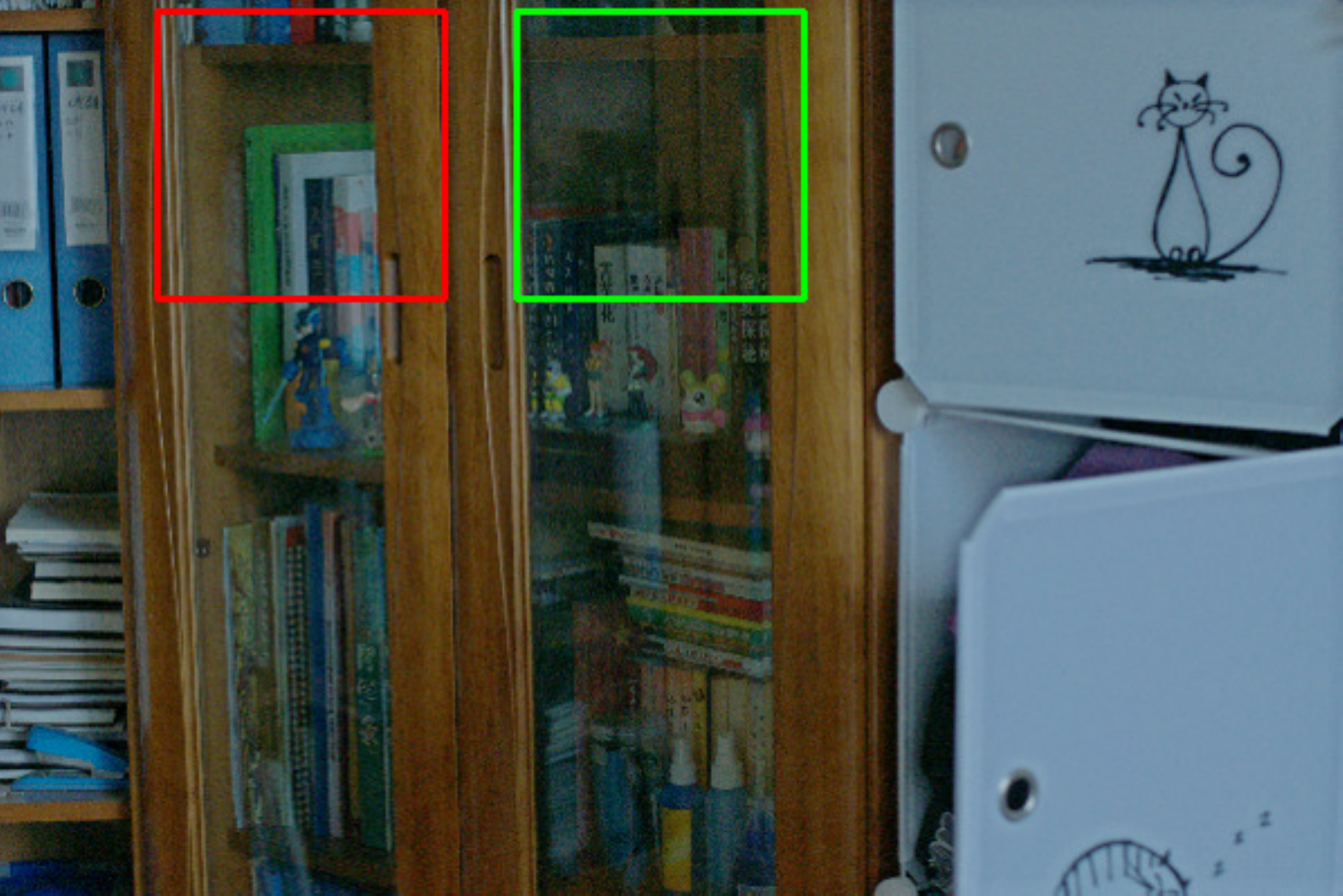}   		\put(70,0){\includegraphics[scale=.28]%
				{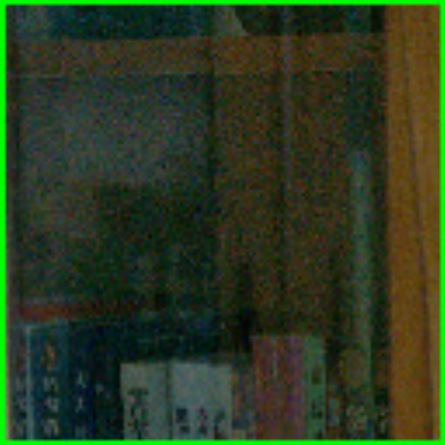}} 
		\end{overpic}
	}%
	\subfigure[DA-DRN]{
		\begin{overpic}[scale=.205]{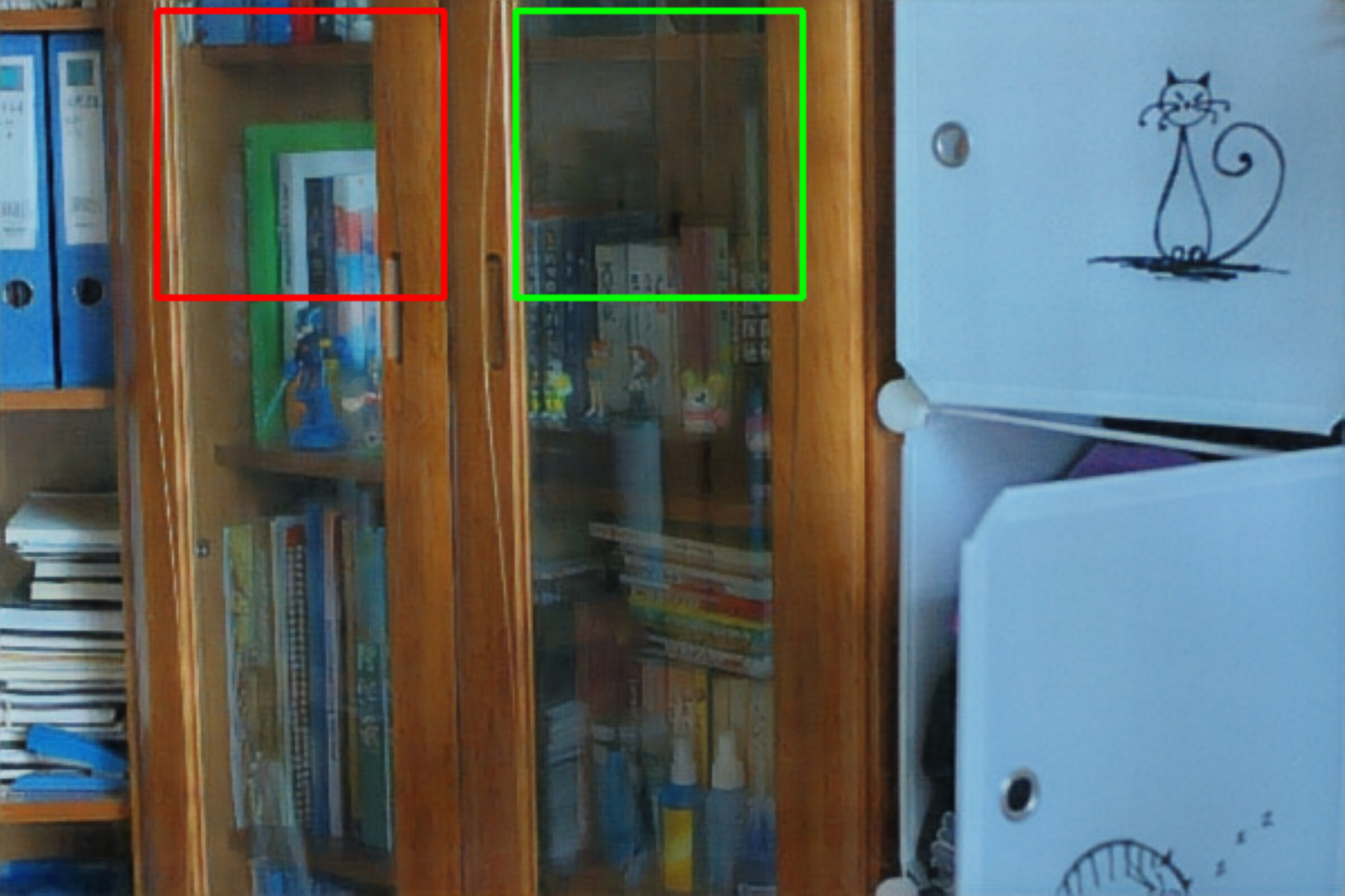}   		\put(70,0){\includegraphics[scale=.28]%
				{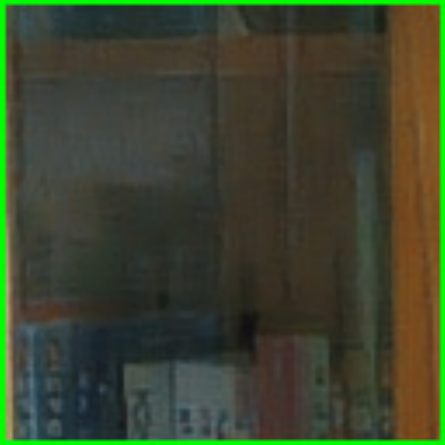}} 
		\end{overpic}
	}%
	\subfigure[Ground-Truth]{
		\begin{overpic}[scale=.205]{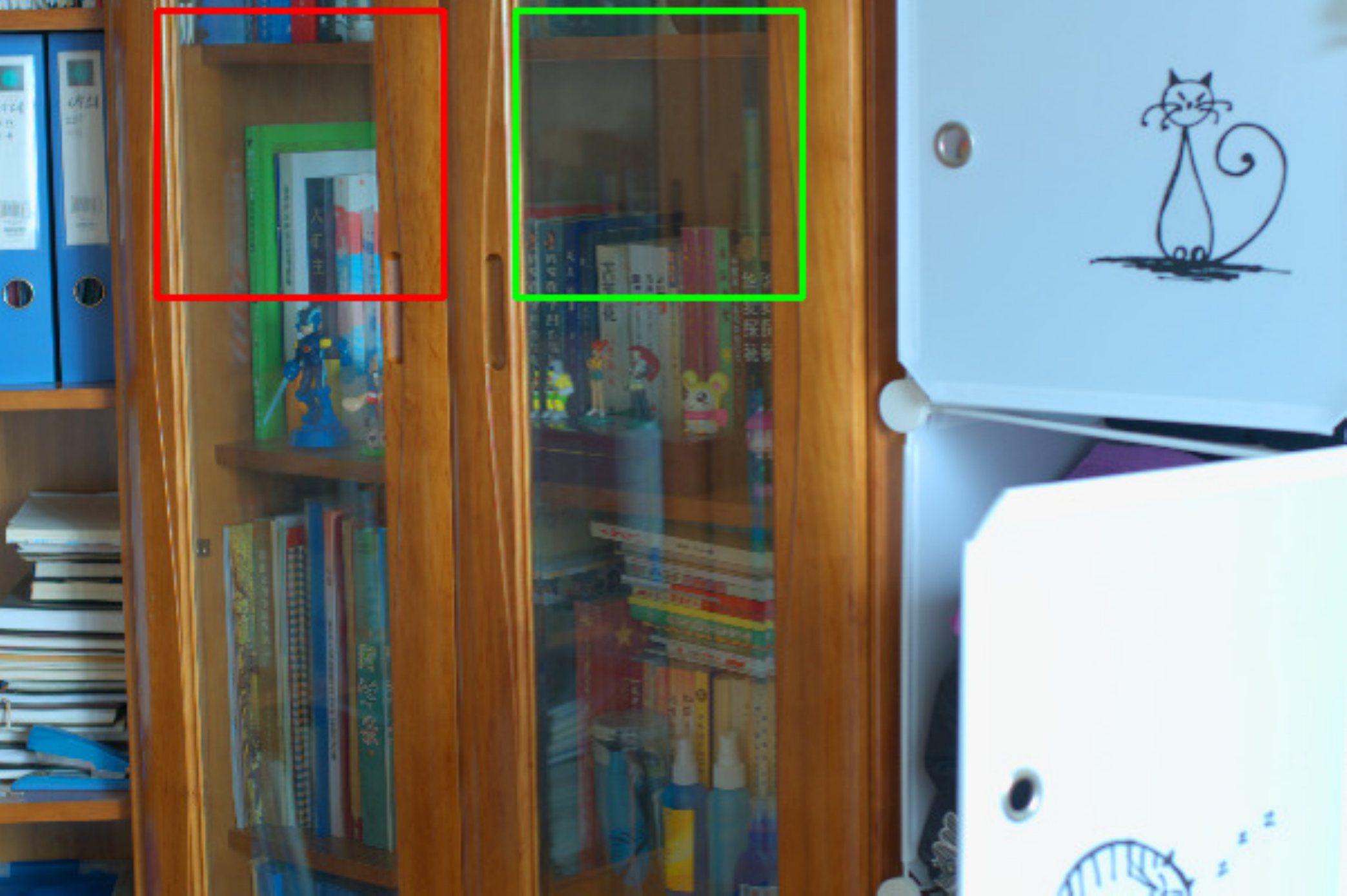}   		\put(70,0){\includegraphics[scale=.28]%
				{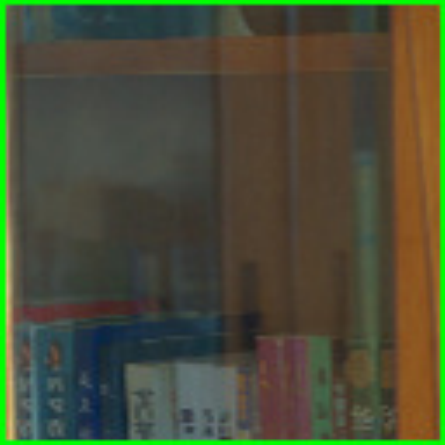}} 
		\end{overpic}
	}%
	
	\flushleft
	\caption{Visual comparison with other state-of-the-art methods on the LOL real-world validation dataset, where the degradation is hidden in darkness.}
	\label{book}
\end{figure*}

\subsection{Visual Performance Analysis}
As shown in Fig. \ref{book}, the low-light images from the LOL dataset are real-world images with very low brightness. The images enhanced by other methods contain serious noise pollution and color distortion, low brightness and local over/under-exposure degradation. In particular, for RetinexNet\cite{wei2018deep}, there is so much amplified noise, color deviation and halo artifacts in the images that they no longer look like real-world images. In contrast, our method can effectively solve these problems. It can be seen that the images enhanced by our method have almost no noise, the most abundant and reasonable color information and no over/under-exposure phenomena.\\
In addition, we test our model on LIME, DICM, MEF and NPE datasets. These four frequently used datasets have no Ground-Truth. As shown in Fig. \ref{LIME}, Fig. \ref{DICM} and Fig \ref{MEF}, the results of other methods have unpleasant problems, such as color distortion, noise and low brightness. In contrast, the output results of our model can effectively suppress noise, reasonably improve brightness, correctly restore color information and eliminate halo artifacts. Our results can also achieve a very good visual effect.

\subsection{Quantitative Performance Analysis}
Higher values of PSNR, SSIM, FSIM, UQI, SRER and SAM and lower value of LPIPS and RMSE indicate better quality of images. We compared our method with other state-of-the-art methods in terms of these metrics, including traditional methods (TD Methods)
and deep learning methods (DL Methods).
As shown in Table.\ref{lol_real} and Table.\ref{lol_syn}, our method achieves state-of-the-art performance in terms of many indicators. 
We also adopted Angular Error \cite{hordley2004re} as a quantitative index to measure the degree of color distortion and bias. Lower values of Angular Error indicate less color distortion and bias and better performance against color distortion. We compared four measures, i.e., DeltaE \cite{sharma2005ciede2000}, the mean and median values of the Angular Error and the average value of them on the LOL real-world and synthetic validation datasets. As shown in Table.\ref{lol_real} and Table.\ref{lol_syn}, compared with other methods, our method achieves the best performance in all four indocators of color distortion, which means our method can effectively solve the color distortion problem after enhancement.
We test DA-DRN on four popular datasets without Ground-Truth in terms of three no-reference image quality assessment metrics, IL-NIQE\cite{zhang2015feature}, PIQE\cite{venkatanath2015blind} and Noise Level\cite{liu2013single}. The lower of them means the better quality of the image and less noise. As shown in Table.\ref{wogt}, compared with other state-of-the-art enhancement methods, our method has almost achieved the best performance in terms of these three non-reference metrics, which means our method has better enhancement effect and stronger ability to suppress noise.\\
The inference speed of our model is very fast. As shown in Fig. \ref{speed}, we tested all of the deep learning models on a TITAN XP GPU with 12G RAM and compared the inference speed and PSNR index of our model with those of other deep learning methods. The closer the method is to the upper left corner in the figure, the faster the method is and the higher the PSNR is. Our model only takes 7 ms to process a real-world low-light image with a resolution of 600x400 from the LOL dataset, and introducing our DA Module does not slow down our framework in the test phase. Zero-DCE \cite{guo2020zero} is the fastest in terms of the inference speed, taking only 2 ms for a 600x400 resolution image. The speed of our model ranks second among all of the compared methods. 

\subsection{Analysis of flexibility and generalization performance}
Through the extensive experiments, we found that any other methods can only work well in certain situations with certain degree of degradation, which limits their generalization and practicability. Our experiments show that KinD can achieve good denoising effect for LOL real-world datasets, but for synthetic datasets with much lower noise level and simpler noise types, the denoising effect becomes worse. Zero-DCE performs very well on those four datasets without Ground-Truth, but it can not handle the LOL datasets well. By contrast, by adjusting the coefficients $\lambda_{TV}$, our method can deal with datasets with different noise levels and degradation levels to achieve the best performance on the corresponding dataset, which makes our model have strong generalization and practicability.\\
As shown in Table.\ref{tv_abla_tb} and Fig.\ref{pic_tv_abla}, we train DA-DRN on the LOL real-world dataset, and when $\lambda_{TV}$ is set to 0.2, the resluts achieve the best performance in terms of SSIM and DeltaE, wihch means the best performance on Structural similarity and color bias. When the coefficient is greater than 0.2, the colors of the decomposed reflection map are too rich, resulting in the phenomenon of color susaturated; when the coefficient is less than 0.2, the color undersaturation will occur due to the loss of too many colors. Therefore, our model can adjust the color saturation of the resulting result by adjusting the coefficient of TV and MSE Loss ($\lambda_{TV}$). But the noise level and structure of the results will also change correspondingly, and we can find a suitable trade-off between them through the adjustment of the $\lambda_{TV}$. This makes our method very flexible and user-friendly.


\section*{Conclusion}

In this paper, inspired by the Retinex theory \cite{land1977retinex} and the idea of RetinexNet \cite{wei2018deep}, we propose a Degradation-Aware Deep Retinex Network named \textbf{DA-DRN} for low-light image enhancement. In the decomposition phase, we use a deep U-Net to decouple images into reflectance and illumination. We propose a Degradation-Aware (DA) Module for the awareness of degradation in the reflectance map decomposed by low-light image and guide the decomposer to directly generate the reflectance map without degradation while preserving details by transferring some of the details information into the illumination map. In the enhancement phase, we use a deep U-Net and introduce perceptual loss to enhance the brightness of the illumination. In addition, our method can be adapted to different kinds of datasets with different degradation degrees and noise levels by adjusting the coefficients $\lambda_{TV}$, which makes our method have better generalization and robustness. Our method can also change the colorfulness of the image by adjusting the coefficient $\lambda_{TV}$, so as to adjust the color saturation of the image flexibly, which makes our method very flexible and user-friendly. Extensive experiments demonstrate the superiority and effectiveness of our method. DA-DRN can also achieve fast inference speed in the test phase, which increases the practicality of our method.

\bibliographystyle{IEEEtran}
\bibliography{ref}

\begin{thebibliography}{10}
\providecommand{\url}[1]{#1}
\csname url@samestyle\endcsname
\providecommand{\newblock}{\relax}
\providecommand{\bibinfo}[2]{#2}
\providecommand{\BIBentrySTDinterwordspacing}{\spaceskip=0pt\relax}
\providecommand{\BIBentryALTinterwordstretchfactor}{4}
\providecommand{\BIBentryALTinterwordspacing}{\spaceskip=\fontdimen2\font plus
\BIBentryALTinterwordstretchfactor\fontdimen3\font minus
  \fontdimen4\font\relax}
\providecommand{\BIBforeignlanguage}[2]{{%
\expandafter\ifx\csname l@#1\endcsname\relax
\typeout{** WARNING: IEEEtran.bst: No hyphenation pattern has been}%
\typeout{** loaded for the language `#1'. Using the pattern for}%
\typeout{** the default language instead.}%
\else
\language=\csname l@#1\endcsname
\fi
#2}}
\providecommand{\BIBdecl}{\relax}
\BIBdecl

\bibitem{wang2018gladnet}
W.~Wang, C.~Wei, W.~Yang, and J.~Liu, ``Gladnet: Low-light enhancement network
  with global awareness,'' in \emph{2018 13th IEEE International Conference on
  Automatic Face \& Gesture Recognition (FG 2018)}.\hskip 1em plus 0.5em minus
  0.4em\relax IEEE, 2018, pp. 751--755.

\bibitem{liu2013single}
X.~Liu, M.~Tanaka, and M.~Okutomi, ``Single-image noise level estimation for
  blind denoising,'' \emph{IEEE transactions on image processing}, vol.~22,
  no.~12, pp. 5226--5237, 2013.

\bibitem{hasler2003measuring}
D.~Hasler and S.~E. Suesstrunk, ``Measuring colorfulness in natural images,''
  in \emph{Human vision and electronic imaging VIII}, vol. 5007.\hskip 1em plus
  0.5em minus 0.4em\relax International Society for Optics and Photonics, 2003,
  pp. 87--95.

\bibitem{lecun1998gradient}
Y.~LeCun, L.~Bottou, Y.~Bengio, and P.~Haffner, ``Gradient-based learning
  applied to document recognition,'' \emph{Proceedings of the IEEE}, vol.~86,
  no.~11, pp. 2278--2324, 1998.

\bibitem{simonyan2014very}
K.~Simonyan and A.~Zisserman, ``Very deep convolutional networks for
  large-scale image recognition,'' \emph{arXiv preprint arXiv:1409.1556}, 2014.

\bibitem{he2016deep}
K.~He, X.~Zhang, S.~Ren, and J.~Sun, ``Deep residual learning for image
  recognition,'' in \emph{Proceedings of the IEEE conference on computer vision
  and pattern recognition}, 2016, pp. 770--778.

\bibitem{redmon2016you}
J.~Redmon, S.~Divvala, R.~Girshick, and A.~Farhadi, ``You only look once:
  Unified, real-time object detection,'' in \emph{Proceedings of the IEEE
  conference on computer vision and pattern recognition}, 2016, pp. 779--788.

\bibitem{girshick2014rich}
R.~Girshick, J.~Donahue, T.~Darrell, and J.~Malik, ``Rich feature hierarchies
  for accurate object detection and semantic segmentation,'' in
  \emph{Proceedings of the IEEE conference on computer vision and pattern
  recognition}, 2014, pp. 580--587.

\bibitem{girshick2015fast}
R.~Girshick, ``Fast r-cnn,'' in \emph{Proceedings of the IEEE international
  conference on computer vision}, 2015, pp. 1440--1448.

\bibitem{chen2017deeplab}
L.-C. Chen, G.~Papandreou, I.~Kokkinos, K.~Murphy, and A.~L. Yuille, ``Deeplab:
  Semantic image segmentation with deep convolutional nets, atrous convolution,
  and fully connected crfs,'' \emph{IEEE transactions on pattern analysis and
  machine intelligence}, vol.~40, no.~4, pp. 834--848, 2017.

\bibitem{he2017mask}
K.~He, G.~Gkioxari, P.~Doll{\'a}r, and R.~Girshick, ``Mask r-cnn,'' in
  \emph{Proceedings of the IEEE international conference on computer vision},
  2017, pp. 2961--2969.

\bibitem{zhao2017pyramid}
H.~Zhao, J.~Shi, X.~Qi, X.~Wang, and J.~Jia, ``Pyramid scene parsing network,''
  in \emph{Proceedings of the IEEE conference on computer vision and pattern
  recognition}, 2017, pp. 2881--2890.

\bibitem{zhang2019kindling}
Y.~Zhang, J.~Zhang, and X.~Guo, ``Kindling the darkness: A practical low-light
  image enhancer,'' in \emph{Proceedings of the 27th ACM International
  Conference on Multimedia}, 2019, pp. 1632--1640.

\bibitem{jobson1997multiscale}
D.~J. Jobson, Z.-u. Rahman, and G.~A. Woodell, ``A multiscale retinex for
  bridging the gap between color images and the human observation of scenes,''
  \emph{IEEE Transactions on Image processing}, vol.~6, no.~7, pp. 965--976,
  1997.

\bibitem{guo2016lime}
X.~Guo, Y.~Li, and H.~Ling, ``Lime: Low-light image enhancement via
  illumination map estimation,'' \emph{IEEE Transactions on image processing},
  vol.~26, no.~2, pp. 982--993, 2016.

\bibitem{wei2018deep}
C.~Wei, W.~Wang, W.~Yang, and J.~Liu, ``Deep retinex decomposition for
  low-light enhancement,'' \emph{arXiv preprint arXiv:1808.04560}, 2018.

\bibitem{jiang2021enlightengan}
Y.~Jiang, X.~Gong, D.~Liu, Y.~Cheng, C.~Fang, X.~Shen, J.~Yang, P.~Zhou, and
  Z.~Wang, ``Enlightengan: Deep light enhancement without paired supervision,''
  \emph{IEEE Transactions on Image Processing}, vol.~30, pp. 2340--2349, 2021.

\bibitem{wang2019rdgan}
J.~Wang, W.~Tan, X.~Niu, and B.~Yan, ``Rdgan: Retinex decomposition based
  adversarial learning for low-light enhancement,'' in \emph{2019 IEEE
  International Conference on Multimedia and Expo (ICME)}.\hskip 1em plus 0.5em
  minus 0.4em\relax IEEE, 2019, pp. 1186--1191.

\bibitem{guo2020zero}
C.~Guo, C.~Li, J.~Guo, C.~C. Loy, J.~Hou, S.~Kwong, and R.~Cong,
  ``Zero-reference deep curve estimation for low-light image enhancement,'' in
  \emph{Proceedings of the IEEE/CVF Conference on Computer Vision and Pattern
  Recognition}, 2020, pp. 1780--1789.

\bibitem{lv2018mbllen}
F.~Lv, F.~Lu, J.~Wu, and C.~Lim, ``Mbllen: Low-light image/video enhancement
  using cnns.'' in \emph{BMVC}, 2018, p. 220.

\bibitem{li2018structure}
M.~Li, J.~Liu, W.~Yang, X.~Sun, and Z.~Guo, ``Structure-revealing low-light
  image enhancement via robust retinex model,'' \emph{IEEE Transactions on
  Image Processing}, vol.~27, no.~6, pp. 2828--2841, 2018.

\bibitem{ren2018joint}
X.~Ren, M.~Li, W.-H. Cheng, and J.~Liu, ``Joint enhancement and denoising
  method via sequential decomposition,'' in \emph{2018 IEEE International
  Symposium on Circuits and Systems (ISCAS)}.\hskip 1em plus 0.5em minus
  0.4em\relax IEEE, 2018, pp. 1--5.

\bibitem{land1977retinex}
E.~H. Land, ``The retinex theory of color vision,'' \emph{Scientific american},
  vol. 237, no.~6, pp. 108--129, 1977.

\bibitem{dabov2006image}
K.~Dabov, A.~Foi, V.~Katkovnik, and K.~Egiazarian, ``Image denoising with
  block-matching and 3d filtering,'' in \emph{Image Processing: Algorithms and
  Systems, Neural Networks, and Machine Learning}, vol. 6064.\hskip 1em plus
  0.5em minus 0.4em\relax International Society for Optics and Photonics, 2006,
  p. 606414.

\bibitem{zhang2017beyond}
K.~Zhang, W.~Zuo, Y.~Chen, D.~Meng, and L.~Zhang, ``Beyond a gaussian denoiser:
  Residual learning of deep cnn for image denoising,'' \emph{IEEE transactions
  on image processing}, vol.~26, no.~7, pp. 3142--3155, 2017.

\bibitem{guo2019toward}
S.~Guo, Z.~Yan, K.~Zhang, W.~Zuo, and L.~Zhang, ``Toward convolutional blind
  denoising of real photographs,'' in \emph{Proceedings of the IEEE/CVF
  Conference on Computer Vision and Pattern Recognition}, 2019, pp. 1712--1722.

\bibitem{ronneberger2015u}
O.~Ronneberger, P.~Fischer, and T.~Brox, ``U-net: Convolutional networks for
  biomedical image segmentation,'' in \emph{International Conference on Medical
  image computing and computer-assisted intervention}.\hskip 1em plus 0.5em
  minus 0.4em\relax Springer, 2015, pp. 234--241.

\bibitem{xiong2020unsupervised}
W.~Xiong, D.~Liu, X.~Shen, C.~Fang, and J.~Luo, ``Unsupervised real-world
  low-light image enhancement with decoupled networks,'' \emph{arXiv preprint
  arXiv:2005.02818}, 2020.

\bibitem{lore2017llnet}
K.~G. Lore, A.~Akintayo, and S.~Sarkar, ``Llnet: A deep autoencoder approach to
  natural low-light image enhancement,'' \emph{Pattern Recognition}, vol.~61,
  pp. 650--662, 2017.

\bibitem{johnson2016perceptual}
J.~Johnson, A.~Alahi, and L.~Fei-Fei, ``Perceptual losses for real-time style
  transfer and super-resolution,'' in \emph{European conference on computer
  vision}.\hskip 1em plus 0.5em minus 0.4em\relax Springer, 2016, pp. 694--711.

\bibitem{pizer1987adaptive}
S.~M. Pizer, E.~P. Amburn, J.~D. Austin, R.~Cromartie, A.~Geselowitz, T.~Greer,
  B.~ter Haar~Romeny, J.~B. Zimmerman, and K.~Zuiderveld, ``Adaptive histogram
  equalization and its variations,'' \emph{Computer vision, graphics, and image
  processing}, vol.~39, no.~3, pp. 355--368, 1987.

\bibitem{pizer1990contrast}
S.~M. Pizer, R.~E. Johnston, J.~P. Ericksen, B.~C. Yankaskas, and K.~E. Muller,
  ``Contrast-limited adaptive histogram equalization: speed and
  effectiveness,'' in \emph{[1990] Proceedings of the First Conference on
  Visualization in Biomedical Computing}.\hskip 1em plus 0.5em minus
  0.4em\relax IEEE Computer Society, 1990, pp. 337--338.

\bibitem{jobson1997properties}
D.~J. Jobson, Z.-u. Rahman, and G.~A. Woodell, ``Properties and performance of
  a center/surround retinex,'' \emph{IEEE transactions on image processing},
  vol.~6, no.~3, pp. 451--462, 1997.

\bibitem{wang2013naturalness}
S.~Wang, J.~Zheng, H.-M. Hu, and B.~Li, ``Naturalness preserved enhancement
  algorithm for non-uniform illumination images,'' \emph{IEEE Transactions on
  Image Processing}, vol.~22, no.~9, pp. 3538--3548, 2013.

\bibitem{fu2016fusion}
X.~Fu, D.~Zeng, Y.~Huang, Y.~Liao, X.~Ding, and J.~Paisley, ``A fusion-based
  enhancing method for weakly illuminated images,'' \emph{Signal Processing},
  vol. 129, pp. 82--96, 2016.

\bibitem{fu2016weighted}
X.~Fu, D.~Zeng, Y.~Huang, X.-P. Zhang, and X.~Ding, ``A weighted variational
  model for simultaneous reflectance and illumination estimation,'' in
  \emph{Proceedings of the IEEE Conference on Computer Vision and Pattern
  Recognition}, 2016, pp. 2782--2790.

\bibitem{ying2017bio}
Z.~Ying, G.~Li, and W.~Gao, ``A bio-inspired multi-exposure fusion framework
  for low-light image enhancement,'' \emph{arXiv preprint arXiv:1711.00591},
  2017.

\bibitem{dong2011fast}
X.~Dong, G.~Wang, Y.~Pang, W.~Li, J.~Wen, W.~Meng, and Y.~Lu, ``Fast efficient
  algorithm for enhancement of low lighting video,'' in \emph{2011 IEEE
  International Conference on Multimedia and Expo}.\hskip 1em plus 0.5em minus
  0.4em\relax IEEE, 2011, pp. 1--6.

\bibitem{ying2017new}
Z.~Ying, G.~Li, Y.~Ren, R.~Wang, and W.~Wang, ``A new low-light image
  enhancement algorithm using camera response model,'' in \emph{Proceedings of
  the IEEE International Conference on Computer Vision Workshops}, 2017, pp.
  3015--3022.

\bibitem{ren2018lecarm}
Y.~Ren, Z.~Ying, T.~H. Li, and G.~Li, ``Lecarm: Low-light image enhancement
  using the camera response model,'' \emph{IEEE Transactions on Circuits and
  Systems for Video Technology}, vol.~29, no.~4, pp. 968--981, 2018.

\bibitem{tao2017llcnn}
L.~Tao, C.~Zhu, G.~Xiang, Y.~Li, H.~Jia, and X.~Xie, ``Llcnn: A convolutional
  neural network for low-light image enhancement,'' in \emph{2017 IEEE Visual
  Communications and Image Processing (VCIP)}.\hskip 1em plus 0.5em minus
  0.4em\relax IEEE, 2017, pp. 1--4.

\bibitem{shen2017msr}
L.~Shen, Z.~Yue, F.~Feng, Q.~Chen, S.~Liu, and J.~Ma, ``Msr-net: Low-light
  image enhancement using deep convolutional network,'' \emph{arXiv preprint
  arXiv:1711.02488}, 2017.

\bibitem{li2018lightennet}
C.~Li, J.~Guo, F.~Porikli, and Y.~Pang, ``Lightennet: A convolutional neural
  network for weakly illuminated image enhancement,'' \emph{Pattern recognition
  letters}, vol. 104, pp. 15--22, 2018.

\bibitem{zhao2021retinexdip}
Z.~Zhao, B.~Xiong, L.~Wang, Q.~Ou, L.~Yu, and F.~Kuang, ``Retinexdip: A unified
  deep framework for low-light image enhancement,'' \emph{IEEE Transactions on
  Circuits and Systems for Video Technology}, 2021.

\bibitem{zhang2020self}
Y.~Zhang, X.~Di, B.~Zhang, and C.~Wang, ``Self-supervised image enhancement
  network: Training with low light images only,'' \emph{arXiv e-prints}, pp.
  arXiv--2002, 2020.

\bibitem{zhao2016loss}
H.~Zhao, O.~Gallo, I.~Frosio, and J.~Kautz, ``Loss functions for image
  restoration with neural networks,'' \emph{IEEE Transactions on computational
  imaging}, vol.~3, no.~1, pp. 47--57, 2016.

\bibitem{wang2004image}
Z.~Wang, A.~C. Bovik, H.~R. Sheikh, and E.~P. Simoncelli, ``Image quality
  assessment: from error visibility to structural similarity,'' \emph{IEEE
  transactions on image processing}, vol.~13, no.~4, pp. 600--612, 2004.

\bibitem{sharma2005ciede2000}
G.~Sharma, W.~Wu, and E.~N. Dalal, ``The ciede2000 color-difference formula:
  Implementation notes, supplementary test data, and mathematical
  observations,'' \emph{Color Research \& Application}, vol.~30, no.~1, pp.
  21--30, 2005.

\bibitem{lee2013contrast}
C.~Lee, C.~Lee, and C.-S. Kim, ``Contrast enhancement based on layered
  difference representation of 2d histograms,'' \emph{IEEE transactions on
  image processing}, vol.~22, no.~12, pp. 5372--5384, 2013.

\bibitem{ma2015perceptual}
K.~Ma, K.~Zeng, and Z.~Wang, ``Perceptual quality assessment for multi-exposure
  image fusion,'' \emph{IEEE Transactions on Image Processing}, vol.~24,
  no.~11, pp. 3345--3356, 2015.

\bibitem{kingma2014adam}
D.~P. Kingma and J.~Ba, ``Adam: A method for stochastic optimization,''
  \emph{arXiv preprint arXiv:1412.6980}, 2014.

\bibitem{zhang2018unreasonable}
R.~Zhang, P.~Isola, A.~A. Efros, E.~Shechtman, and O.~Wang, ``The unreasonable
  effectiveness of deep features as a perceptual metric,'' in \emph{Proceedings
  of the IEEE conference on computer vision and pattern recognition}, 2018, pp.
  586--595.

\bibitem{zhang2011fsim}
L.~Zhang, L.~Zhang, X.~Mou, and D.~Zhang, ``Fsim: A feature similarity index
  for image quality assessment,'' \emph{IEEE transactions on Image Processing},
  vol.~20, no.~8, pp. 2378--2386, 2011.

\bibitem{wang2002universal}
Z.~Wang and A.~C. Bovik, ``A universal image quality index,'' \emph{IEEE signal
  processing letters}, vol.~9, no.~3, pp. 81--84, 2002.

\bibitem{de2000spectral}
O.~A. De~Carvalho and P.~R. Meneses, ``Spectral correlation mapper (scm): an
  improvement on the spectral angle mapper (sam),'' in \emph{Summaries of the
  9th JPL Airborne Earth Science Workshop, JPL Publication 00-18},
  vol.~9.\hskip 1em plus 0.5em minus 0.4em\relax JPL publication Pasadena, CA,
  2000.

\bibitem{hordley2004re}
S.~D. Hordley and G.~D. Finlayson, ``Re-evaluating colour constancy
  algorithms,'' in \emph{Proceedings of the 17th International Conference on
  Pattern Recognition, 2004. ICPR 2004.}, vol.~1.\hskip 1em plus 0.5em minus
  0.4em\relax IEEE, 2004, pp. 76--79.

\bibitem{zhang2015feature}
L.~Zhang, L.~Zhang, and A.~C. Bovik, ``A feature-enriched completely blind
  image quality evaluator,'' \emph{IEEE Transactions on Image Processing},
  vol.~24, no.~8, pp. 2579--2591, 2015.

\bibitem{venkatanath2015blind}
N.~Venkatanath, D.~Praneeth, M.~C. Bh, S.~S. Channappayya, and S.~S. Medasani,
  ``Blind image quality evaluation using perception based features,'' in
  \emph{2015 Twenty First National Conference on Communications (NCC)}.\hskip
  1em plus 0.5em minus 0.4em\relax IEEE, 2015, pp. 1--6.

\end{thebibliography}

\appendix
\section{Appendix}

\begin{figure*}
	\centering
	
	
	\subfigure[Input]{
		\begin{minipage}[b]{0.13\textwidth}
			\includegraphics[width=2.5cm]{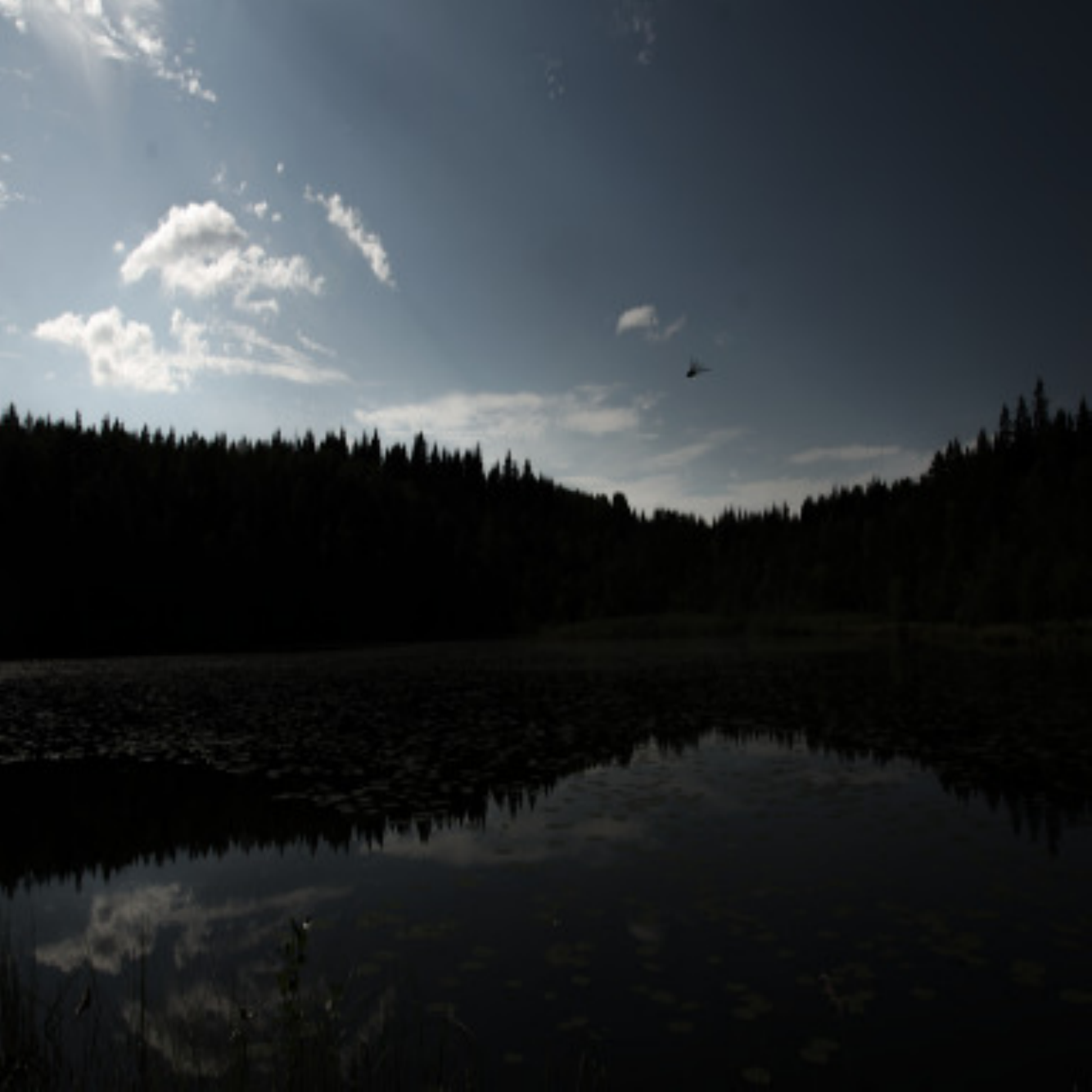}\vspace{2pt} \\
			\includegraphics[width=2.5cm]{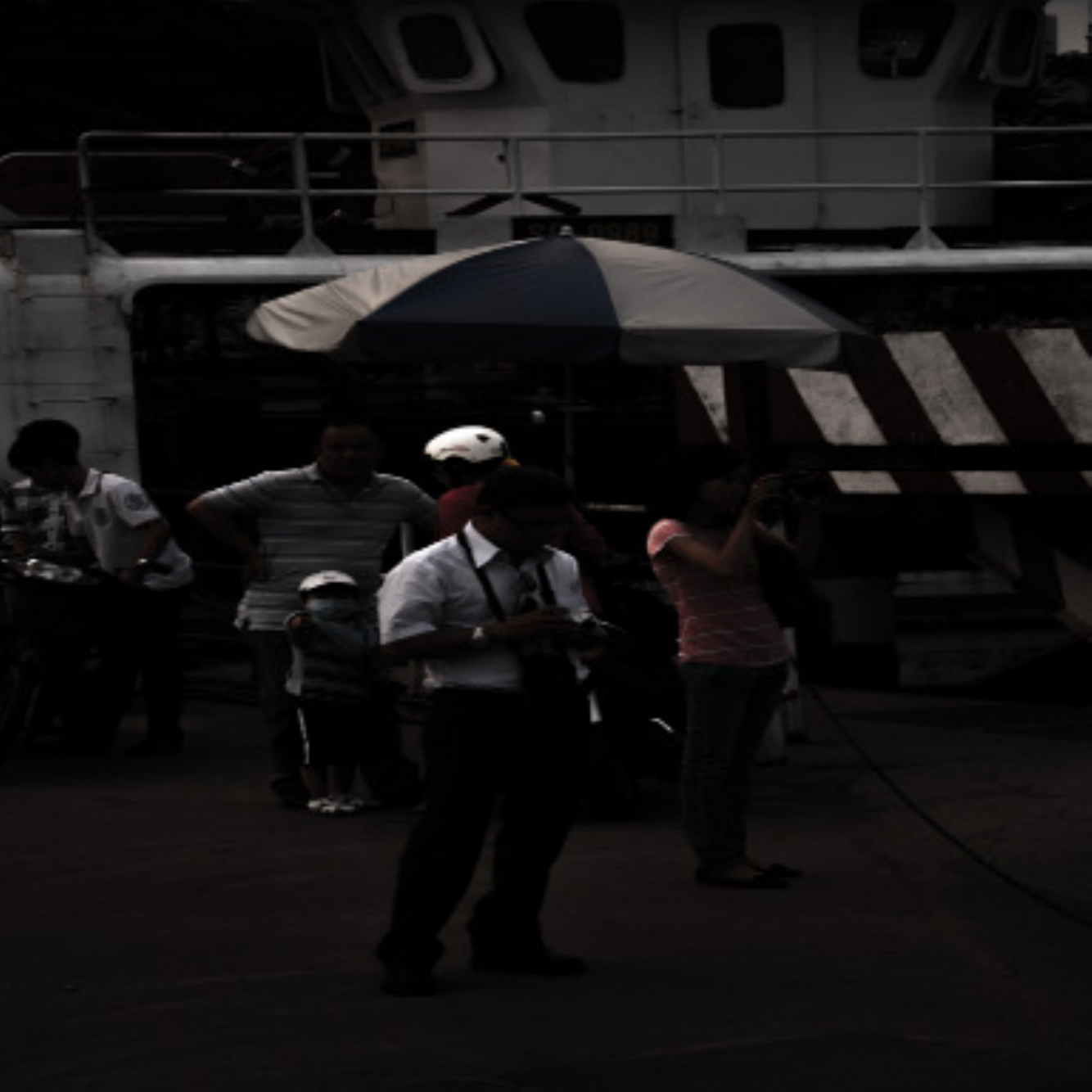}\vspace{2pt}
			\includegraphics[width=2.5cm]{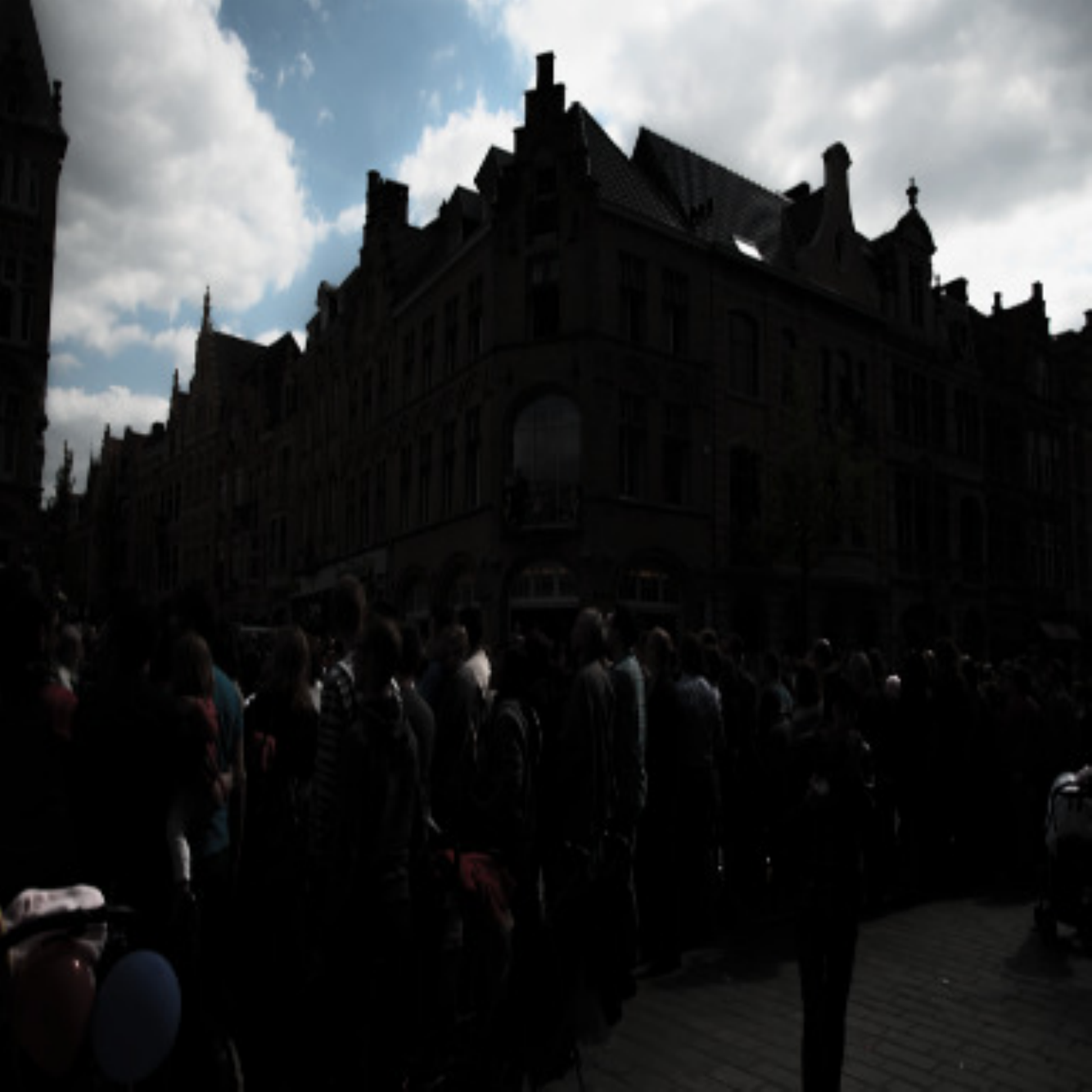}\vspace{2pt}
			\includegraphics[width=2.5cm]{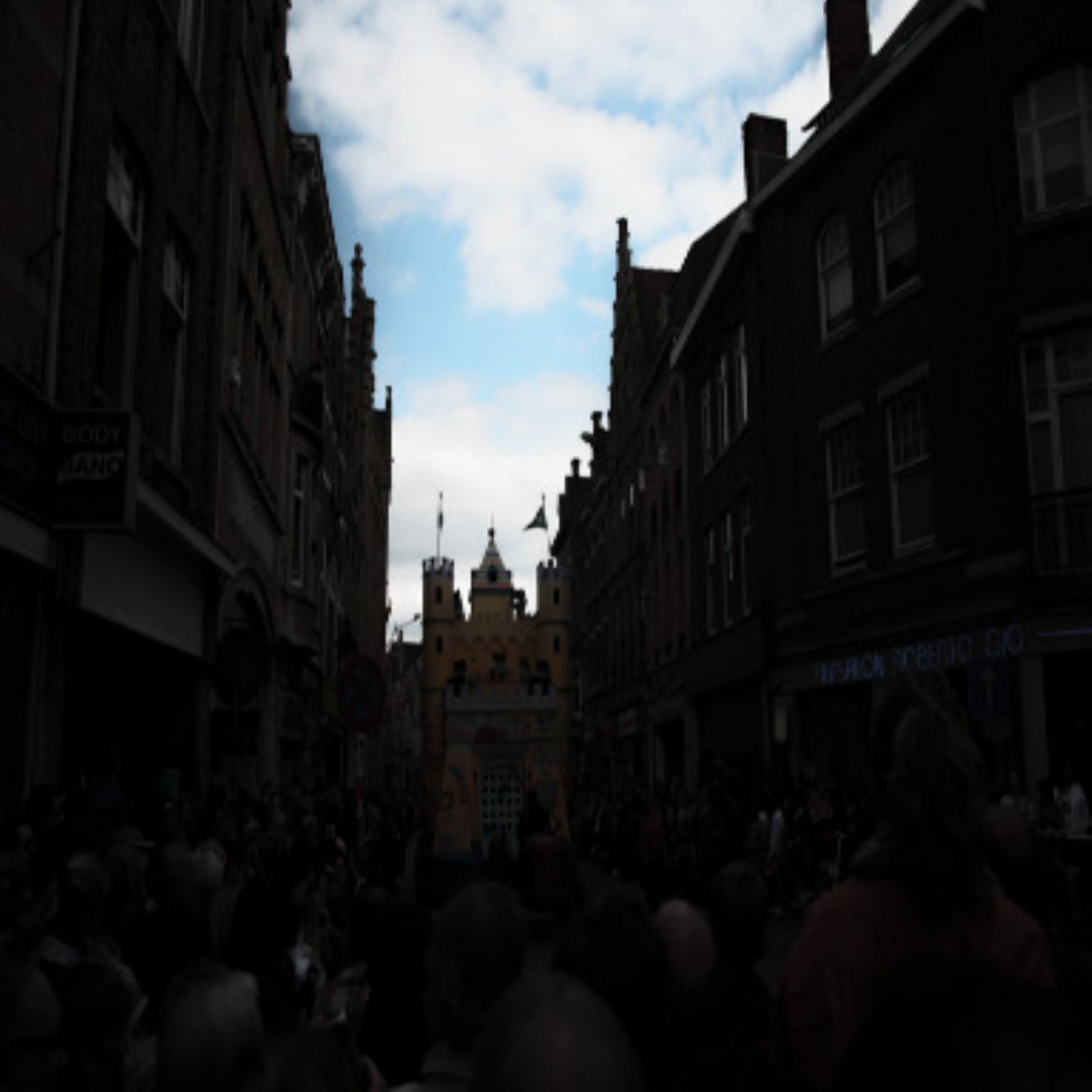}\vspace{2pt} 
		\end{minipage}
	}\hspace{-5pt}
	\subfigure[CRM\cite{ying2017new}]{
		\begin{minipage}[b]{0.13\textwidth}
			\includegraphics[width=2.5cm]{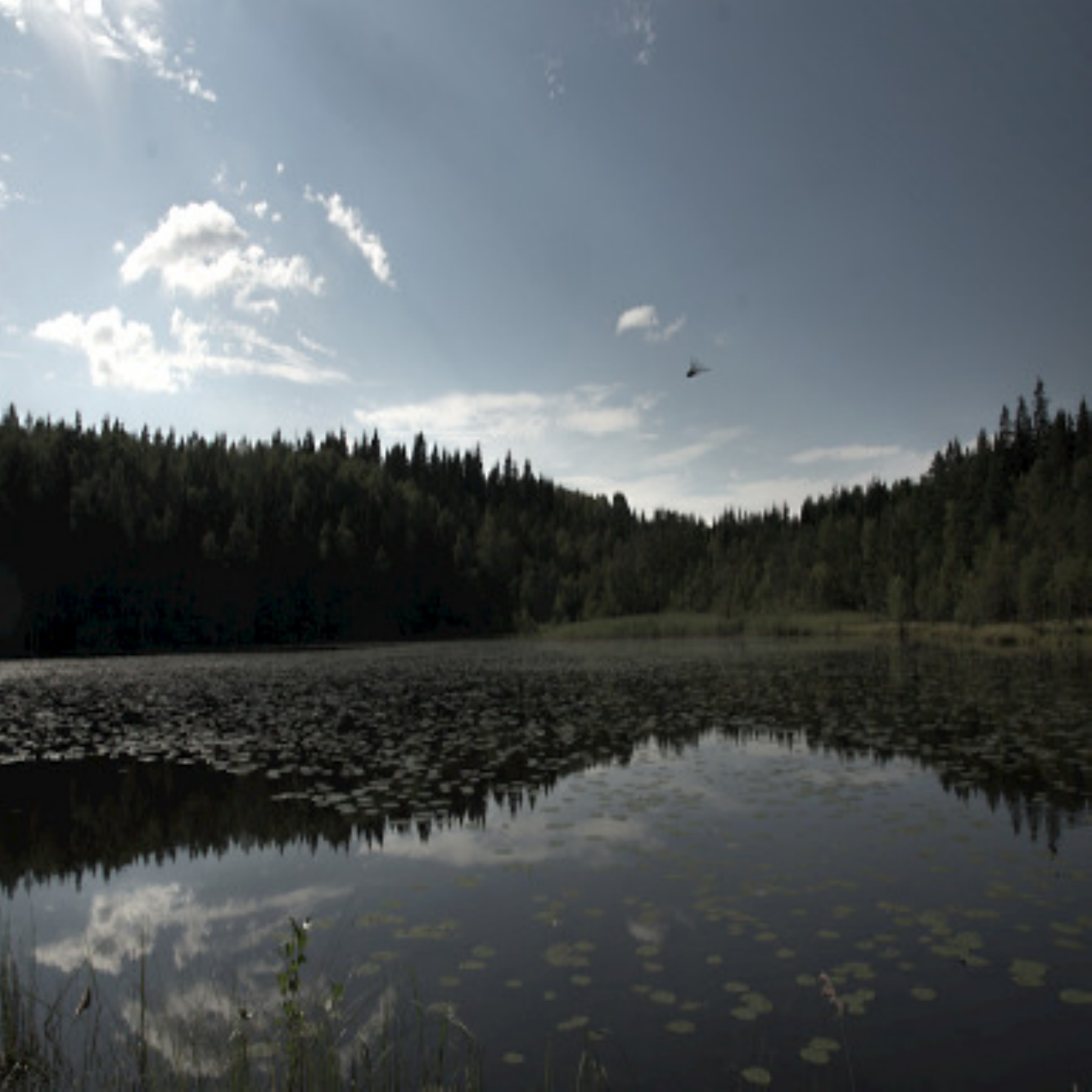}\vspace{2pt} \\
			\includegraphics[width=2.5cm]{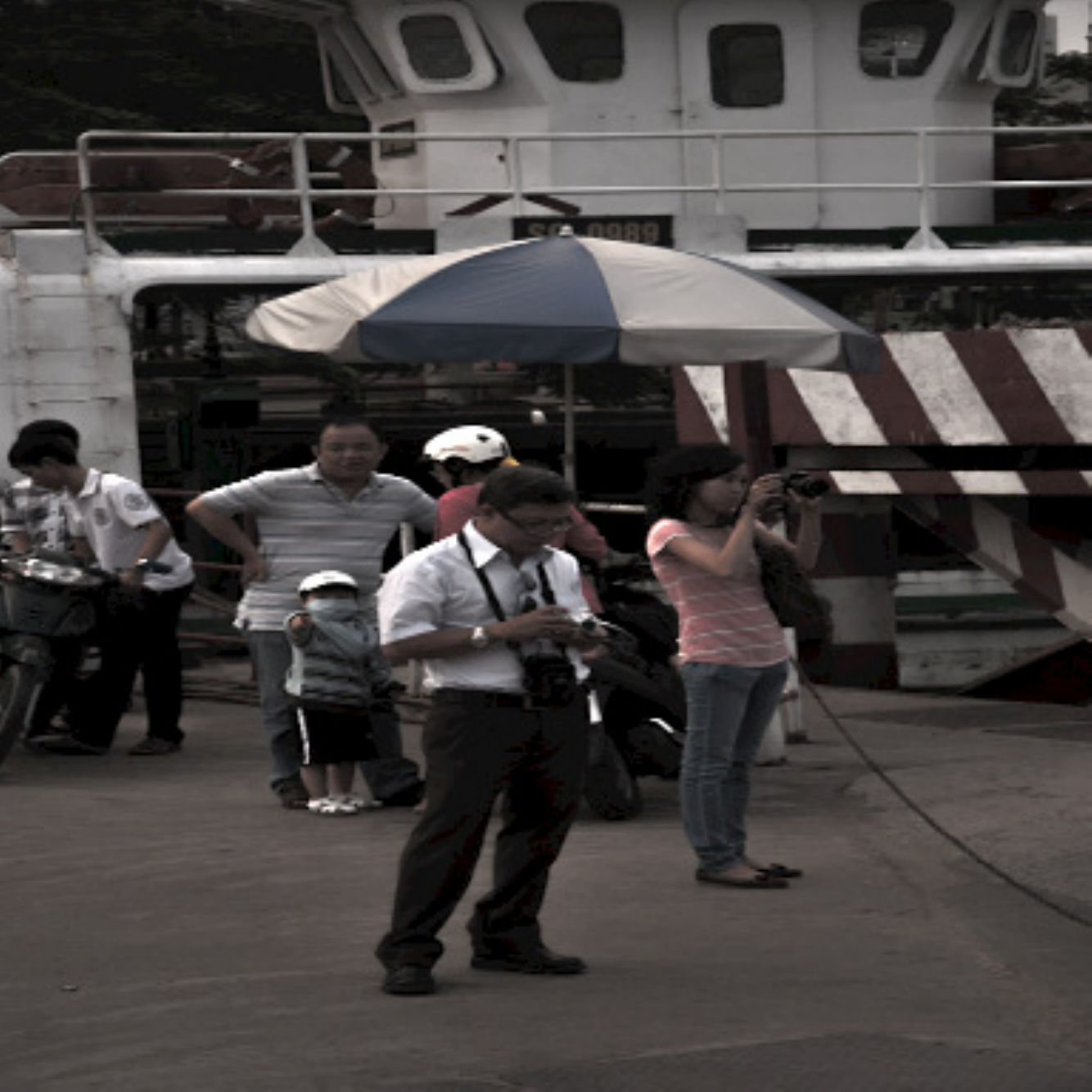}\vspace{2pt}
			\includegraphics[width=2.5cm]{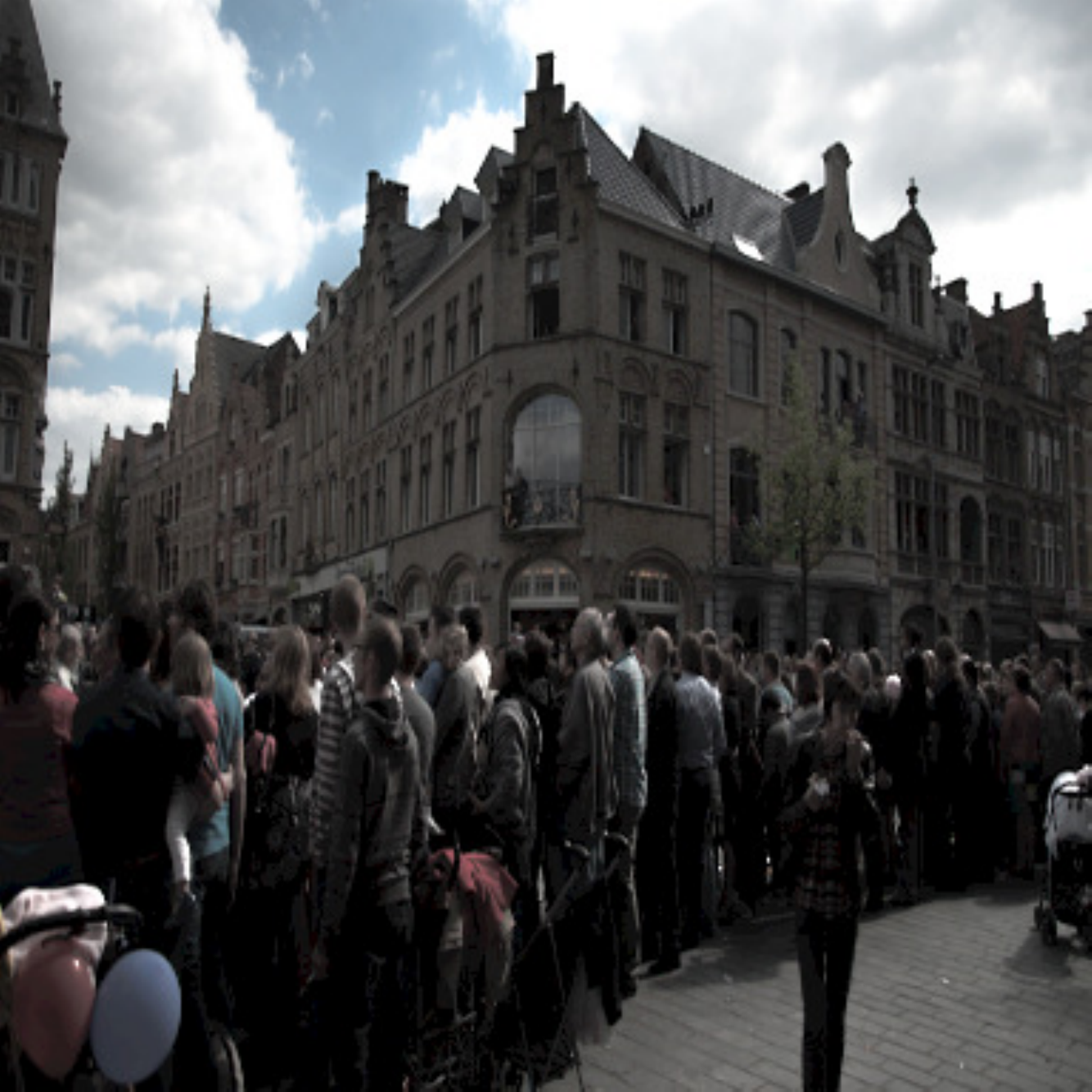}\vspace{2pt}
			\includegraphics[width=2.5cm]{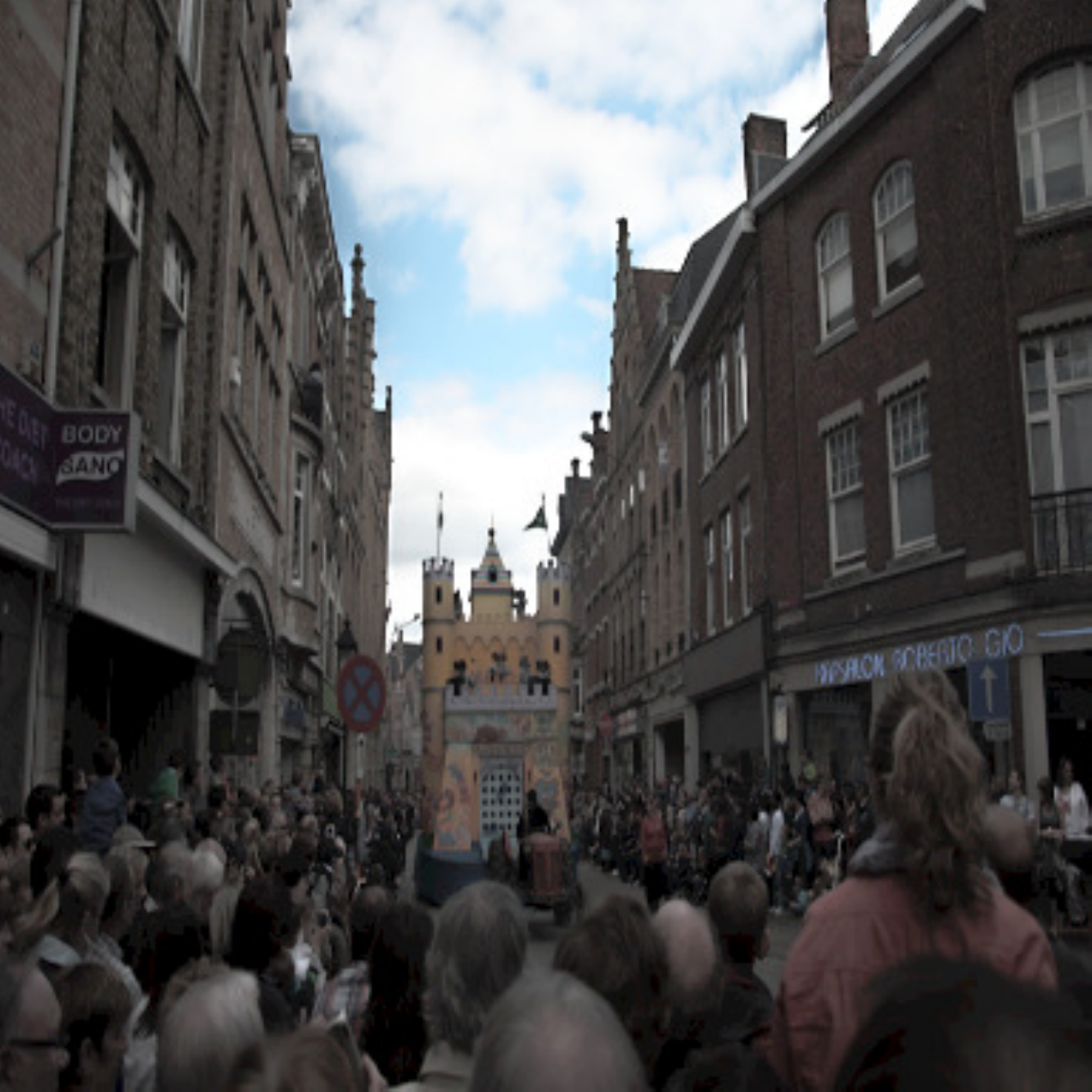}\vspace{2pt} 
		\end{minipage}
	}\hspace{-5pt}
	\subfigure[MBLLEN\cite{lv2018mbllen}]{
		\begin{minipage}[b]{0.13\textwidth}
			\includegraphics[width=2.5cm]{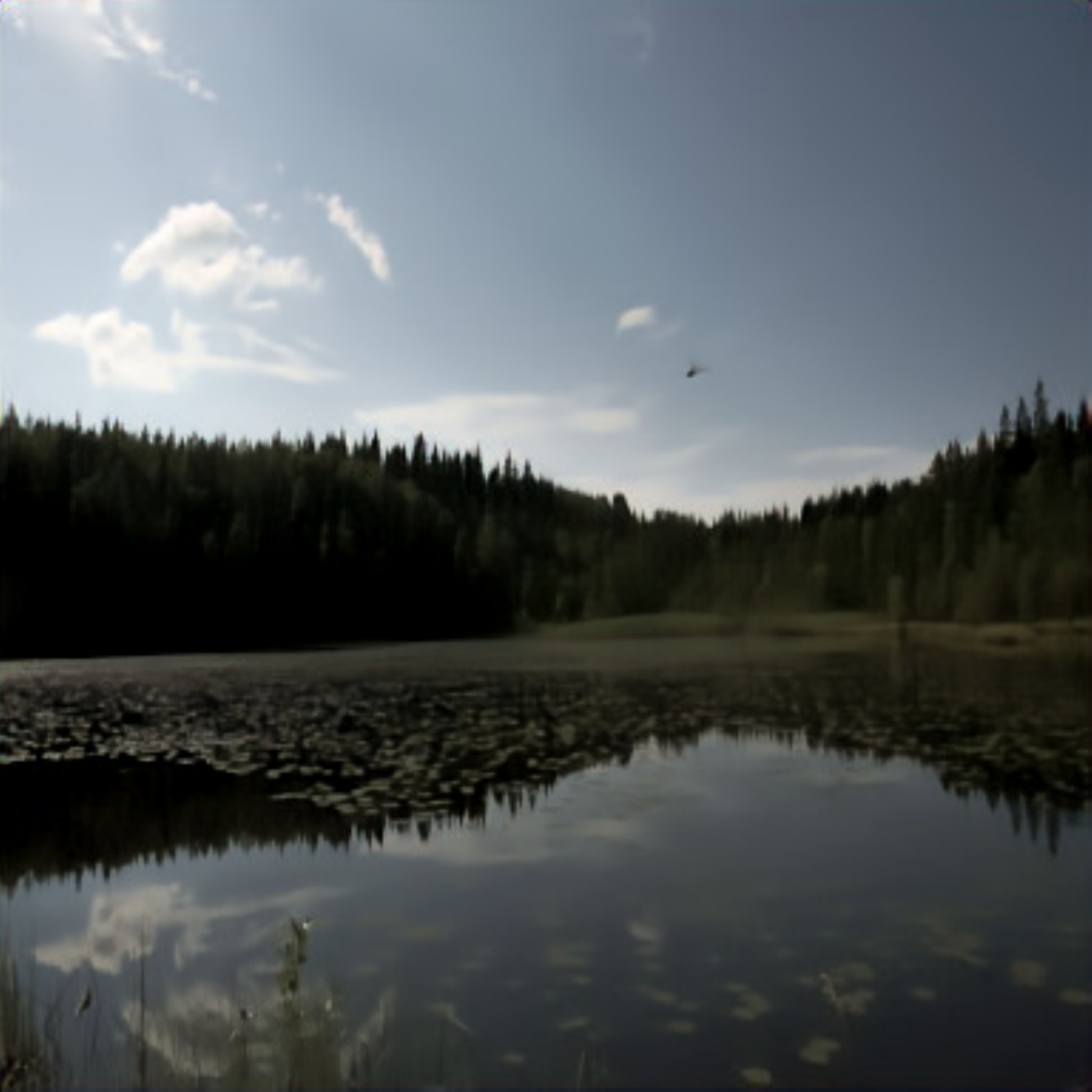}\vspace{2pt} \\
			\includegraphics[width=2.5cm]{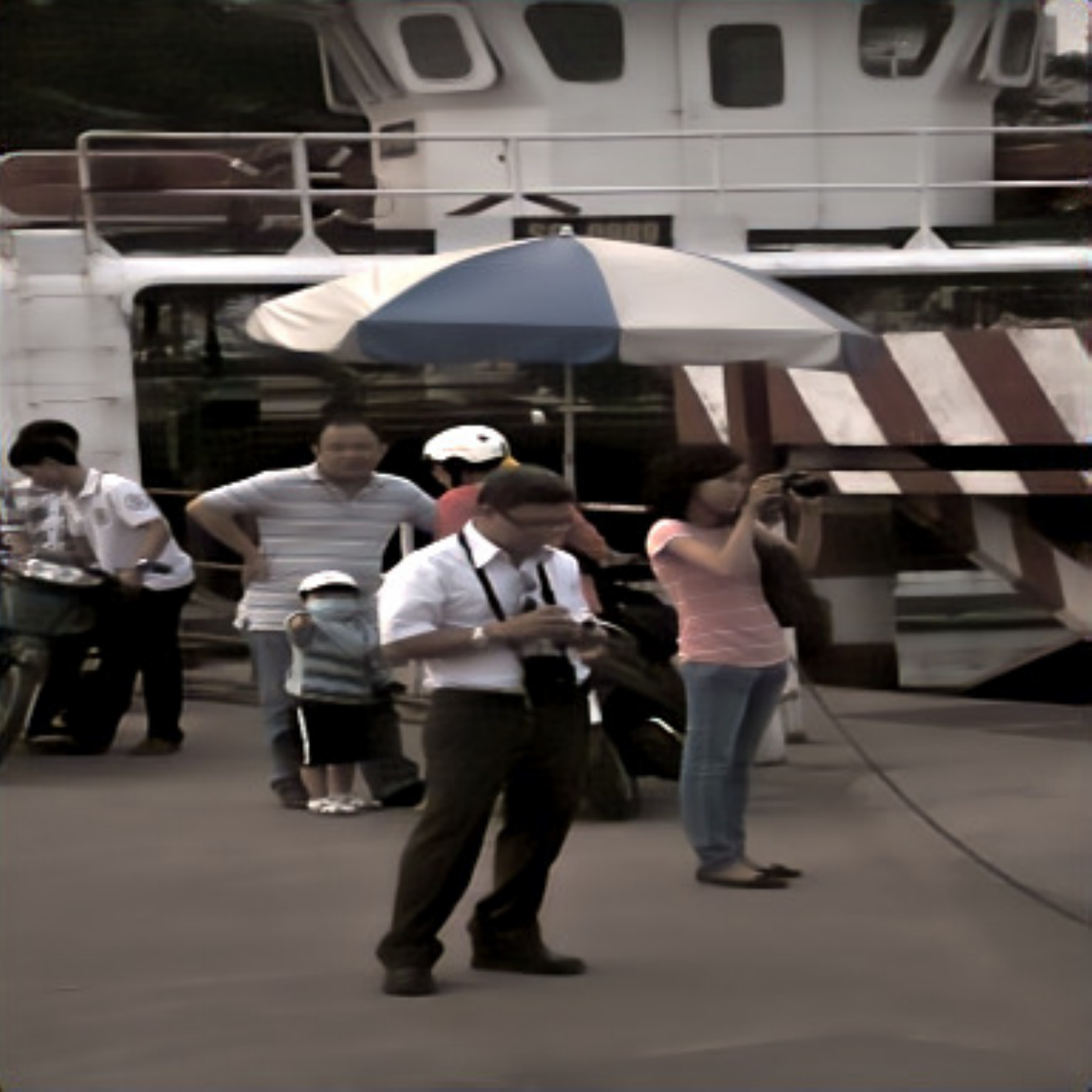}\vspace{2pt}
			\includegraphics[width=2.5cm]{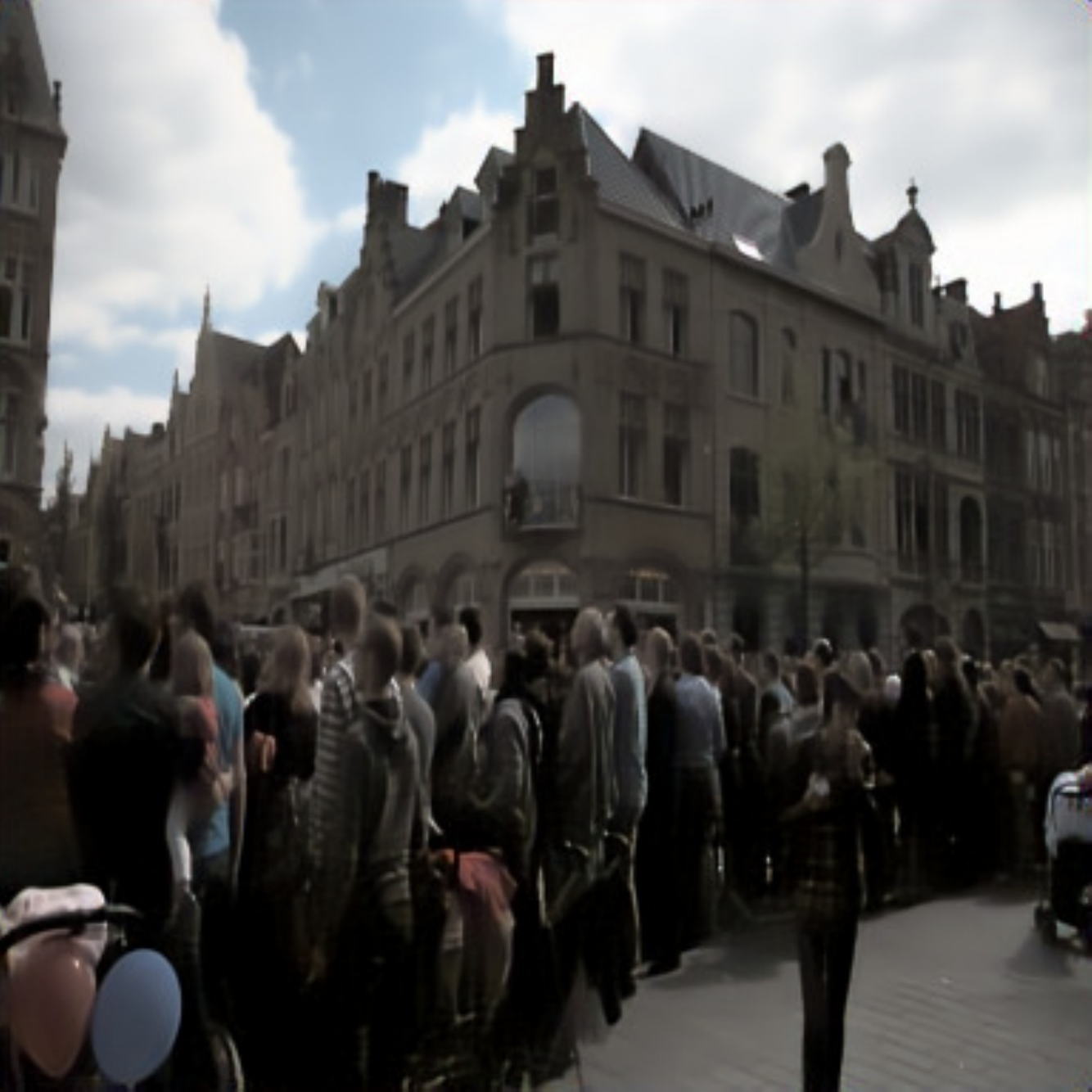}\vspace{2pt}
			\includegraphics[width=2.5cm]{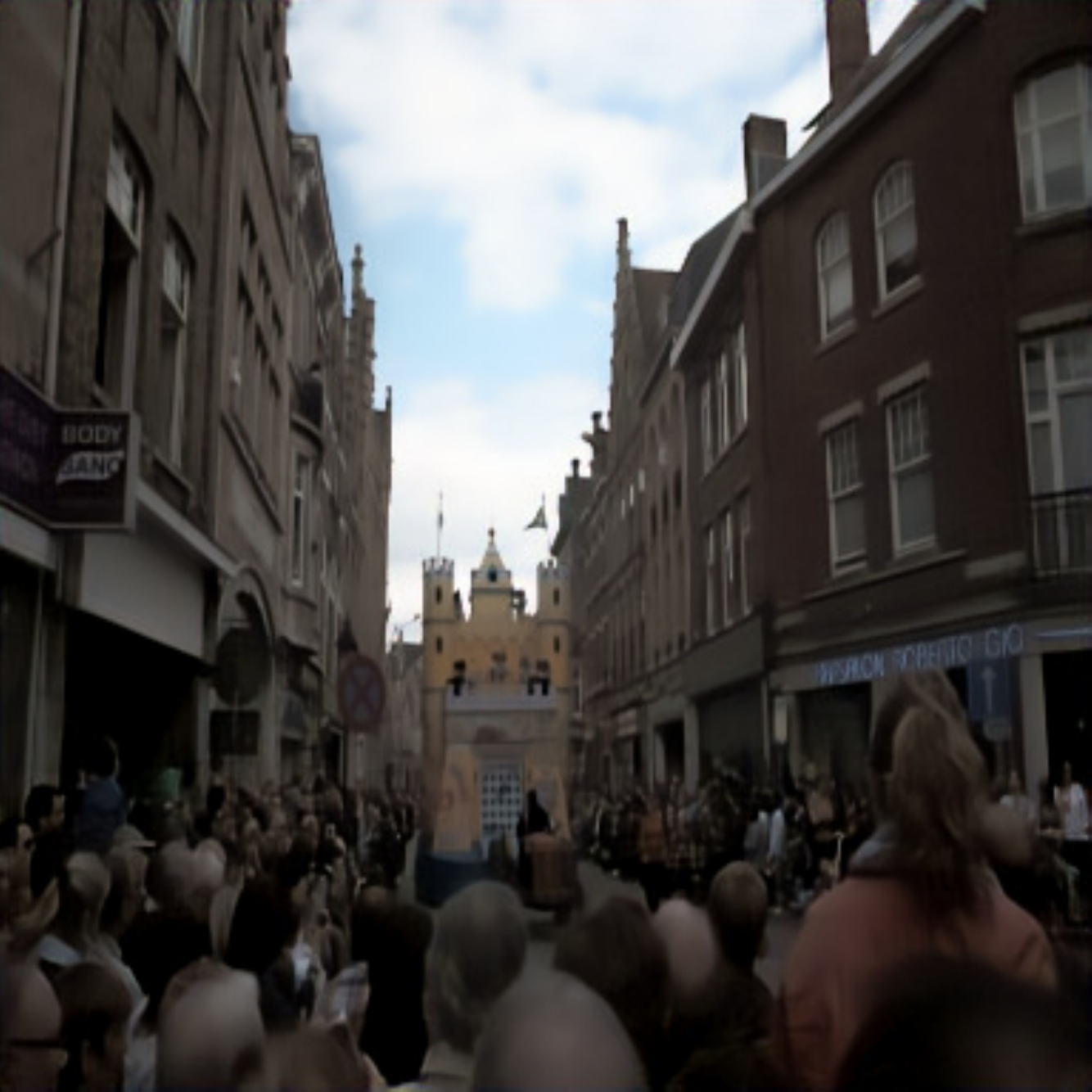}\vspace{2pt} 
		\end{minipage}
	}\hspace{-5pt}
	\subfigure[Zero-DCE\cite{guo2020zero}]{
		\begin{minipage}[b]{0.13\textwidth}
			\includegraphics[width=2.5cm]{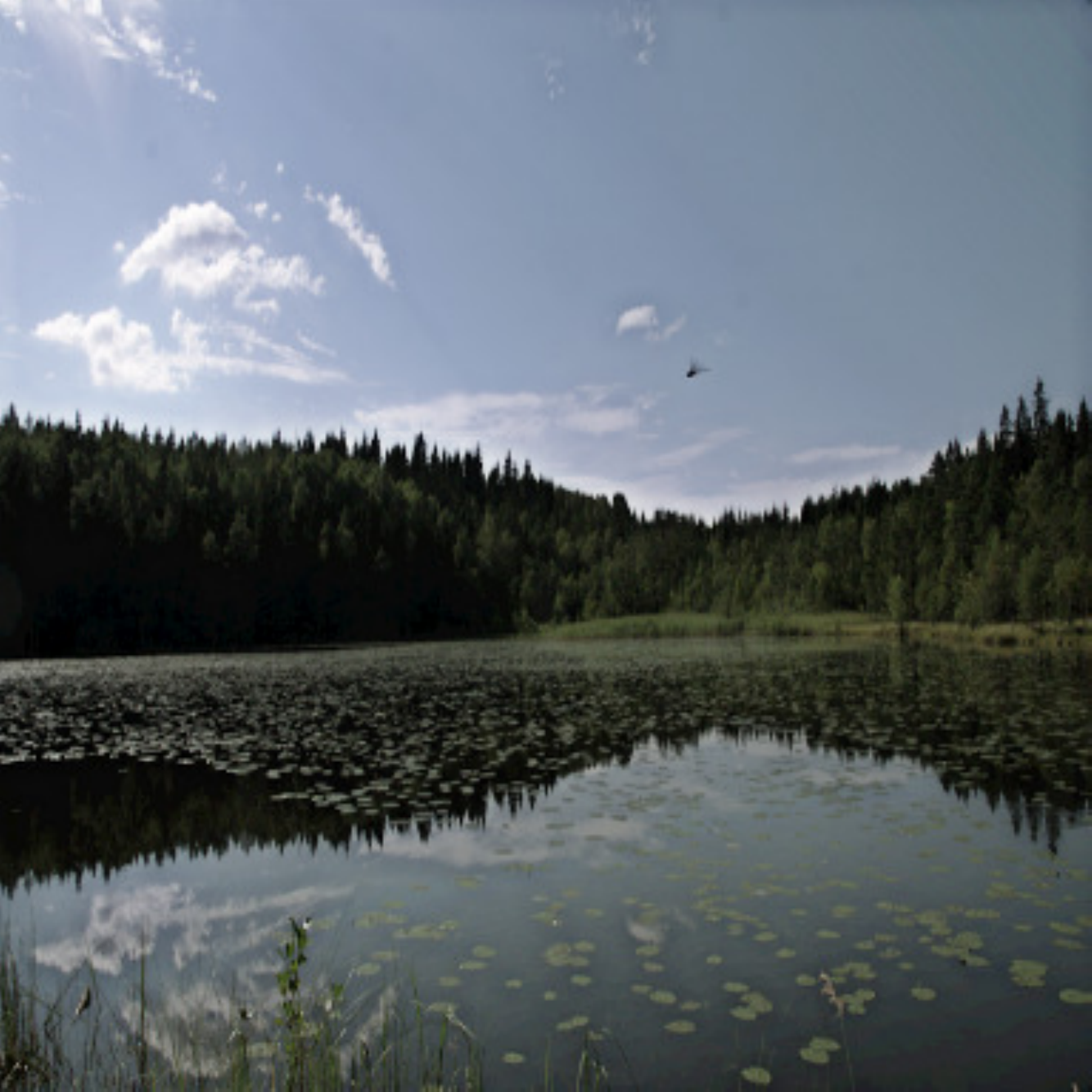}\vspace{2pt} \\
			\includegraphics[width=2.5cm]{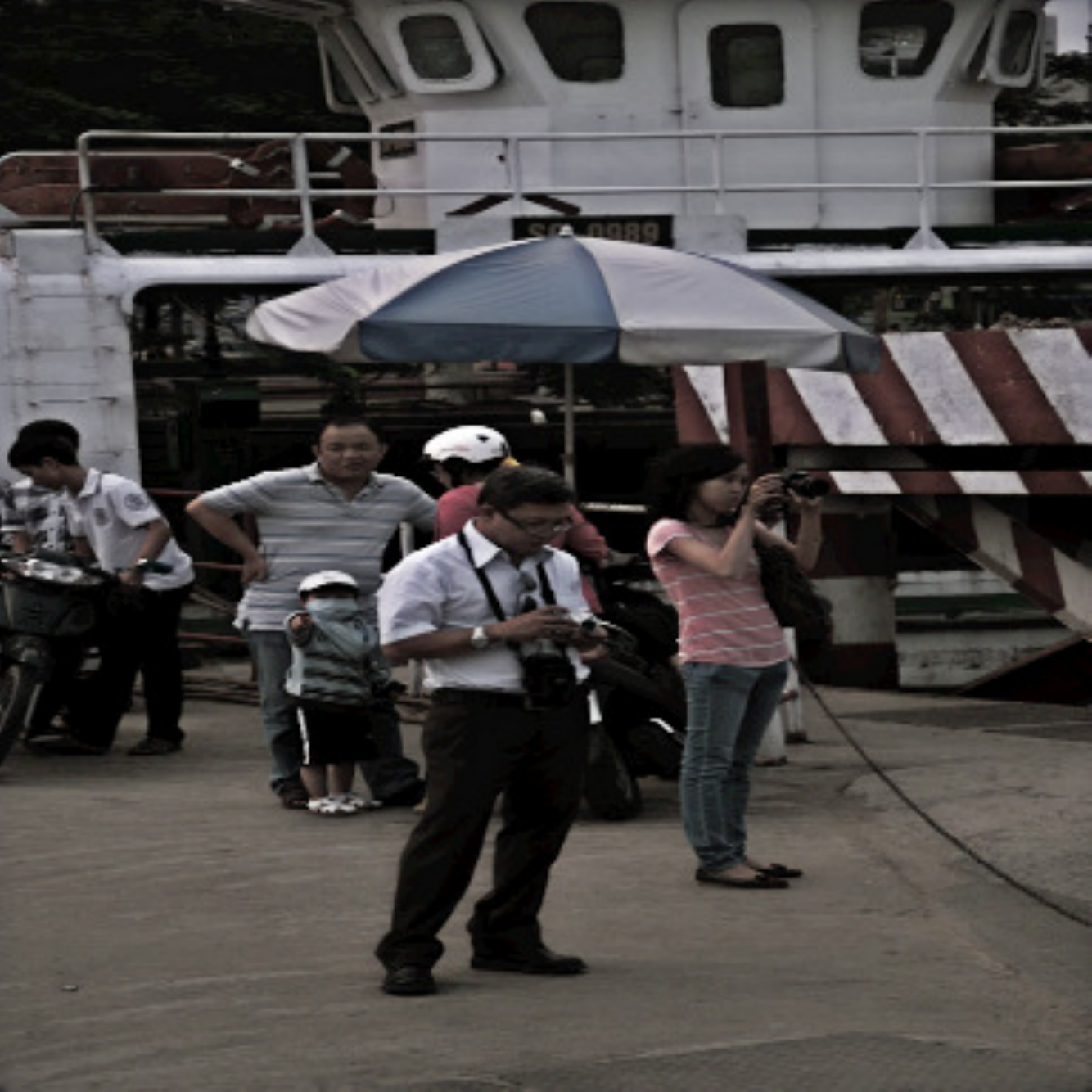}\vspace{2pt}
			\includegraphics[width=2.5cm]{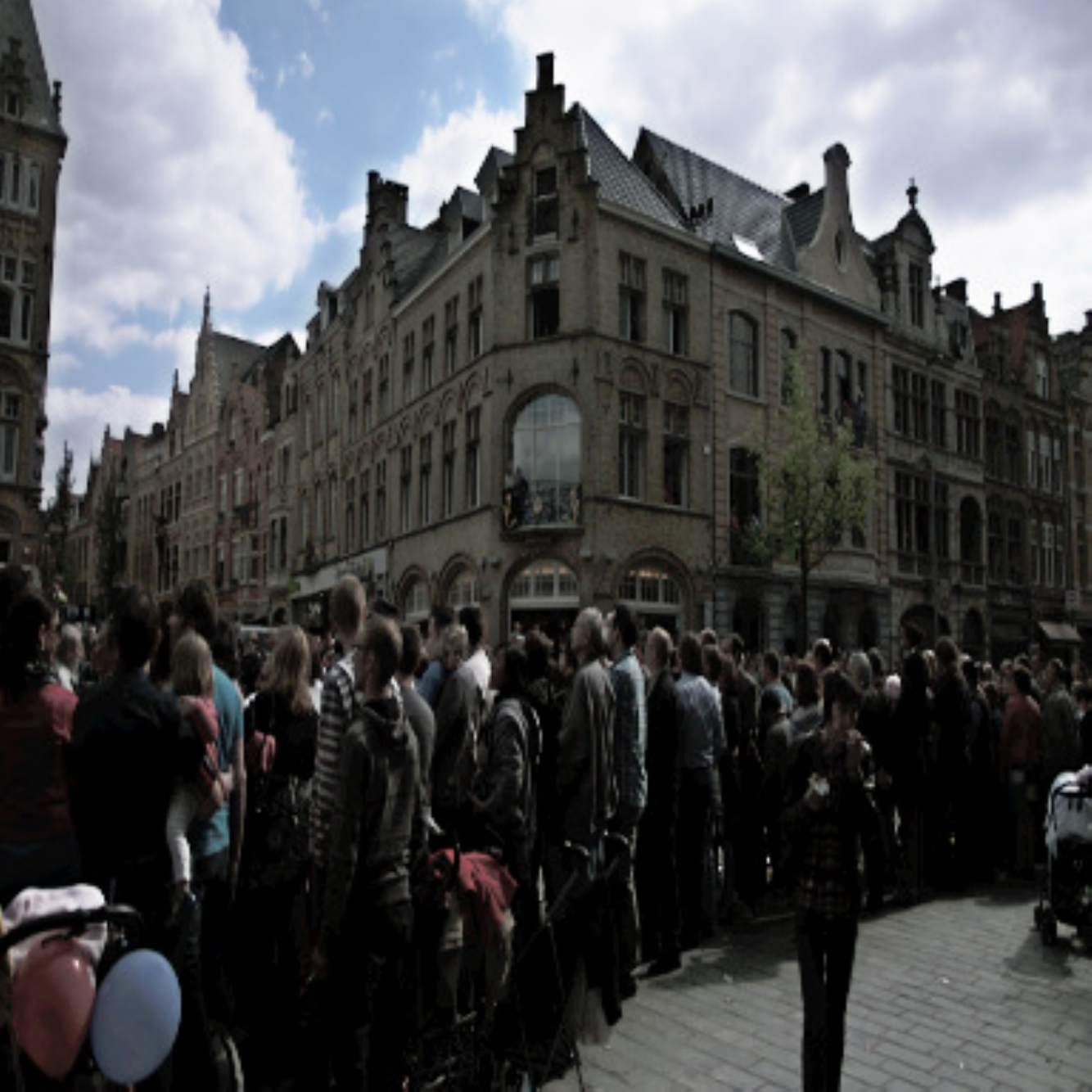}\vspace{2pt}
			\includegraphics[width=2.5cm]{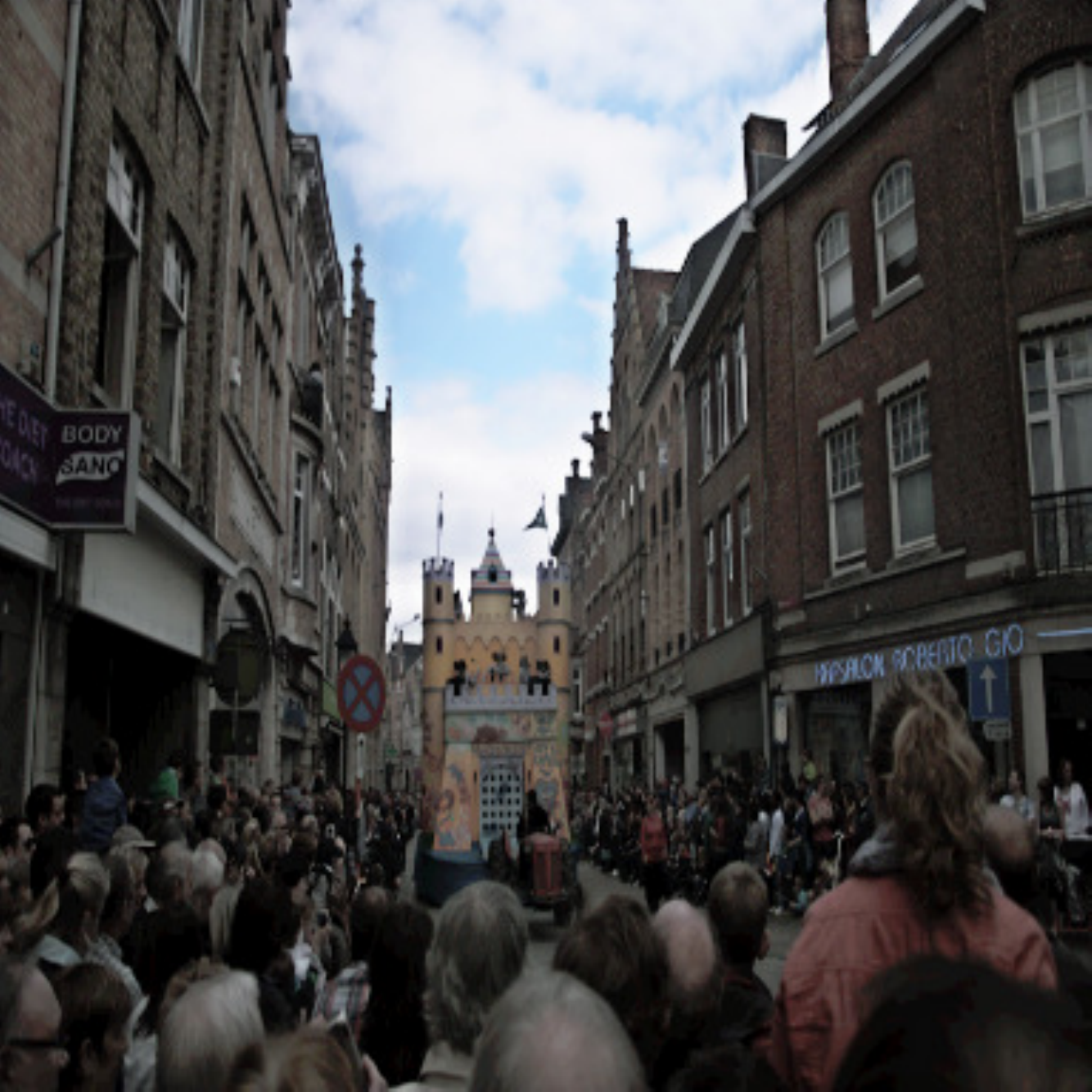}\vspace{2pt} 
		\end{minipage}
	}\hspace{-5pt}
	\subfigure[EnlightenGan\cite{jiang2021enlightengan}]{
		\begin{minipage}[b]{0.13\textwidth}
			\includegraphics[width=2.5cm]{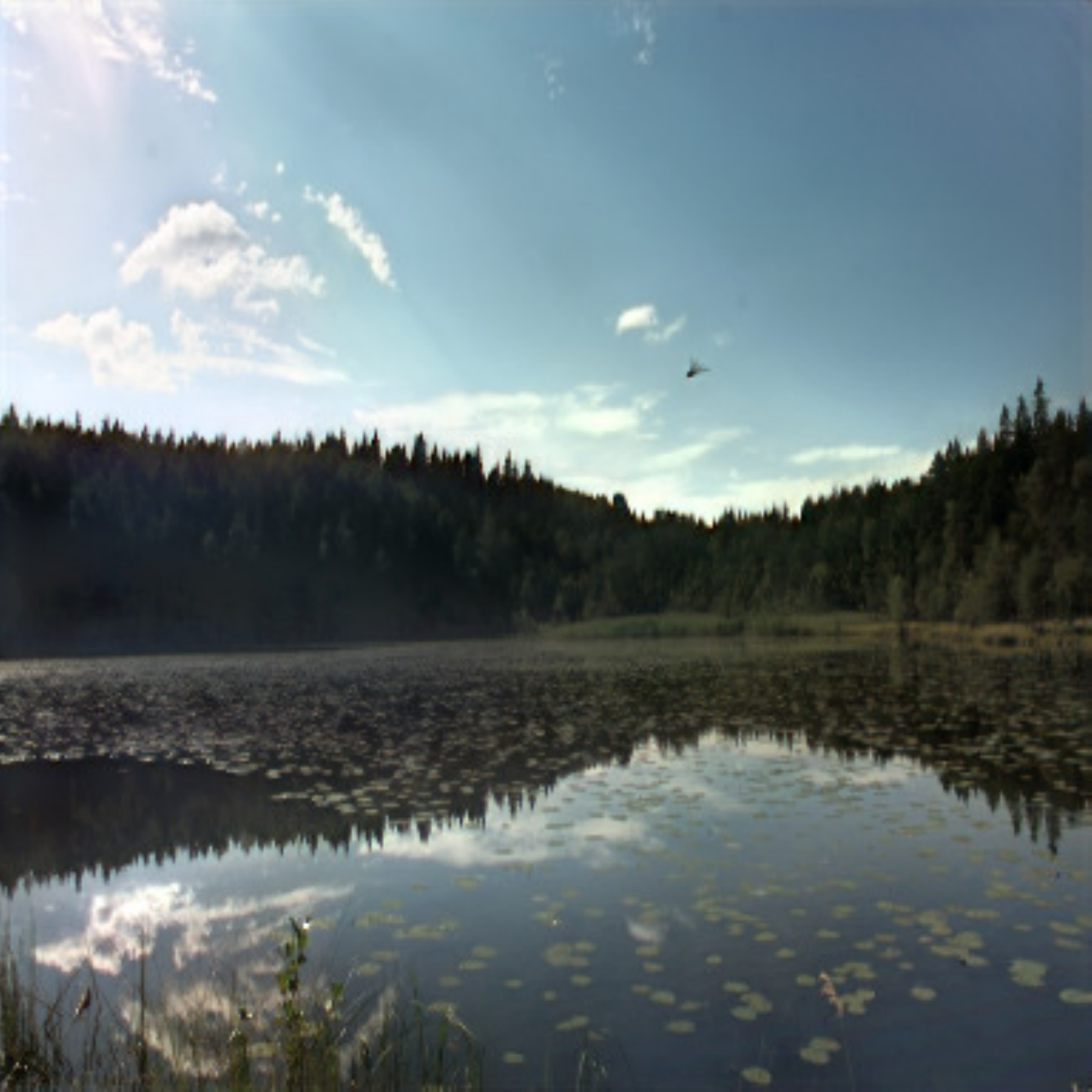}\vspace{2pt} \\
			\includegraphics[width=2.5cm]{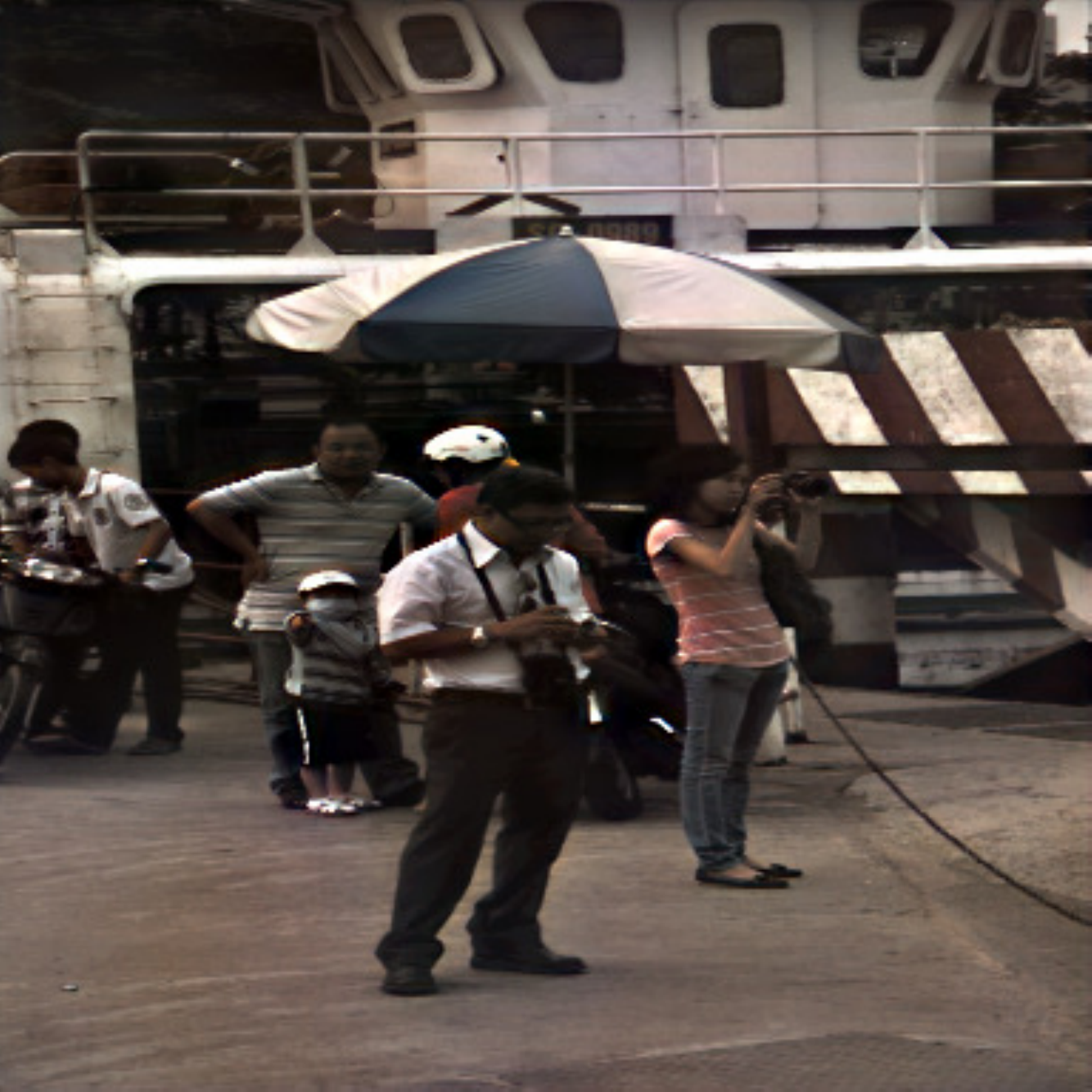}\vspace{2pt}
			\includegraphics[width=2.5cm]{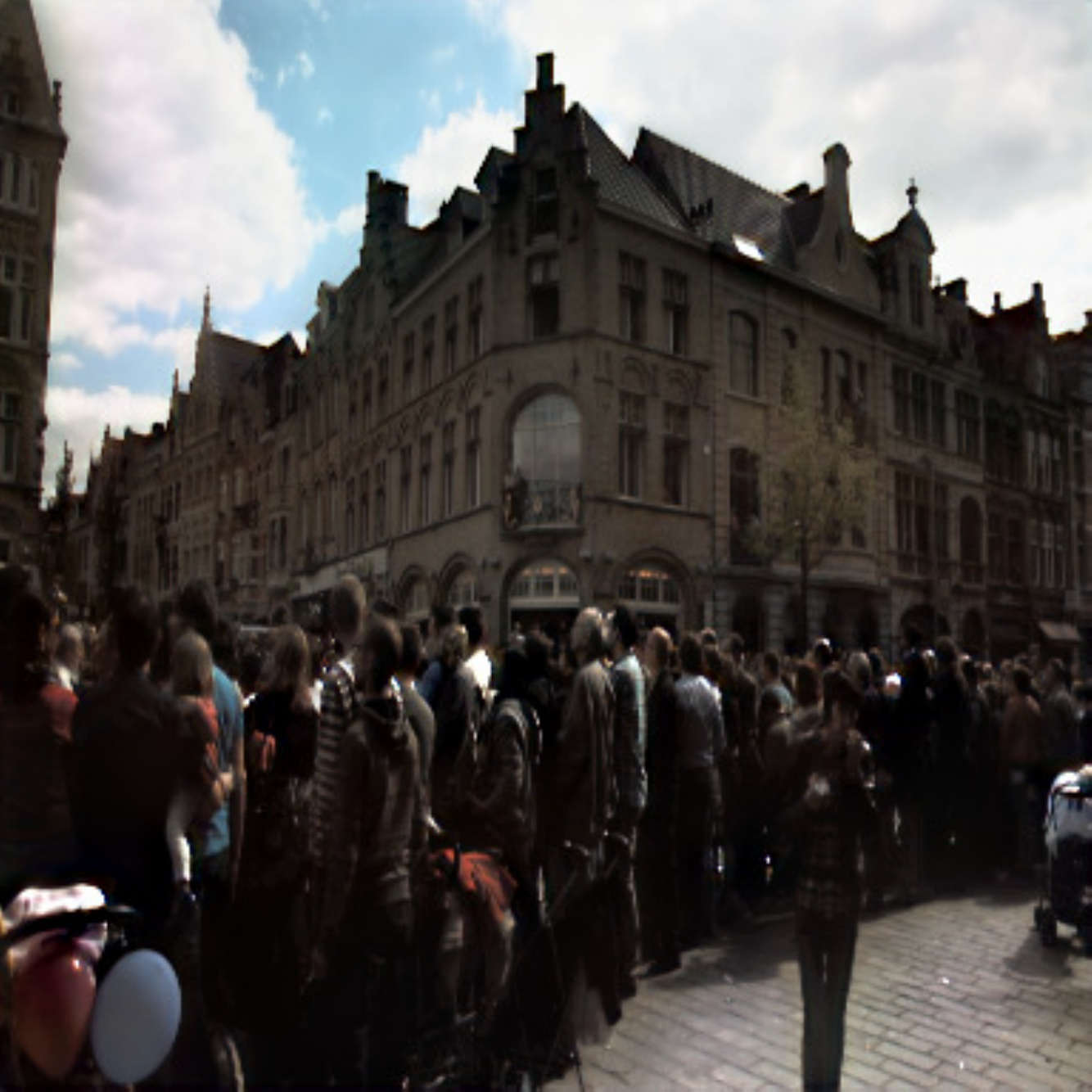}\vspace{2pt}
			\includegraphics[width=2.5cm]{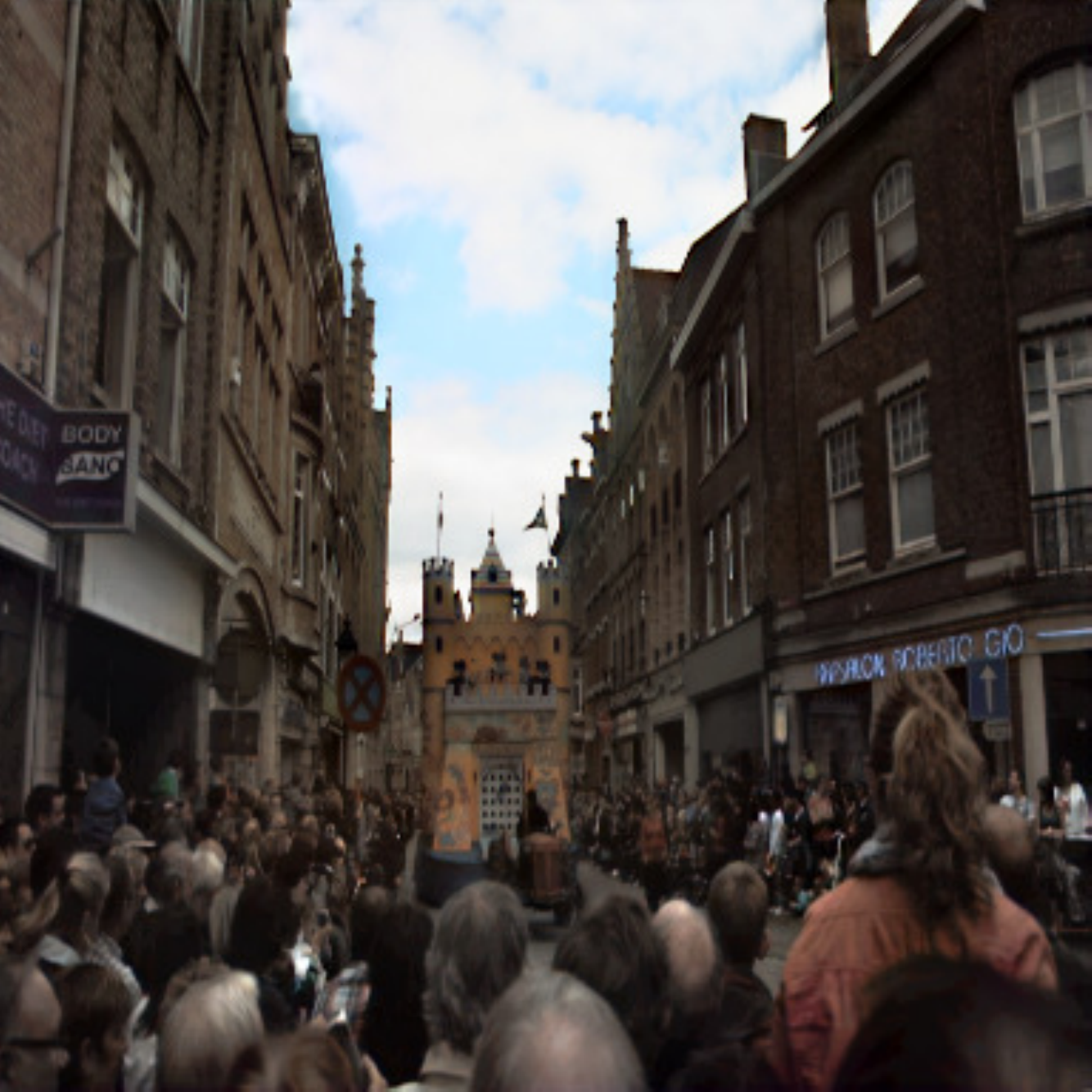}\vspace{2pt} 
		\end{minipage}
	}\hspace{-5pt}
	\subfigure[DA-DRN]{
		\begin{minipage}[b]{0.13\textwidth}
			\includegraphics[width=2.5cm]{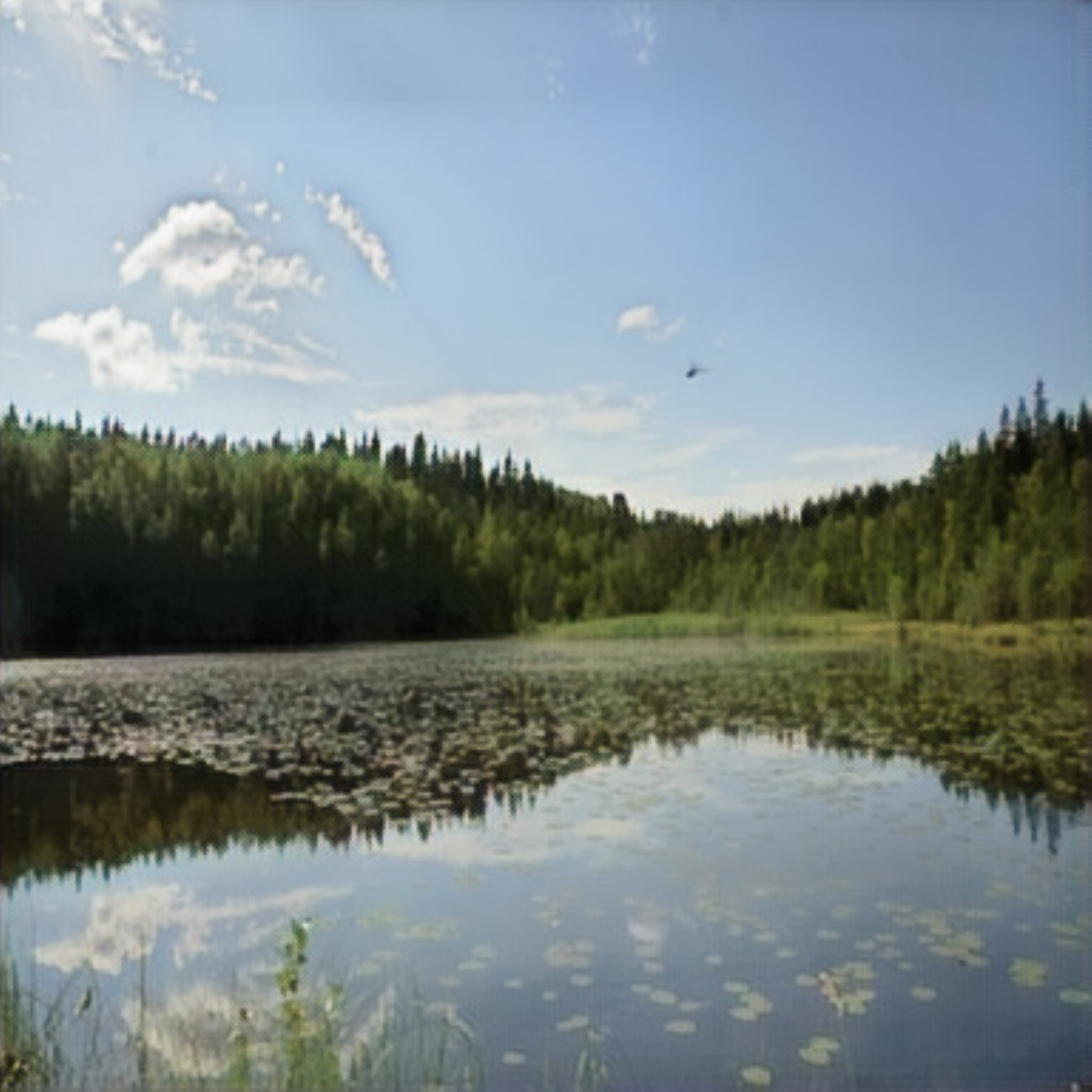}\vspace{2pt} \\
			\includegraphics[width=2.5cm]{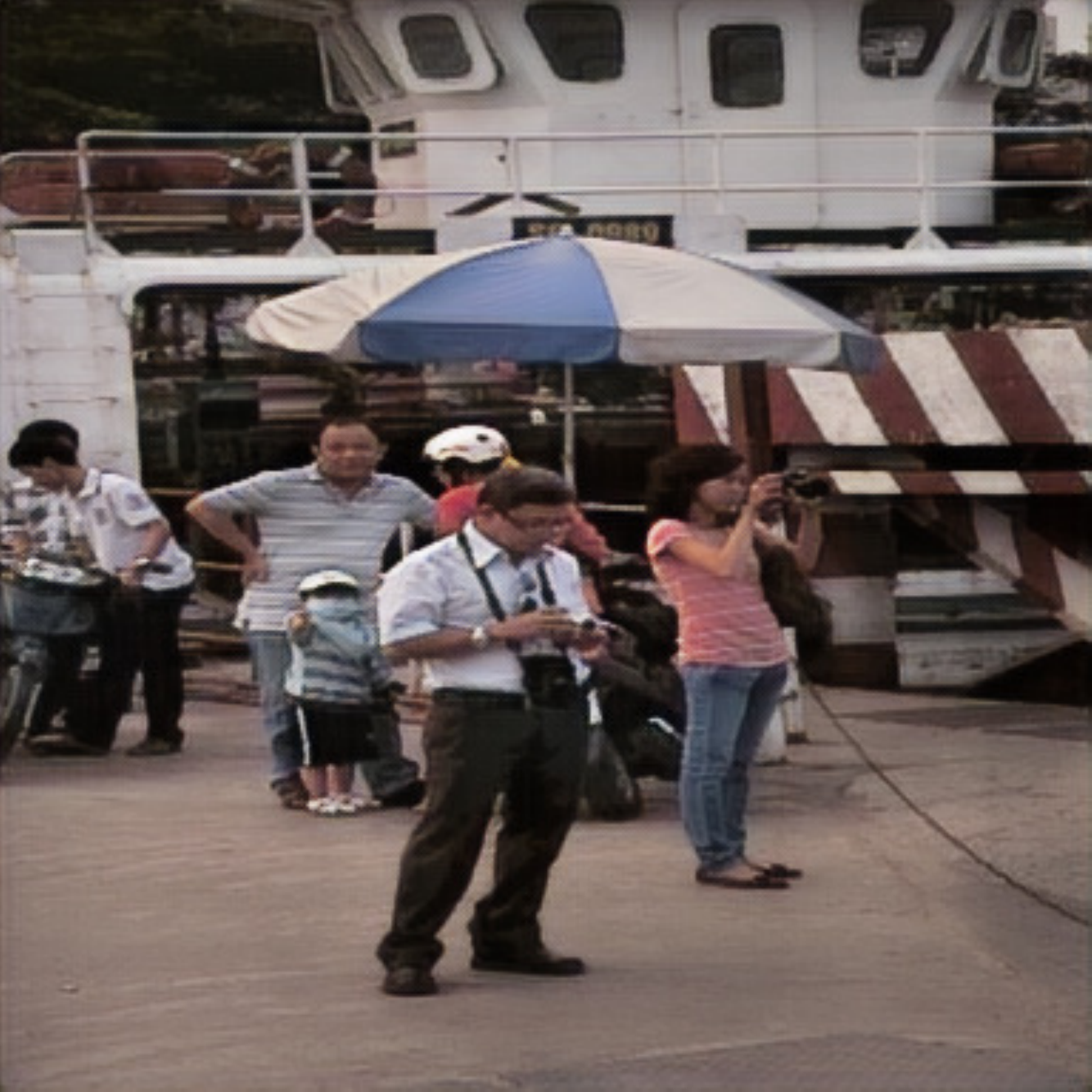}\vspace{2pt}
			\includegraphics[width=2.5cm]{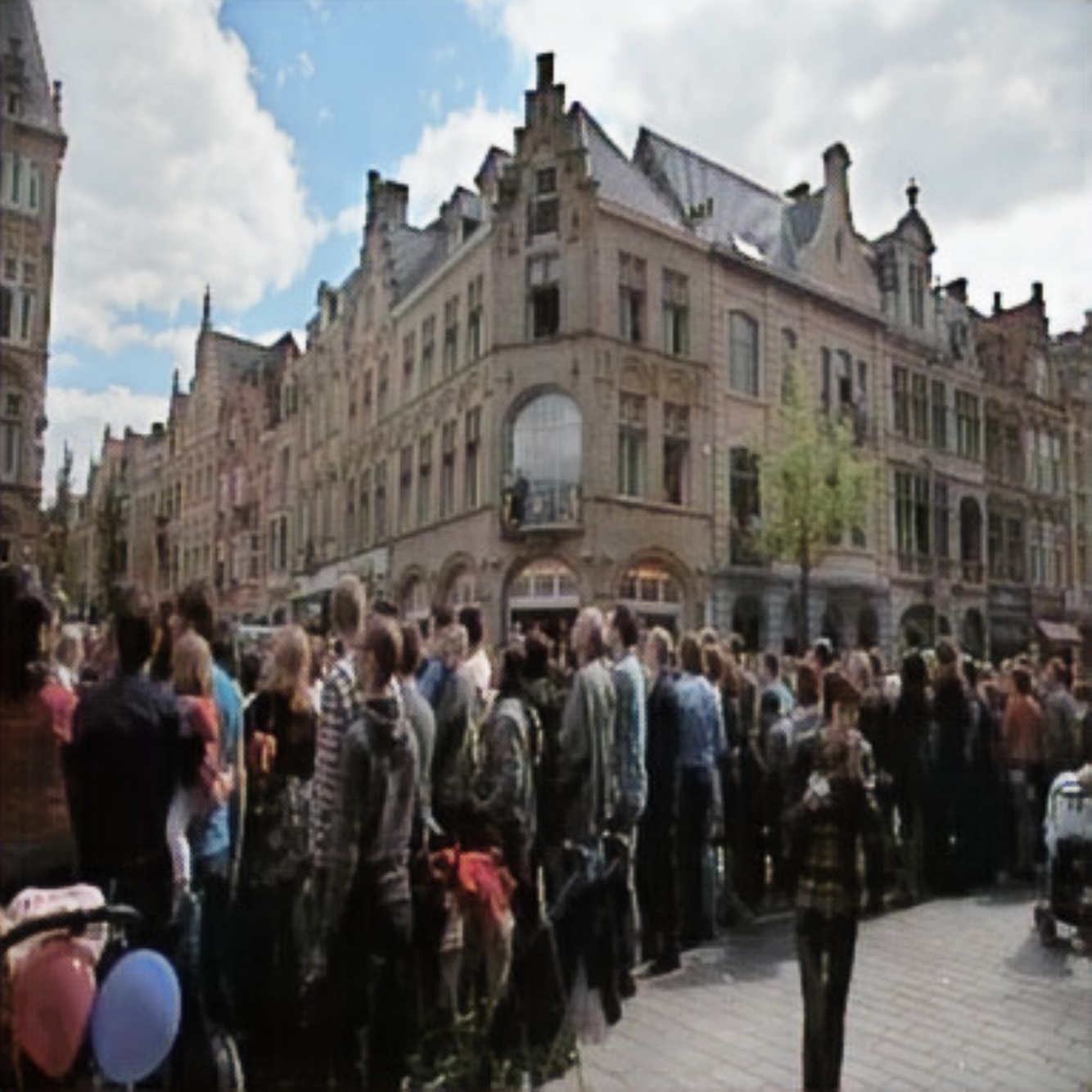}\vspace{2pt}
			\includegraphics[width=2.5cm]{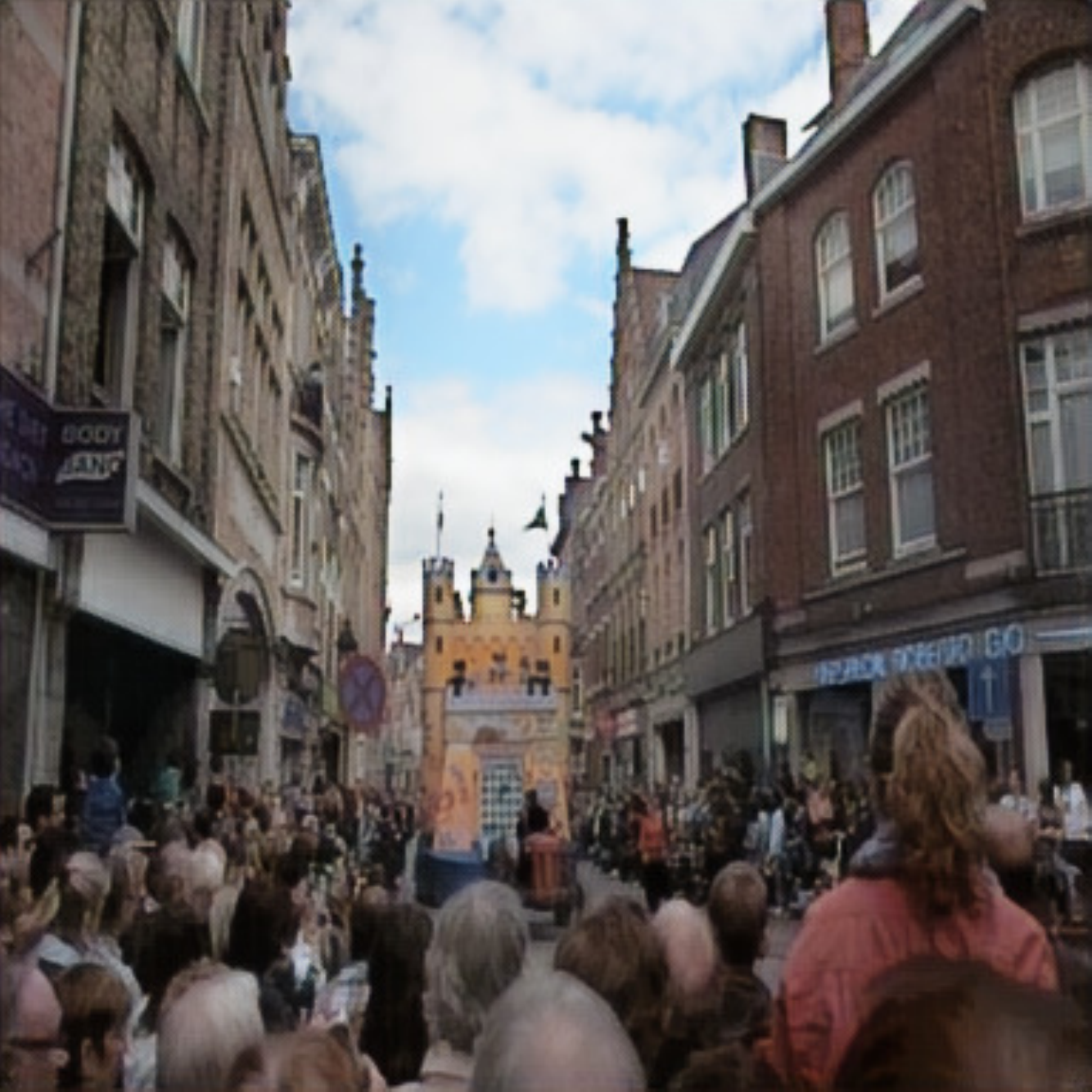}\vspace{2pt} 
		\end{minipage}
	}\hspace{-5pt}
	\subfigure[Ground-Truth]{
		\begin{minipage}[b]{0.13\textwidth}
			\includegraphics[width=2.5cm]{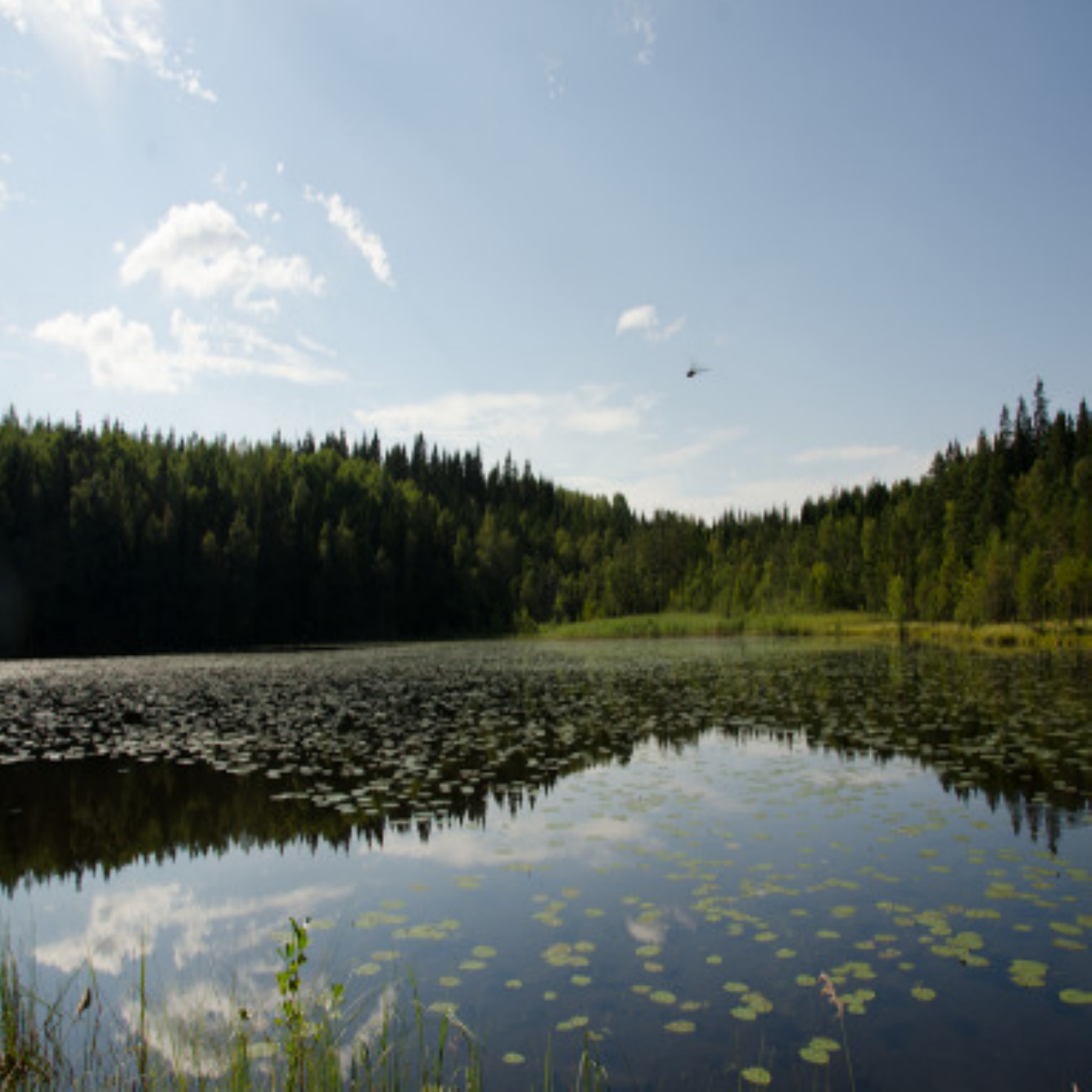}\vspace{2pt} \\
			\includegraphics[width=2.5cm]{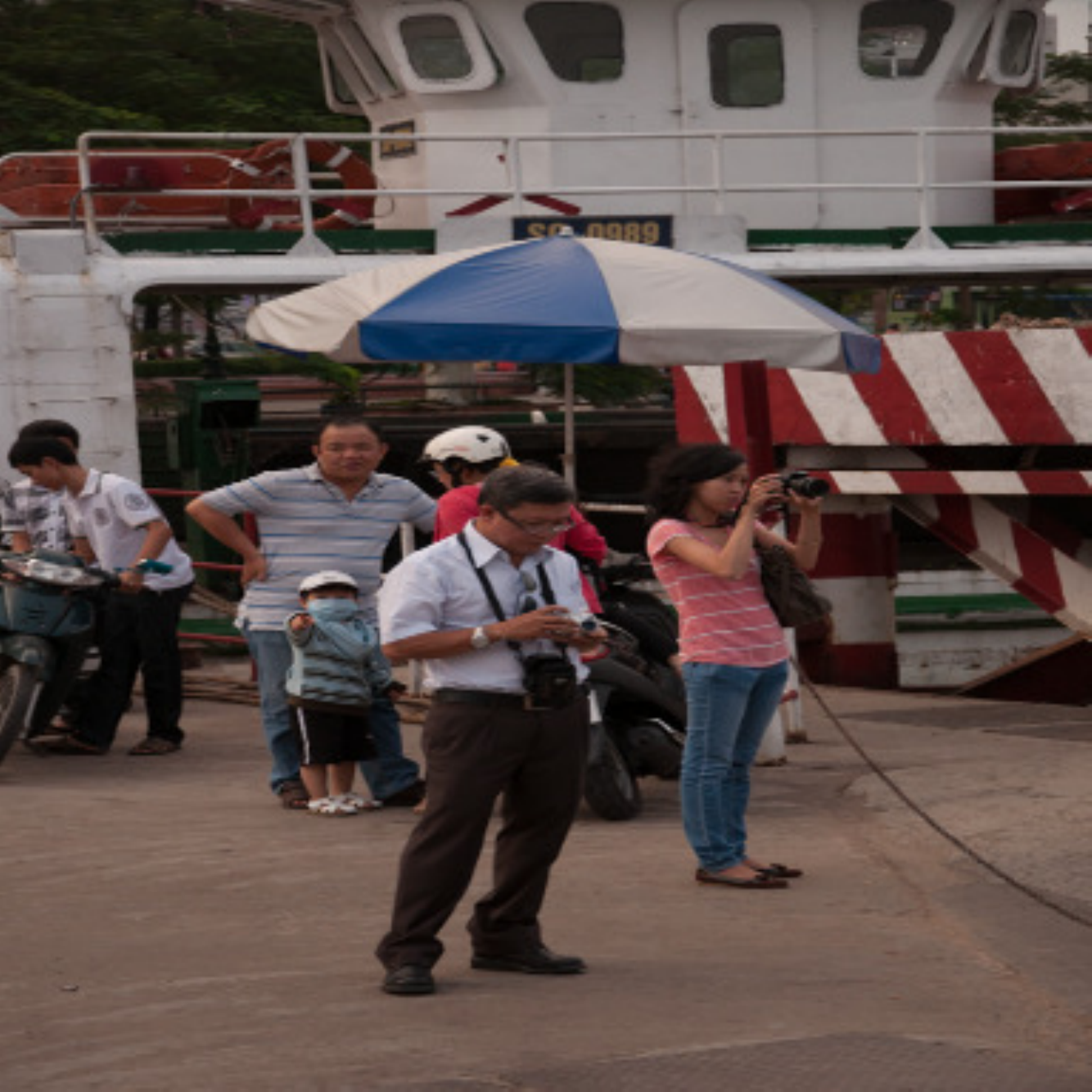}\vspace{2pt}
			\includegraphics[width=2.5cm]{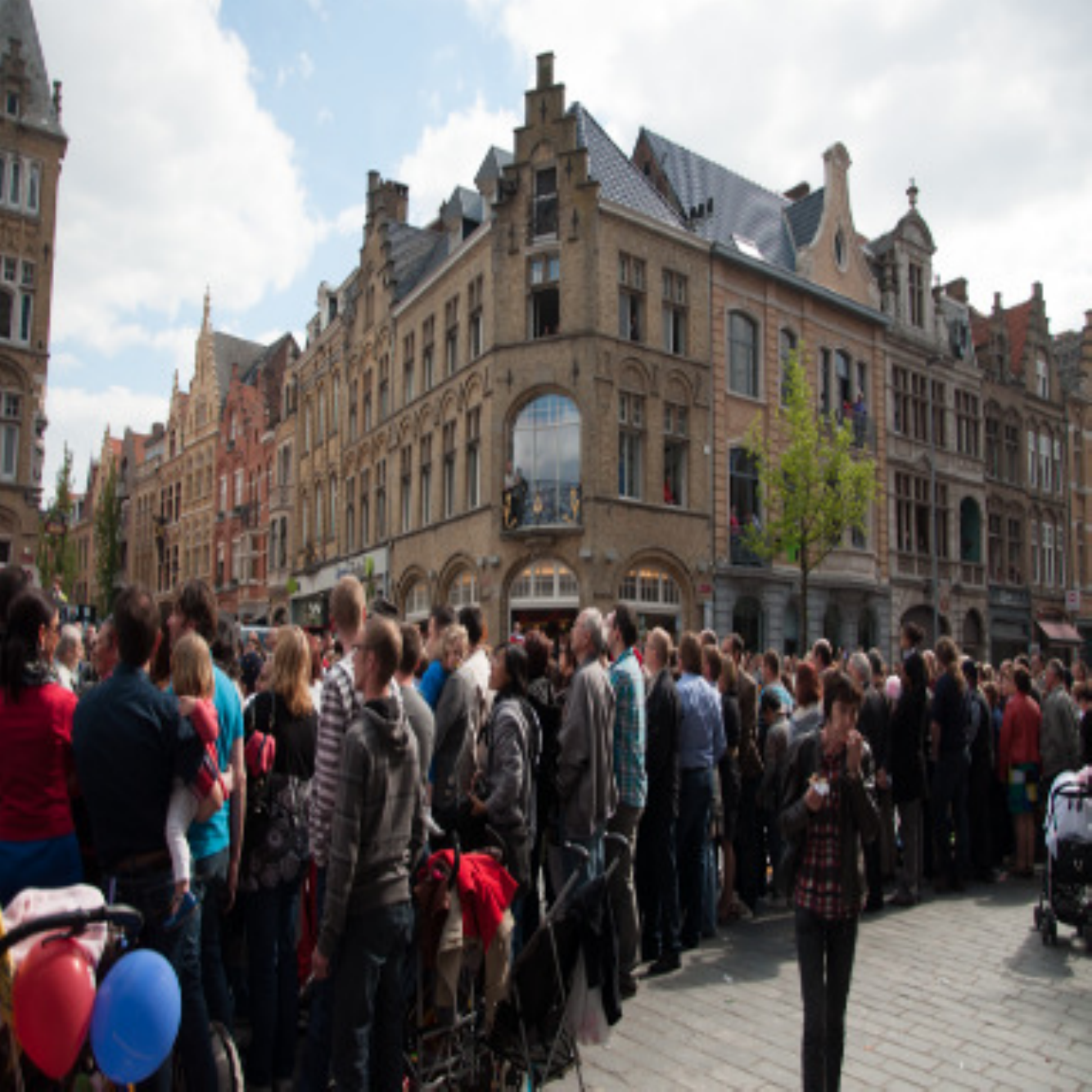}\vspace{2pt}
			\includegraphics[width=2.5cm]{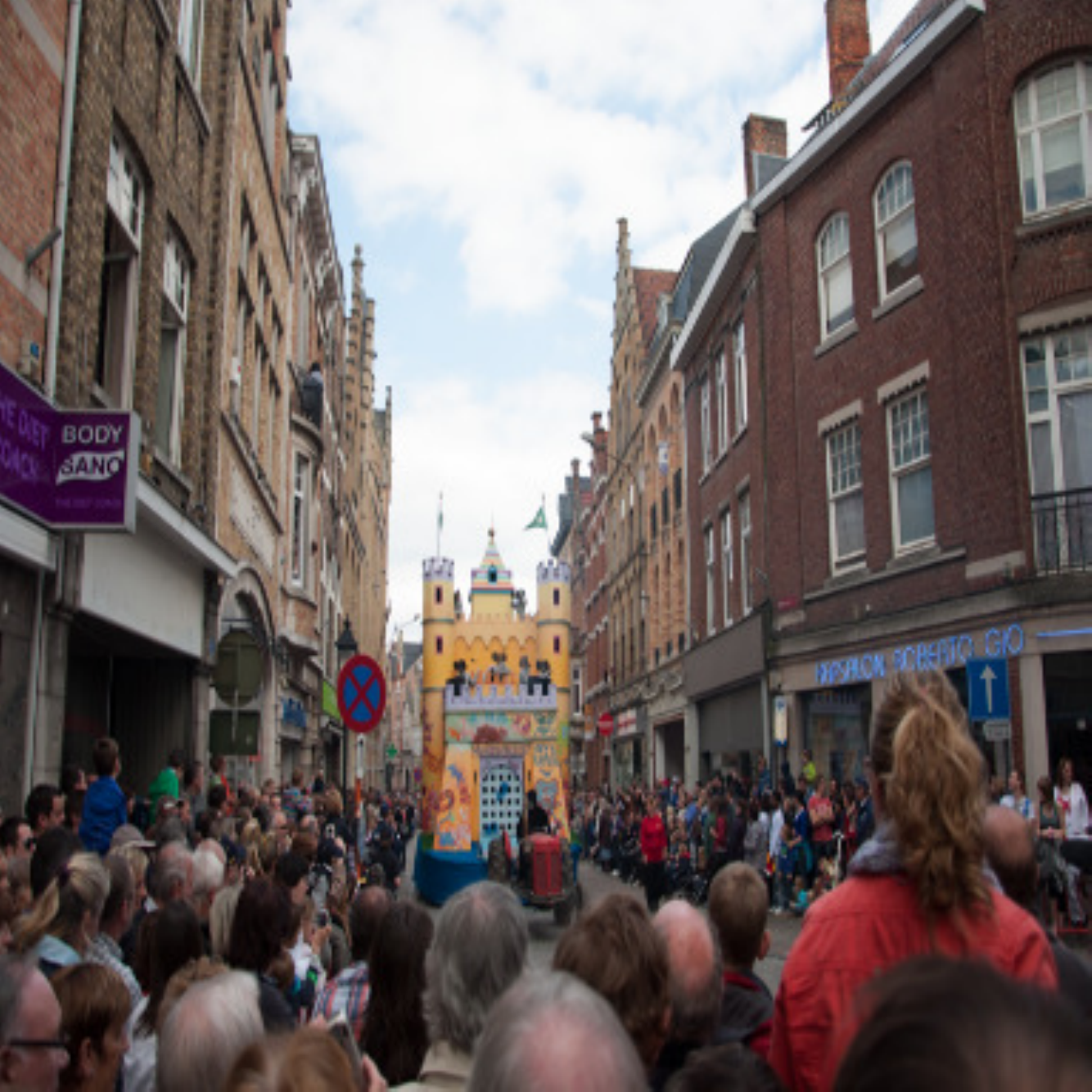}\vspace{2pt} 
		\end{minipage}
	}
	\caption{Visual comparison with other state-of-the-art methods on the LOL synthetic validation dataset.}
	\label{syn}
\end{figure*}

\begin{figure*}
	\flushleft
	
	
	\subfigure[Input]{
		\begin{minipage}[b]{0.155\textwidth}
			\includegraphics[width=2.8cm]{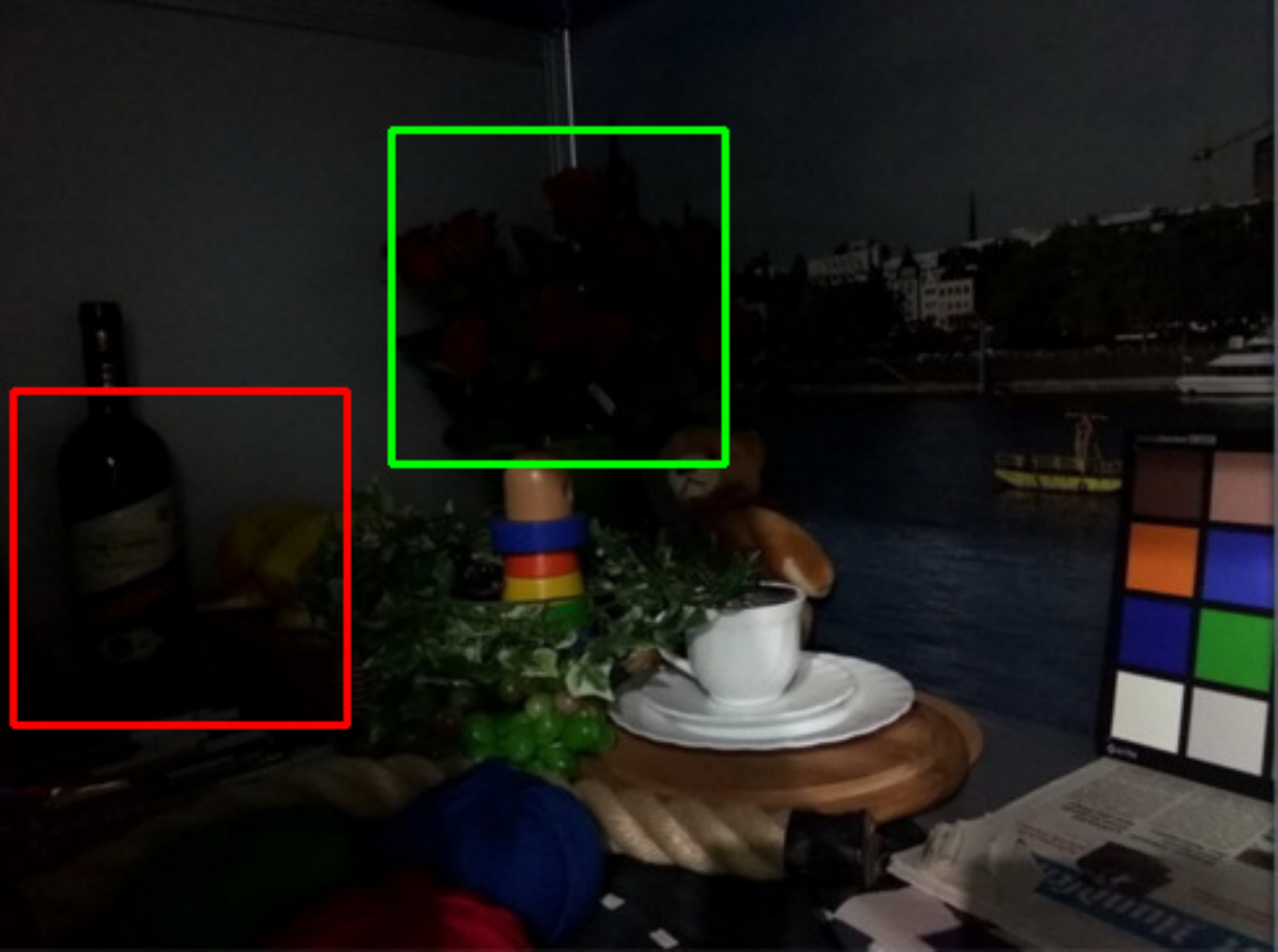}\vspace{1pt} \\
			\includegraphics[width=1.35cm]{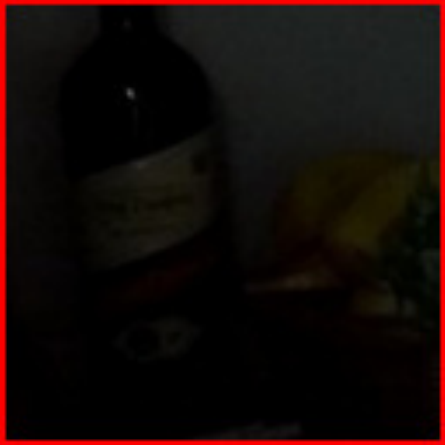}
			\includegraphics[width=1.35cm]{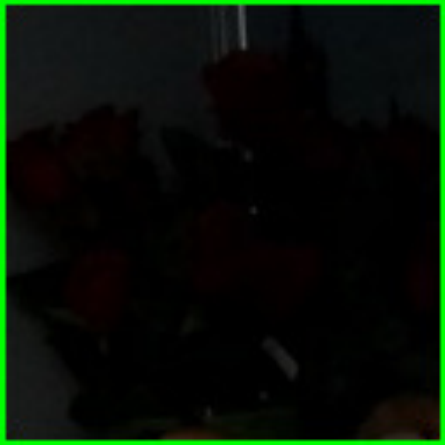}\vspace{5pt}
			\includegraphics[width=2.8cm]{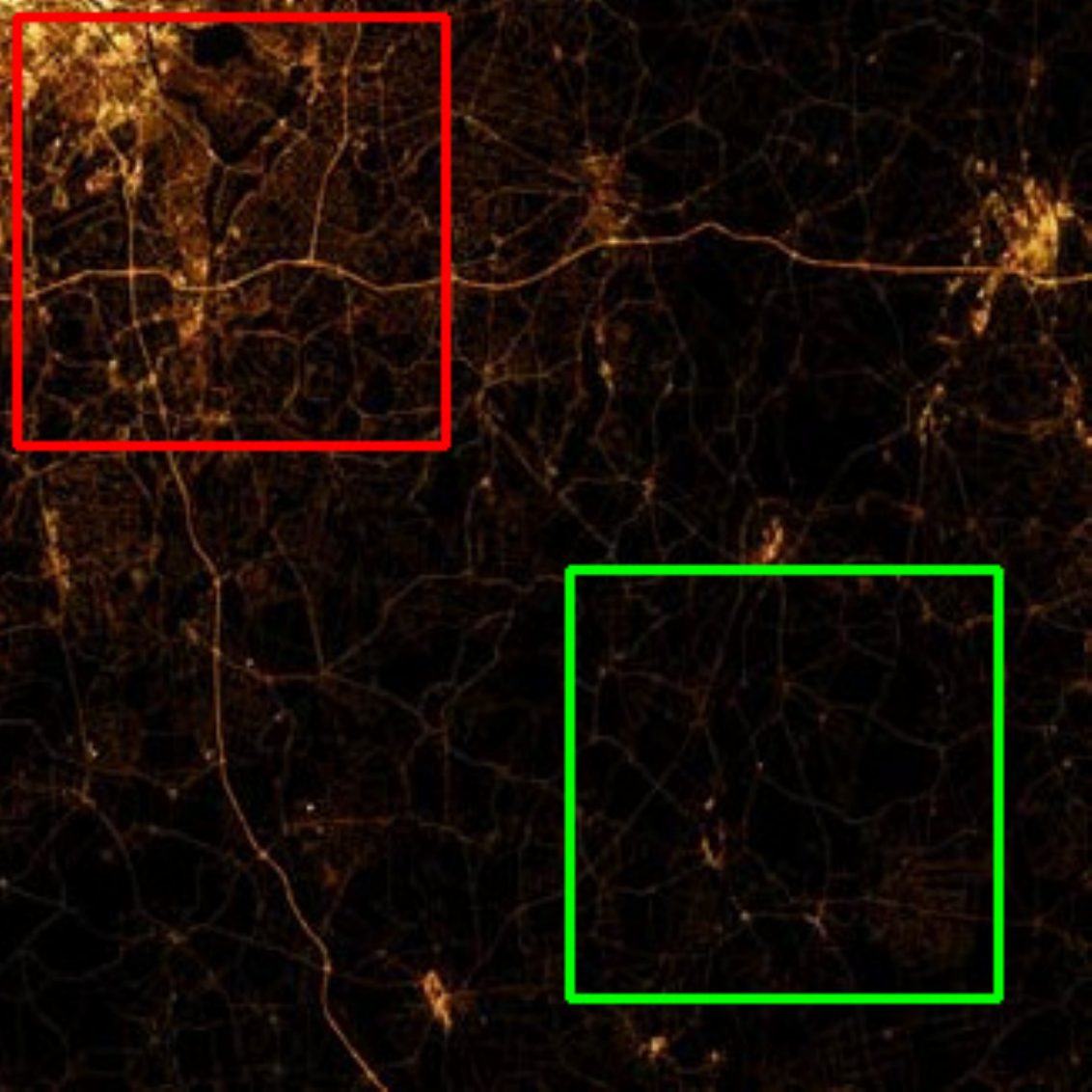}\vspace{1.5pt} \\
			\includegraphics[width=1.35cm]{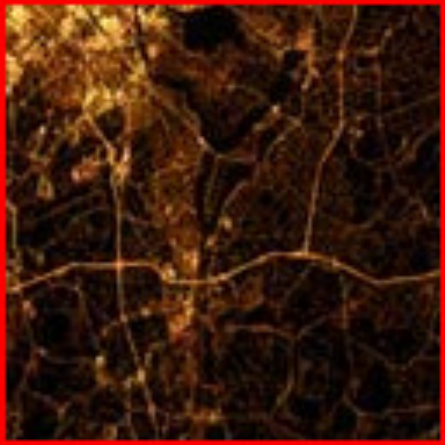}
			\includegraphics[width=1.35cm]{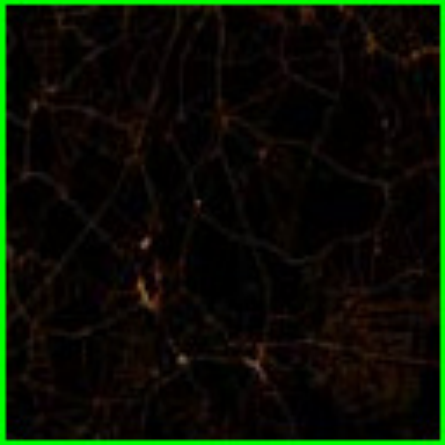}
		\end{minipage}
	}\hspace{-5pt}
	\subfigure[MSRCR\cite{jobson1997multiscale}]{
		\begin{minipage}[b]{0.155\textwidth}
			\includegraphics[width=2.8cm]{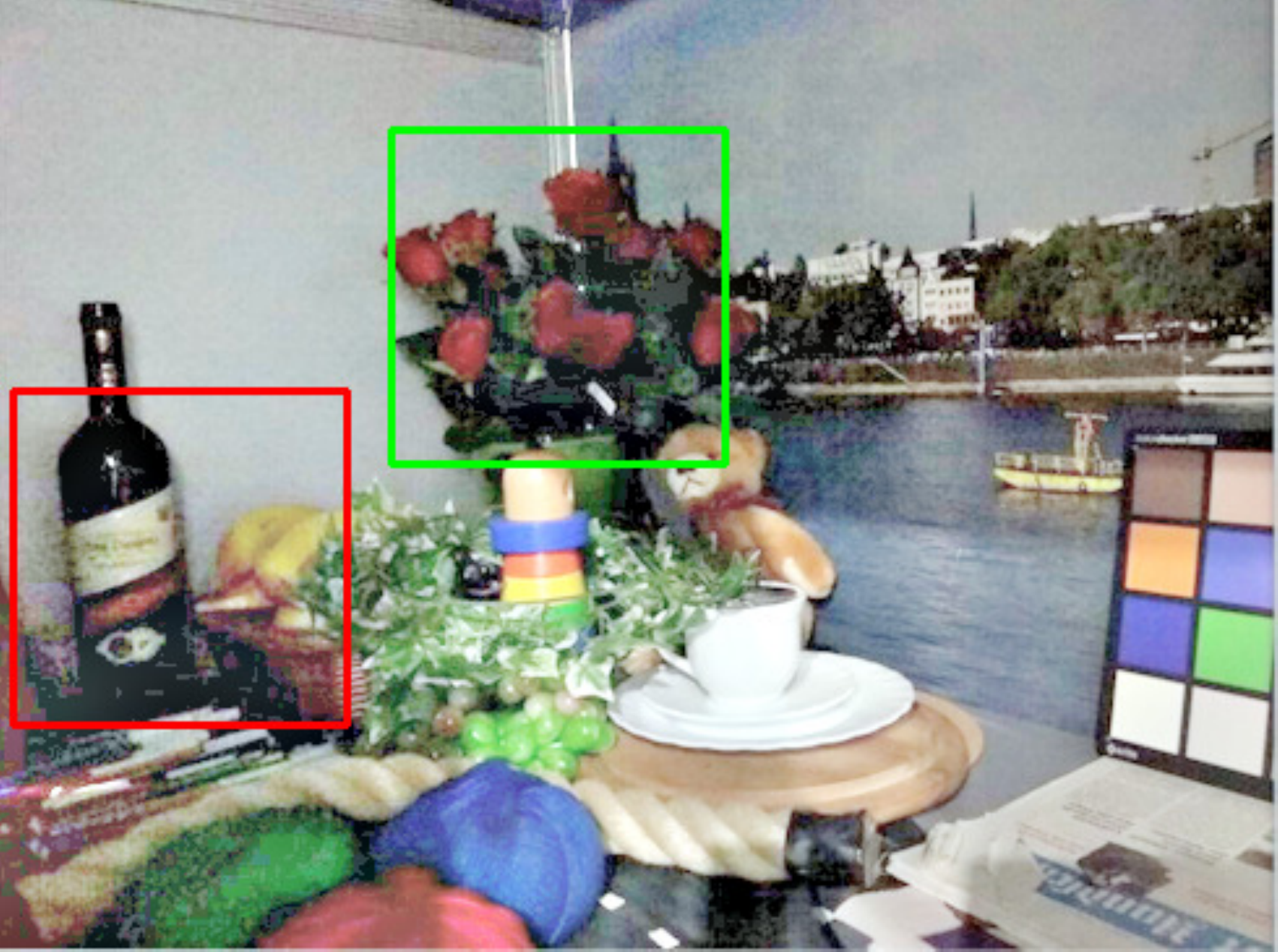}\vspace{1pt} \\
			\includegraphics[width=1.35cm]{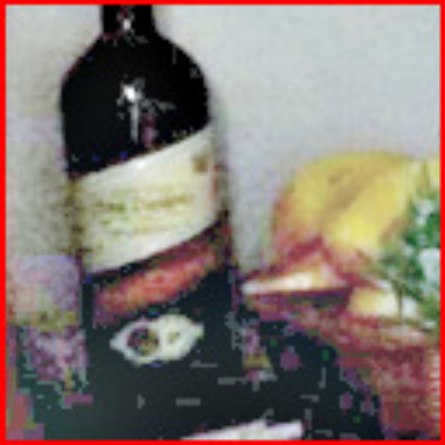}
			\includegraphics[width=1.35cm]{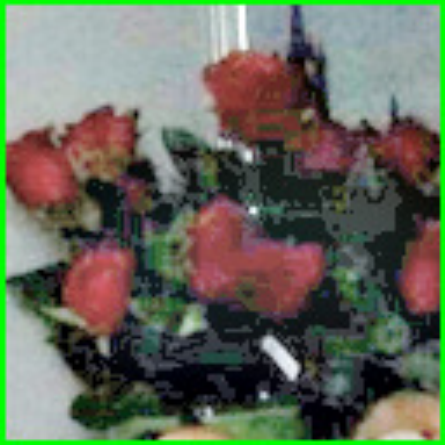}\vspace{5pt}
			\includegraphics[width=2.8cm]{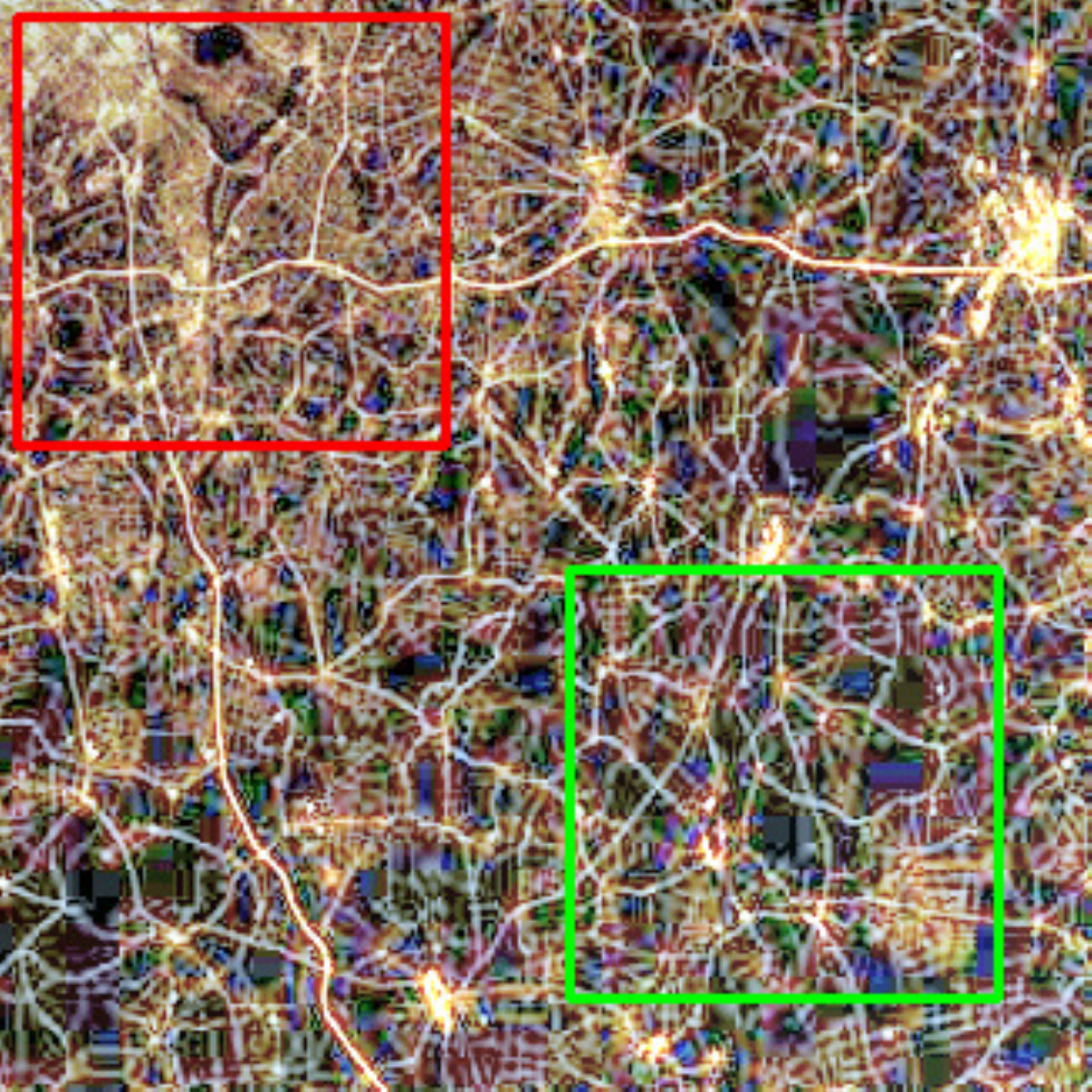}\vspace{1.5pt} \\
			\includegraphics[width=1.35cm]{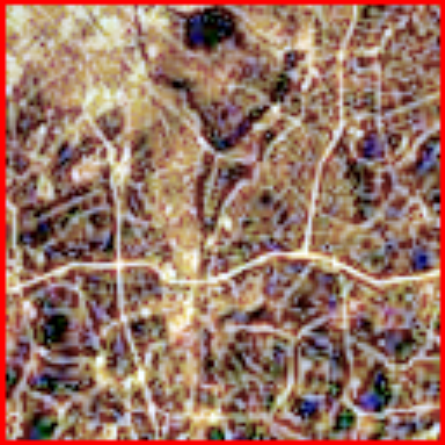}
			\includegraphics[width=1.35cm]{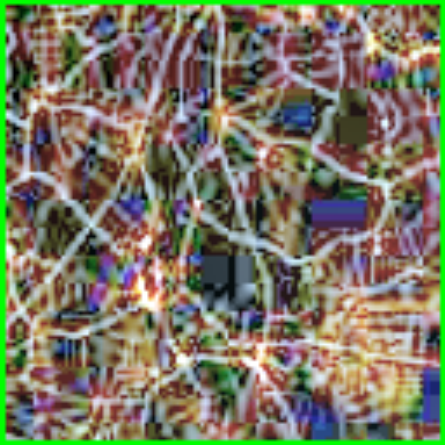}
		\end{minipage}
	}\hspace{-5pt}
	\subfigure[BIMEF\cite{ying2017bio}]{
		\begin{minipage}[b]{0.155\textwidth}
			\includegraphics[width=2.8cm]{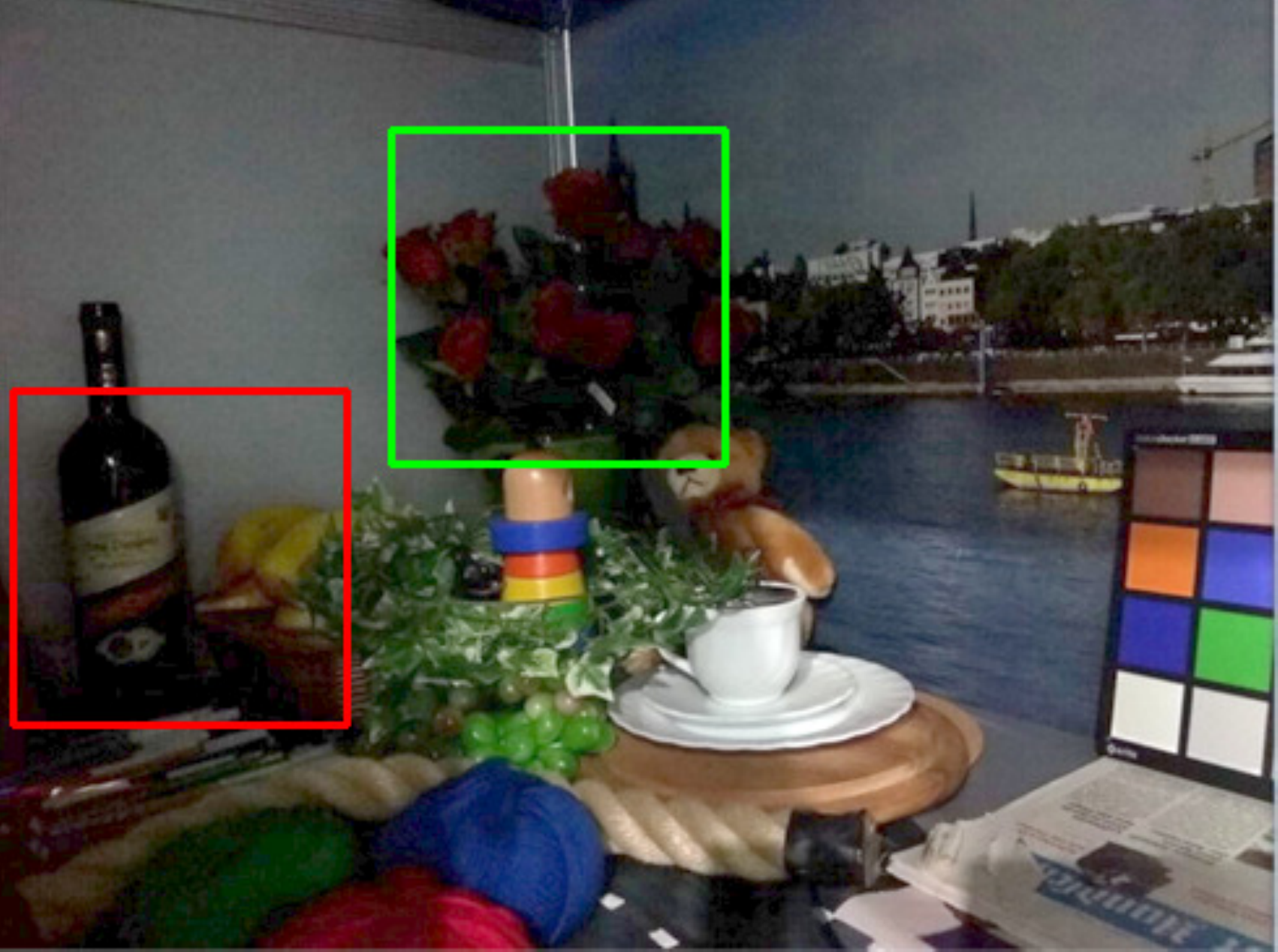}\vspace{1pt} \\
			\includegraphics[width=1.35cm]{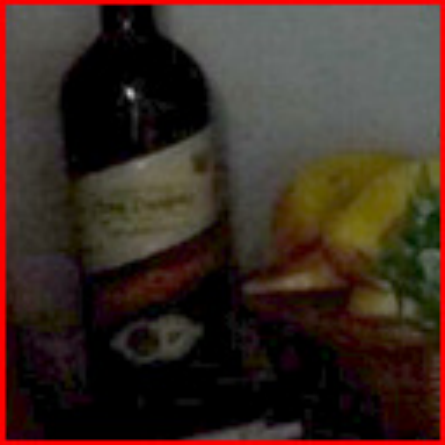}
			\includegraphics[width=1.35cm]{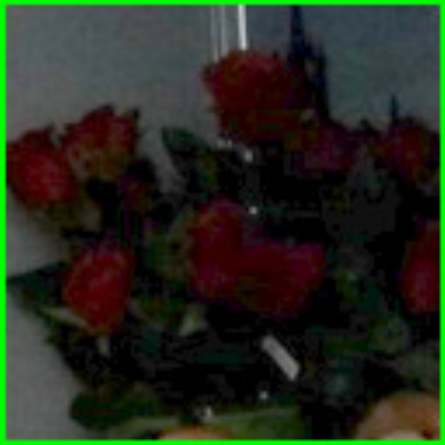}\vspace{5pt}
			\includegraphics[width=2.8cm]{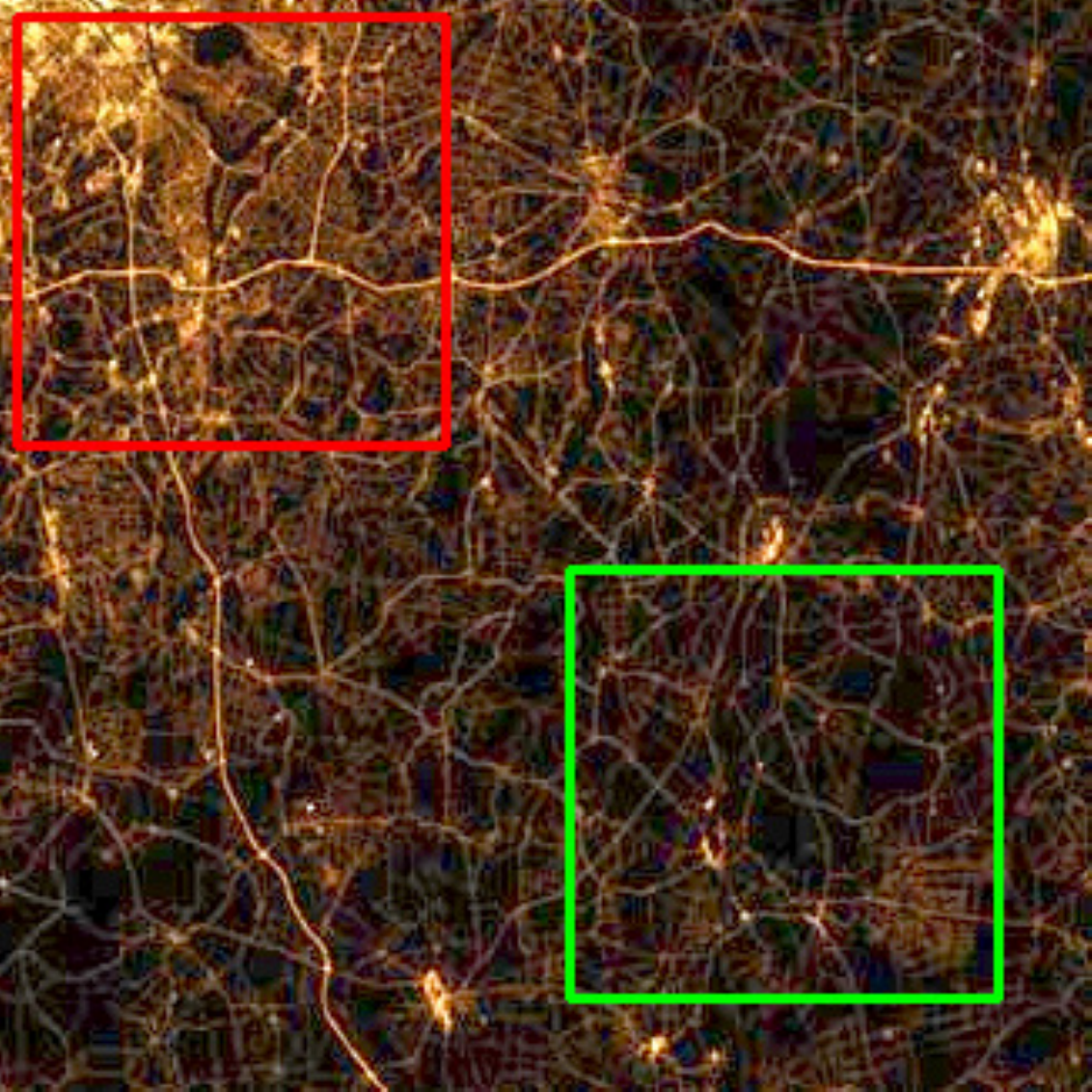}\vspace{1.5pt} \\
			\includegraphics[width=1.35cm]{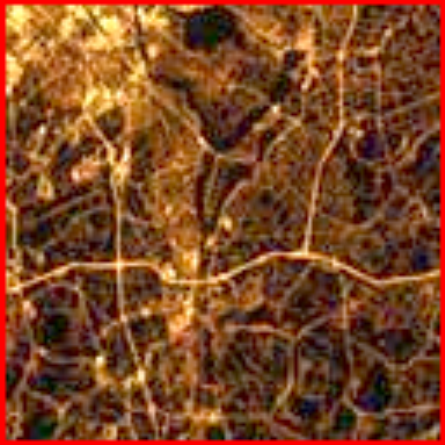}
			\includegraphics[width=1.35cm]{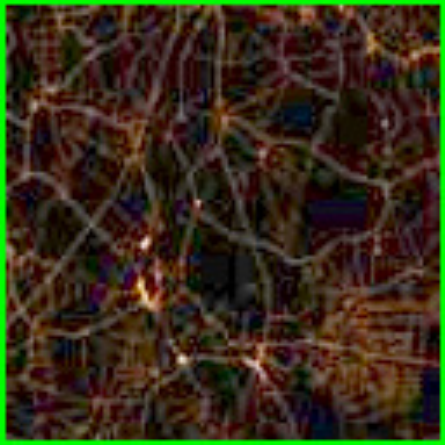}
		\end{minipage}
	}\hspace{-5pt}
	\subfigure[Dong\cite{dong2011fast}]{
		\begin{minipage}[b]{0.155\textwidth}
			\includegraphics[width=2.8cm]{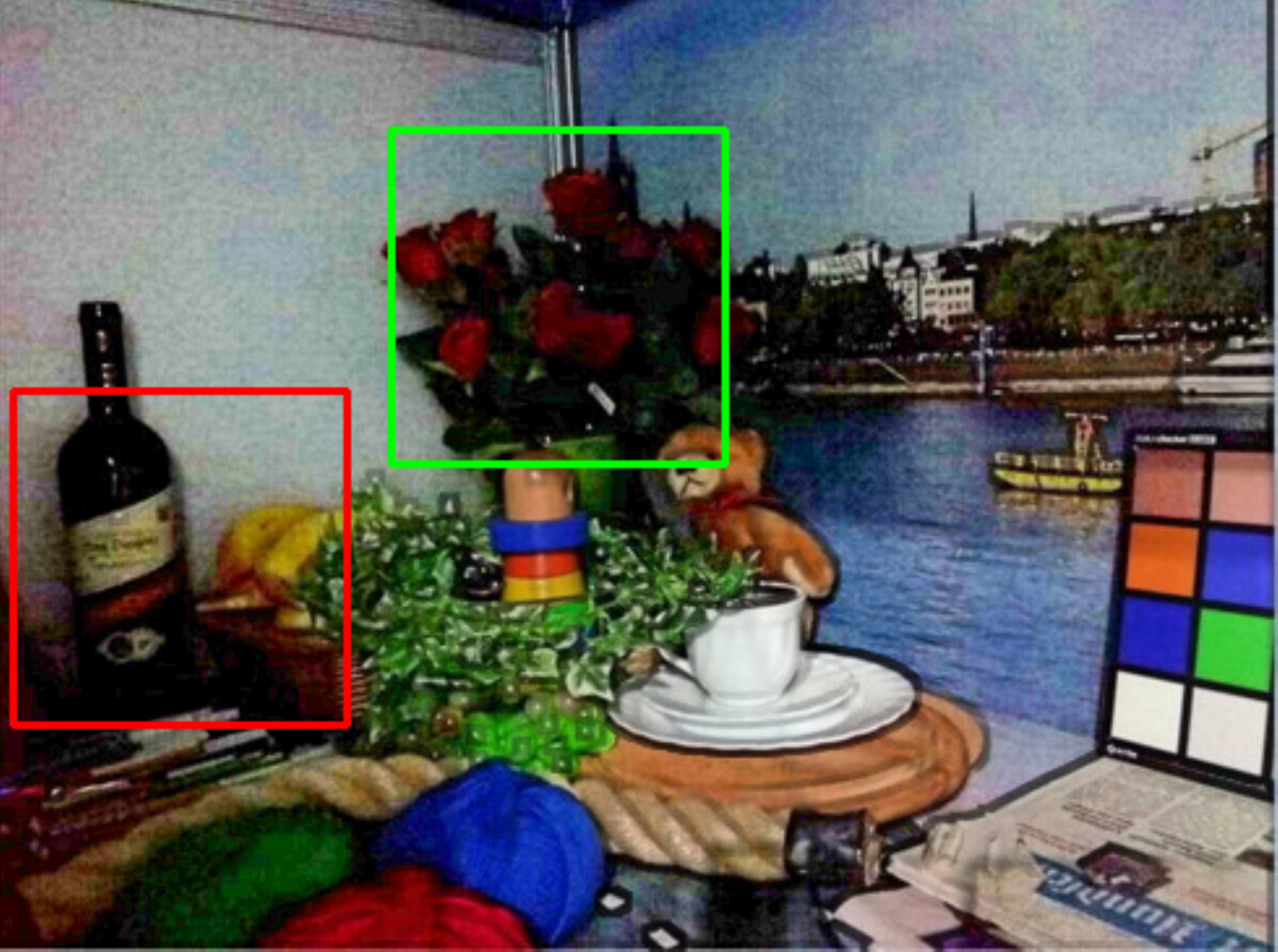}\vspace{1pt} \\
			\includegraphics[width=1.35cm]{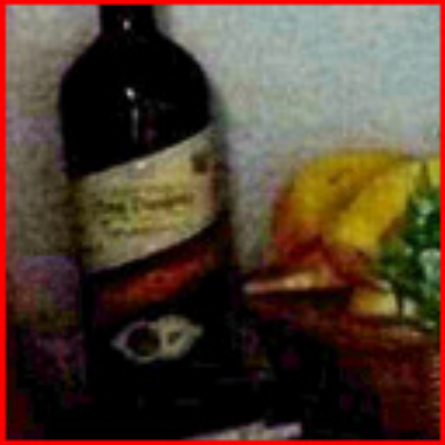}
			\includegraphics[width=1.35cm]{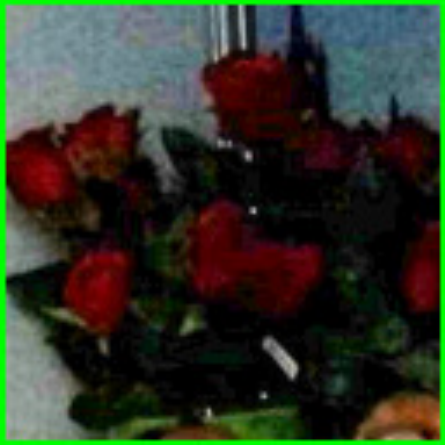}\vspace{5pt}
			\includegraphics[width=2.8cm]{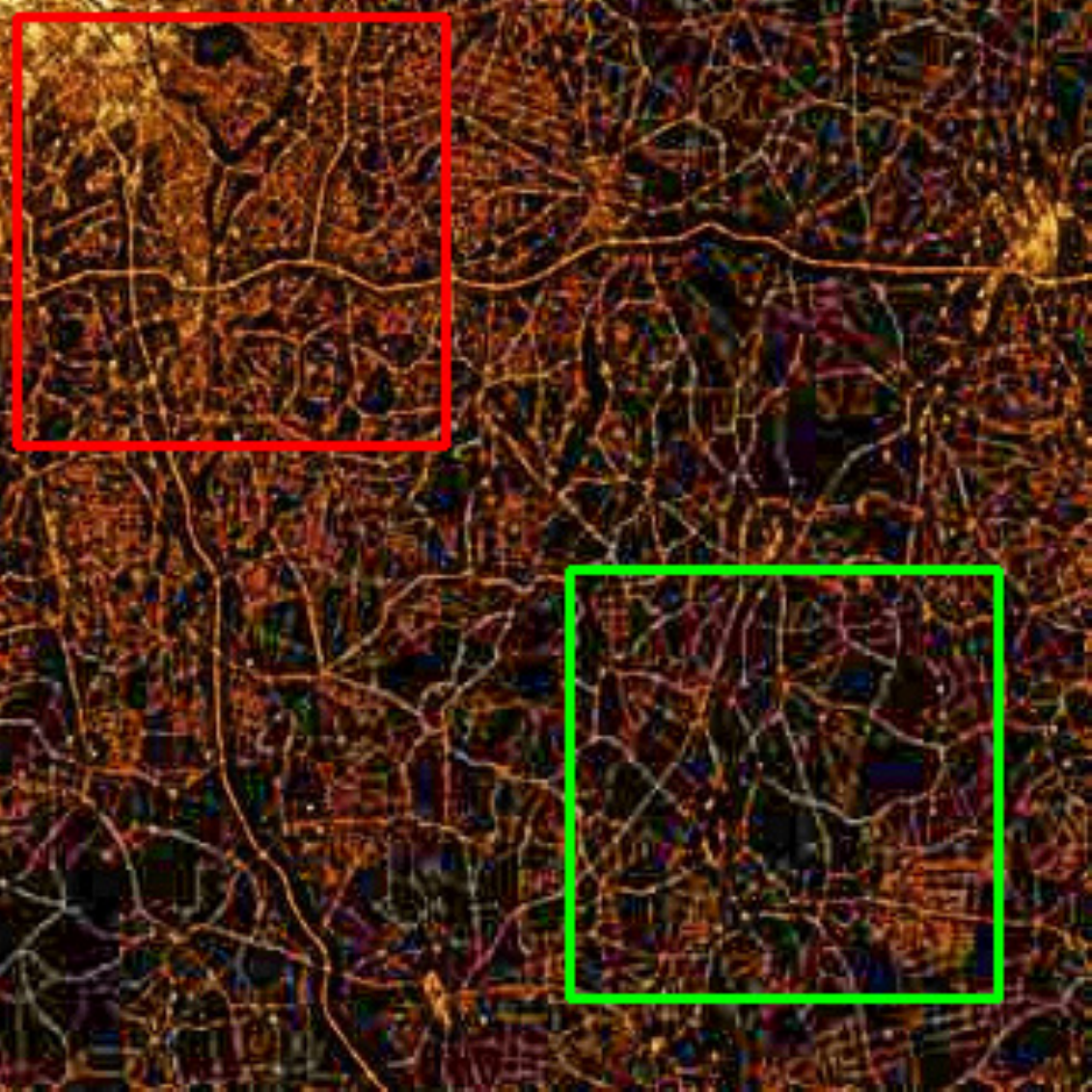}\vspace{1.5pt} \\
			\includegraphics[width=1.35cm]{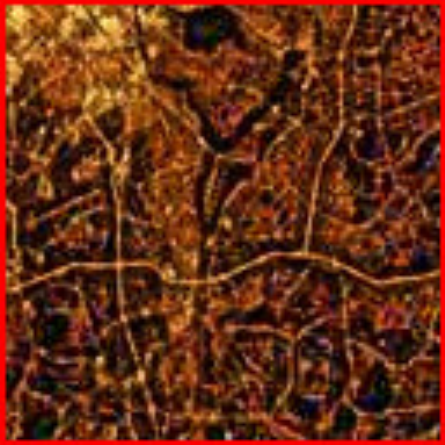}
			\includegraphics[width=1.35cm]{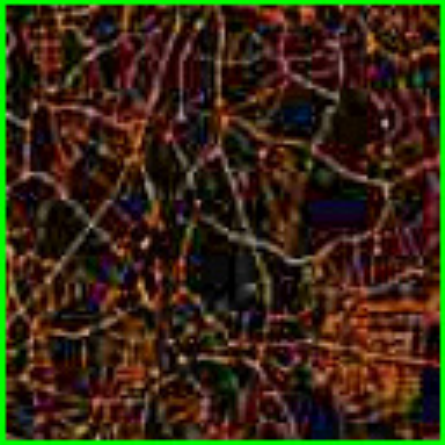}
		\end{minipage}
	}\hspace{-5pt}
	\subfigure[KinD\cite{zhang2019kindling}]{
		\begin{minipage}[b]{0.155\textwidth}
			\includegraphics[width=2.8cm]{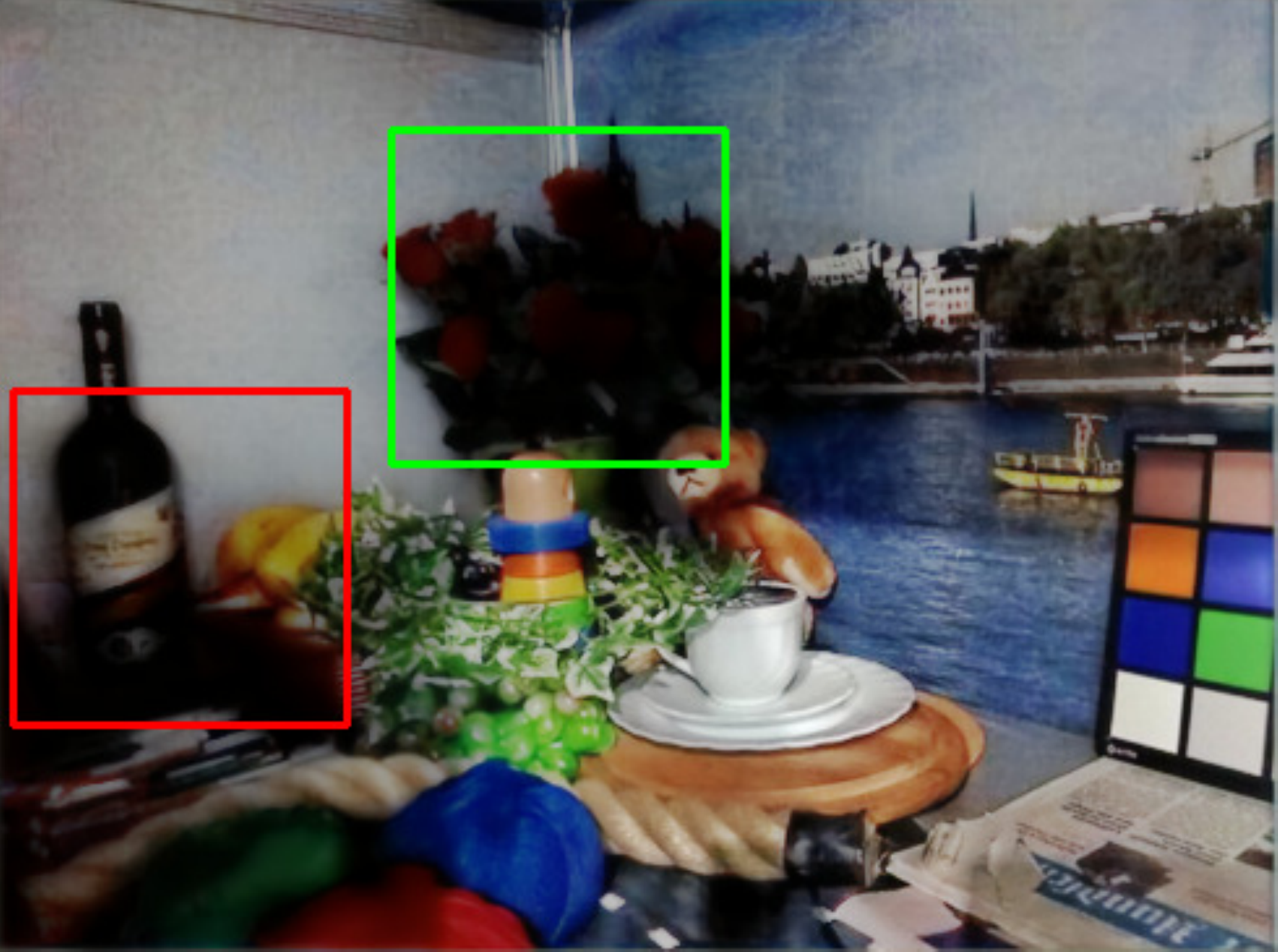}\vspace{1pt} \\
			\includegraphics[width=1.35cm]{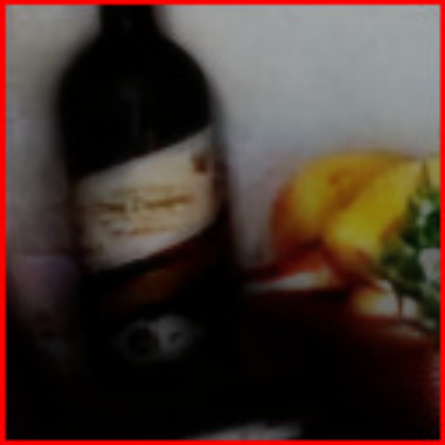}
			\includegraphics[width=1.35cm]{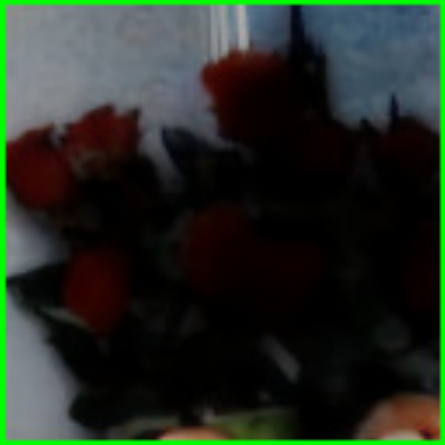}\vspace{5pt}
			\includegraphics[width=2.8cm]{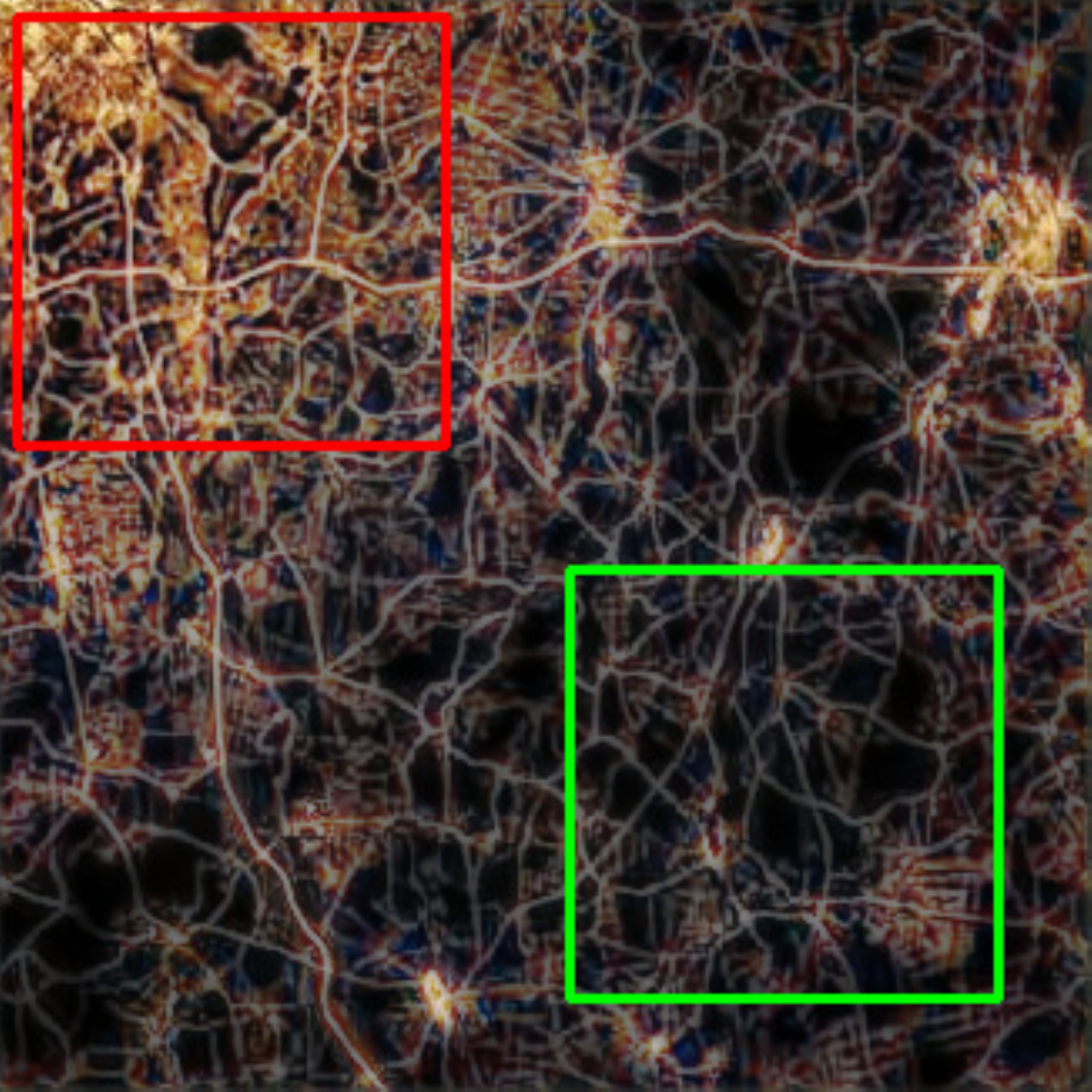}\vspace{1.5pt} \\
			\includegraphics[width=1.35cm]{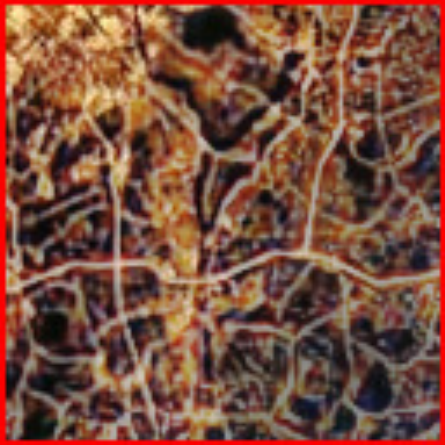}
			\includegraphics[width=1.35cm]{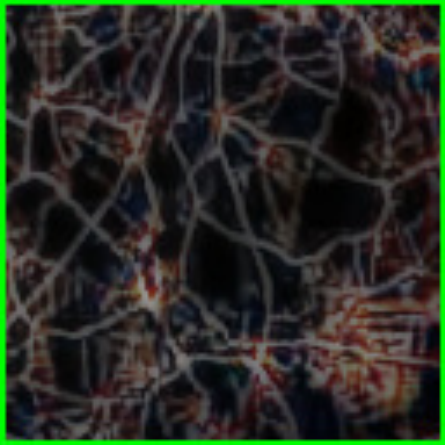}
		\end{minipage}
	}\hspace{-5pt}
	\subfigure[DA-DRN]{
		\begin{minipage}[b]{0.155\textwidth}
			\includegraphics[width=2.8cm]{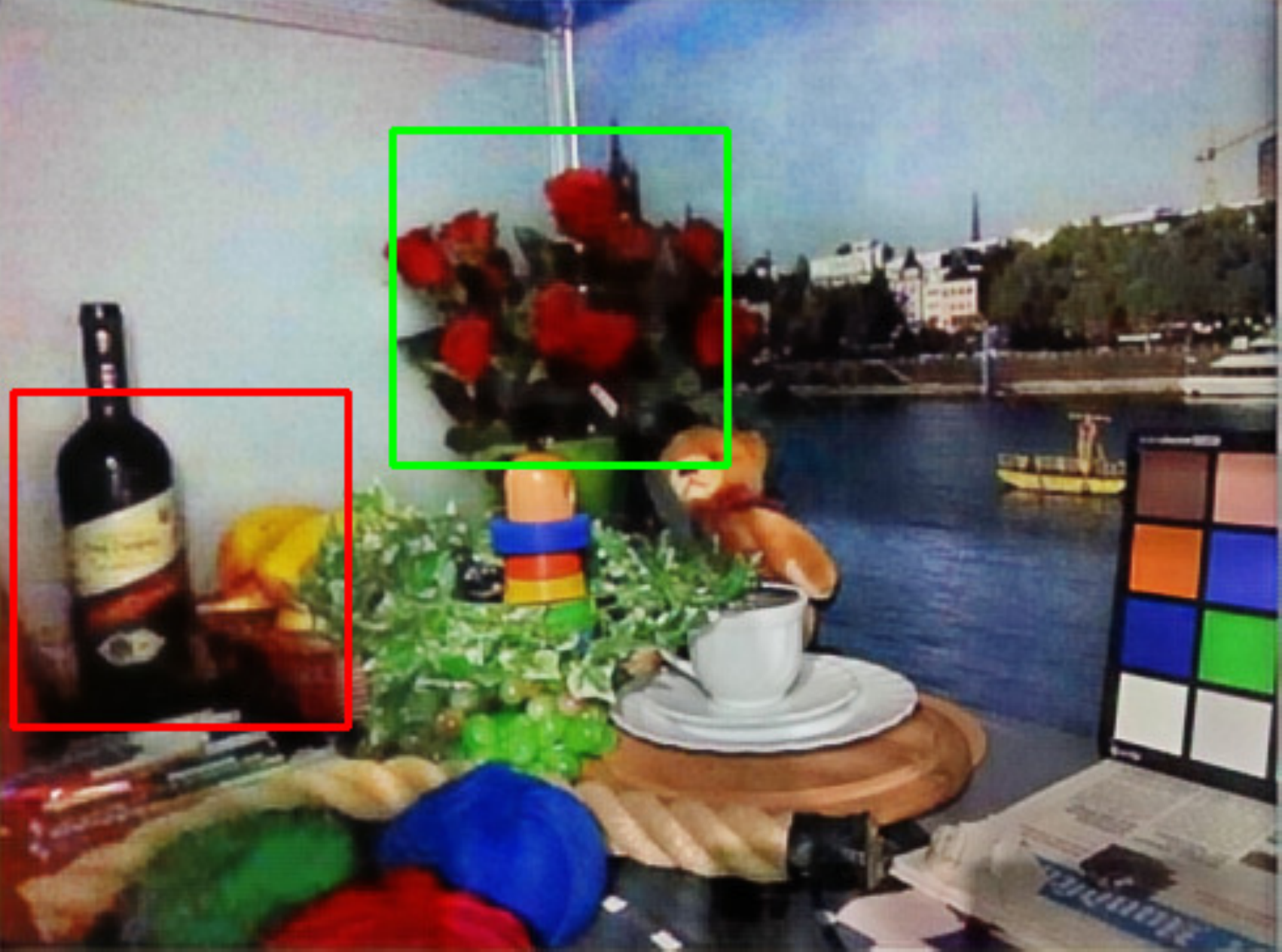}\vspace{1pt} \\
			\includegraphics[width=1.35cm]{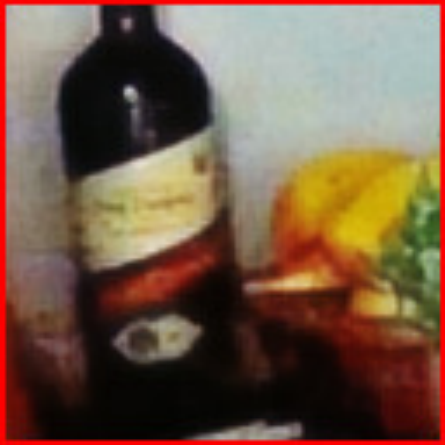}
			\includegraphics[width=1.35cm]{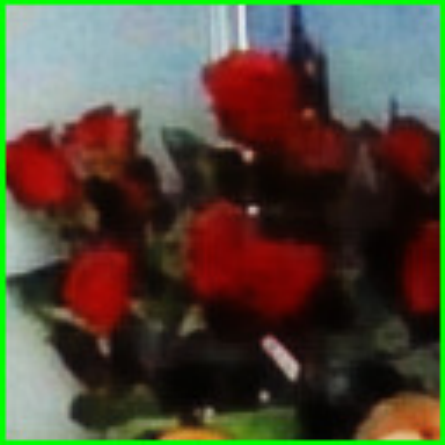}\vspace{5pt}
			\includegraphics[width=2.8cm]{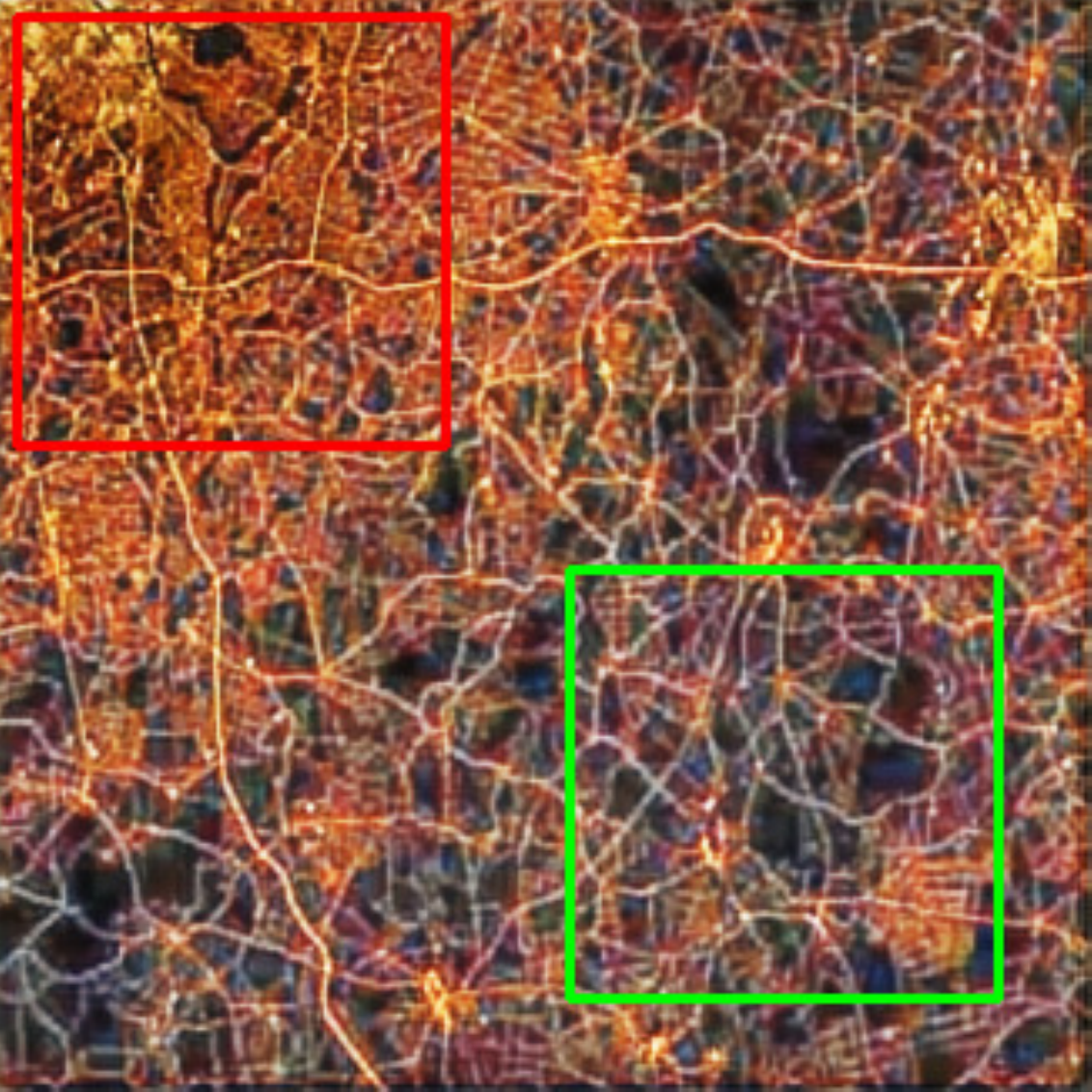}\vspace{1.5pt} \\
			\includegraphics[width=1.35cm]{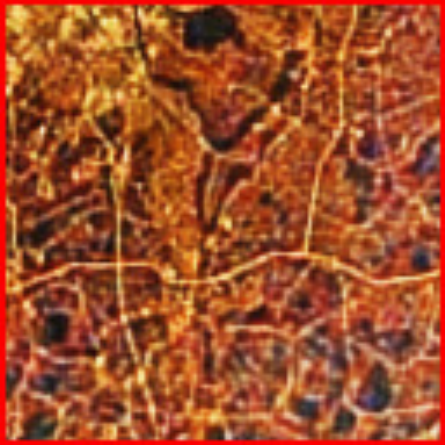}
			\includegraphics[width=1.35cm]{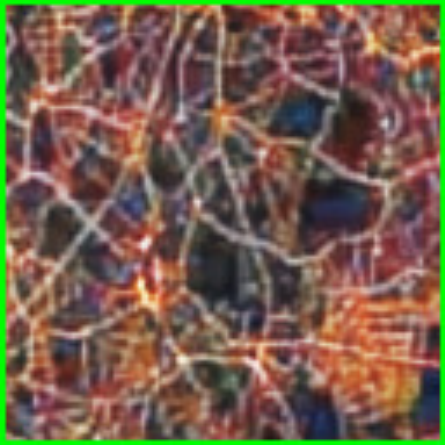}
		\end{minipage}
	}
	\caption{Visual comparison with other state-of-the-art methods on the LIME dataset. There is no Ground-Truth in LIME dataset.}
	\label{LIME}
\end{figure*}

\begin{figure*}
	\flushleft
	
	
	\subfigure[Input]{
		\begin{minipage}[b]{0.155\textwidth}
			\includegraphics[width=2.8cm]{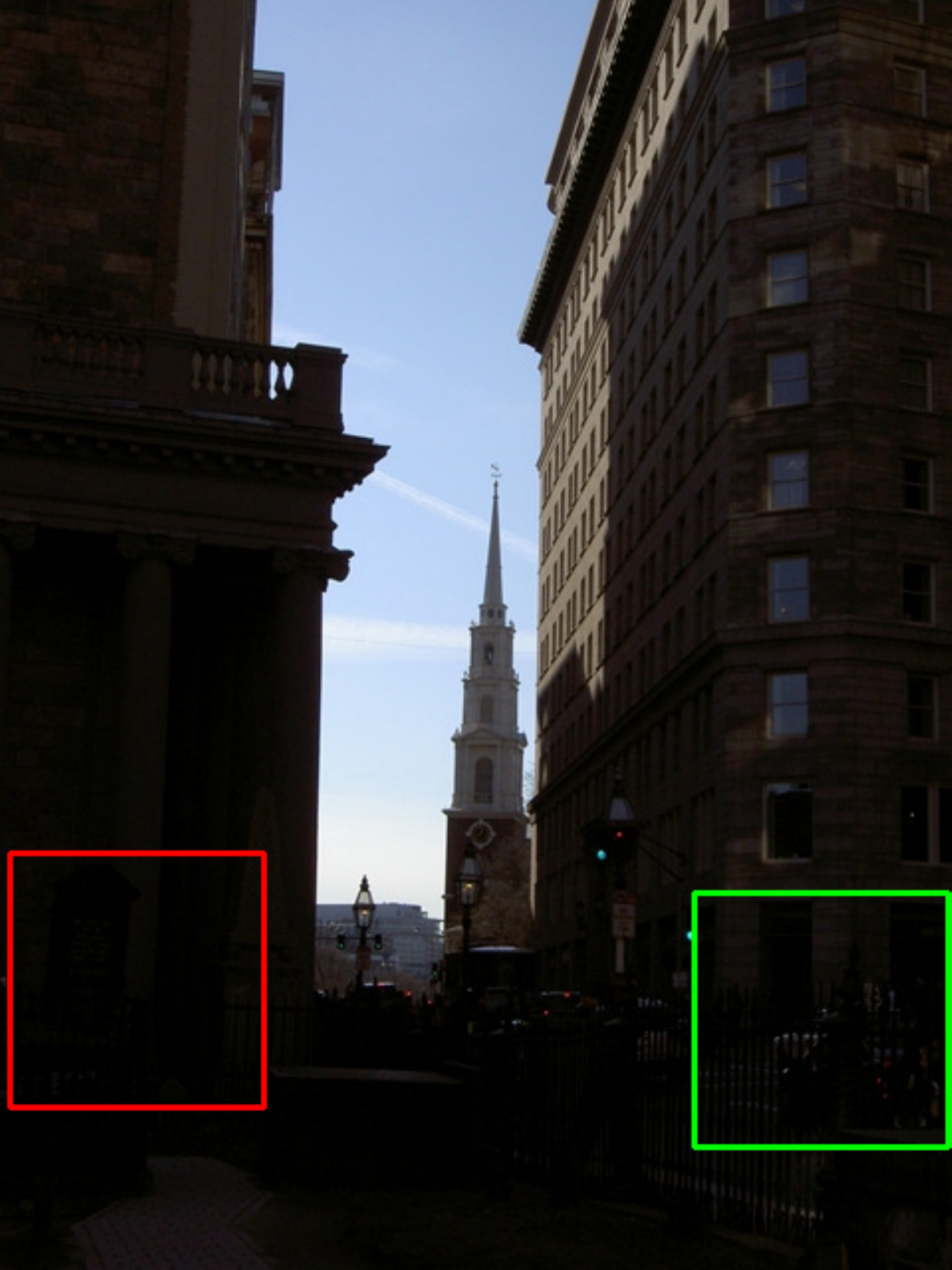}\vspace{1pt} \\
			\includegraphics[width=1.35cm]{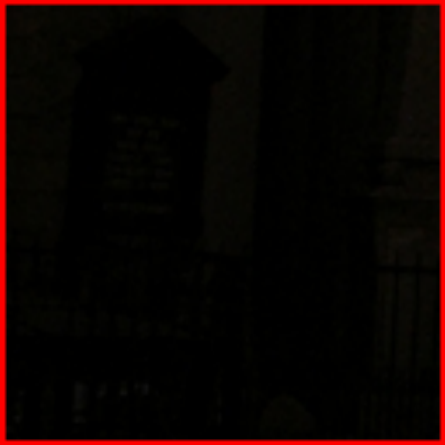}
			\includegraphics[width=1.35cm]{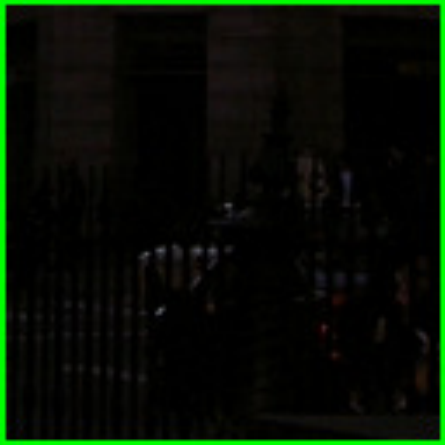}\vspace{5pt}
			\includegraphics[width=2.8cm]{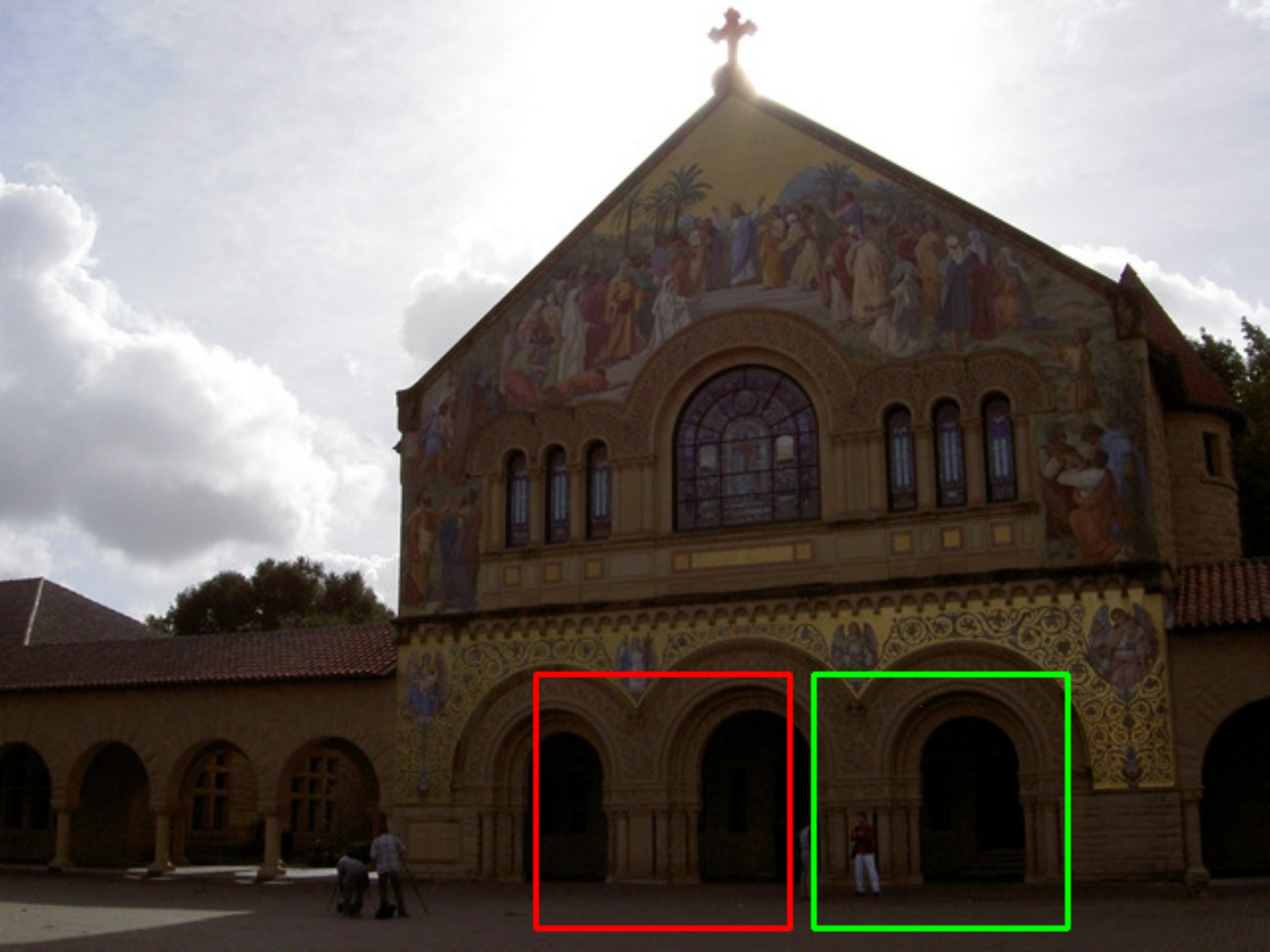}\vspace{1.5pt} \\
			\includegraphics[width=1.35cm]{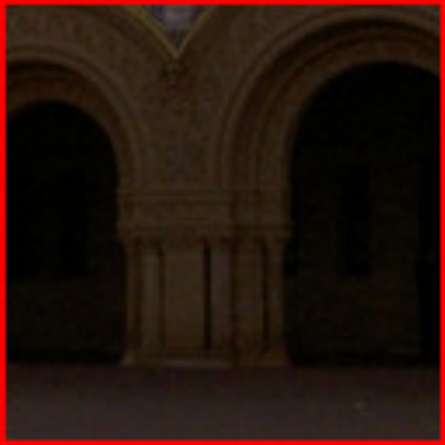}
			\includegraphics[width=1.35cm]{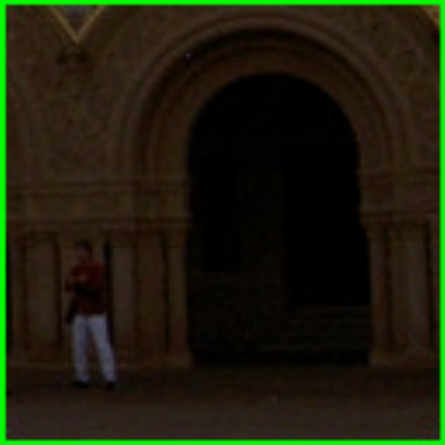}\vspace{5pt}
			\includegraphics[width=2.8cm]{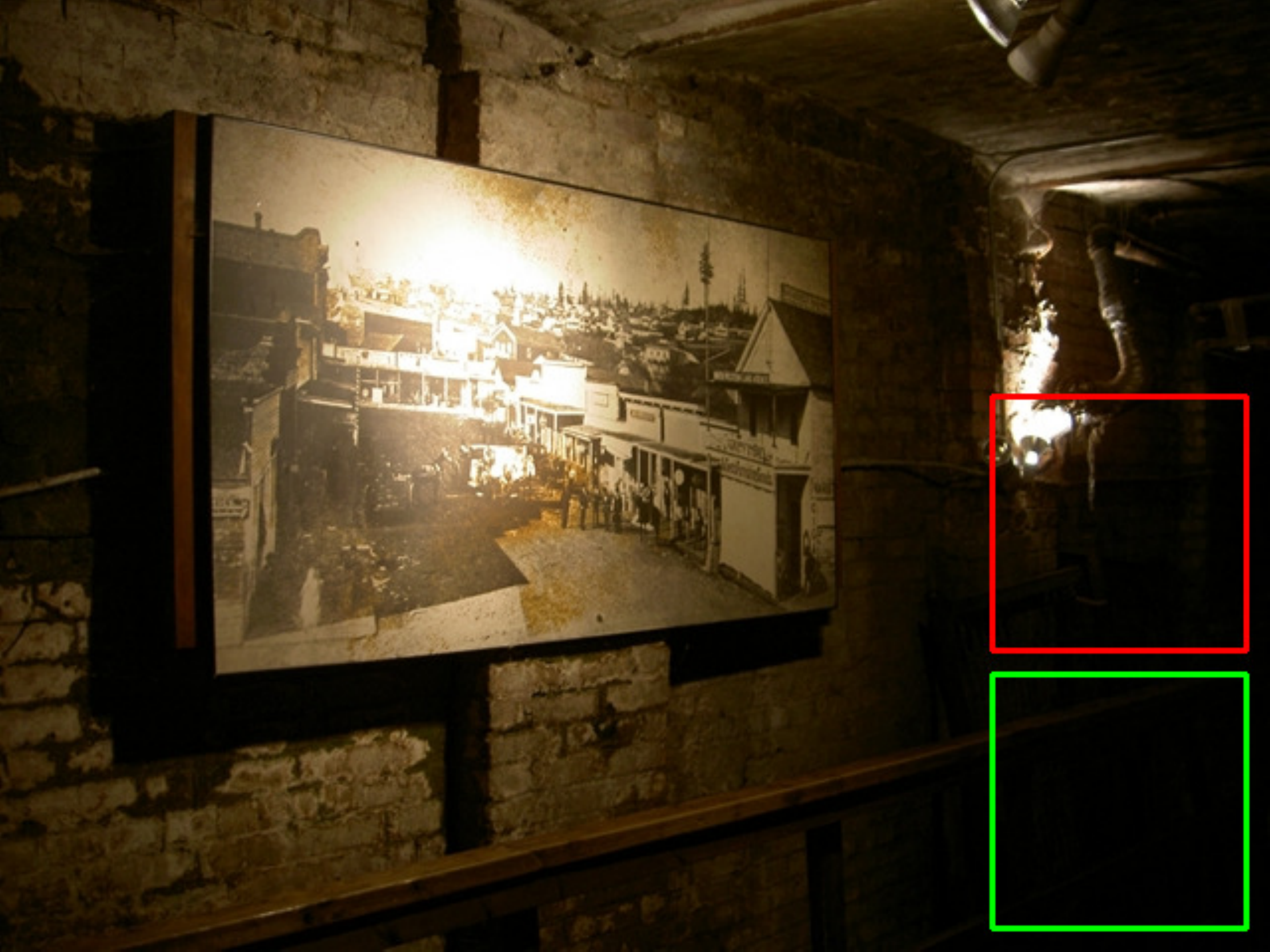}\vspace{1pt} \\
			\includegraphics[width=1.35cm]{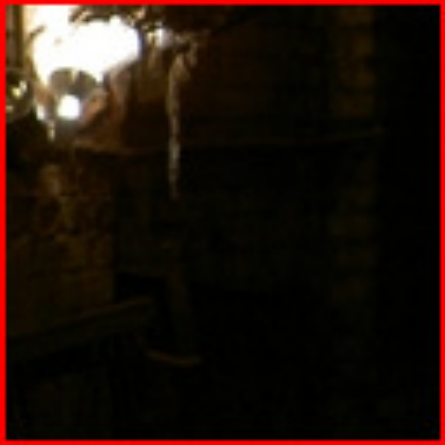}
			\includegraphics[width=1.35cm]{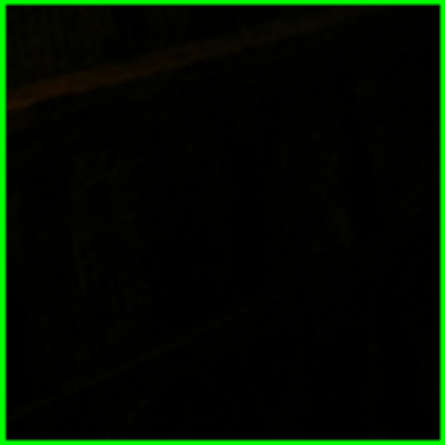}
		\end{minipage}
	}\hspace{-5pt}
	\subfigure[NPE\cite{wang2013naturalness}]{
		\begin{minipage}[b]{0.155\textwidth}
			\includegraphics[width=2.8cm]{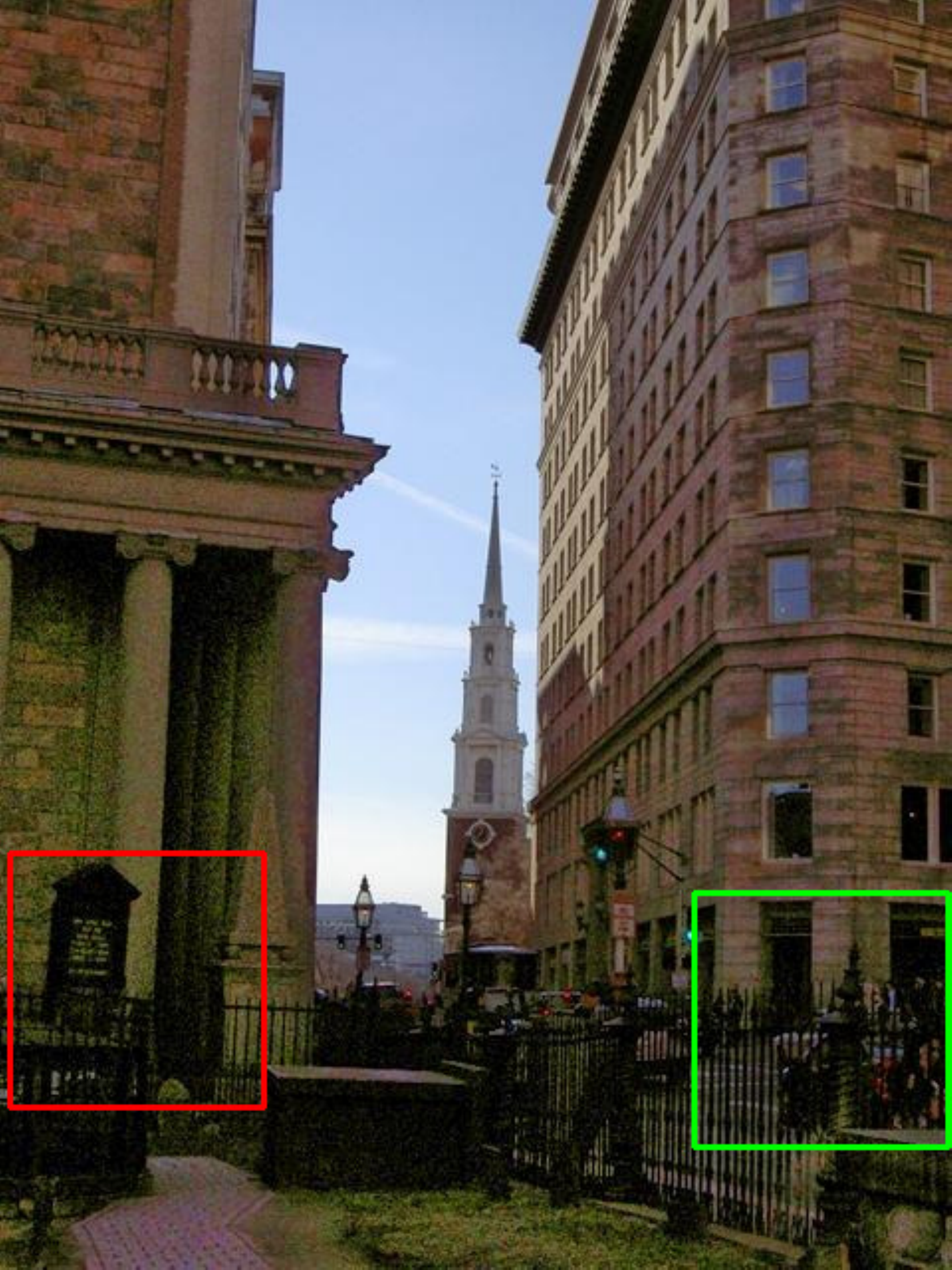}\vspace{1pt} \\
			\includegraphics[width=1.35cm]{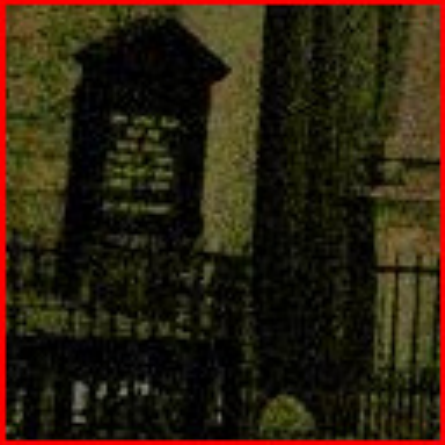}
			\includegraphics[width=1.35cm]{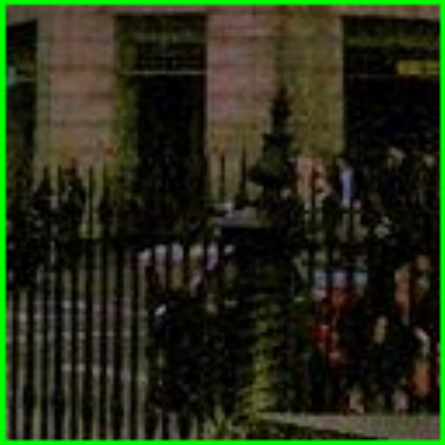}\vspace{5pt}
			\includegraphics[width=2.8cm]{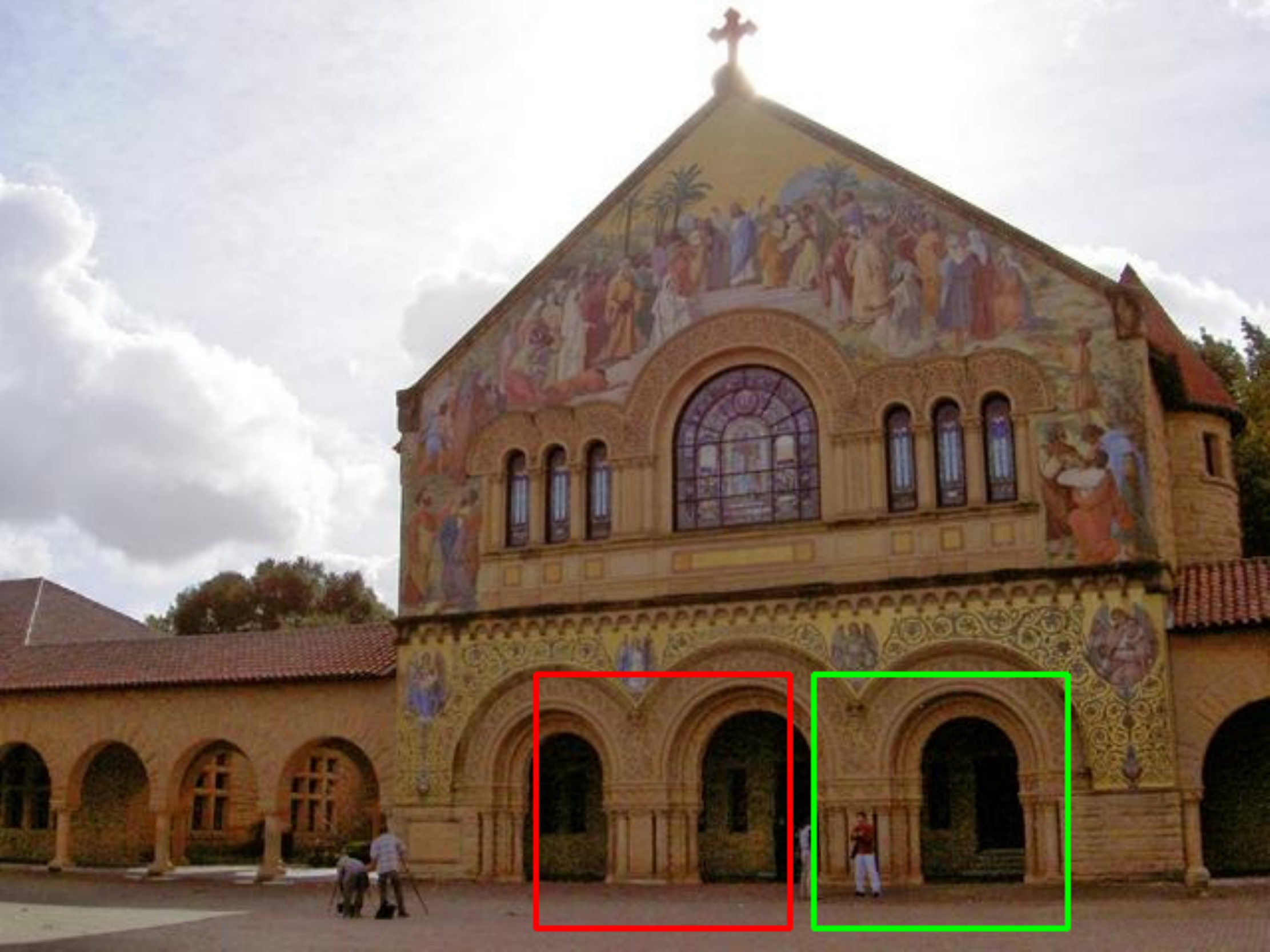}\vspace{1.5pt} \\
			\includegraphics[width=1.35cm]{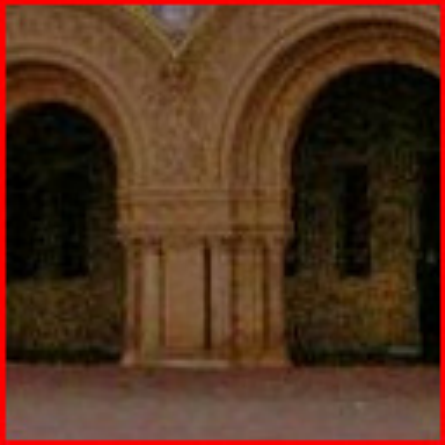}
			\includegraphics[width=1.35cm]{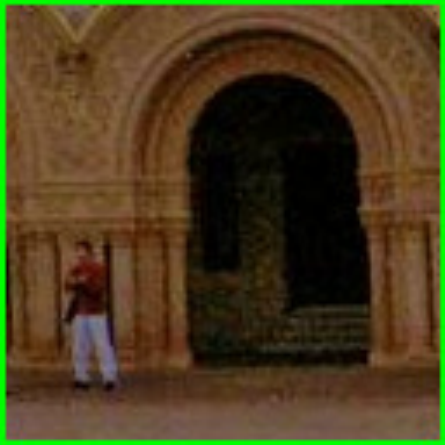}\vspace{5pt}
			\includegraphics[width=2.8cm]{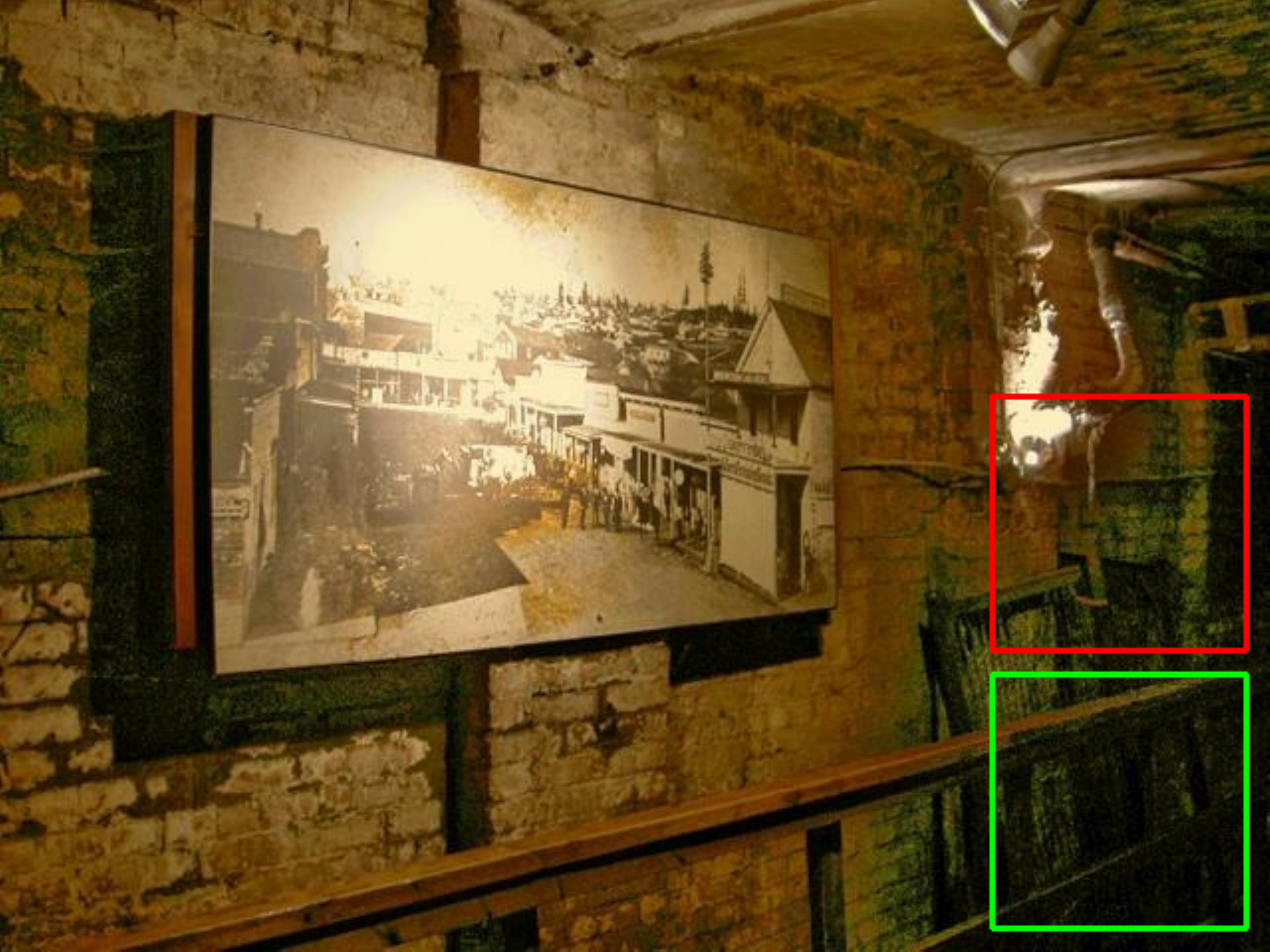}\vspace{1pt} \\
			\includegraphics[width=1.35cm]{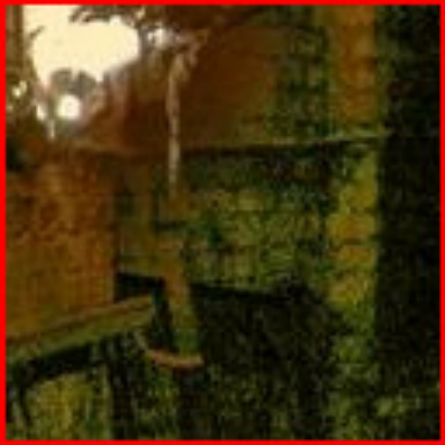}
			\includegraphics[width=1.35cm]{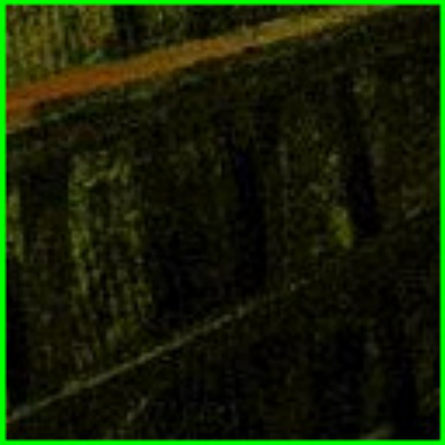}
		\end{minipage}
	}\hspace{-5pt}
	\subfigure[GLAD\cite{wang2018gladnet}]{
		\begin{minipage}[b]{0.155\textwidth}
			\includegraphics[width=2.8cm]{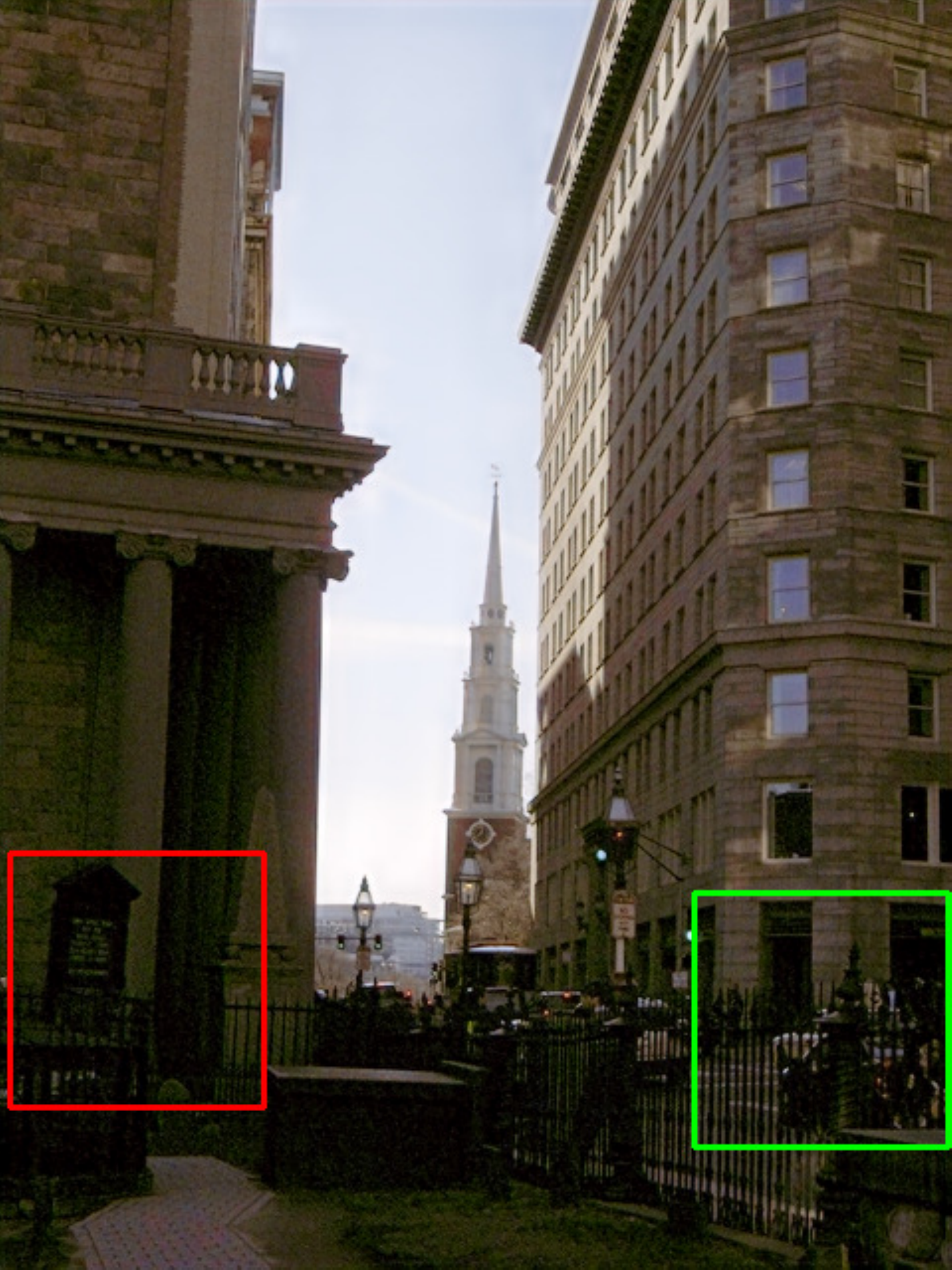}\vspace{1pt} \\
			\includegraphics[width=1.35cm]{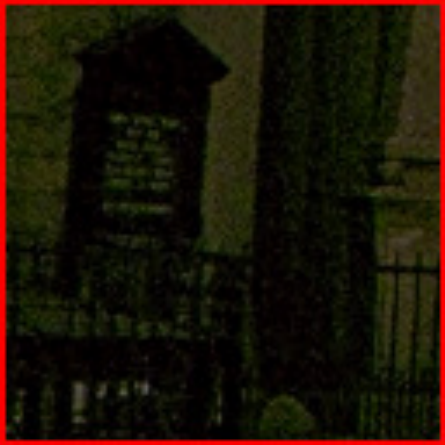}
			\includegraphics[width=1.35cm]{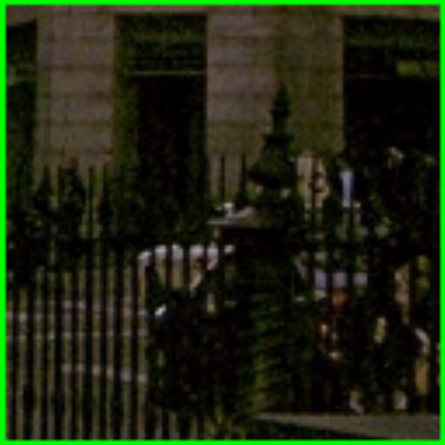}\vspace{5pt}
			\includegraphics[width=2.8cm]{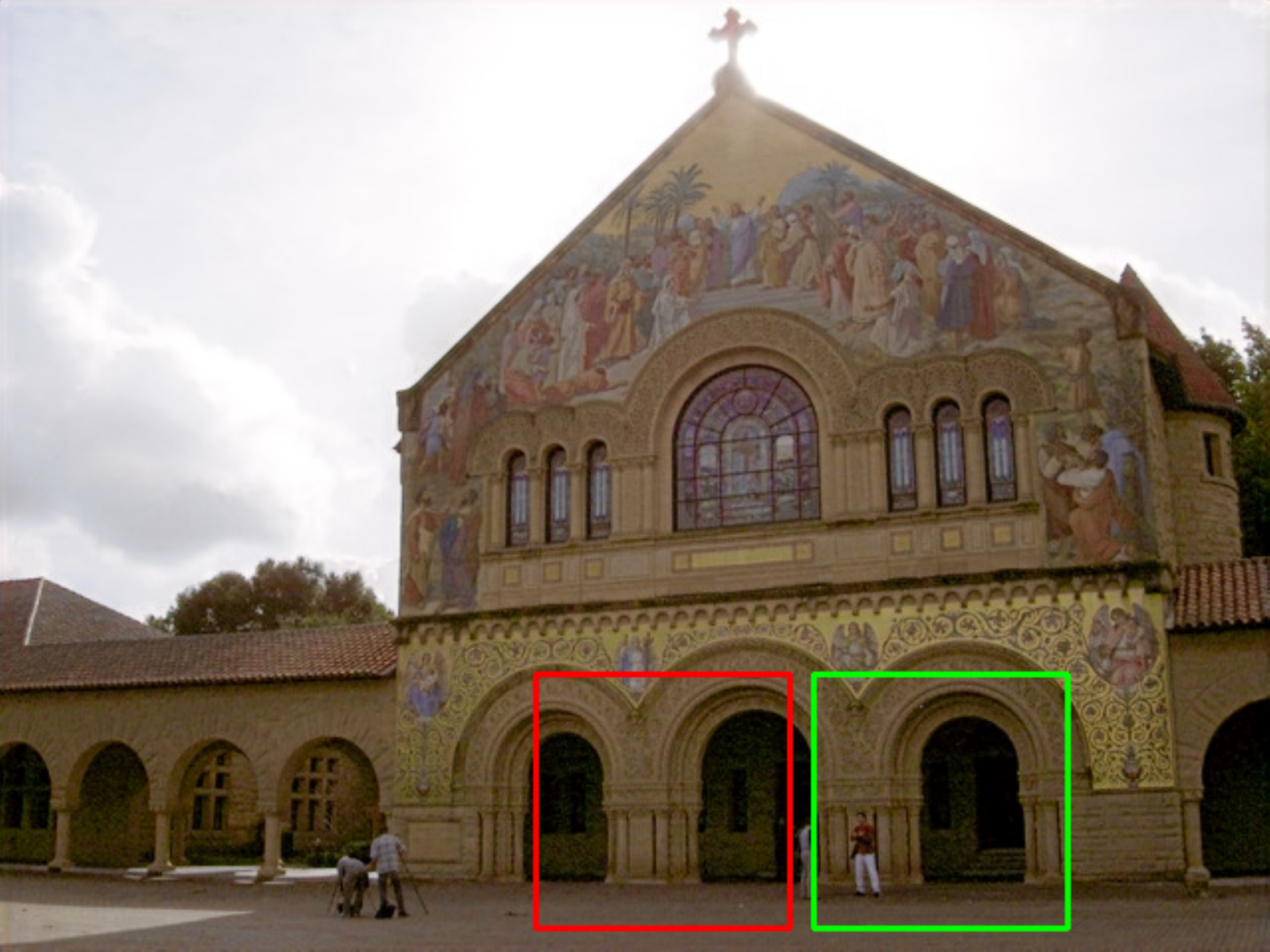}\vspace{1.5pt} \\
			\includegraphics[width=1.35cm]{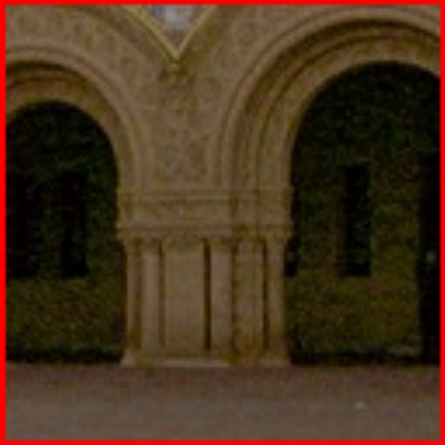}
			\includegraphics[width=1.35cm]{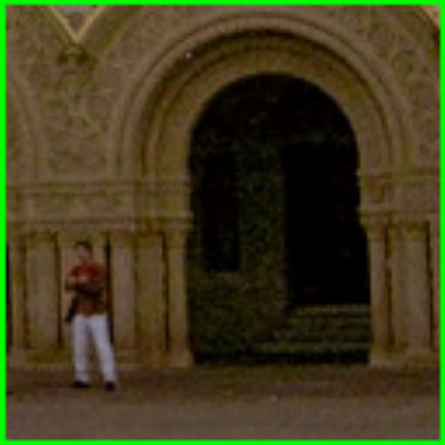}\vspace{5pt}
			\includegraphics[width=2.8cm]{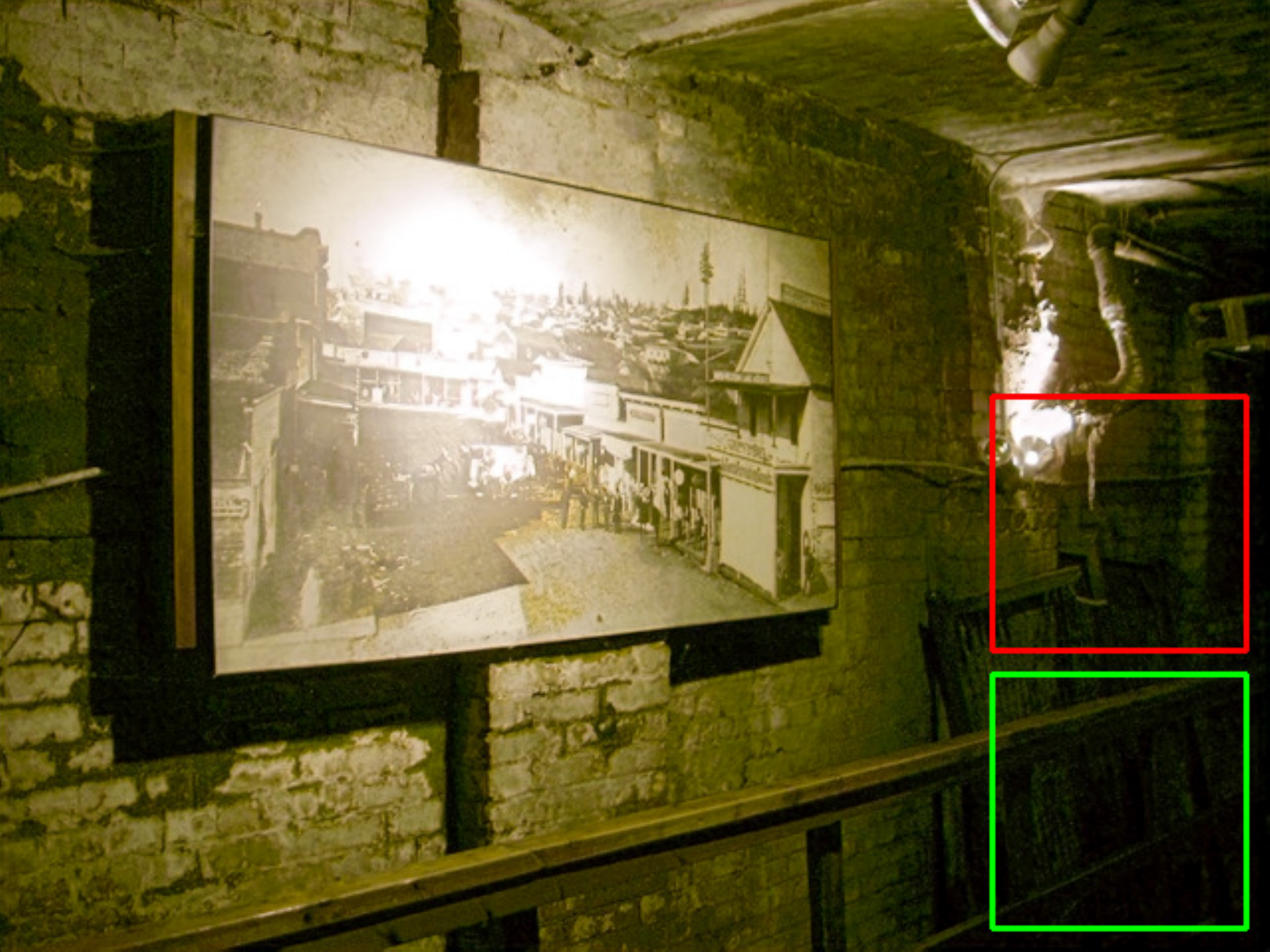}\vspace{1pt} \\
			\includegraphics[width=1.35cm]{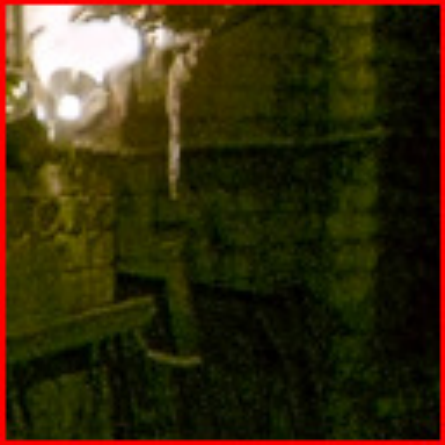}
			\includegraphics[width=1.35cm]{pic/test/DICM03_crop/low_1-eps-converted-to.pdf}
		\end{minipage}
	}\hspace{-5pt}
	\subfigure[RetinexNet\cite{wei2018deep}]{
		\begin{minipage}[b]{0.155\textwidth}
			\includegraphics[width=2.8cm]{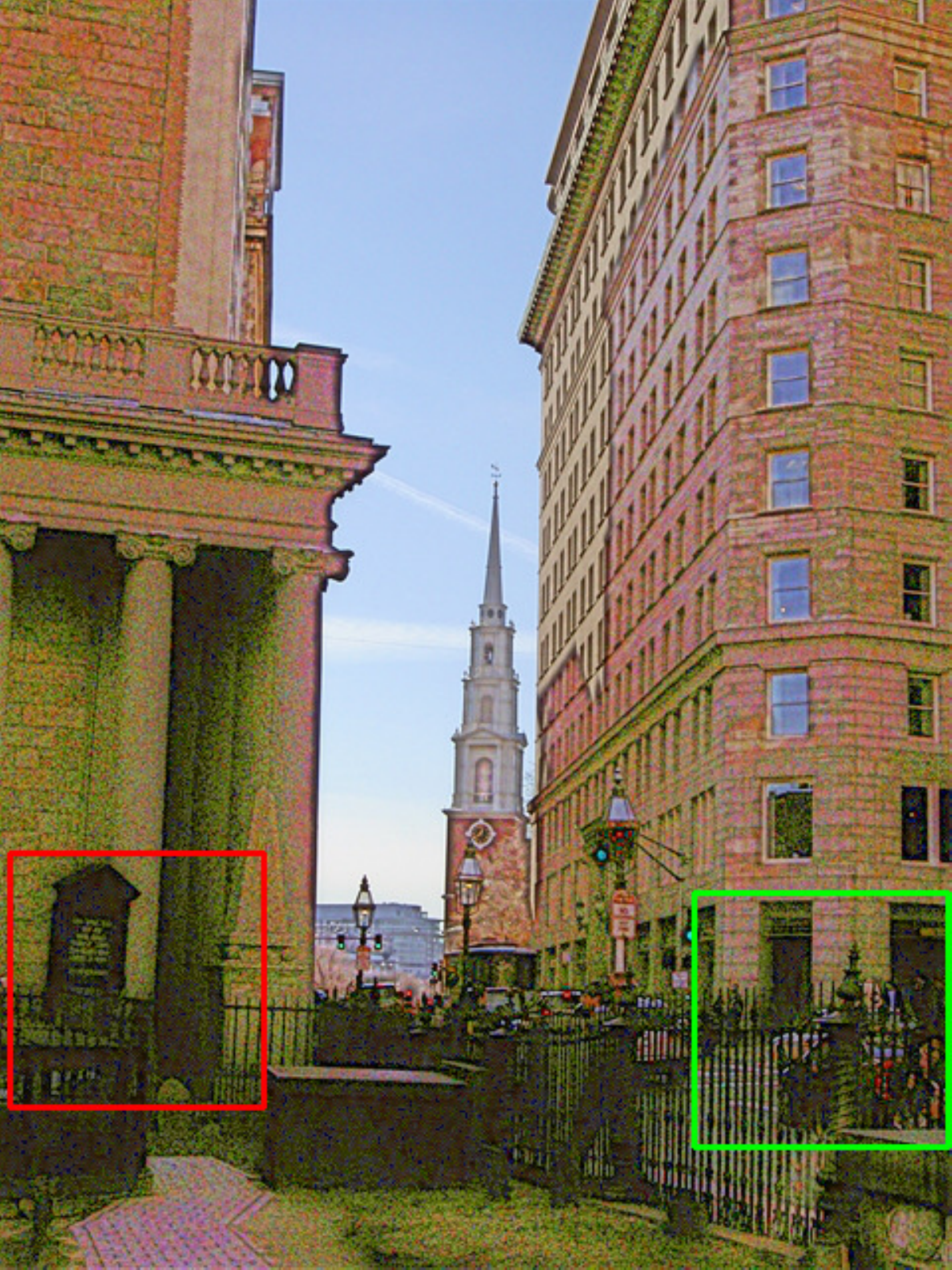}\vspace{1pt} \\
			\includegraphics[width=1.35cm]{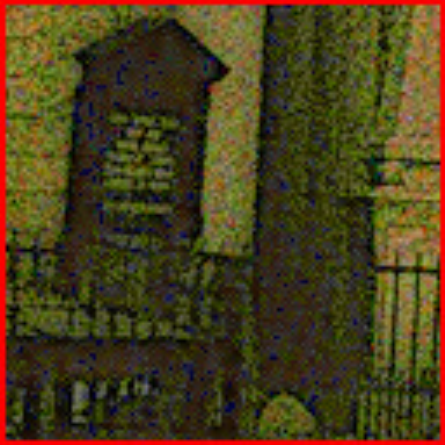}
			\includegraphics[width=1.35cm]{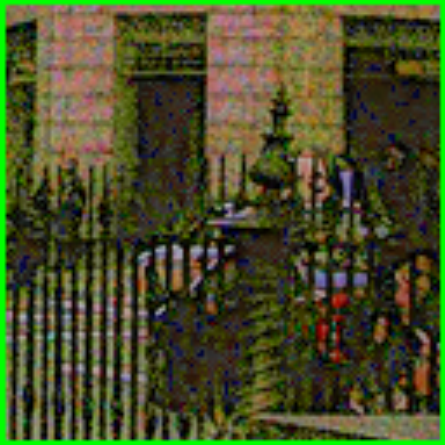}\vspace{5pt}
			\includegraphics[width=2.8cm]{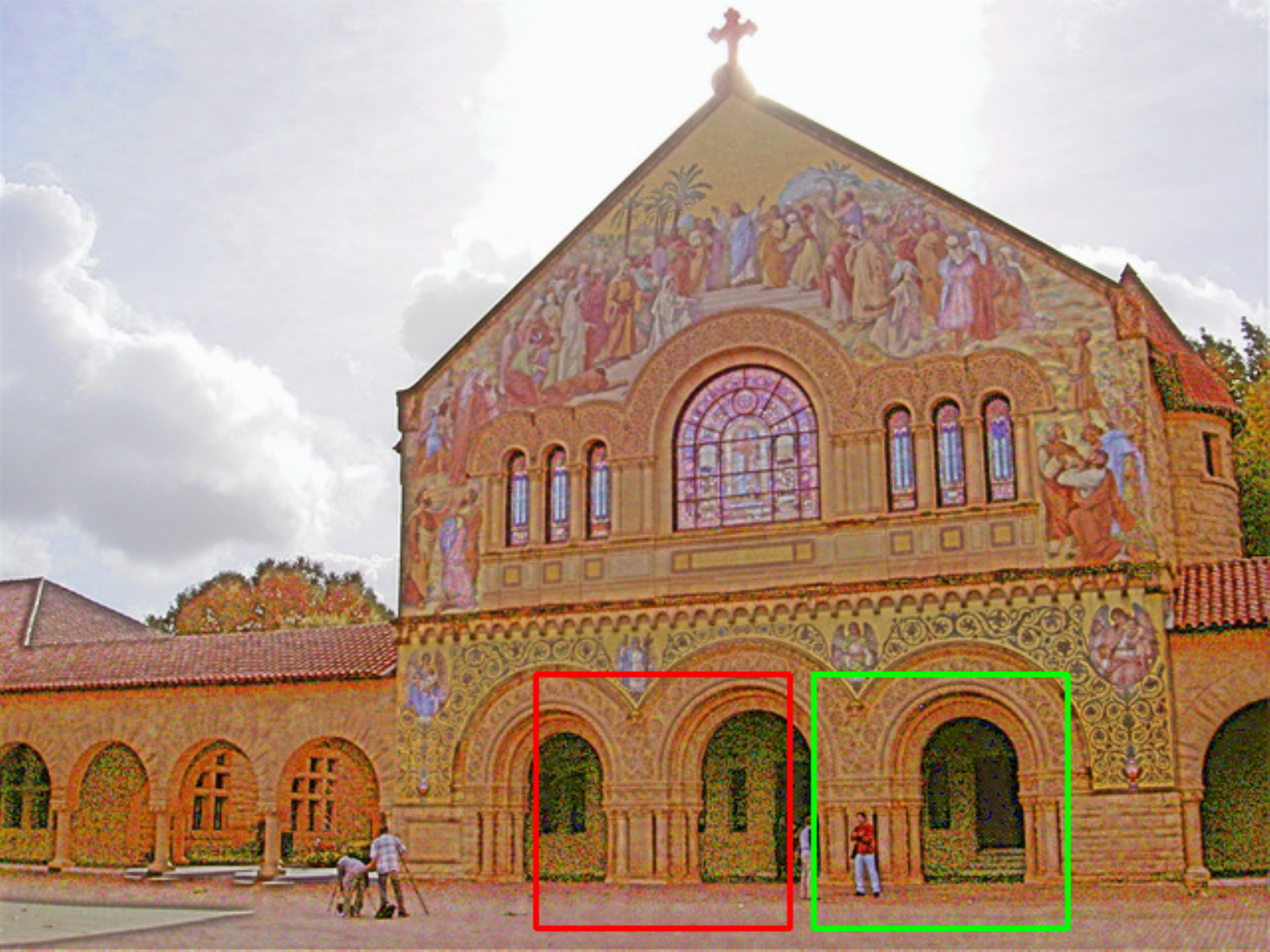}\vspace{1.5pt} \\
			\includegraphics[width=1.35cm]{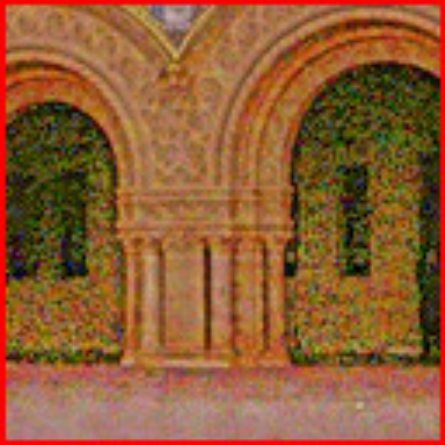}
			\includegraphics[width=1.35cm]{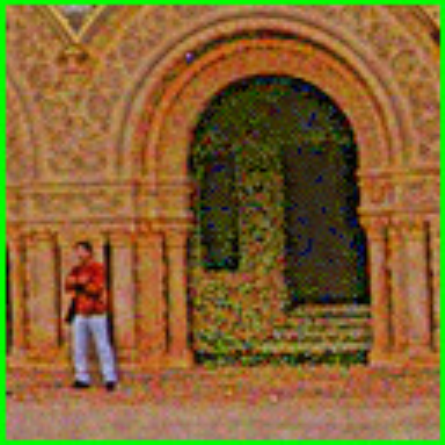}\vspace{5pt}
			\includegraphics[width=2.8cm]{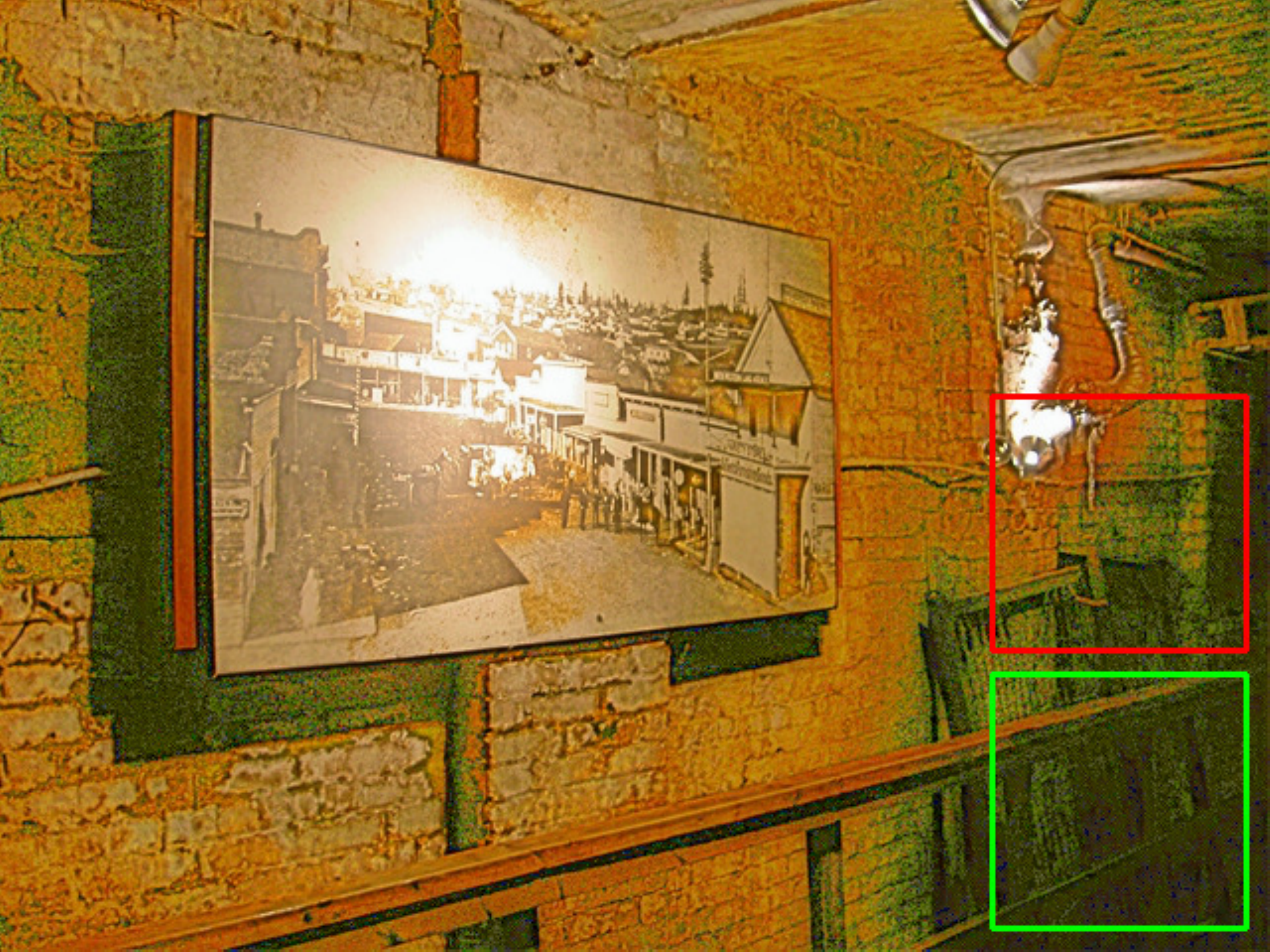}\vspace{1pt} \\
			\includegraphics[width=1.35cm]{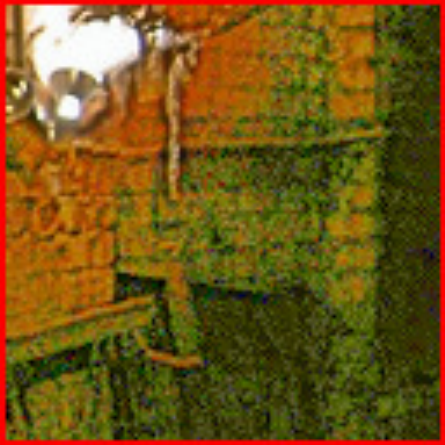}
			\includegraphics[width=1.35cm]{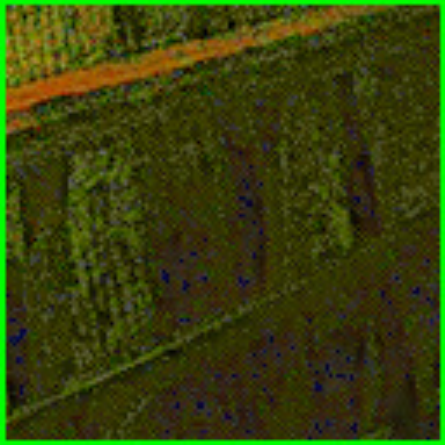}
		\end{minipage}
	}\hspace{-5pt}
	\subfigure[MBLLEN\cite{lv2018mbllen}]{
		\begin{minipage}[b]{0.155\textwidth}
			\includegraphics[width=2.8cm]{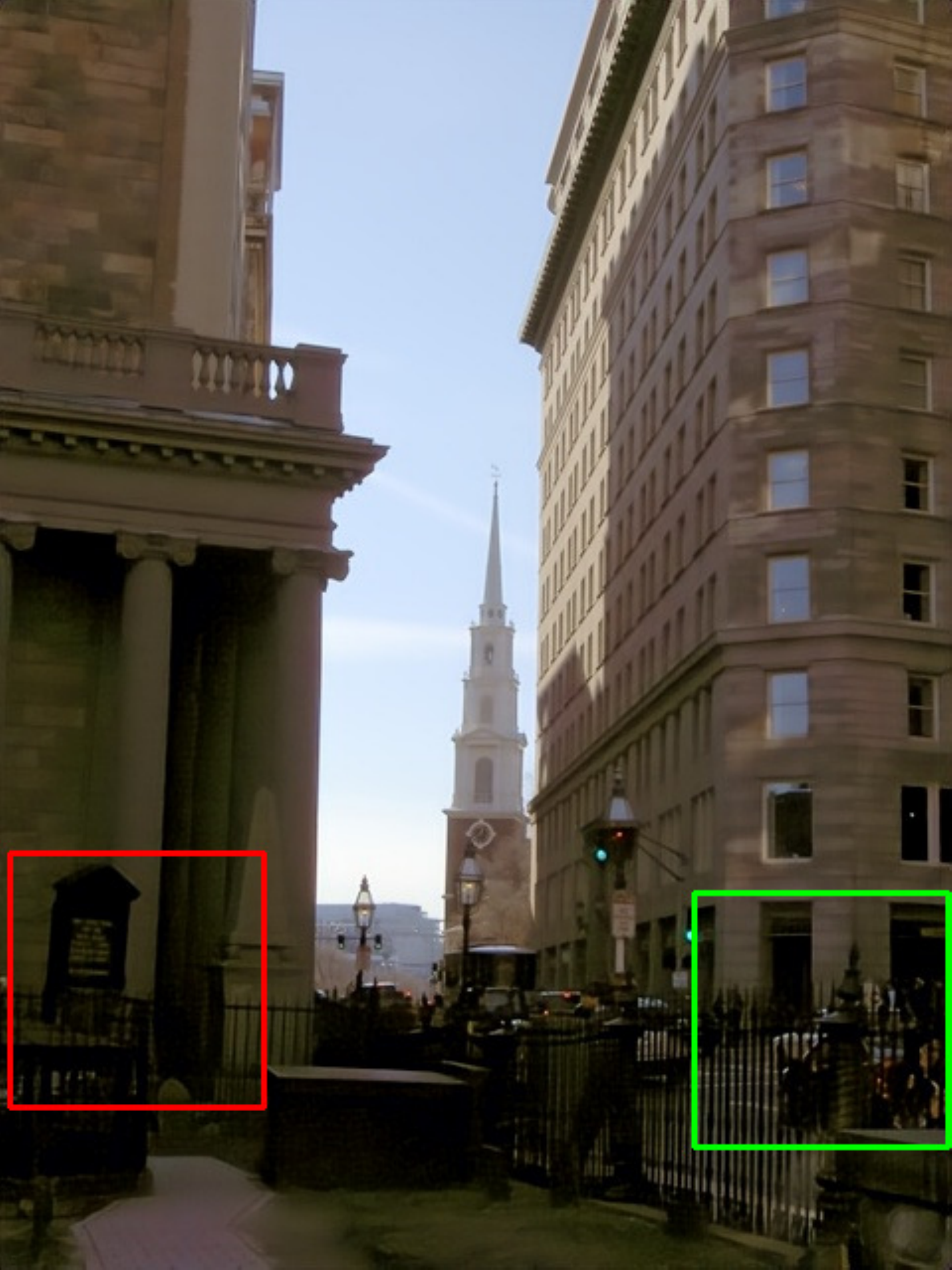}\vspace{1pt} \\
			\includegraphics[width=1.35cm]{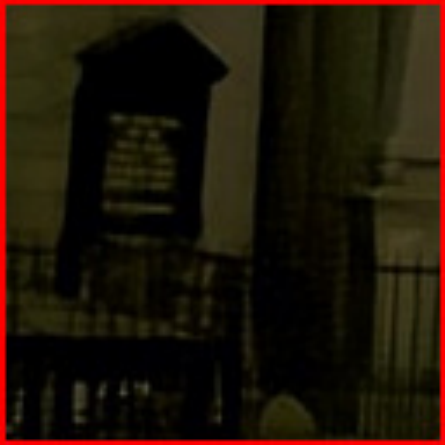}
			\includegraphics[width=1.35cm]{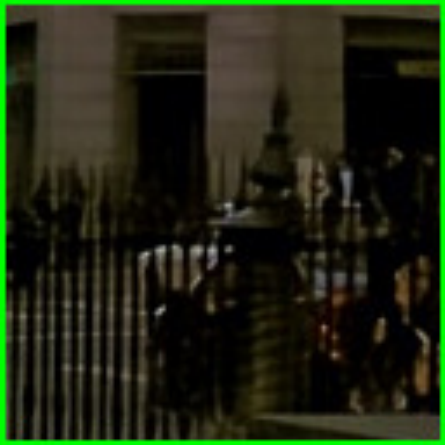}\vspace{5pt}
			\includegraphics[width=2.8cm]{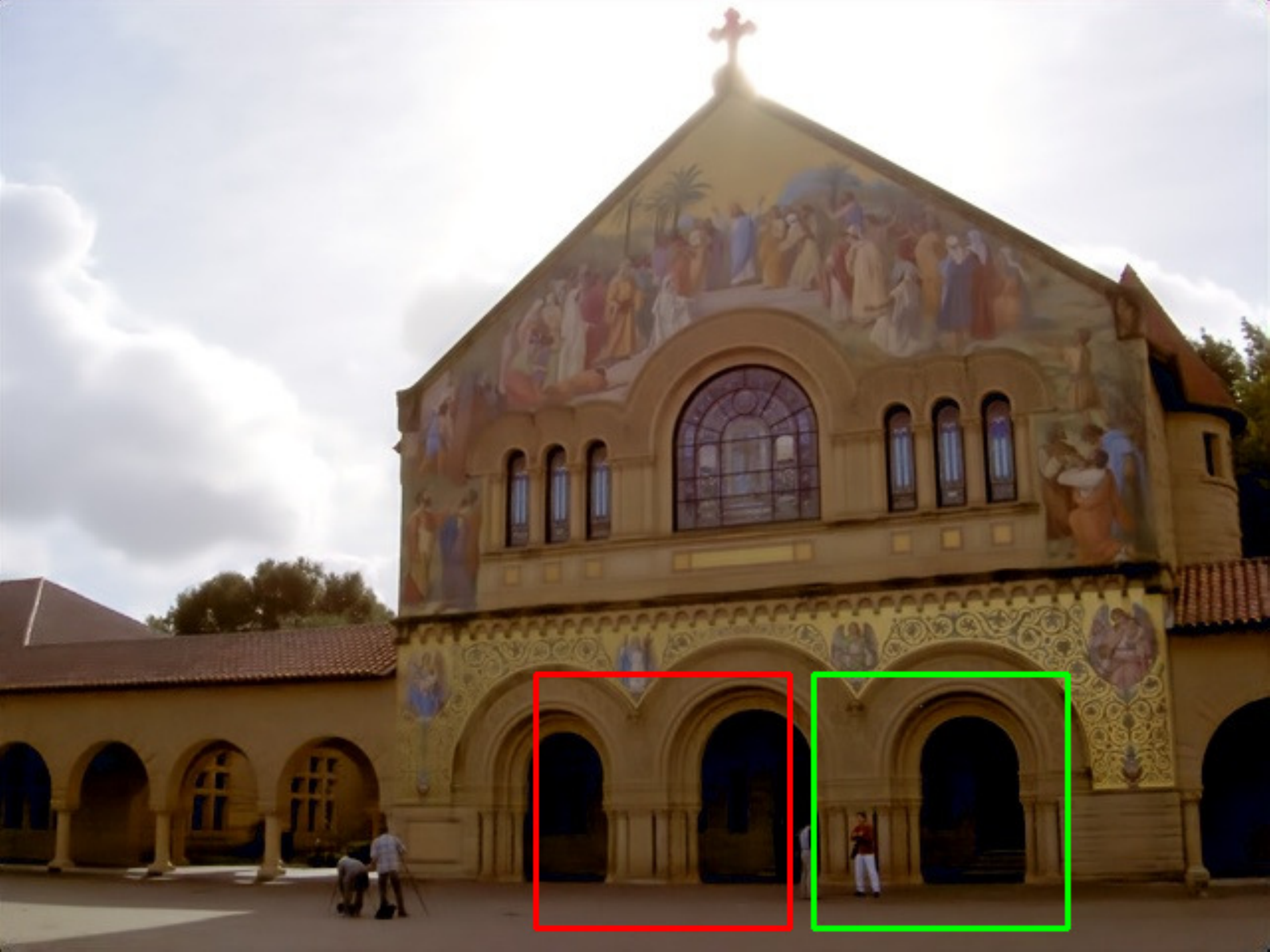}\vspace{1.5pt} \\
			\includegraphics[width=1.35cm]{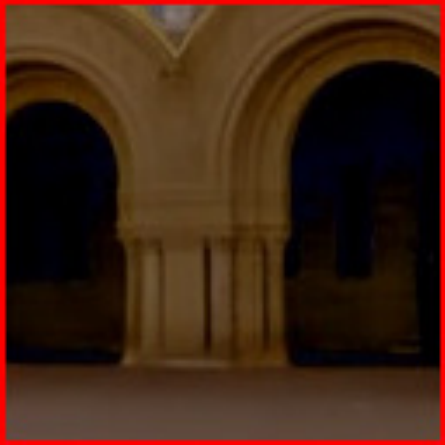}
			\includegraphics[width=1.35cm]{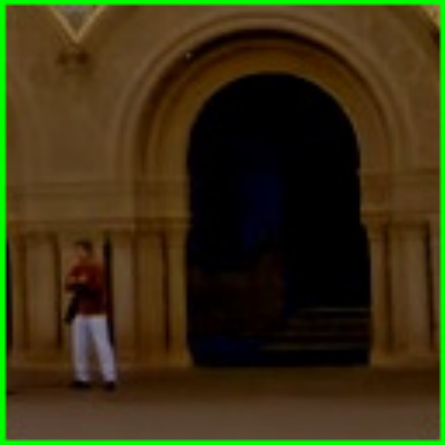}\vspace{5pt}
			\includegraphics[width=2.8cm]{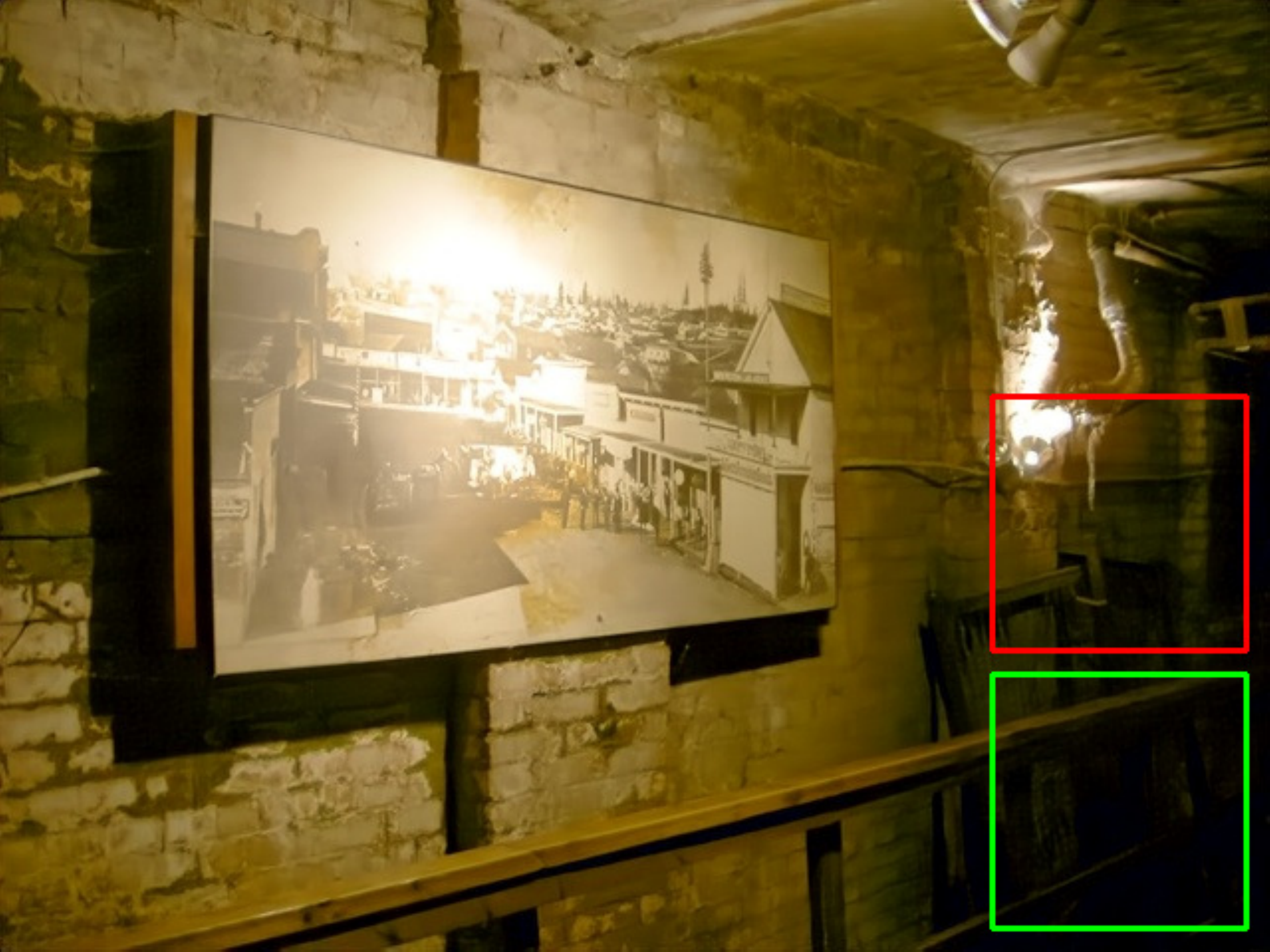}\vspace{1pt} \\
			\includegraphics[width=1.35cm]{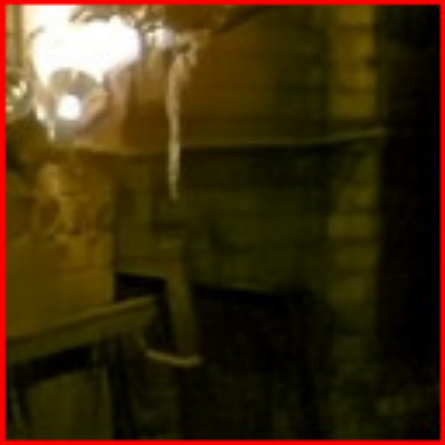}
			\includegraphics[width=1.35cm]{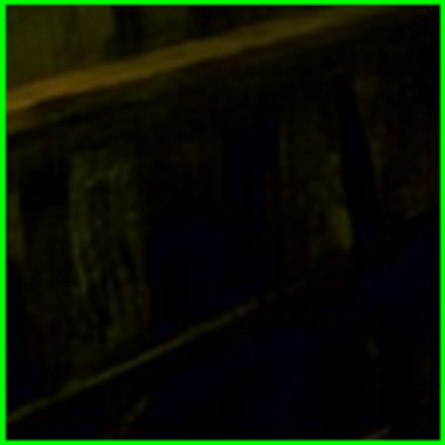}
		\end{minipage}
	}\hspace{-5pt}
	\subfigure[DA-DRN]{
		\begin{minipage}[b]{0.155\textwidth}
			\includegraphics[width=2.8cm]{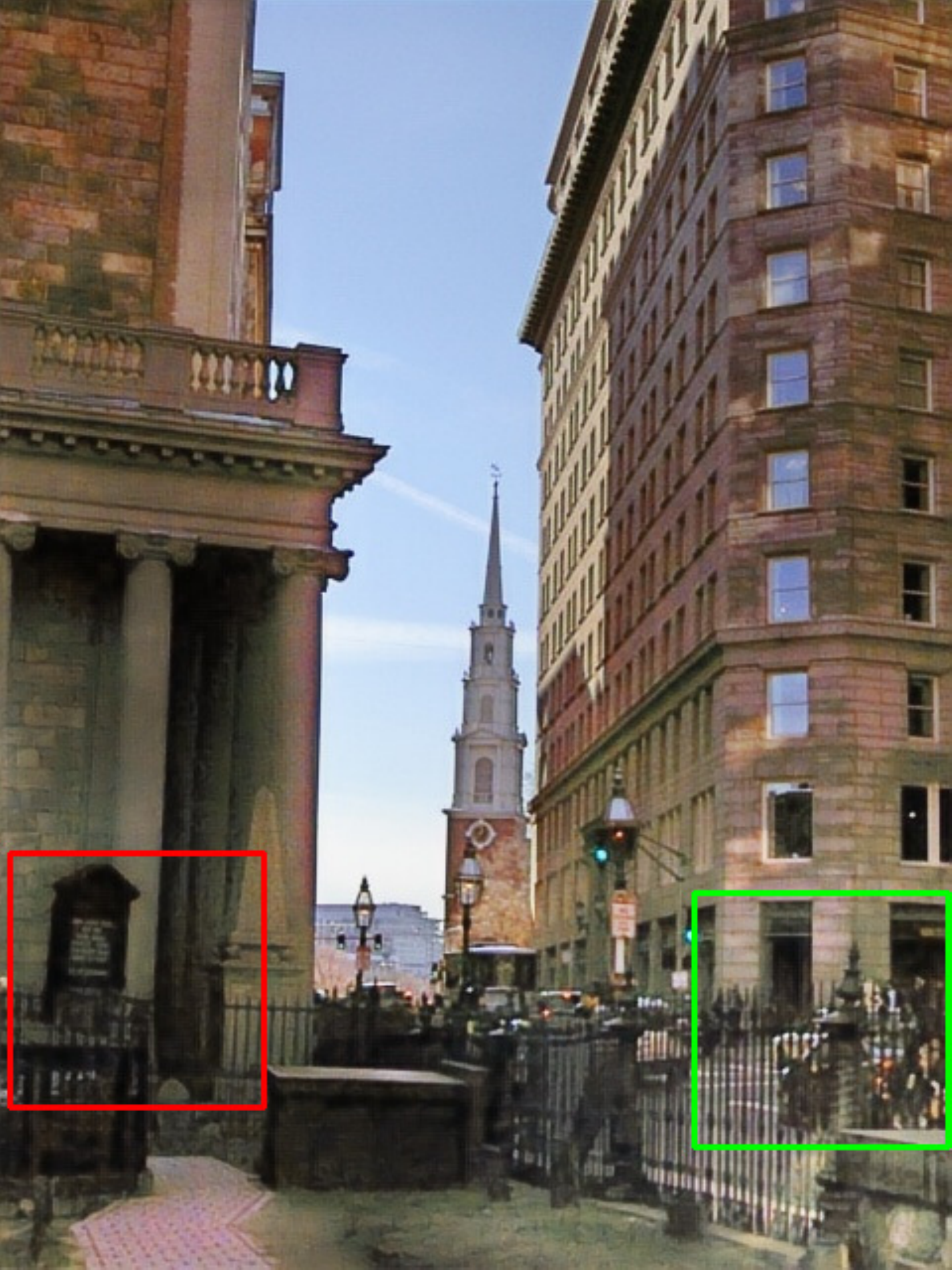}\vspace{1pt} \\
			\includegraphics[width=1.35cm]{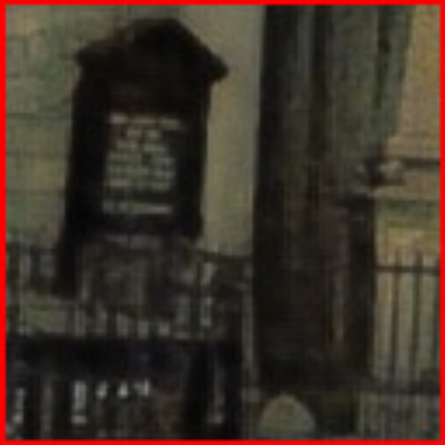}
			\includegraphics[width=1.35cm]{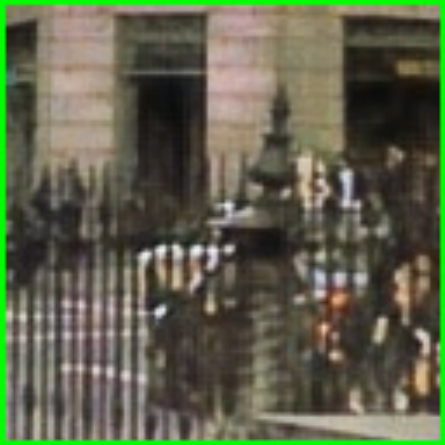}\vspace{5pt}
			\includegraphics[width=2.8cm]{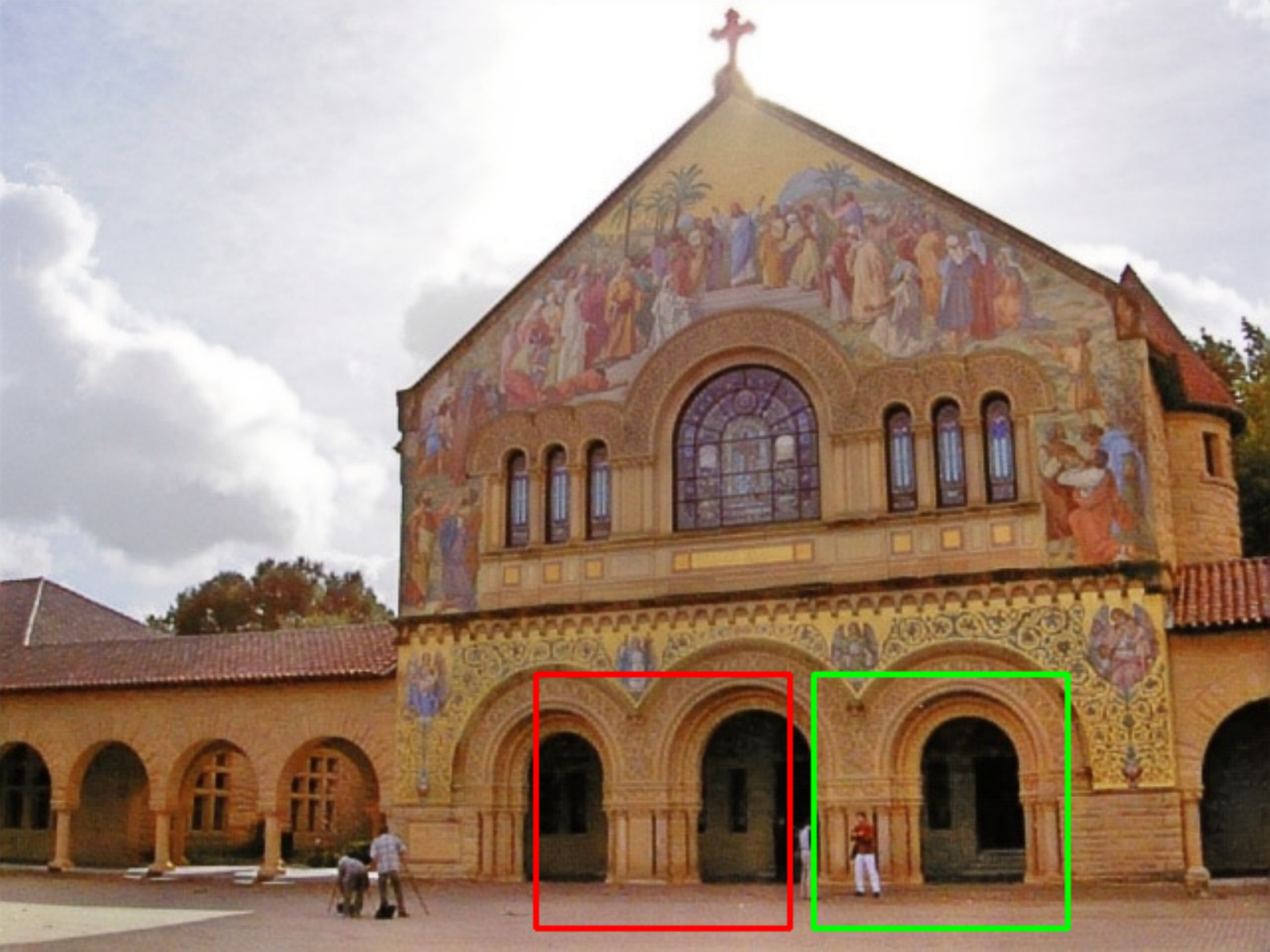}\vspace{1.5pt} \\
			\includegraphics[width=1.35cm]{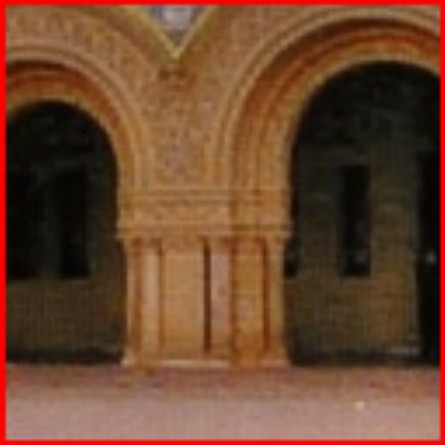}
			\includegraphics[width=1.35cm]{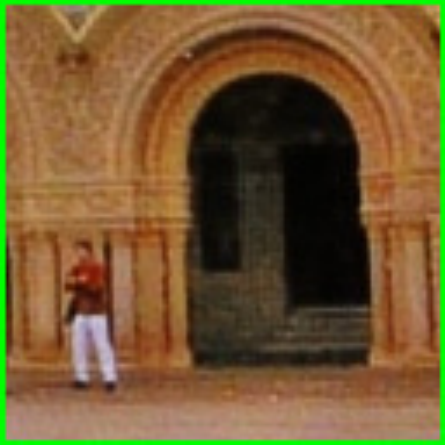}\vspace{5pt}
			\includegraphics[width=2.8cm]{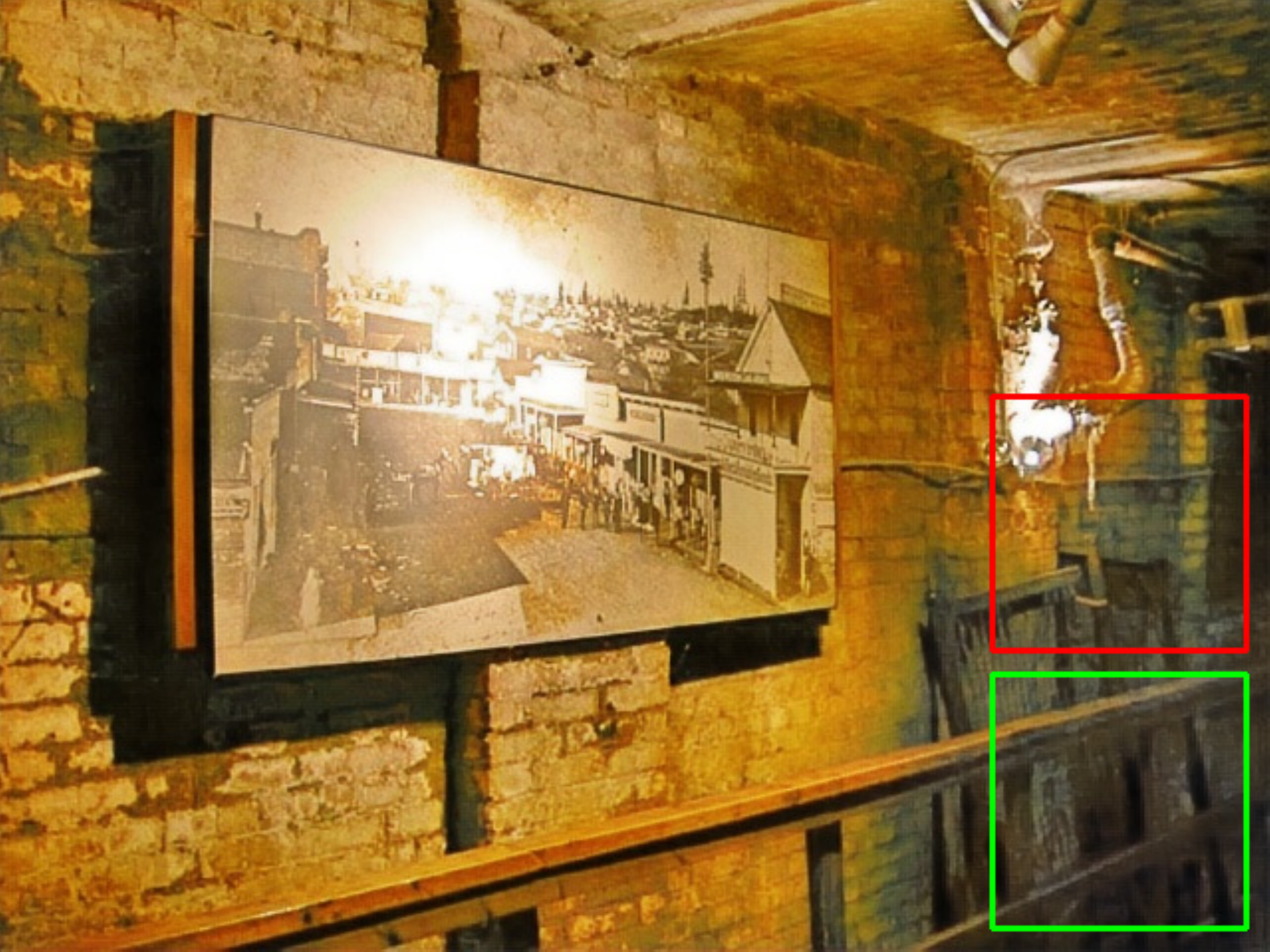}\vspace{1pt} \\
			\includegraphics[width=1.35cm]{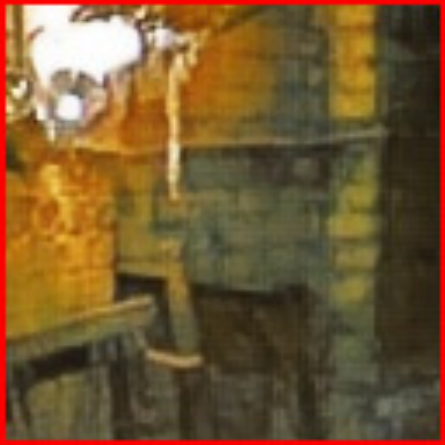}
			\includegraphics[width=1.35cm]{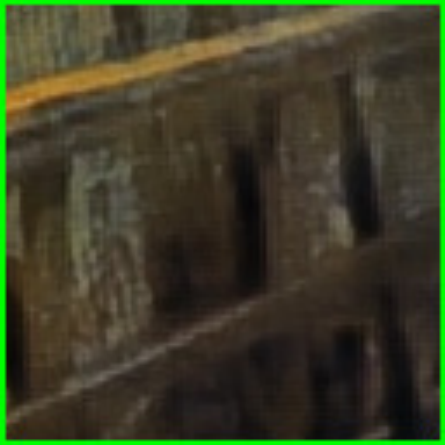}
		\end{minipage}
	}
	\caption{Visual comparison with other state-of-the-art methods on the DICM dataset. There is no Ground-Truth in DICM dataset.}
	\label{DICM}
\end{figure*}

\begin{figure*}
	\flushleft
	
	
	\subfigure[Input]{
		\begin{minipage}[b]{0.155\textwidth}
			\includegraphics[width=2.8cm]{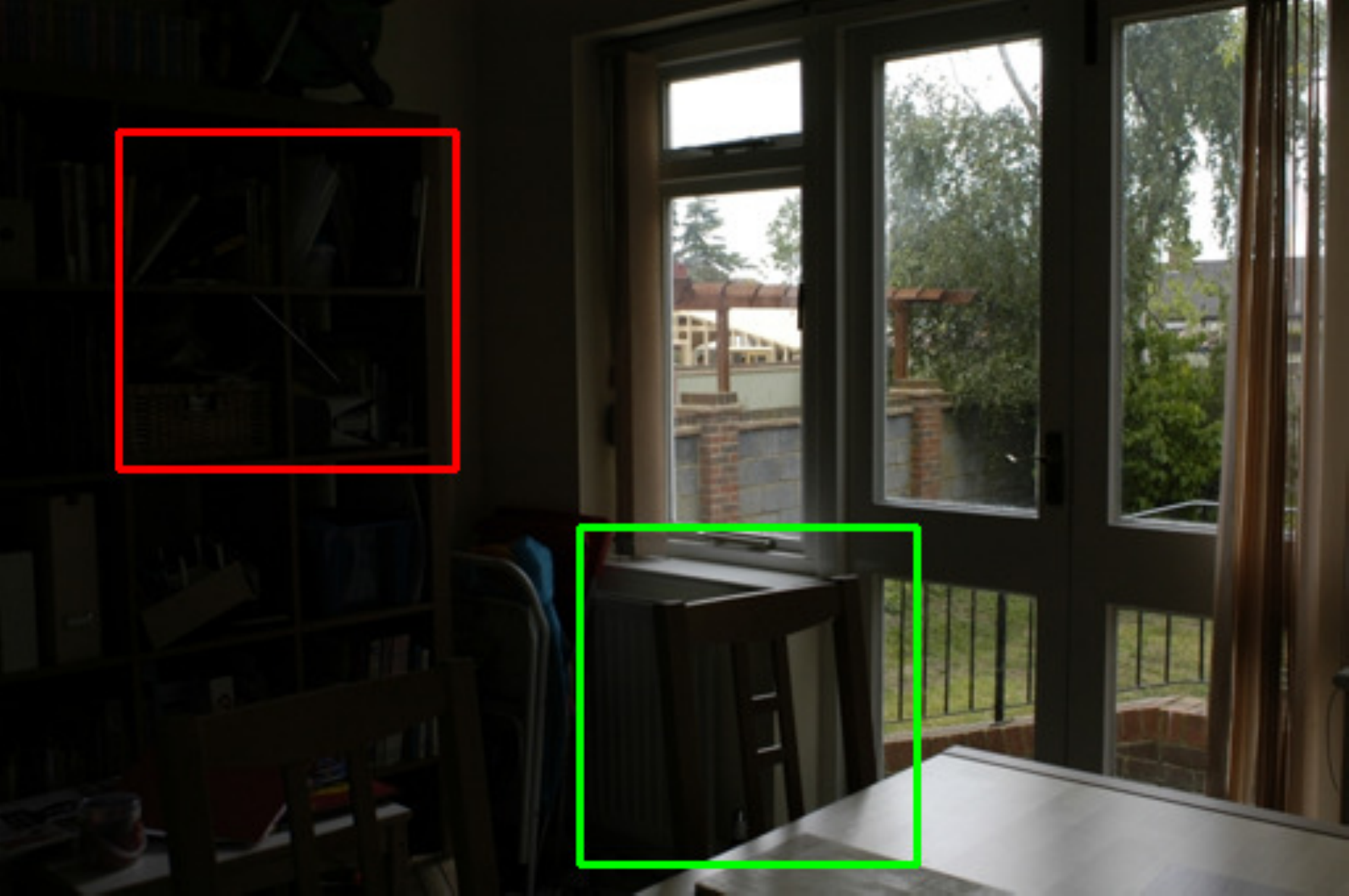}\vspace{1pt} \\
			\includegraphics[width=1.35cm]{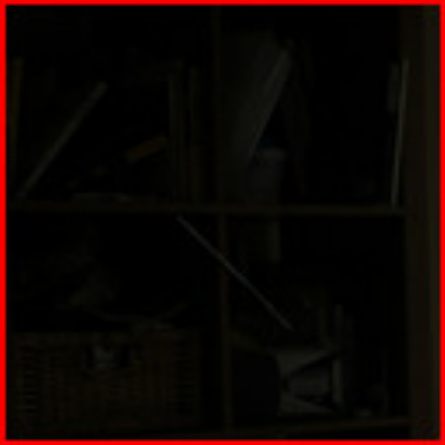}
			\includegraphics[width=1.35cm]{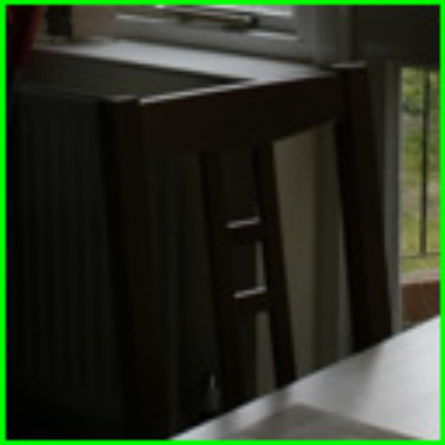}\vspace{5pt}
			\includegraphics[width=2.8cm]{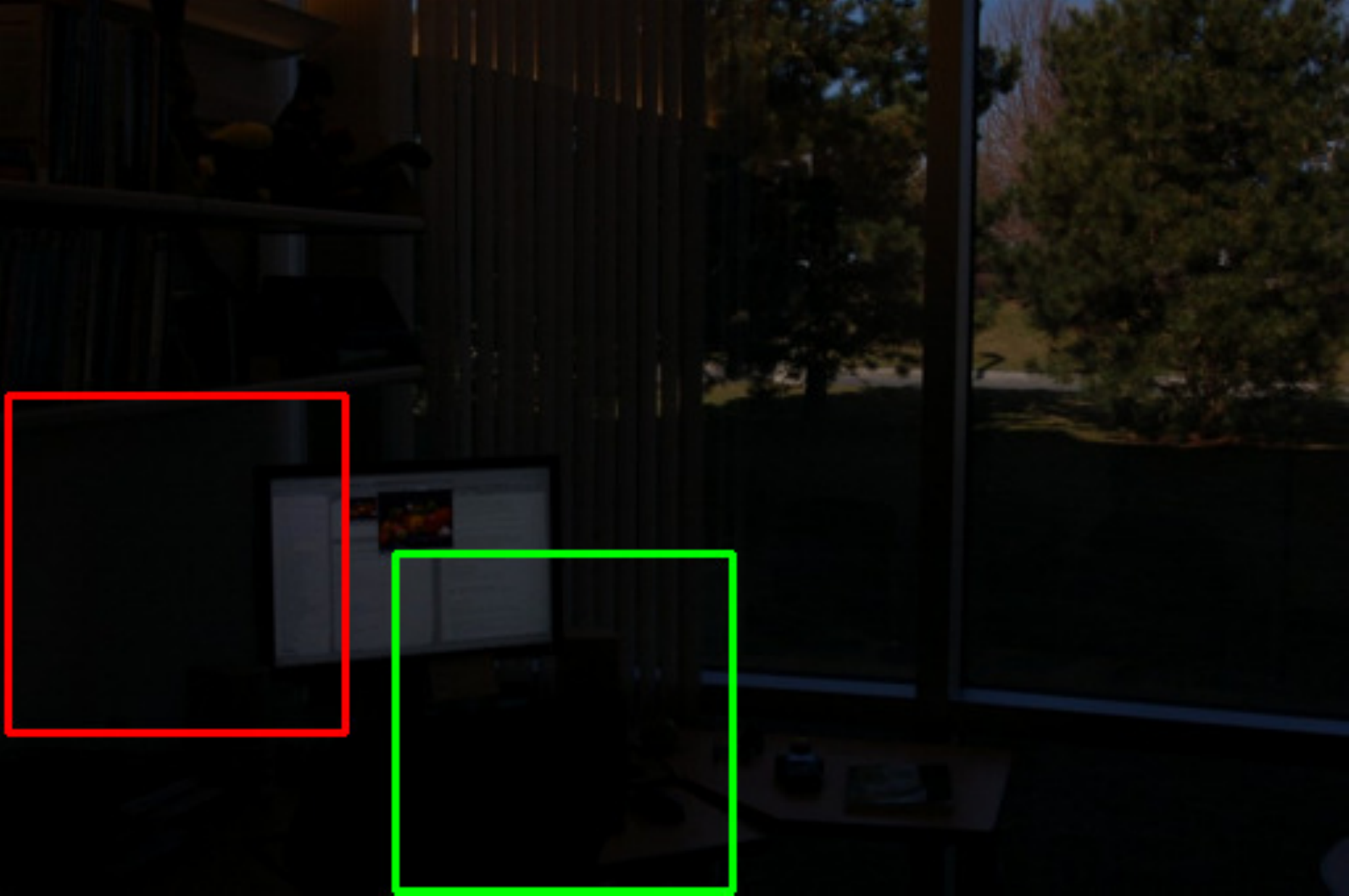}\vspace{1.5pt} \\
			\includegraphics[width=1.35cm]{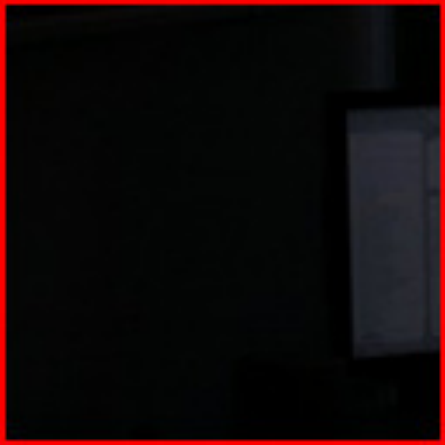}
			\includegraphics[width=1.35cm]{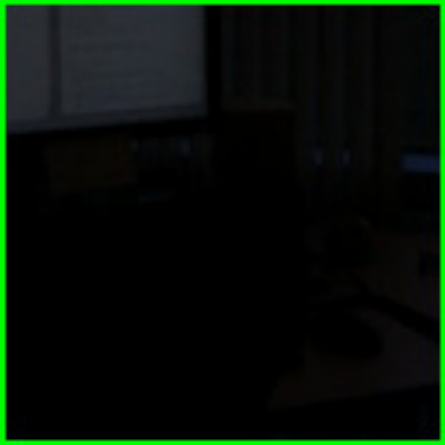}
		\end{minipage}
	}\hspace{-5pt}
	\subfigure[NPE\cite{wang2013naturalness}]{
		\begin{minipage}[b]{0.155\textwidth}
			\includegraphics[width=2.8cm]{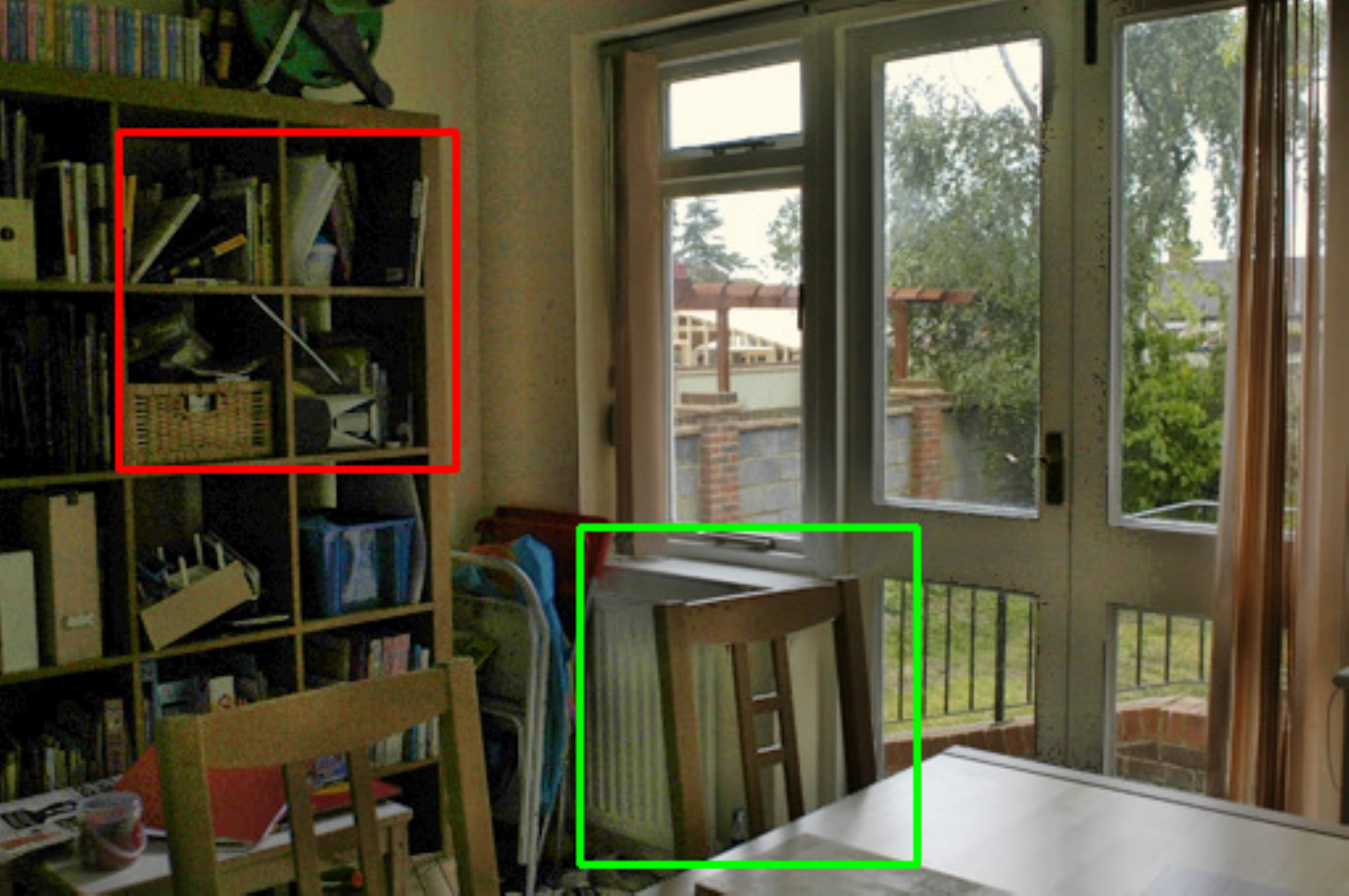}\vspace{1pt} \\
			\includegraphics[width=1.35cm]{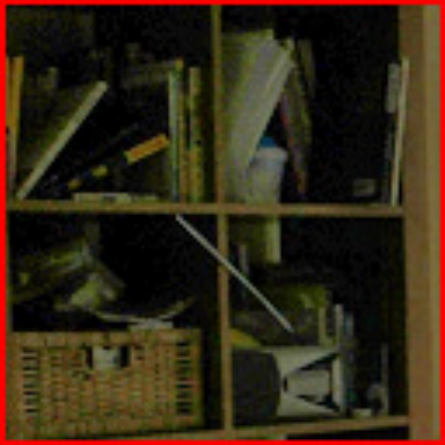}
			\includegraphics[width=1.35cm]{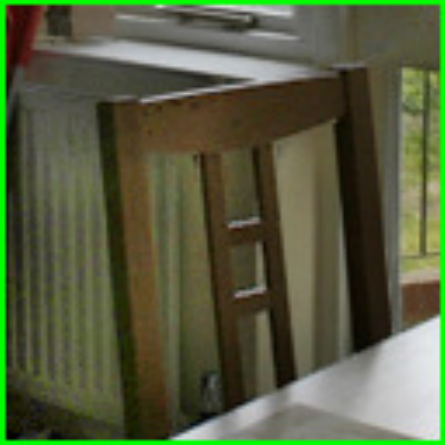}\vspace{5pt}
			\includegraphics[width=2.8cm]{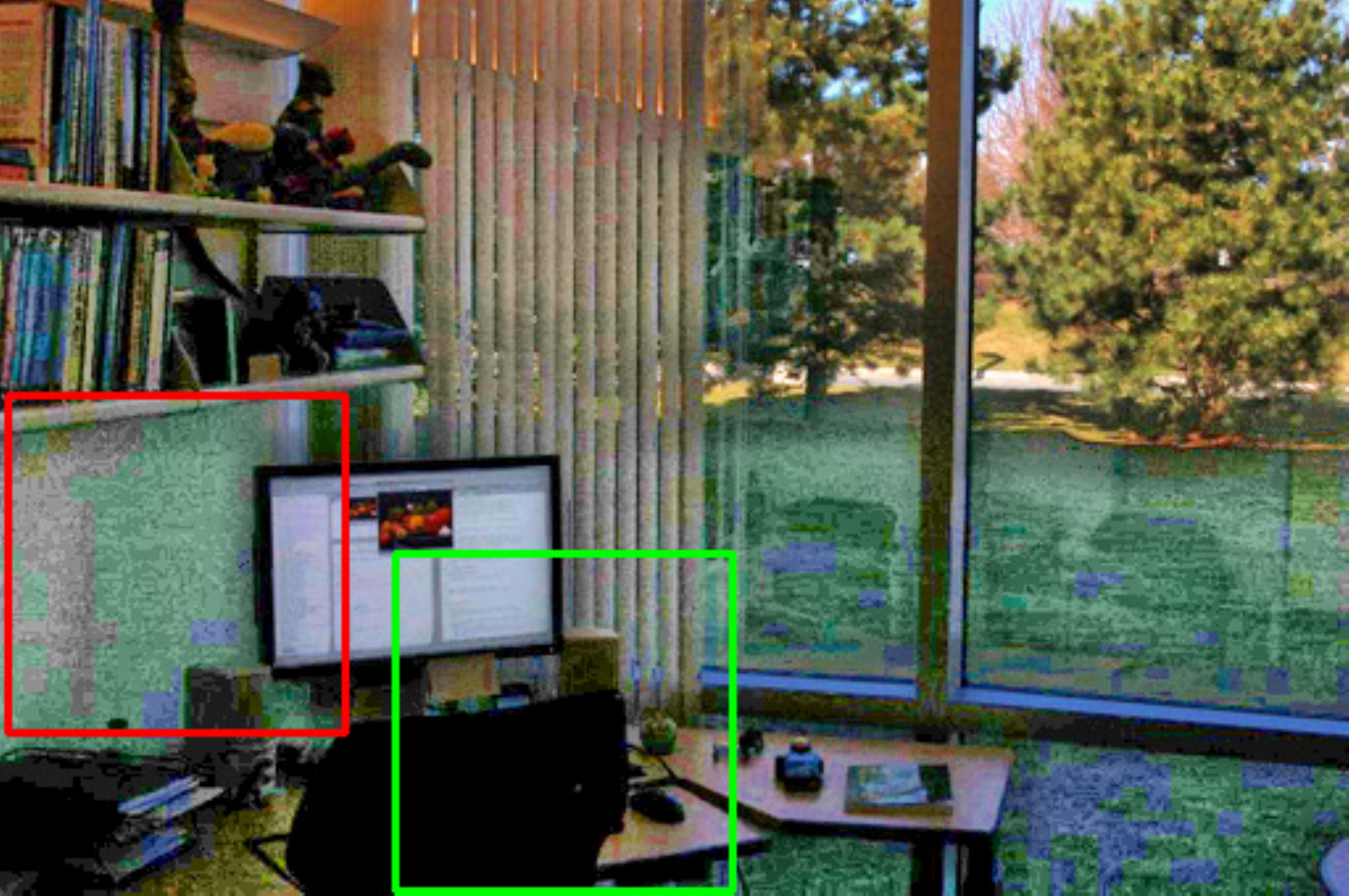}\vspace{1.5pt} \\
			\includegraphics[width=1.35cm]{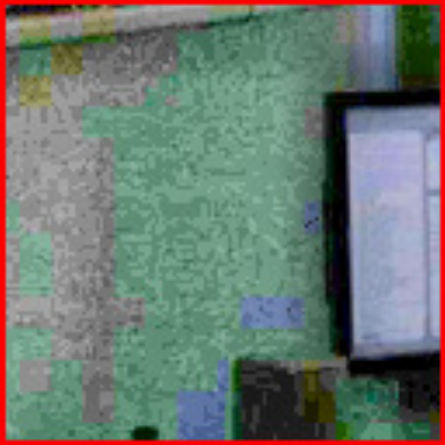}
			\includegraphics[width=1.35cm]{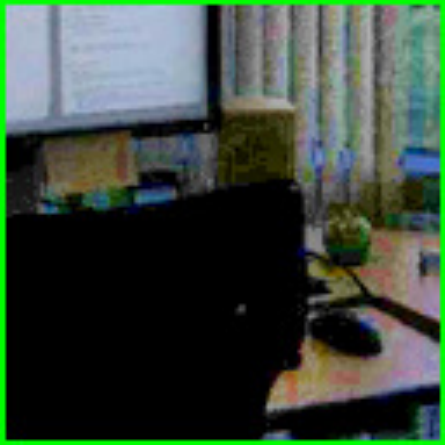}
		\end{minipage}
	}\hspace{-5pt}
	\subfigure[MF\cite{fu2016fusion}]{
		\begin{minipage}[b]{0.155\textwidth}
			\includegraphics[width=2.8cm]{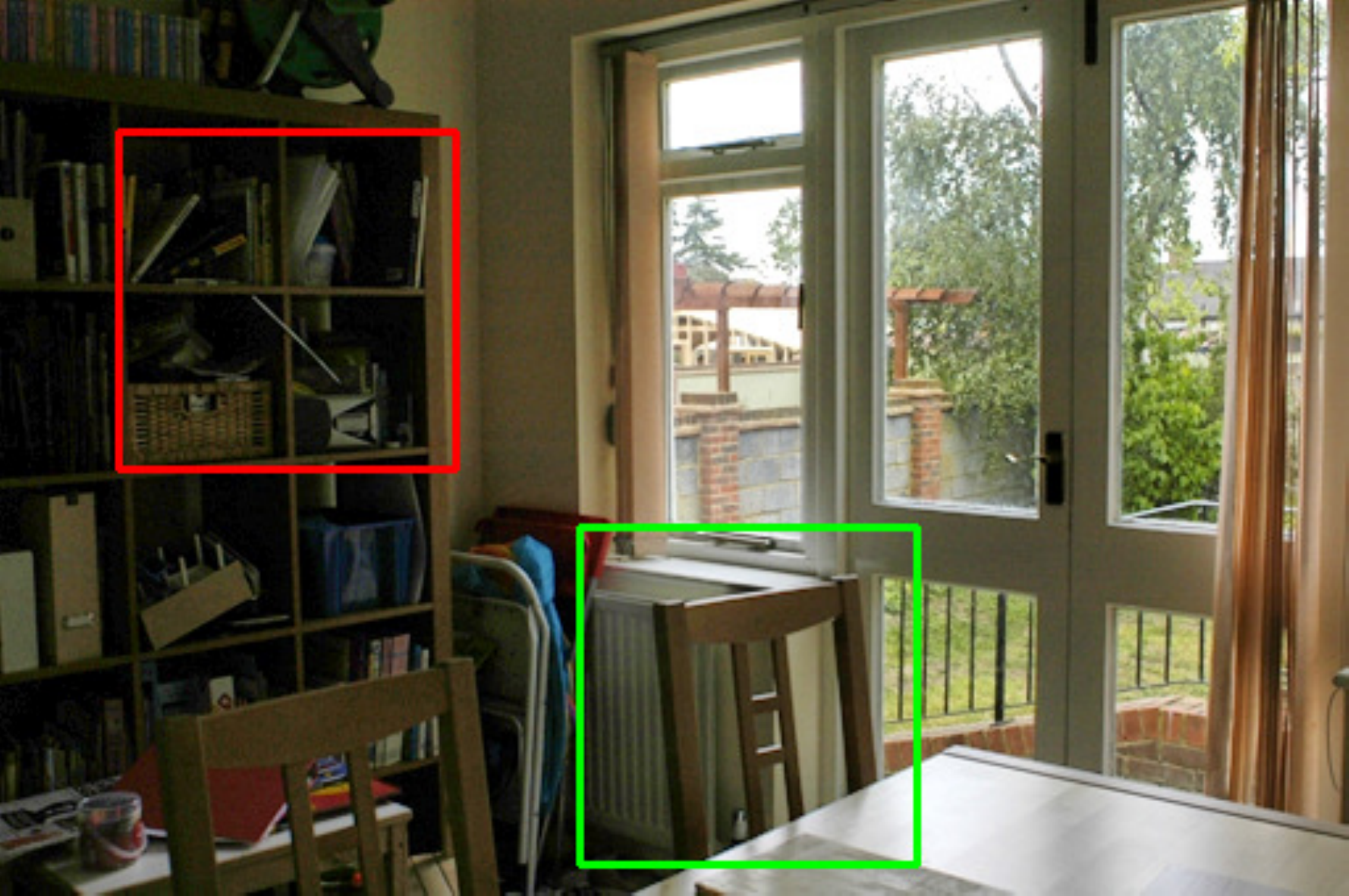}\vspace{1pt} \\
			\includegraphics[width=1.35cm]{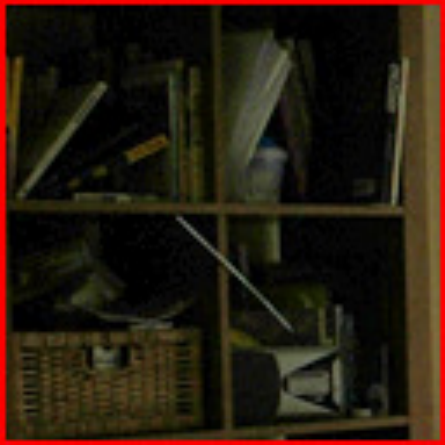}
			\includegraphics[width=1.35cm]{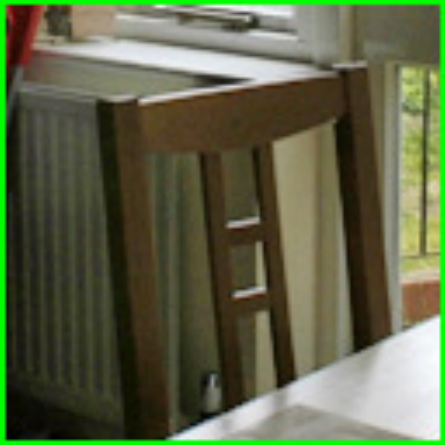}\vspace{5pt}
			\includegraphics[width=2.8cm]{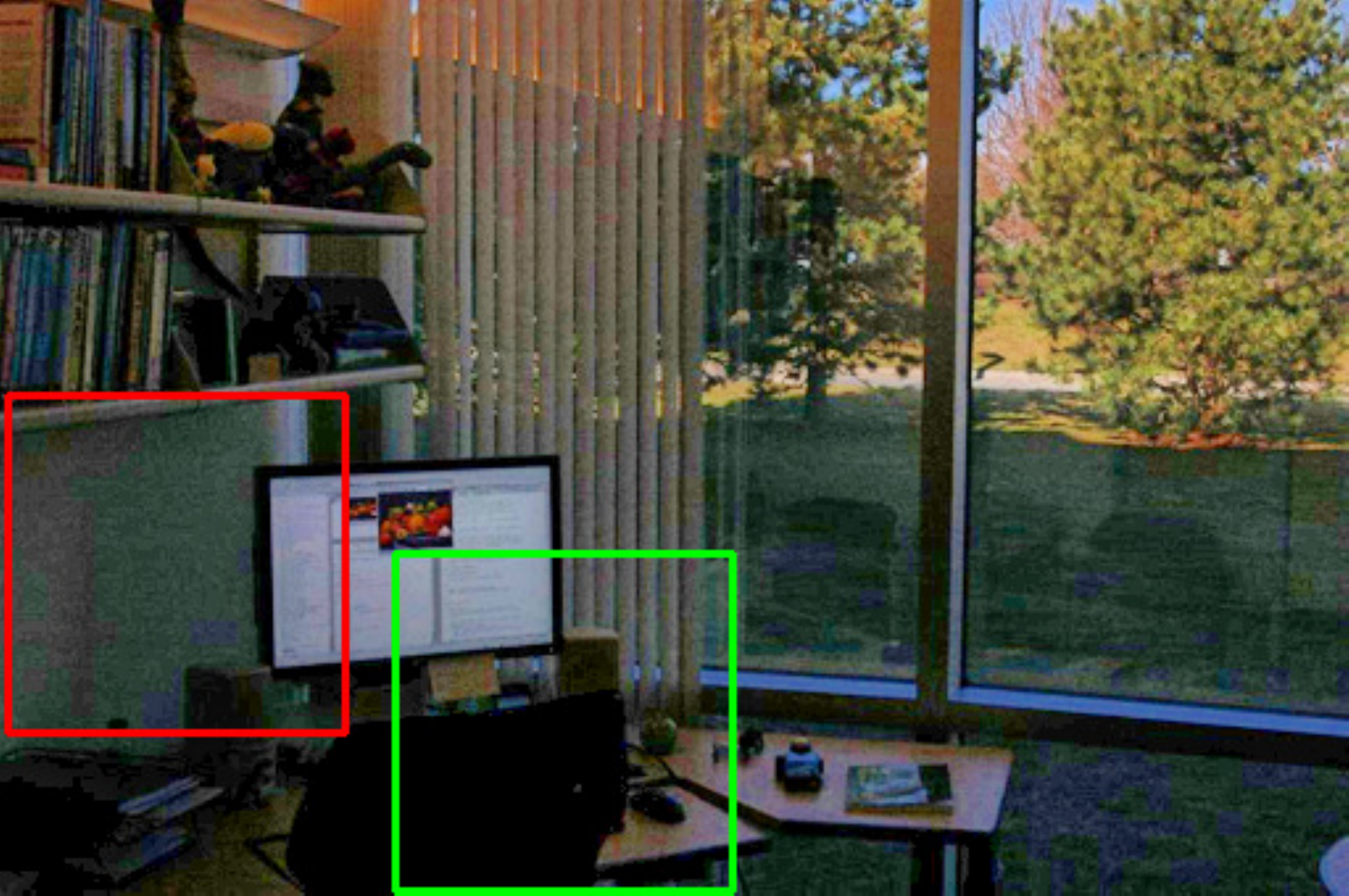}\vspace{1.5pt} \\
			\includegraphics[width=1.35cm]{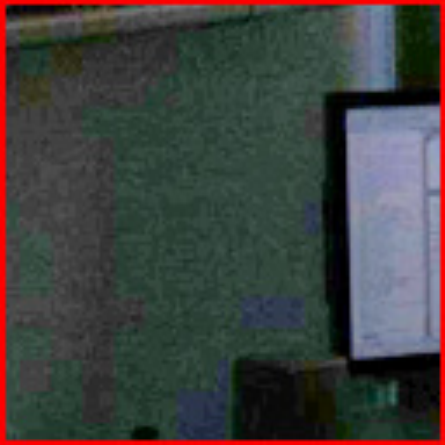}
			\includegraphics[width=1.35cm]{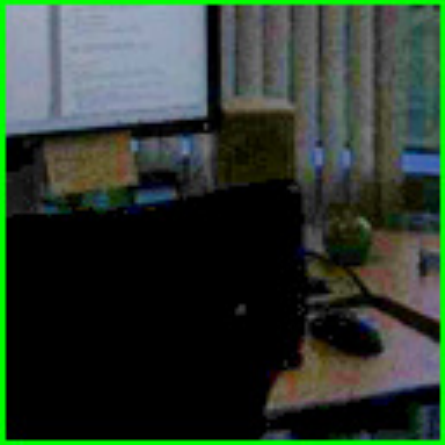}
		\end{minipage}
	}\hspace{-5pt}
	\subfigure[Zero-DCE\cite{guo2020zero}]{
		\begin{minipage}[b]{0.155\textwidth}
			\includegraphics[width=2.8cm]{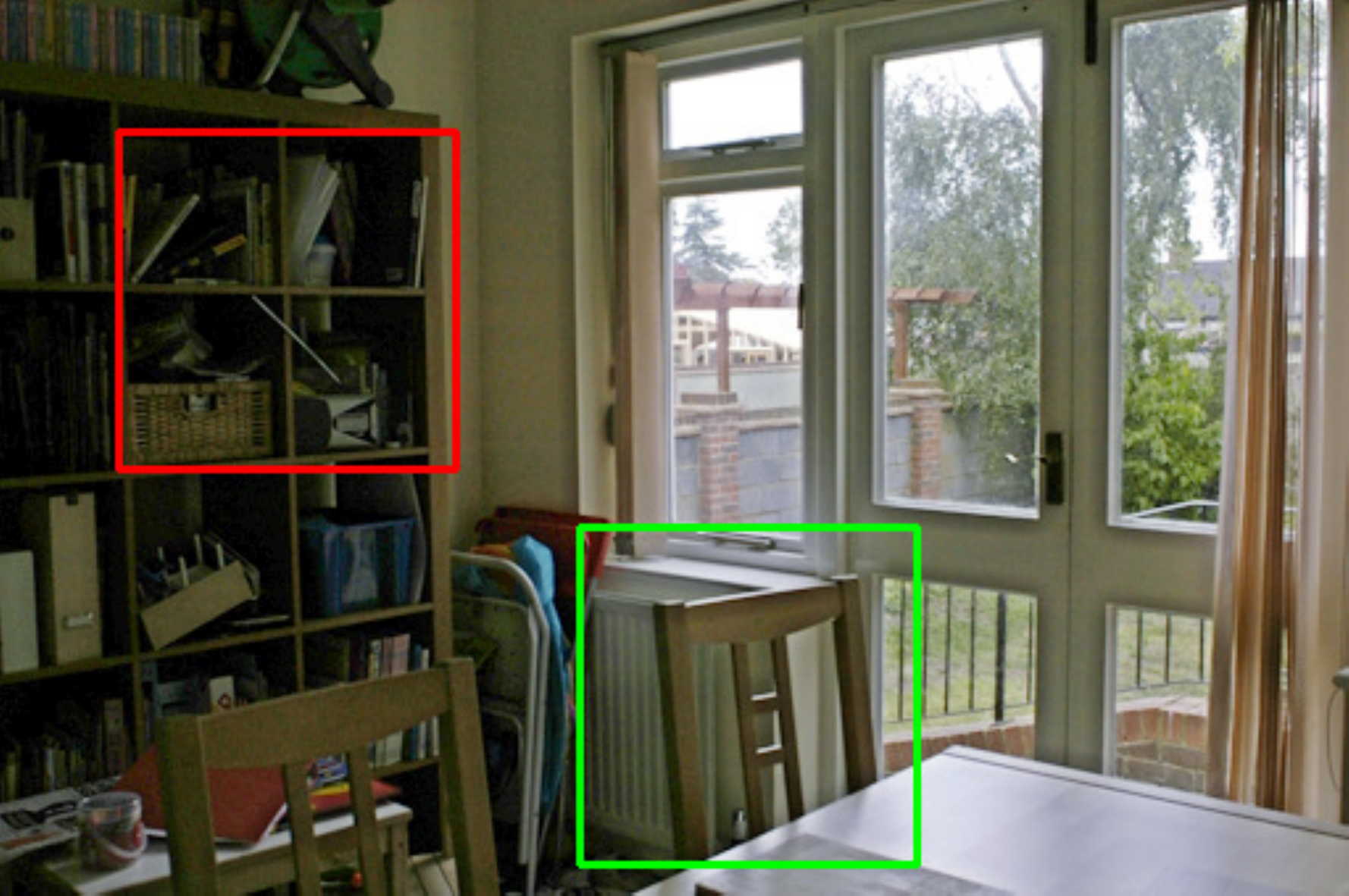}\vspace{1pt} \\
			\includegraphics[width=1.35cm]{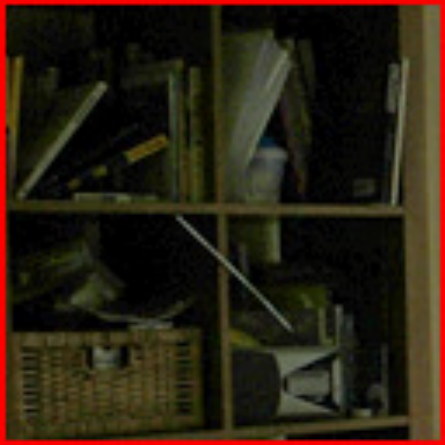}
			\includegraphics[width=1.35cm]{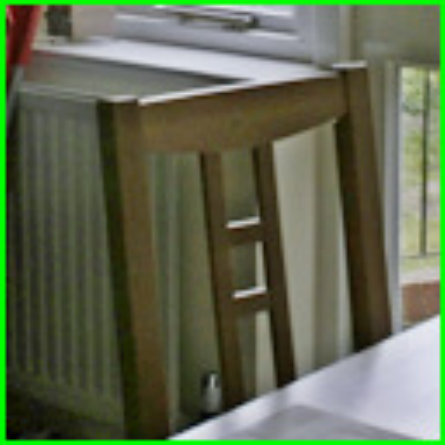}\vspace{5pt}
			\includegraphics[width=2.8cm]{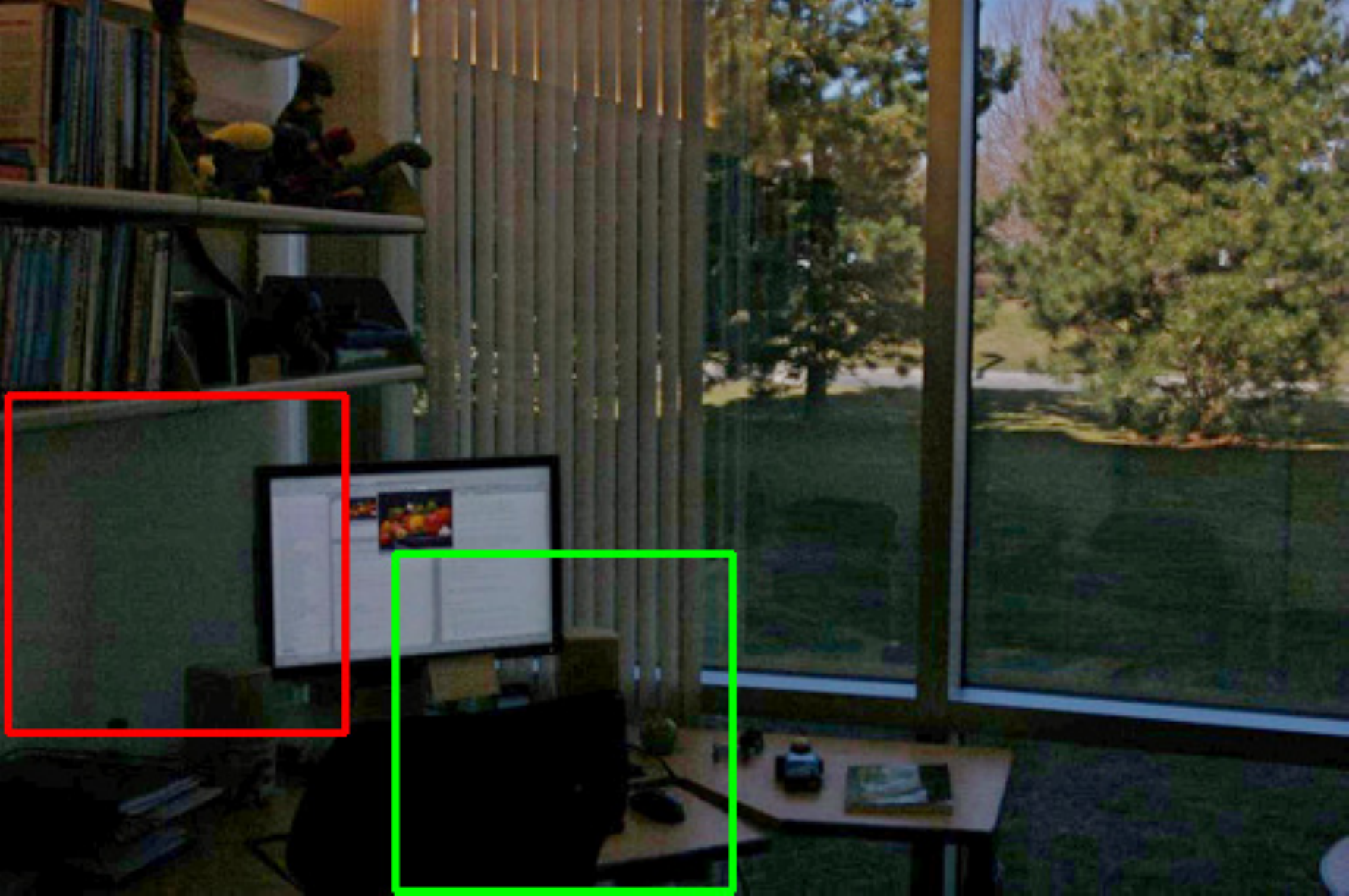}\vspace{1.5pt} \\
			\includegraphics[width=1.35cm]{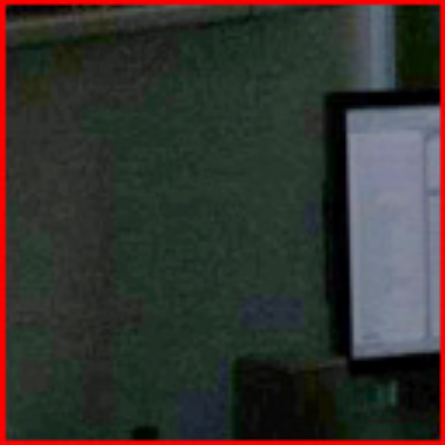}
			\includegraphics[width=1.35cm]{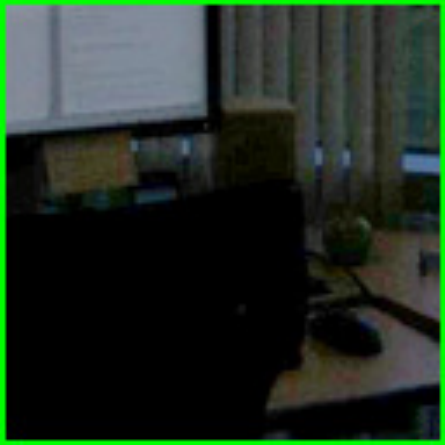}
		\end{minipage}
	}\hspace{-5pt}
	\subfigure[SRIE\cite{fu2016weighted}]{
		\begin{minipage}[b]{0.155\textwidth}
			\includegraphics[width=2.8cm]{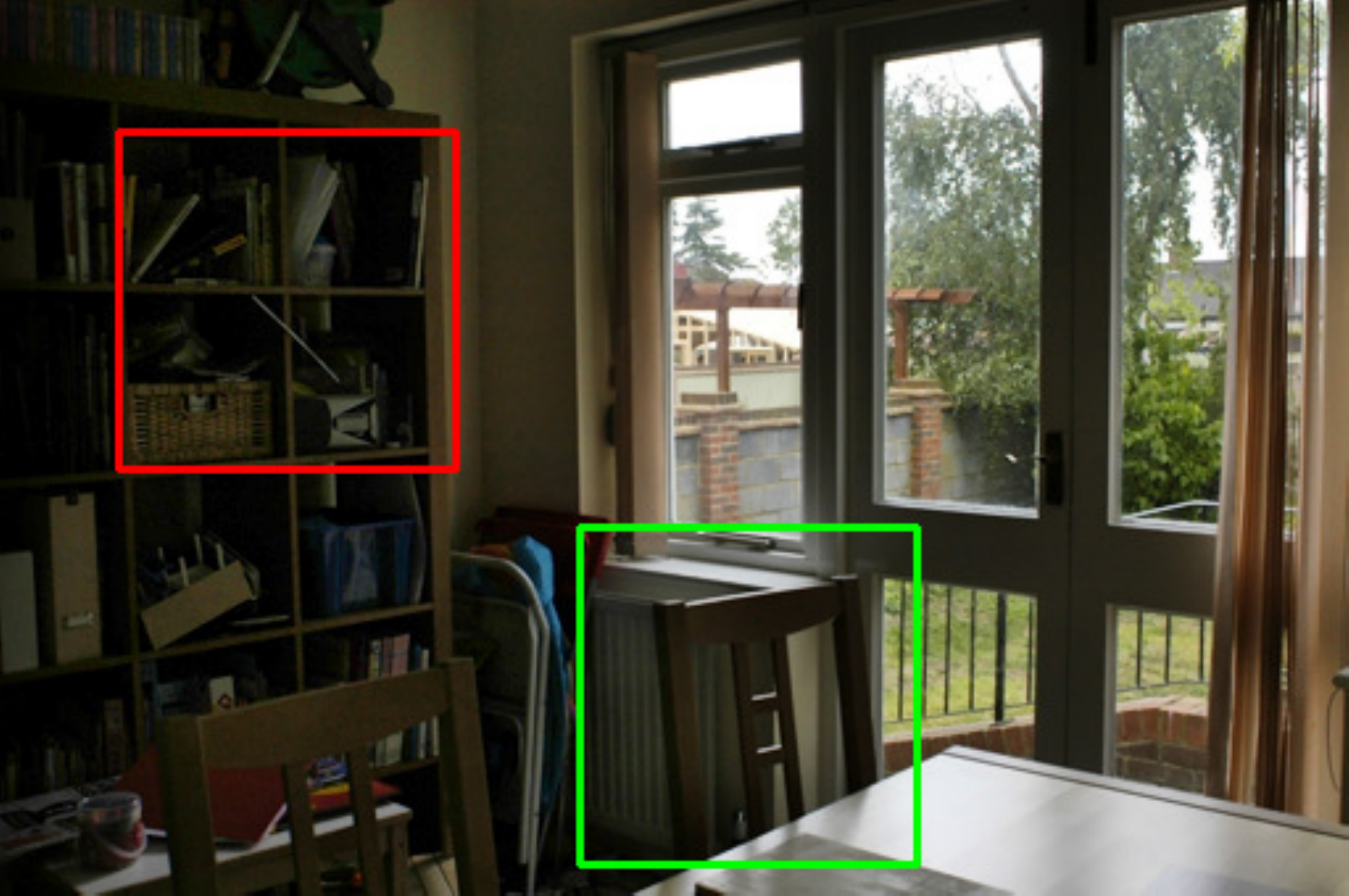}\vspace{1pt} \\
			\includegraphics[width=1.35cm]{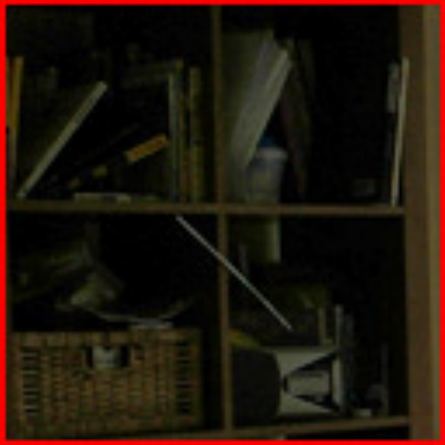}
			\includegraphics[width=1.35cm]{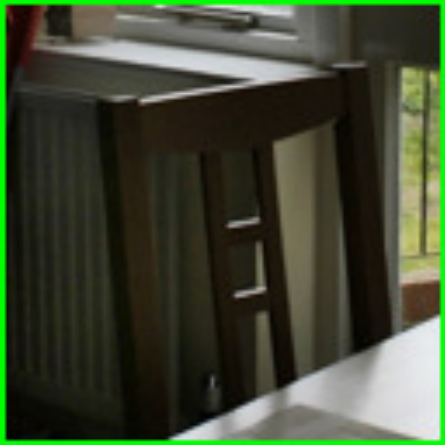}\vspace{5pt}
			\includegraphics[width=2.8cm]{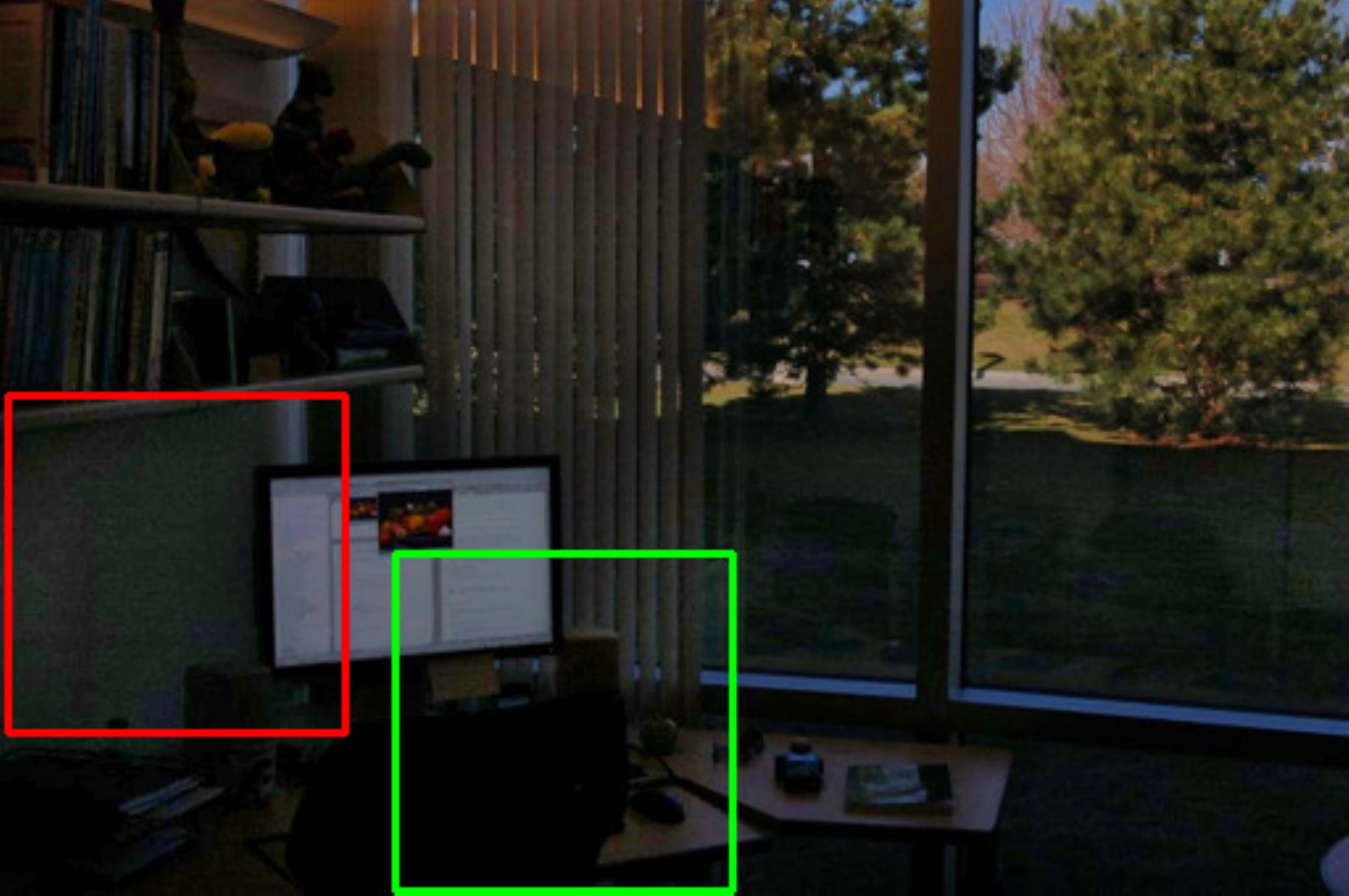}\vspace{1.5pt} \\
			\includegraphics[width=1.35cm]{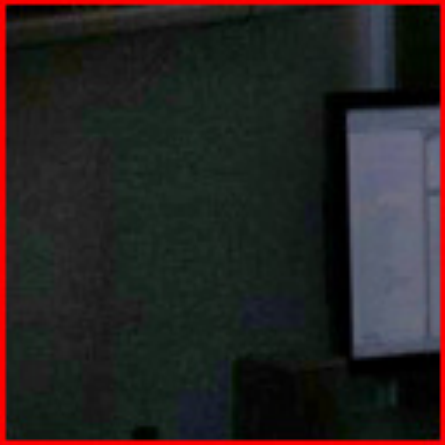}
			\includegraphics[width=1.35cm]{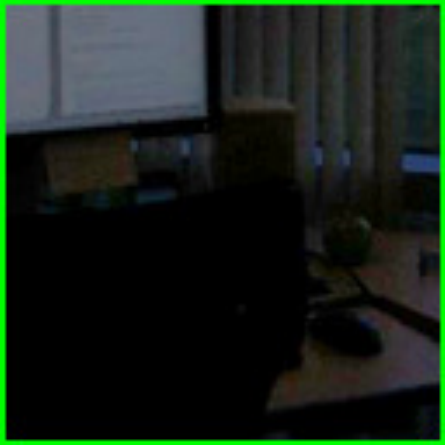}
		\end{minipage}
	}\hspace{-5pt}
	\subfigure[DA-DRN]{
		\begin{minipage}[b]{0.155\textwidth}
			\includegraphics[width=2.8cm]{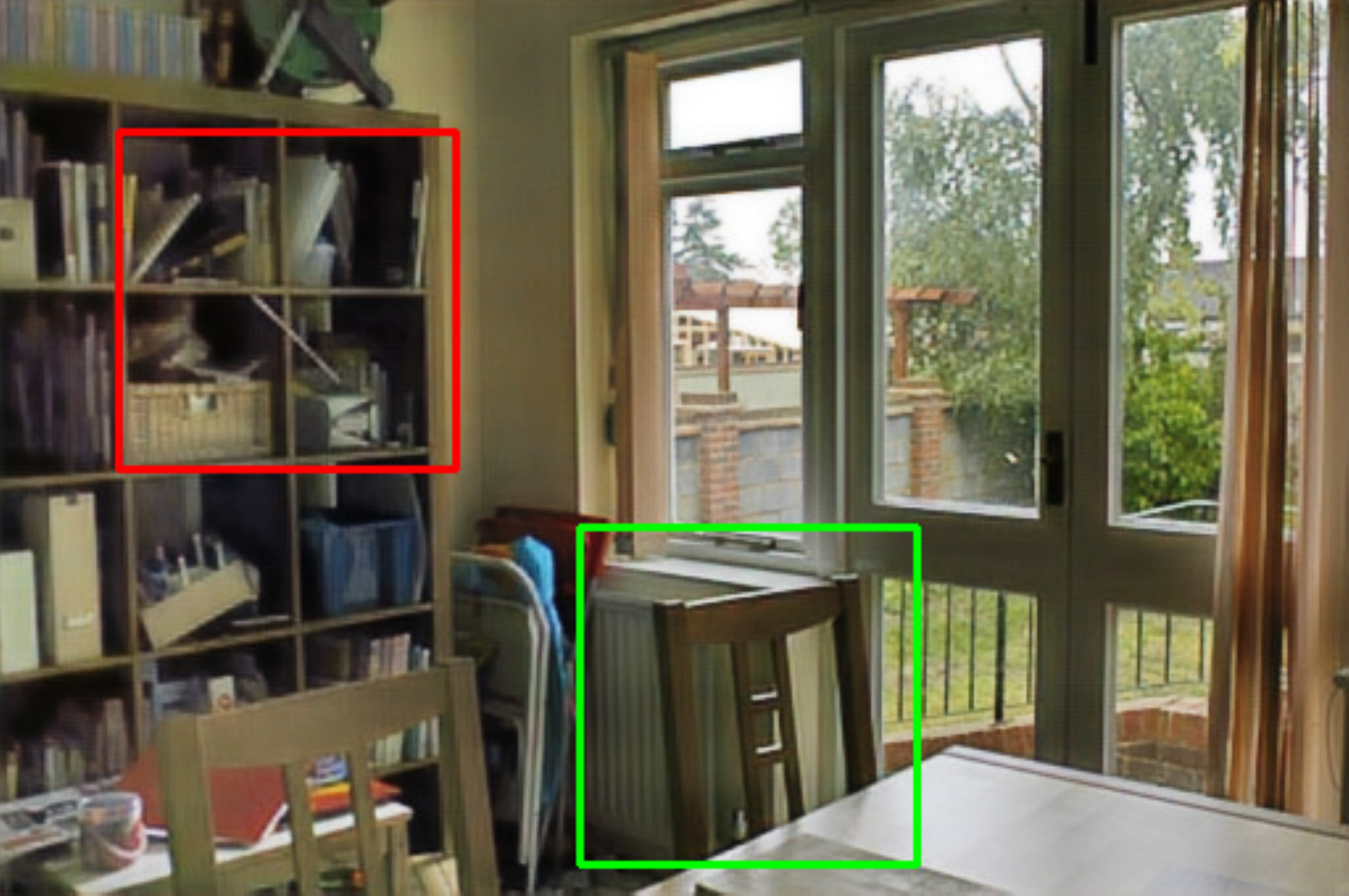}\vspace{1pt} \\
			\includegraphics[width=1.35cm]{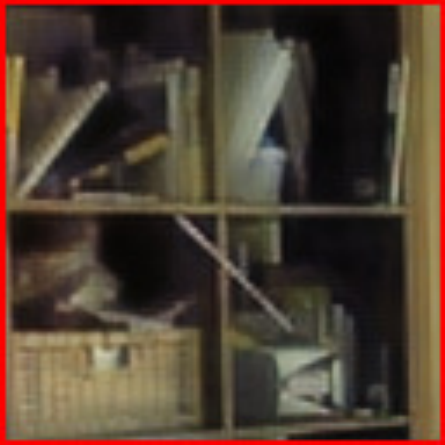}
			\includegraphics[width=1.35cm]{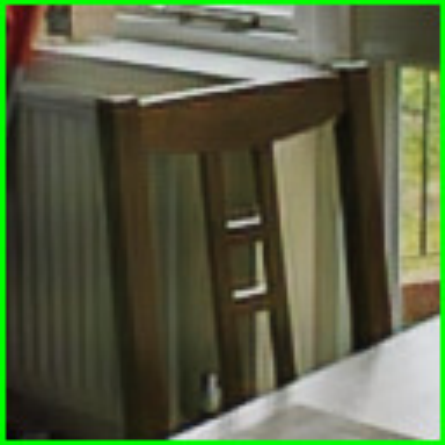}\vspace{5pt}
			\includegraphics[width=2.8cm]{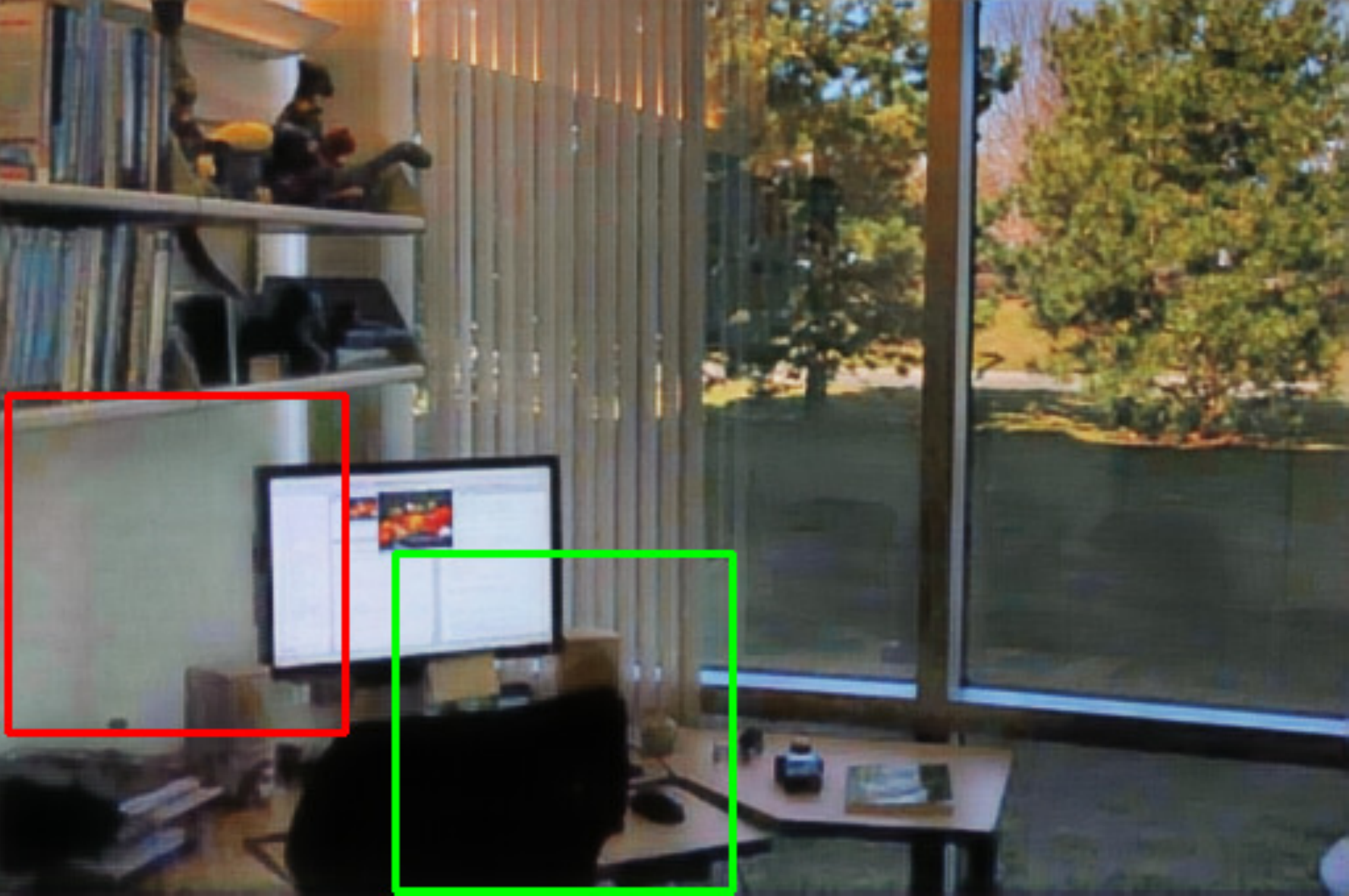}\vspace{1.5pt} \\
			\includegraphics[width=1.35cm]{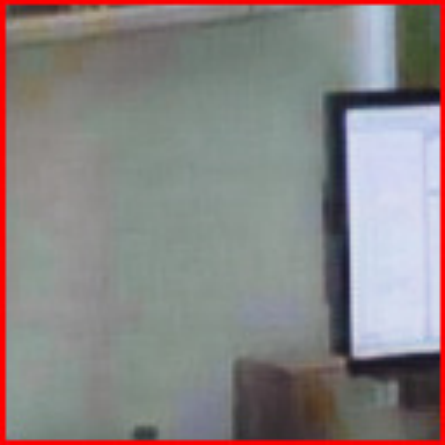}
			\includegraphics[width=1.35cm]{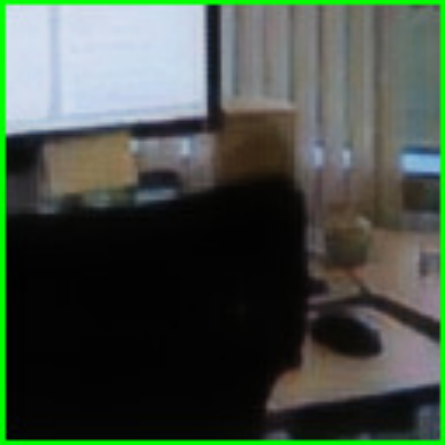}
		\end{minipage}
	}
	\caption{Visual comparison with other state-of-the-art methods on the MEF dataset. There is no Ground-Truth in MEF dataset.}
	\label{MEF}
\end{figure*}

As shown in Fig.\ref{syn},\ref{LIME},\ref{DICM},\ref{MEF}, In the enhanced results of previous methods, if the image has uneven brightness, there will be a problem of the unreasonable enhancement degree of brightness, such as some dark regions do not get enough enhancement, lead to the brightness in these regions after enhancement still dark, other areas that are already bright are overenhanced, resulting in overexposure in these areas. 
In contrast, in the enhanced results of our method, the brightness distribution of each region is more reasonable and there is no overexposure or underexposure problem. 
In addition, in the previous methods, noise hidden in the dark regions are amplified after enhancement, and the color is seriously distorted. However, in our enhanced results, noise is well eliminated and color distortion is corrected.

\end{document}